\documentclass[aps,prd,showpacs,showkeys,amsmath,amssymb,
twocolumn,
floatfix,
superscriptaddress
]{revtex4-1}
\usepackage{graphicx}
\usepackage{times}
\usepackage{verbatim}
\usepackage{color}
\usepackage{cancel}
\usepackage[normalem]{ulem}
\usepackage{multirow}
\usepackage{braket}
\usepackage{bm}
\usepackage{hyperref}
\usepackage{tabularx}

\newcommand*\syst{\mathrm{ (syst.)}}
\newcommand*\stat{\mathrm{ (stat.)}}

\newcommand{\nuebar}{$\overline{\nu}_{e}$}
\newcommand{\TRSNOERROR}{0.0841} 
\newcommand{\TRS}{\TRSNOERROR \pm 0.0027(\mathrm{stat.}) \pm 0.0019(\mathrm{syst.})} 
\newcommand{\DMEE}{(2.50 \pm 0.06(\mathrm{stat.}) \pm 0.06(\mathrm{syst.})) \times 10^{-3}\ {\rm eV}^2} 
\newcommand{\DMNH}{(2.45 \pm 0.06(\mathrm{stat.}) \pm 0.06(\mathrm{syst.})) \times 10^{-3}\ {\rm eV}^2} 
\newcommand{\DMIH}{(-2.56 \pm 0.06(\mathrm{stat.}) \pm 0.06(\mathrm{syst.})) \times 10^{-3}\ {\rm eV}^2} 

\graphicspath{{figures/}}

\begin{document}

\title{Measurement of electron antineutrino oscillation based on
  1230~days of operation of the Daya Bay experiment}

\newcommand{\ECUST}{\affiliation{Institute of Modern Physics, East China University of Science and Technology, Shanghai}}
\newcommand{\IHEP}{\affiliation{Institute~of~High~Energy~Physics, Beijing}}
\newcommand{\Wisconsin}{\affiliation{University~of~Wisconsin, Madison, Wisconsin 53706, USA}}
\newcommand{\Yale}{\affiliation{Department~of~Physics, Yale~University, New~Haven, Connecticut 06520, USA}}
\newcommand{\BNL}{\affiliation{Brookhaven~National~Laboratory, Upton, New York 11973, USA}}
\newcommand{\NTU}{\affiliation{Department of Physics, National~Taiwan~University, Taipei}}
\newcommand{\NUU}{\affiliation{National~United~University, Miao-Li}}
\newcommand{\Dubna}{\affiliation{Joint~Institute~for~Nuclear~Research, Dubna, Moscow~Region}}
\newcommand{\CalTech}{\affiliation{California~Institute~of~Technology, Pasadena, California 91125, USA}}
\newcommand{\CUHK}{\affiliation{Chinese~University~of~Hong~Kong, Hong~Kong}}
\newcommand{\NCTU}{\affiliation{Institute~of~Physics, National~Chiao-Tung~University, Hsinchu}}
\newcommand{\NJU}{\affiliation{Nanjing~University, Nanjing}}
\newcommand{\TsingHua}{\affiliation{Department~of~Engineering~Physics, Tsinghua~University, Beijing}}
\newcommand{\SZU}{\affiliation{Shenzhen~University, Shenzhen}}
\newcommand{\NCEPU}{\affiliation{North~China~Electric~Power~University, Beijing}}
\newcommand{\Siena}{\affiliation{Siena~College, Loudonville, New York  12211, USA}}
\newcommand{\IIT}{\affiliation{Department of Physics, Illinois~Institute~of~Technology, Chicago, Illinois  60616, USA}}
\newcommand{\LBNL}{\affiliation{Lawrence~Berkeley~National~Laboratory, Berkeley, California 94720, USA}}
\newcommand{\UIUC}{\affiliation{Department of Physics, University~of~Illinois~at~Urbana-Champaign, Urbana, Illinois 61801, USA}}
\newcommand{\RPI}{\affiliation{Department~of~Physics, Applied~Physics, and~Astronomy, Rensselaer~Polytechnic~Institute, Troy, New~York  12180, USA}}
\newcommand{\SJTU}{\affiliation{Department of Physics and Astronomy, Shanghai Jiao Tong University, Shanghai Laboratory for Particle Physics and Cosmology, Shanghai}}
\newcommand{\BNU}{\affiliation{Beijing~Normal~University, Beijing}}
\newcommand{\WM}{\affiliation{College~of~William~and~Mary, Williamsburg, Virginia  23187, USA}}
\newcommand{\Princeton}{\affiliation{Joseph Henry Laboratories, Princeton~University, Princeton, New~Jersey 08544, USA}}
\newcommand{\VirginiaTech}{\affiliation{Center for Neutrino Physics, Virginia~Tech, Blacksburg, Virginia  24061, USA}}
\newcommand{\CIAE}{\affiliation{China~Institute~of~Atomic~Energy, Beijing}}
\newcommand{\SDU}{\affiliation{Shandong~University, Jinan}}
\newcommand{\NanKai}{\affiliation{School of Physics, Nankai~University, Tianjin}}
\newcommand{\UC}{\affiliation{Department of Physics, University~of~Cincinnati, Cincinnati, Ohio 45221, USA}}
\newcommand{\DGUT}{\affiliation{Dongguan~University~of~Technology, Dongguan}}
\newcommand{\XJTU}{\affiliation{Xi'an Jiaotong University, Xi'an}}
\newcommand{\UCB}{\affiliation{Department of Physics, University~of~California, Berkeley, California  94720, USA}}
\newcommand{\HKU}{\affiliation{Department of Physics, The~University~of~Hong~Kong, Pokfulam, Hong~Kong}}
\newcommand{\UH}{\affiliation{Department of Physics, University~of~Houston, Houston, Texas  77204, USA}}
\newcommand{\Charles}{\affiliation{Charles~University, Faculty~of~Mathematics~and~Physics, Prague, Czech~Republic}} 
\newcommand{\USTC}{\affiliation{University~of~Science~and~Technology~of~China, Hefei}}
\newcommand{\TempleUniversity}{\affiliation{Department~of~Physics, College~of~Science~and~Technology, Temple~University, Philadelphia, Pennsylvania  19122, USA}}
\newcommand{\CUC}{\affiliation{Instituto de F\'isica, Pontificia Universidad Cat\'olica de Chile, Santiago, Chile}} 
\newcommand{\CGNPG}{\affiliation{China General Nuclear Power Group}}
\newcommand{\NUDT}{\affiliation{College of Electronic Science and Engineering, National University of Defense Technology, Changsha}} 
\newcommand{\IowaState}{\affiliation{Iowa~State~University, Ames, Iowa  50011, USA}}
\newcommand{\ZSU}{\affiliation{Sun Yat-Sen (Zhongshan) University, Guangzhou}}
\newcommand{\CQU}{\affiliation{Chongqing University, Chongqing}} 
\newcommand{\BCC}{\altaffiliation[Now at ]{Department of Chemistry and Chemical Technology, Bronx Community College, Bronx, New York  10453, USA}} 
\author{F.~P.~An}\ECUST
\author{A.~B.~Balantekin}\Wisconsin
\author{H.~R.~Band}\Yale
\author{M.~Bishai}\BNL
\author{S.~Blyth}\NTU\NUU
\author{D.~Cao}\NJU
\author{G.~F.~Cao}\IHEP
\author{J.~Cao}\IHEP
\author{W.~R.~Cen}\IHEP
\author{Y.~L.~Chan}\CUHK
% ...

\collaboration{The Daya Bay Collaboration}\noaffiliation
\date{\today}

\begin{abstract}
\noindent
A measurement of electron antineutrino oscillation by the Daya Bay
Reactor Neutrino Experiment is described in detail.
Six 2.9-GW$_{\rm th}$ nuclear power reactors of the Daya Bay and Ling
Ao nuclear power facilities served as intense sources of \nuebar{}'s.
Comparison of the \nuebar{} rate and energy spectrum measured by
antineutrino detectors far from the nuclear reactors
($\sim$1500-1950~m) relative to detectors near the reactors
($\sim$350-600~m) allowed a precise measurement of \nuebar{}
disappearance.
More than 2.5~million \nuebar{} inverse beta decay interactions were
observed, based on the combination of 217 days of operation of six
antineutrino detectors (Dec.~2011--Jul.~2012) with a subsequent
1013~days using the complete configuration of eight detectors
(Oct.~2012--Jul.~2015).
The \nuebar{} rate observed at the far detectors relative to the near
detectors showed a significant deficit, $R=0.949 \pm
0.002(\mathrm{stat.}) \pm 0.002(\mathrm{syst.})$.
The energy dependence of \nuebar{} disappearance showed the distinct
variation predicted by neutrino oscillation.
Analysis using an approximation for the three-flavor oscillation
probability yielded the flavor-mixing angle
$\sin^22\theta_{13}=\TRS{}$ and the effective neutrino mass-squared
difference of $\left|{\Delta}m^2_{\mathrm{ee}}\right|=\DMEE{}$\@.
Analysis using the exact three-flavor probability found
${\Delta}m^2_{32}=\DMNH{}$ assuming the normal neutrino mass hierarchy
and ${\Delta}m^2_{32}=\DMIH{}$ for the inverted hierarchy.
\end{abstract}

\maketitle
\tableofcontents
\clearpage

\section{\label{sec:intro} Introduction}

Recent experiments have resulted in significant advances in our
understanding of neutrinos.
Although neutrinos were considered massless within the Standard Model,
abundant evidence of lepton flavor violation by neutrinos strongly
implies small but non-zero masses.
A long-standing disparity between measurement and models of the solar
$\nu_e$ flux was corroborated by successive
radiochemical~\cite{Cleveland:1998nv,Kaether:2010ag,Abdurashitov:2009tn}
and water-Cherenkov~\cite{Hirata:1990xa,Fukuda:1998fd} experiments.
Variation of the ratio of atmospheric $\nu_\mu$ to $\nu_e$ provided
initial evidence for distance-dependent neutrino
disappearance~\cite{Fukuda:1998mi}.
Subsequent observation of the disappearance of $\nu_\mu$ produced in
particle accelerators confirmed atmospheric $\nu$
measurements~\cite{Ahn:2002up}.
A comparison of the solar $\nu_e$ to the total solar $\nu$ flux showed
that the apparent disappearance was a consequence of the conversion of
$\nu_e$'s to other neutrino flavors~\cite{Ahmad:2001an,Ahmad:2002jz}.
Disappearance of $\overline{\nu}_e$'s emitted by nuclear reactors
demonstrated a distinct dependence on the ratio of propagation
distance to antineutrino energy, $L/E_{\nu}$, cementing neutrino
flavor oscillation as the explanation for the observed flavor
violation~\cite{Abe:2008aa}.

The rich phenomena of neutrino flavor oscillation arise from two
remarkable characteristics of neutrinos: small differences between the
masses of the three neutrino states, $m_1 \neq m_2 \neq m_3$, and an
inequivalence between neutrino flavor and mass eigenstates.
Produced in a flavor eigenstate by the weak interaction, a neutrino
state evolves as a coherent superposition of mass eigenstates.
Interference between the phases of each mass component results in the
oscillation of the neutrino flavor.
The flavor oscillates with phases given by
${\Delta}m^2_{ji}L/4E_{\nu}$, where $L$ is the propagation distance,
$E_{\nu}$ is the neutrino energy, and ${\Delta}m^2_{ji}=m^2_j-m^2_i$
is the difference of the squared masses.
The amplitude of flavor oscillation is determined by the amount of
mixing between the flavor and mass eigenstates.
Using the unitary Pontecorvo-Maki-Nakagawa-Sakata (PMNS) matrix,
$U_{\rm PMNS}$, a neutrino with flavor $\alpha$ can be expressed as a
combination of mass states,
$\ket{\nu_\alpha}=\sum_iU^*_{{\alpha}i}\ket{\nu_i}$.
In the three-flavor model, $U_{\rm PMNS}$ is commonly parameterized
using three mixing angles, $\theta_{12}$, $\theta_{23}$,
$\theta_{13}$, and an off-diagonal CP-violating phase $\delta_{\rm
  CP}$.
With sensitivity to the small neutrino mass separations, oscillation
experiments have provided strong evidence for three distinct neutrino
mass states $\nu_{i}$ with masses $m_1$, $m_2$, and $m_3$\@.
Evidence for matter-enhanced resonant flavor conversion of solar
neutrinos has shown that ${\Delta}m^2_{21}\cos(2\theta_{21})>0$\@.
Whether $m_3$ is much lighter or heavier than $m_1$ and $m_2$, also
referred to as the neutrino mass hierarchy, is currently unknown and
is the focus of a broad experimental program~\cite{deGouvea:2013onf}.
Direct measurements of decay kinematics and indirect cosmological
observations are currently consistent with massless neutrinos,
implying that the absolute masses are less than $\sim$1~eV\@.
Neutrino mass qualitatively alters the Standard Model, for example by
inhibiting renormalization or by requiring a new global
symmetry~\cite{DeGouvea:2005gd,Mohapatra:2005wg}.

The Daya Bay Reactor Neutrino Experiment set out to answer the
question: Does the $\nu_3$ mass eigenstate mix with the electron
neutrino state $\nu_e$?
This is equivalent to asking whether the parameter $\theta_{13}$ is
non-zero.
Solar and reactor experiments have established significant mixing
between the $\nu_e$ and $\nu_{1,2}$ states, given by
$\sin^2(2\theta_{12})=0.846\pm0.021$~\cite{PDG2015}.
Atmospheric and accelerator experiments yielded nearly maximal mixing
of the $\nu_\mu$ and $\nu_{2,3}$ states, with
$\sin^2(2\theta_{23})=0.999^{+0.001}_{-0.018}$~\cite{PDG2015}.
Previous searches found no evidence of \nuebar{} disappearance at
$\sim$1~km from reactors, limiting $\sin^22\theta_{13}\leq0.17$ at the
90\% C.L.\@~\cite{Apollonio:2002gd, Boehm:2001ik}.
Measurement of $\theta_{13}$ provides a key parameter of a new
Standard Model which incorporates massive neutrinos, and may allow a
deeper insight into the flavor and mass structure of nature.
A non-zero value for $\theta_{13}$ also makes it possible for future
experiments to determine the neutrino mass hierarchy and to search for
neutrino CP-violation~\cite{deGouvea:2013onf}.



%
Nuclear reactors produce an intense and pure flux of \nuebar{}'s,
which is useful for experimental searches for $\theta_{13}$.
Approximately $2\times10^{20}$~\nuebar{}'s per second are emitted per
gigawatt of thermal power, with a steeply-falling energy spectrum
showing minuscule flux above 10~MeV\@.
Section~\ref{sec:osc_analysis} gives further details of~\nuebar{}
emission by nuclear reactors.
Reactor \nuebar{} are most commonly detected via inverse beta decay
(IBD),
\begin{equation} \label{eq:inverseBetaDecay}
  \overline{\nu}_{e} + p \rightarrow e^+ + n.
\end{equation}
Convolving the energy spectrum with the IBD
cross-section~\cite{Vogel:1999zy} results in an expected spectrum
which rises with neutrino energy from the 1.8~MeV interaction
threshold, peaks at $\sim$4~MeV, and falls to a very low rate above
8~MeV\@.
Charge-current interactions of $\overline{\nu}_{\mu}$ or
$\overline{\nu}_{\tau}$ at these energies are forbidden by energy
conservation, hence oscillation is observed as a reduction, or {\em
  disappearance}, of the expected \nuebar{} signal.
In the three-flavor model of neutrino oscillation, the survival
probability of detecting an \nuebar{} of energy $E_\nu$ at a distance
$L$ from the production source can be expressed as
\begin{equation}\label{eq:survProb_31}
  \begin{split}
  P_{\rm sur} = & 1 - \cos^4\theta_{13}\sin^22\theta_{12}\sin^2\Delta_{21} \\
  & -  \sin^22\theta_{13}(\cos^2\theta_{12}\sin^2\Delta_{31} + \sin^2\theta_{12}\sin^2\Delta_{32}),
  \end{split}
\end{equation}
where $\Delta_{ji} \simeq 1.267 {\Delta}m^2_{ji}{\rm (eV^2)}L{\rm
  (m)}/E_{\nu}{\rm (MeV)}$.
The KamLAND experiment measured the first term, demonstrating
large-amplitude disappearance of reactor \nuebar{} with an oscillation
length of $\sim$60~km.
Atmospheric and accelerator $\nu$ measurements of
$\left|{\Delta}m^2_{32}\right|$ predict an oscillation length of
$\sim$1.6~km for the latter terms.
At this distance, the two oscillation phases $\Delta_{31}$ and
$\Delta_{32}$ are indistinguishable.
Therefore, the expression can be approximated using a single effective
\nuebar{} disappearance phase $\Delta_{\mathrm{ee}}$,
\begin{equation} \label{eq:survProb_ee}
  \begin{split}
  P_{\mathrm{sur}} \simeq 1 & - \cos^4\theta_{13}\sin^22\theta_{12}\sin^2\Delta_{21} \\
                      & - \sin^22\theta_{13}\sin^2\Delta_{\mathrm{ee}},
  \end{split}
\end{equation}
which is independent of the neutrino mass hierarchy.
Here the definition of ${\Delta}m^2_{\mathrm{ee}}$ is empirical; it is
the mass-squared difference obtained by modeling the observed reactor
\nuebar{} disappearance using Eq.~\ref{eq:survProb_ee}.
The mass-squared differences obtained by modeling an observation using
either Eq.~\ref{eq:survProb_31} or Eq.~\ref{eq:survProb_ee} are
expected to follow the relation ${\Delta}m^2_{\mathrm{ee}} \simeq
\cos^2\theta_{12}\left|{\Delta}m^2_{\mathrm{31}}\right| +
\sin^2\theta_{12}\left|{\Delta}m^2_{\mathrm{32}}\right|$~\cite{Minakata:2006gq}.
Based on current measurements, $\left|{\Delta}m^2_{31}\right| \simeq
\left|{\Delta}m^2_{\mathrm{ee}}\right| \pm 2.3 \times 10^{-5}$~eV$^2$
and $\left|{\Delta}m^2_{32}\right| \simeq
\left|{\Delta}m^2_{\mathrm{ee}}\right| \mp 5.2 \times 10^{-5}$~eV$^2$,
assuming the normal (upper sign) or inverted (lower sign) mass
hierarchy.

Previous searches for oscillation due to $\theta_{13}$ were limited by
uncertainty in the \nuebar{} flux emitted by
reactors~\cite{Apollonio:2002gd, Boehm:2001ik}.
A differential comparison with an additional detector located near the
reactor was proposed to overcome this
uncertainty~\cite{Mikaelyan:1998yg}.
With a far-versus-near detector arrangement, sensitivity to neutrino
oscillation depends on relative uncertainties between detectors in the
number of target protons $N_{\rm p}$, \nuebar{} detection efficiency
$\epsilon$, and distances from the reactor $L$.
If these relative uncertainties are well-controlled, small differences
in the oscillation survival probability $P_{\mathrm{sur}}$ become
detectable in the ratio of the number of \nuebar{} interactions in the
far relative to near detector,
\begin{equation}\label{eq:nearFarRatio}
  \frac{N_{\rm f}}{N_{\rm n}} = \left(\frac{N_{\rm p,f}}{N_{\rm p,n}}\right) \left(\frac{L_{\rm n}}{L_{\rm f}}\right)^2 \left(\frac{\epsilon_{\rm f}}{\epsilon_{\rm n}}\right) \left[\frac{P_{\rm sur}(E_{\nu},L_{\rm f})}{P_{\rm sur}(E_{\nu},L_{\rm n})}\right].
\end{equation}
Three experiments were constructed based on this technique: the Daya
Bay~\cite{DYBProposal}, RENO~\cite{Ahn:2010vy}, and Double CHOOZ
experiments~\cite{DCProposal}.
In Mar.~2012, the Daya Bay experiment reported the discovery of
\nuebar{} disappearance due to a non-zero value of
$\theta_{13}$~\cite{DYB2012}.
Oscillation due to $\theta_{13}$ has since been confirmed by the other
experiments~\cite{RENO2012,DC2012}, as well as by other
techniques~\cite{Abe:2015awa,Adamson:2016tbq}.
The relatively large $\theta_{13}$ mixing has also allowed measurement
of $\left|{\Delta}m^2_{\mathrm{ee}}\right|$ from the variation of the
disappearance probability versus \nuebar{} energy~\cite{An:2015rpe}.
Compatibility of the mass-squared difference with that obtained from
the disappearance of accelerator and atmospheric $\nu_\mu$'s with
\mbox{GeV-energies} firmly establishes the three-flavor model of
neutrino mass and mixing.

This paper provides a detailed review of the Daya Bay measurement of
neutrino oscillation.
Section~\ref{sec:experiment} gives an overview of the experiment.
The calibration and characterization of the experiment are presented
in Section~\ref{sec:calibration}.
Identification of reactor \nuebar{} interactions, signal efficiencies,
and assessment of backgrounds are discussed in
Section~\ref{sec:background}.
Section~\ref{sec:osc_analysis} presents an analysis of neutrino
oscillation using the measured \nuebar{} rate and spectra, while
Section~\ref{sec:conclusions} contains concluding remarks.

\section{\label{sec:experiment} Experiment Description}

The relative measurement of oscillation, as summarized in
Eq.~\ref{eq:nearFarRatio}, motivated much of the design of the Daya
Bay experiment.
The disappearance signal is most pronounced at the first oscillation
minimum of $P_{\mathrm{sur}}$.
Based on existing accelerator and atmospheric $\nu_\mu$ measurements
of ${\Delta}m^2_{32}$, this corresponded to a distance $L_{\rm
  f}$$\approx$1.6~km for reactor \nuebar{} with a mean energy of
$\sim$4~MeV\@.
Significant \nuebar{} disappearance in the near detectors would have
reduced the overall sensitivity of the far-to-near comparison, so
$L_{\rm n}$ was kept to $\sim$500~m or less.
The use of identically-designed modular detectors limited variations
in relative number of target protons $N_{\rm p}$ and efficiency
$\epsilon$ between detectors.
Situating detectors at a sufficient depth underground reduced
muon-induced neutrons and short-lived isotopes, the most prominent
backgrounds for reactor \nuebar{} detection.
Statistical sensitivity increases with \nuebar{} flux, target size,
and detector efficiency, arguing for the use of intense reactors and
large detectors.
%

The campus of the Daya Bay nuclear power plant near Shenzhen, China
was well-suited for this purpose.
At the time of this measurement the facility consisted of six
2.9~GW$_{\mathrm{th}}$ pressurized water reactors and produced roughly
$3.5\times10^{21}$~\nuebar{}/s, making it one of the most intense
\nuebar{} sources on Earth.
Steep mountains adjacent to the reactors provided ample shielding from
muons produced by cosmic ray showers.
Underground experimental halls were excavated to accommodate 160~tons
of fiducial target mass for \nuebar{} interactions, equally divided
between locations near and far from the reactors.
With this arrangement, a total of $\sim$2000 \nuebar{} interactions
per day were detected near to, and $\sim$250 far from, the reactors,
with muon-induced backgrounds contributing less than 0.5\%.
The target mass was divided between 8 identically-designed modular
antineutrino detectors (ADs).
Installing at least two ADs in each experimental hall allowed
side-by-side demonstration of $<$0.2\% variation in \nuebar{}
detection efficiency between detectors.
A confirmation of the side-by-side performance of the first two ADs
was given in~\cite{DayaBay:2012aa}.
These basic characteristics have yielded measurements of
$\sin^22\theta_{13}$ with $\sim$4\% precision and
$\left|{\Delta}m^2_{\mathrm{ee}}\right|$ with $\sim$3\% precision, as
will be discussed in this paper.
This section provides an abbreviated description of the Daya Bay
experiment, while a more detailed description is given
in~\cite{An:2015qga}.

\begin{figure}[!htb]
\includegraphics[width=.45\textwidth]{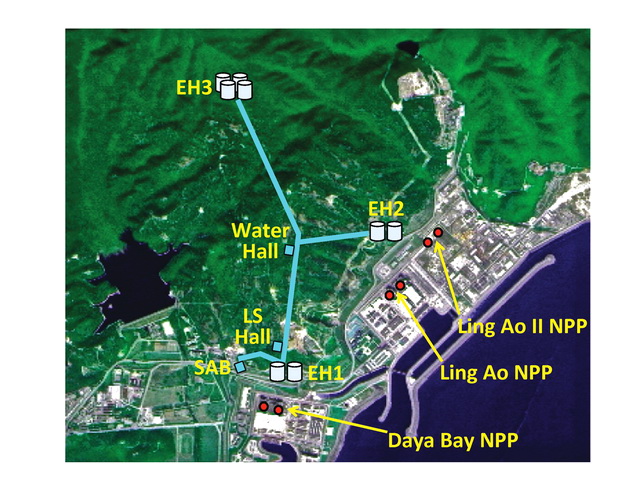}
\caption{\label{fig:dayabay_map} Layout of the Daya Bay experiment.
  The Daya Bay and Ling Ao nuclear power plant (NPP) reactors (red
  circles) were situated on a narrow coastal shelf between the Daya
  Bay coastline and inland mountains.  Two antineutrino detectors
  installed in each underground experimental hall near to the reactors
  (EH1 and EH2) measured the \nuebar{} flux emitted by the reactors,
  while four detectors in the far experimental hall (EH3) measured a
  deficit in the \nuebar{} flux due to oscillation.  The detectors
  were built and initially tested in a surface assembly building
  (SAB), transported to a liquid scintillator hall (LS Hall) for
  filling, and then installed in an experimental hall.}
\end{figure}

%
%
%
\begin{table*}[!htb]
  \caption[Site Survey]{The surveyed coordinates of the geometric
  centers of the nuclear reactor cores and antineutrino detectors of
  the Daya Bay experiment.  The detectors are labeled as AD1 through
  AD8, according to their order of assembly and installation.  The $X$
  coordinate is due north, while the $Z$ coordinate is vertical at the
  survey origin.  Coordinates were determined from a combination of
  total station electronic theodolite and GPS measurements, with a
  precision of 18~mm.  The corresponding neutrino oscillation
  baselines for each reactor-detector pair are provided.  The
  approximate rock overburden of each experimental hall and the mass
  of the GdLS antineutrino target in each detector are also given in
  both meters and meters-water-equivalent.  The average thermal power
  of each reactor core, in gigawatts, is given separately for the six
  detector and eight detector periods.}
\label{tab:site_survey}
  \begin{minipage}[c]{\textwidth}
%
 \begin{ruledtabular}
 \begin{tabular}{lcccrrrrrrrrr}
  \multicolumn{7}{c}{} & \multicolumn{6}{c}{Reactor} \\
  \multicolumn{7}{c}{} & \multicolumn{1}{c}{D1} & \multicolumn{1}{c}{D2} & \multicolumn{1}{c}{L1} & \multicolumn{1}{c}{L2} & \multicolumn{1}{c}{L3} & \multicolumn{1}{c}{L4} \\
  \cline{7-13}
  \multicolumn{6}{c}{} & \multicolumn{1}{c}{$\overline{W}^{\mathrm{6AD}}_{\mathrm{th}}$} & 2.082 & 2.874 & 2.516 & 2.554 & 2.825 & 1.976 \rule{0pt}{2.6ex}\\ 
  \multicolumn{6}{c}{} & \multicolumn{1}{c}{$\overline{W}^{\mathrm{8AD}}_{\mathrm{th}}$} & 2.514 & 2.447 & 2.566 & 2.519 & 2.519 & 2.550 \rule{0pt}{2.6ex}\\ 
  \multicolumn{6}{c}{} & \multicolumn{1}{c}{X [m]} & 359.20 & 448.00 & -319.67 & -267.06 & -543.28 & -490.69 \\ 
  \multicolumn{6}{c}{} & \multicolumn{1}{c}{Y [m]} & 411.49 & 411.00 & -540.75 & -469.21 & -954.70 & -883.15 \\
  \multicolumn{6}{c}{} & \multicolumn{1}{c}{Z [m]} & -40.23 & -40.24 & -39.73 & -39.72 & -39.80 & -39.79 \\ 
  \hline
  Hall & Depth [m(mwe)] & Detector & Target [kg] & \multicolumn{1}{c}{X [m]} & \multicolumn{1}{c}{Y [m]} & \multicolumn{1}{c}{Z [m]} & \multicolumn{6}{c}{Baseline [m]} \\
  \hline
  \multirow{2}{*}{EH1} & 93 & AD1 & 19941 $\pm$ 3 & 362.83 & 50.42 & -70.82 & 362.38 & 371.76 & 903.47 & 817.16 & 1353.62 & 1265.32 \\
  & (250) & AD2 & 19967 $\pm$ 3 & 358.80 & 54.86 & -70.81 & 357.94 & 368.41 & 903.35 & 816.90 & 1354.23 & 1265.89 \\ 
  \multirow{2}{*}{EH2} & 100 & AD3 & 19891 $\pm$ 4 & 7.65 & -873.49 & -67.52 & 1332.48 & 1358.15 & 467.57 & 489.58 & 557.58 & 499.21 \\ 
  & (265) & AD8 & 19944 $\pm$ 5 & 9.60 & -879.15 & -67.52 & 1337.43 & 1362.88 & 472.97 & 495.35 & 558.71 & 501.07 \\ 
  \multirow{4}{*}{EH3} & & AD4 & 19917 $\pm$ 4 & 936.75 & -1419.01 & -66.49 & 1919.63 & 1894.34 & 1533.18 & 1533.63 & 1551.38 & 1524.94 \\ 
  & 324 & AD5 & 19989 $\pm$ 3 & 941.45 & -1415.31 & -66.50 & 1917.52 & 1891.98 & 1534.92 & 1535.03 & 1554.77 & 1528.05 \\ 
  & (860) & AD6 & 19892 $\pm$ 3 & 940.46 & -1423.74 & -66.50 & 1925.26 & 1899.86 & 1538.93 & 1539.47 & 1556.34 & 1530.08 \\
  & & AD7 & 19931 $\pm$ 3 & 945.17 & -1420.03 & -66.49 & 1923.15 & 1897.51 & 1540.67 & 1540.87 & 1559.72 & 1533.18 \\   
 \end{tabular}
 \end{ruledtabular}
%
  \end{minipage}
\end{table*}

The reactors at Daya Bay were arranged in two clusters: the Daya Bay
cluster hosted two reactors (D1 and D2), while the Ling Ao cluster
hosted four (L1, L2, L3 and L4).
Correspondingly, four near detectors were divided between two near
experimental halls (EH1 and EH2) near the two clusters.
The remaining four detectors were installed in a single far hall
(EH3).
The locations of the experimental halls were determined to optimize
sensitivity to $\theta_{13}$, considering reactor locations and
mountain topography.
While uncertainties in reactor flux were not completely canceled as
would happen for the case of a single reactor, this arrangement of
detectors reduced the far-to-near flux ratio uncertainty to
$\leq$0.1\% (see Sec.~\ref{sec:osc_analysis}).
The layout of the six reactors and three experimental halls is shown
in Fig.~\ref{fig:dayabay_map}.

When comparing the measurements between near and far detectors, the
largest relative correction was from the baselines of the detectors,
as seen in Eq.~\ref{eq:nearFarRatio}.
Accurate surveys of the experiment site allowed precise correction for
this effect.
Surveys consisted of total station electronic theodolite measurements
combined with supplemental global positioning system (GPS)
measurements.
Lacking GPS reception underground, surveys of the experimental halls
and access tunnels relied on redundant total station measurements.
Table~\ref{tab:site_survey} provides the surveyed reactor and detector
coordinates, where $X$ is due north and $Z$ is vertical at the survey
origin.
Uncertainties in the survey results were 18~mm in each coordinate,
dominated by the precision of the GPS measurements and the tension
between GPS and total station survey results.
\nuebar{} emission was distributed throughout the fuel elements of
each reactor core, spanning a region 3.7~m in height and 3~m in
diameter.
Reactor models established the horizontal centroid of \nuebar{}
emission to within 2~cm of the geometric center of each core.
With the centroid determined, the spatial variation of the
distribution of \nuebar{} emission within the core had negligible
impact to the oscillation measurement.

%
The combination of organic liquid scintillator with photomultiplier
tubes (PMTs) results in a powerful technique for reactor \nuebar{}
detection.
Scintillator contains protons (as $^1$H) which serve as targets for
\nuebar{} inverse beta decay interactions (see
Eq.~\ref{eq:inverseBetaDecay}).
Scintillators simultaneously function as a sensitive medium, emitting
photons in response to ionization by the products of IBD interactions.
Detection of the photons using PMTs allows a calorimetric measurement
of the prompt positron energy deposition.
This energy is the sum of the IBD positron kinetic plus annihilation
energy, $E_{\mathrm{prompt}} = T_{e^+} + 2m_{e}$, where $m_{e}$ is the
mass of the electron.
The initial \nuebar{} energy can be accurately estimated using
$E_{\nu} \simeq E_{\mathrm{prompt}} + 0.8$~MeV, based on the
kinematics of inverse beta decay.
The IBD neutron generally carries only a small fraction of the initial
kinetic energy, O(10~keV).
The neutron thermalizes and is then captured on a nucleus within the
scintillator in a time of O(100~$\mu$s).
The resulting nucleus rapidly de-excites by emitting one or more
characteristic $\gamma$-rays.
Detection of this subsequent pulse of scintillation light from the
delayed neutron capture $\gamma$-rays efficiently discriminates
\nuebar{} interactions from background.

\begin{figure}[!htb]
\includegraphics[width=.45\textwidth]{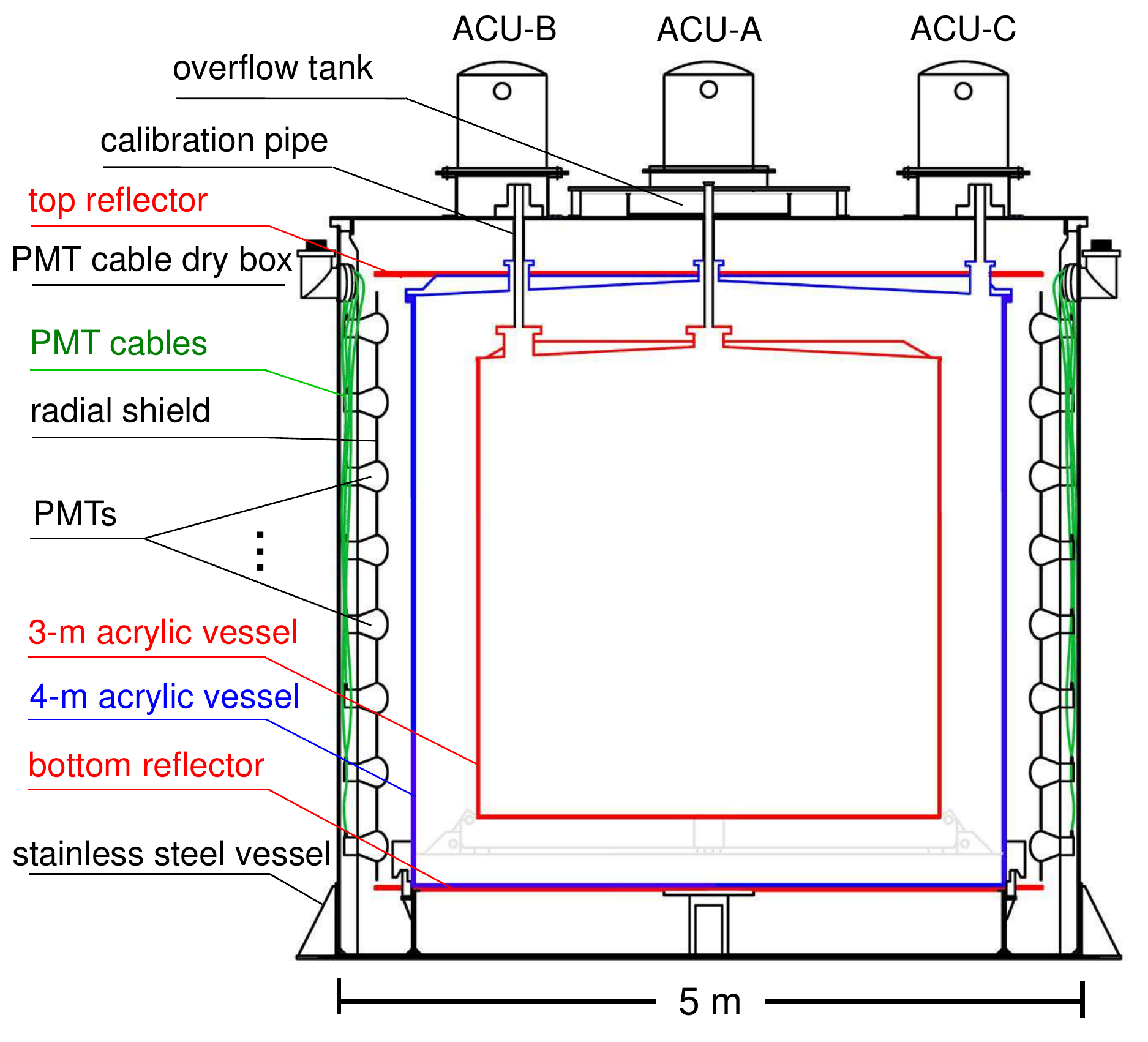}
\caption{\label{fig:antineutrinoDetector} Cross-sectional diagram of
  an antineutrino detector (AD)\@. Scintillation light was produced
  when a reactor \nuebar{} interacted within the central 20-ton GdLS
  target, which was contained in a transparent acrylic vessel.  The
  target was nested within an additional 20-tons of pure LS to
  increase efficiency for detection of gamma rays produced within the
  target.  Scintillation light was detected by 192 photomultipliers
  mounted on the inner circumference of a 5~m by 5~m stainless steel
  vessel, which was filled with mineral oil (MO).}
\end{figure}

%
The eight antineutrino detectors of the Daya Bay experiment relied on
this technique, and were designed to specifically limit potential
variations in response and efficiency between detectors.
Each detector consisted of a nested three-zone structure, as shown in
Fig.~\ref{fig:antineutrinoDetector}.
The central \nuebar{} target was 20-tons of linear-alkyl-benzene-based
liquid scintillator, loaded with 0.1\% of $^{\rm nat}$Gd by mass
(GdLS)\@.
Details of the production and composition of the scintillator are
discussed in~\cite{Beriguete:2014gua}.
Gadolinium (Gd) efficiently captures thermalized neutrons, emitting a
few $\gamma$-rays with a total energy of $\sim$8~MeV per capture.
The relatively high capture energy enhanced discrimination of the
signal from backgrounds produced by natural radioactivity, primarily
at energies $\lesssim$5~MeV\@.
Gd-loading also provided a physical method to fiducialize the
detector, allowing efficient rejection of \nuebar{} interactions which
occurred outside the target volume.
The target scintillator was contained within a 3~m by 3~m cylindrical
tank, referred to as the Inner Acrylic Vessel (IAV), which was made of
UV-transparent acrylic.
This was nested within a similar 4~m by 4~m acrylic tank, refered to
as the Outer Acrylic Vessel (OAV), which was filled with scintillator
without Gd-loading (LS).
This outer scintillating region significantly increased the efficiency
for detection of gamma rays produced in the target region, reducing
systematic uncertainties from effects at the target boundary.
Hence, this region was referred to as the gamma catcher.
Both regions were nested within a 5~m by 5~m stainless steel vessel
(SSV), which was filled with mineral oil (MO).
The MO had a density matching that of the LS and GdLS, which balanced
stresses across the thin-walled ($\sim$1.5~cm) acrylic vessels.
It also shielded the scintillating regions from gamma rays from
radioactivity in the SSV and PMTs, and provided a transparent medium
for propagation of scintillation light to the PMTs.

Scintillation light was detected using 192 8-inch PMTs (Hamamatsu
R5912) which were immersed in the MO, and mounted in 8 rings of 24 on
the inner cylindrical surface of each SSV\@.
Specular reflectors located above and below the OAV improved the
uniformity of light collection versus position within the
scintillating regions.
In the radial direction, a black light-absorbing radial shield masked
all but the photocathode of the PMTs, simplifying and unifying the
optical characteristics of the eight detectors.
Liquid overflow tanks on the top of the detector allowed for small
changes in liquid volume of each region in response to changes of
temperature and pressure.
Three automated calibration units (ACUs) were used to deploy
radioactive sources ($^{60}$Co, $^{68}$Ge, and $^{241}$Am-$^{13}$C)
and light-emitting diodes (LEDs) through narrow teflon-bellow
penetrations into the GdLS and LS regions.
Details of the calibration system are provided in~\cite{Liu:2013ava}.
%

%
Small differences ($<$0.5\%) in the total number of protons within
each AD target region was the next most significant correction when
comparing the measurements of the far versus near detectors
(i.e.~$N_{\mathrm{p}}$ in Eq.~\ref{eq:nearFarRatio}).
After mechanical assembly and testing within a surface assembly
building (SAB), each dry AD was transported to an underground liquid
scintillator hall for filling.
All GdLS was produced in advance and divided equally between the eight
ADs in order to ensure a consistent proton and Gd density, as well as
optical performance.
Each AD target was filled with GdLS from a reservoir mounted on
precision weigh-bridge load cells, whose performance was confirmed
using calibrated test masses.
Drift in the load cell readings over several days provided the
dominant systematic uncertainty of $\pm$2~kg.
An independent measurement used a coriolis flow meter to confirm the
relative differences in mass delivered to each AD with few-kg
precision, although this instrument measured the absolute mass with
far less precision than the load cells.
A 0.13\% correction accounted for the weight of nitrogen gas which
displaced the GdLS within the reservoir during filling.
After filling, another correction was made for the small fraction of
GdLS present within the calibration tubes and overflow tank, and hence
outside of the IAV target volume.
Table~\ref{tab:site_survey} summarizes the measured GdLS mass within
each IAV target.
The 5-kg precision of the target mass correction corresponded to a
negligible 0.03\% systematic uncertainty in the comparison of
antineutrino interaction rates among the ADs.

\begin{figure}[!htb]
\includegraphics[width=.45\textwidth]{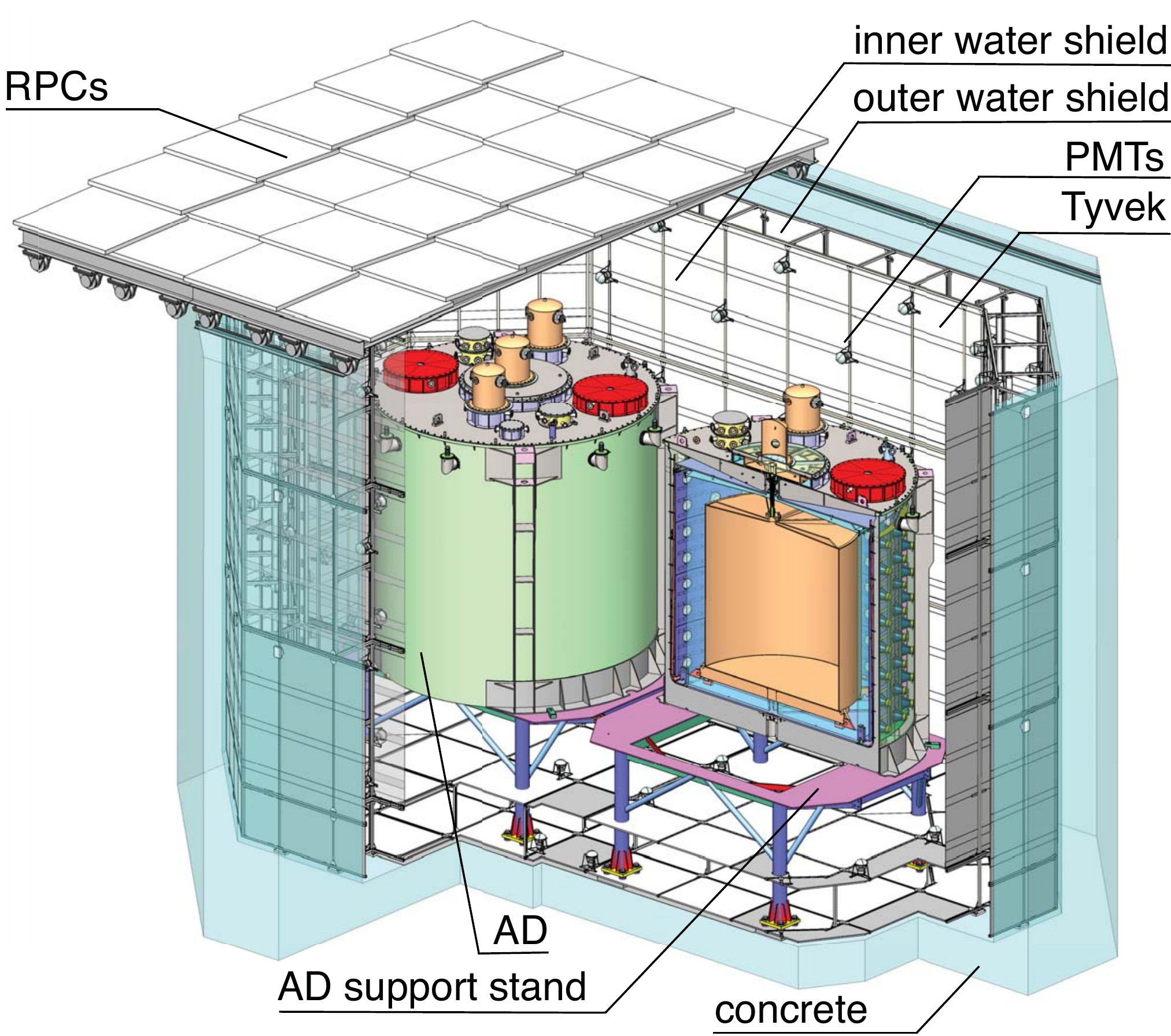}
\caption{\label{fig:vetoAndAD} Diagram of a near site detector
  system. Two ADs were immersed in a water Cherenkov muon detector
  which functioned as both a passive radiation shield and an active
  muon tag.  Tyvek sheets divided the pool into two optically separate
  detectors, the inner and outer water shields.  An RPC system covered
  the pool, providing additional muon identification.}
\end{figure}

%
After filling, the antineutrino detectors were installed in a 10~m
deep water pool in each underground experimental hall, as shown in
Fig.~\ref{fig:vetoAndAD}.
The water shielded the detectors from $\gamma$-rays arising from
natural radioactivity and muon-induced neutrons, which were primarily
emanating from the cavern rock walls.
The pool was optically separated into two independent regions, the
inner (IWS) and outer water shield (OWS).
Both regions were instrumented with PMTs to detect the Cherenkov light
produced by muons.
A 4-layer modular resistive plate chamber (RPC) system was installed
over the pool, which served in studies of muons and muon-induced
backgrounds.
Identification of muons which passed through the IWS, OWS, and RPC
system enhanced the rejection of background from neutrons generated by
muon interactions in the immediate vicinity of the antineutrino
detectors.
A detailed description of the muon system was given
in~\cite{Dayabay:2014vka} and muon-induced backgrounds are discussed
in Sec.~\ref{sec:background}.
%

%
A single coaxial cable delivered positive high-voltage to, and
returned the signal from, each PMT\@.
A passive circuit AC-decoupled the fast PMT signal from the HV, and
the signal was input to a channel of the front-end electronics (FEE).
The HV for each PMT was tuned for detecting single photoelectrons
(PE), with gains matched at $\sim$$1.0\times10^7$ to within 5\%.
After an initial fast amplification, the FEE split the signal for
separate measurements of the charge and relative arrival time.
One copy of the signal was passed to a discriminator with a threshold
of $\sim$0.25~pe, which served as the start signal for a TDC with
1.6~ns resolution.
The other copy was sent to a CR-(RC)$^4$ pulse shaping circuit which
provided an integral measure of the incoming signal charge with a
$\sim$100~ns time constant~\cite{Knoll:2010xta}.
The signal was $\times$10 amplified and then sampled by a 40~MHz
12-bit ADC, which provided better than 0.1~pe resolution.
To increase the dynamic charge range for processing very large
signals, an additional copy of the shaper output was passed to a
$\times$0.5 amplifier and sampled by an equivalent ADC\@.
The peak sample value obtained by each ADC, as well as a measure of
each ADC pedestal preceding the signal, was buffered and awaited
triggering of the detector.

%
Each detector (ADs, IWS, OWS) operated as an independently triggered
system using a Local Trigger Board (LTB).
Each FEE board accepted signals from up to 16 PMTs, and transmitted a
count of the number of channels over threshold, as well as an analog
sum of all signals, to the LTB\@.
A trigger was issued for each detector under the following conditions:
\begin{itemize}
\item AD: The total count of channels over threshold (NHIT) was
  $\geq$45 or analog sum (ESUM) was $\geq$65~PE ($\sim$0.4~MeV).
\item IWS: The NHIT was $\geq$6 for an IWS\@.
\item OWS: The NHIT was $\geq$7 for a near-hall OWS, or $\geq$8 for
  the far-hall OWS\@.
\item RPC: 3 of 4 layers of a module were above threshold.
\item Random: Randomly issued at a rate of $\sim$10~Hz in order to
  monitor the level of sub-threshold or accidental activity in each
  detector.
\item Calibration: Simultaneous with each pulse of light emitted from
  a light-emitting diode (LED).
\item Cross-detector: A Master Trigger Board (MTB) at each site could
  forward triggers from one detector to another.  An intended use was
  to capture activity within the muon systems when an AD detected a
  potential reactor \nuebar{} (e.g.\ two AD triggers separated by
  $\leq$200~$\mu$s).
\end{itemize}
When a detector received a trigger, it served as a stop signal for the
TDCs\@.
The TDC, peak ADC, and pedestal ADC values for each channel over
threshold within the past 1.2~$\mu$s were then recorded.
A digital hit map was recorded for each RPC module which satisfied the
3 of 4 layer trigger threshold.
A GPS-synchronized time stamp (25~ns resolution) provided a measure of
the absolute time for each triggered event.

%
The analysis presented here relied on the combination of data from two
periods of operation.
Extending from Dec.~24,~2011, to Jul.~28,~2012, the first period
consisted of 217~days of operation with the first 6 ADs: 2 in EH1, 1
in EH2, and 3 in EH3.
The final two ADs, AD7 and AD8, were completed and installed in EH3
and EH2 respectively during the Summer of 2012.
An additional 1013~days of data were collected from Oct.~19,~2012, to
Jul.~28,~2015.
For these two periods, 189~days (87\%) and 920~days (91\%) of livetime
were accepted for the oscillation analysis, with the majority of the
downtime attributed to weekly detector calibration.

\section{\label{sec:calibration}Detector Calibration}

%
As a first step in the analysis, the recorded digital information was
converted to time and charge.
From the converted values we established the energy scale, and studied
the temporal and spatial response of the detectors to particle
interactions.
The details of the calibration process are discussed in this section.
Descriptions of the calibration systems are given
in~\cite{Liu:2013ava} and~\cite{Dayabay:2014vka}.

\subsection{\label{sec:time_calib}Time Calibration}

%
As discussed in the previous section, the time at which each detector
triggered was recorded with 25~ns precision.
%
%
%
Calibration LEDs were used to measure the relative time responses of
the PMTs within a single detector.
The time delays observed in each channel were corrected for LED-to-PMT
distances and were fitted as a function of light intensity.
The results were recorded to a database and used to correct TDC values
during data analysis.
The timing calibration was repeated whenever a modification was made
to a detector system (e.g.\ replacement of FEE, LTB, or MTB board).

\subsection{\label{sec:energy_calib}Energy Calibration}

%
The most critical calibration task was to reduce potential differences
in reactor \nuebar{} detection efficiency between ADs, as shown in
Eq.~\ref{eq:nearFarRatio}.
Therefore, a calibration process was implemented to reduce
detector-to-detector variations in the energy estimated for equivalent
particle interactions.
At the lowest level, the uncalibrated charge from each PMT $i$ was
determined from the difference between the ADC peak value and the ADC
pedestal value reported by the FEE, $Q_{i} = ADC^{\rm peak}_{i} -
ADC^{\rm ped}_{i}$\@.
Each AD was principally a calorimetric detector, hence the estimate of
the total energy deposited by a particle interaction, $E_{\rm rec}$,
was proportional to a calibrated sum of the charges measured by each
PMT, $Q_{i}$.
This sum can be expressed as
\begin{equation} \label{eq:energyCalibration} 
  E_{\rm rec} = \left(\sum_i \frac{Q_{i}}{\overline{Q}^{\rm SPE}_{i}(t)}\right) \frac{f_{\rm act}(t)}{N^{\rm PE}(t)} f_{\rm pos}(\mathbf{r}_{\rm rec}, t),
\end{equation}
including the following calibration terms:
\begin{itemize}
\item PMT charge scale, $\overline{Q}^{\rm SPE}_{i}(t)$: a scale
  conversion from charge to detected light unique for each PMT plus
  electronics channel, roughly 19 ADC counts per single photoelectron
  (SPE).
\item Active PMT correction, $f_{\rm act}(t)$: a unitless factor which
  compensated for the reduced light collected when a PMT channel was
  temporarily disabled.  The factor is defined as the ratio of total
  to active AD PMTs, and amounts to a minor correction of
  (1/192)$\simeq$0.5\% per inactive PMT in an AD\@.
\item Light yield, $N^{\rm PE}(t)$: a scale conversion from total AD
  detected light to mean particle interaction energy, approximately
  170 photoelectrons per MeV of deposited energy.
\item AD nonuniformity correction, $f_{\rm pos}(\mathbf{r}_{\rm rec},
  t)$: a unitless factor which compensated for the observed variation
  in collected light versus the estimated position $\mathbf{r}_{\rm
    rec}$ of a particle interaction in the AD\@.  The correction was
  $\pm$5\% within the target region, and from -6\% to +15\% including
  interactions in the gamma catcher.
\end{itemize}
The following sections discuss how these calibration factors were
determined and validated.

\subsubsection{\label{sec:pmt_calib}PMT Charge Calibration}

%
The first step in the energy scale calibration chain consisted of
correcting for the few-percent differences that exist in the gain of
each PMT and the associated electronics versus time,
$\overline{Q}^{\rm SPE}(t)$.
The operating voltages necessary to achieve a common gain of
$1\times10^7$, $\pm$5\%, were determined for each PMT prior to
installation.
Each electronics channel introduced an additional 3\% variation in
gain.
Since the response of each channel drifts with changing environmental
conditions as well as with hardware replacements, a calibration method
that operated concurrently with regular antineutrino data collection
was developed.
In this method, the gain was determined using individual PMT signals
uncorrelated with particle interactions within the scintillator.
These signals were primarily single photoelectrons from thermal
emission, also referred to as PMT dark noise, and were captured by the
data acquisition system in the few hundred nanoseconds prior to a
particle interaction which triggered a detector.
The baseline subtracted charge distributions of these uncorrelated
signals for each channel were used to estimate the SPE gain versus
time.
The gain was re-estimated every $\sim$6~hours, as this was the minimum
time required to collect sufficient uncorrelated signals from each PMT
channel.

\begin{figure}[h!]
\includegraphics[width=.45\textwidth]{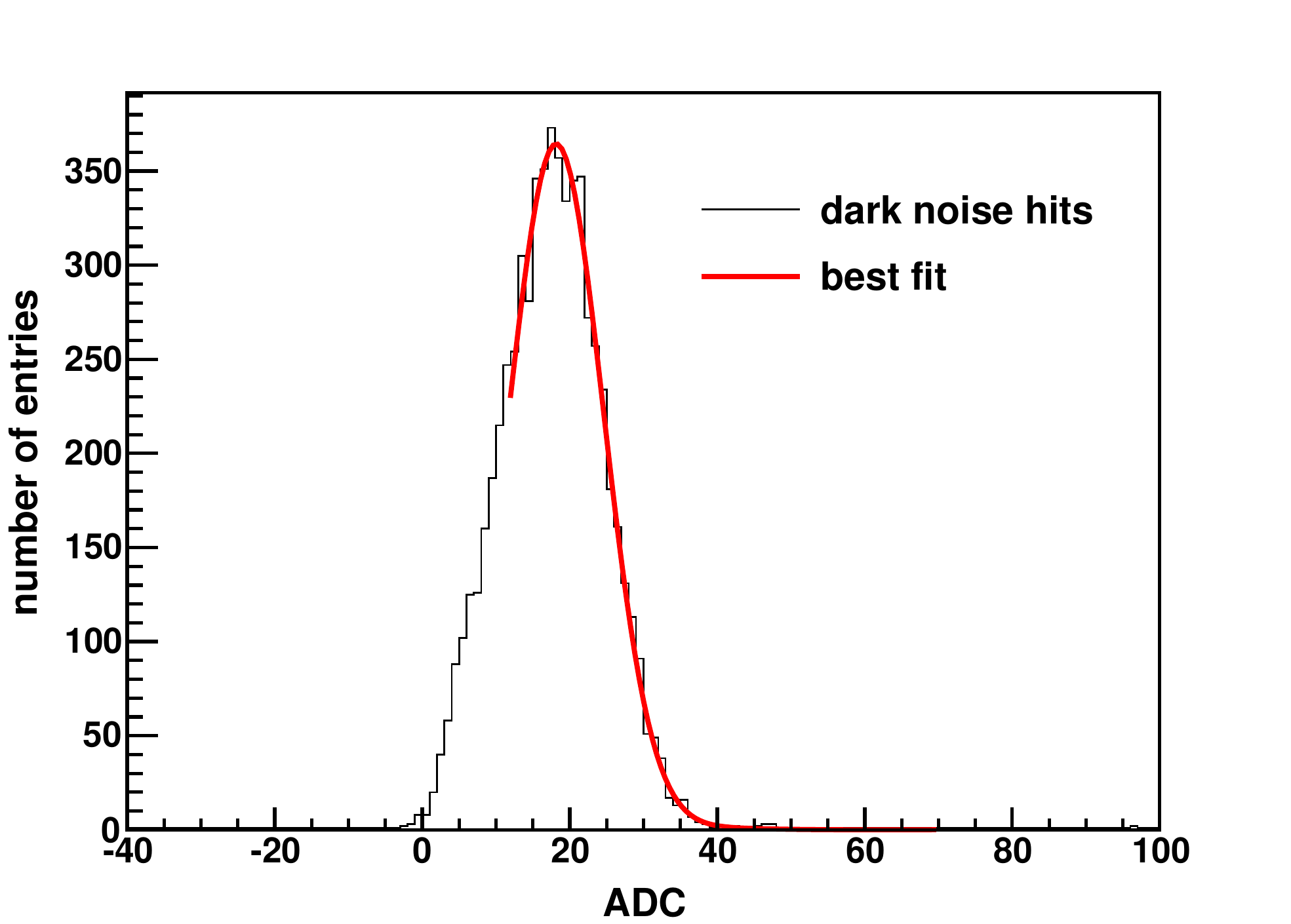}
\caption{\label{fig:spe_fit_example} Example of the
  baseline-subtracted ADC charge distribution of uncorrelated PMT
  signals modeled using Eq.~\ref{eq:spe_eq_fit}.}
\end{figure}

\begin{figure}[h!]
\includegraphics[width=.45\textwidth]{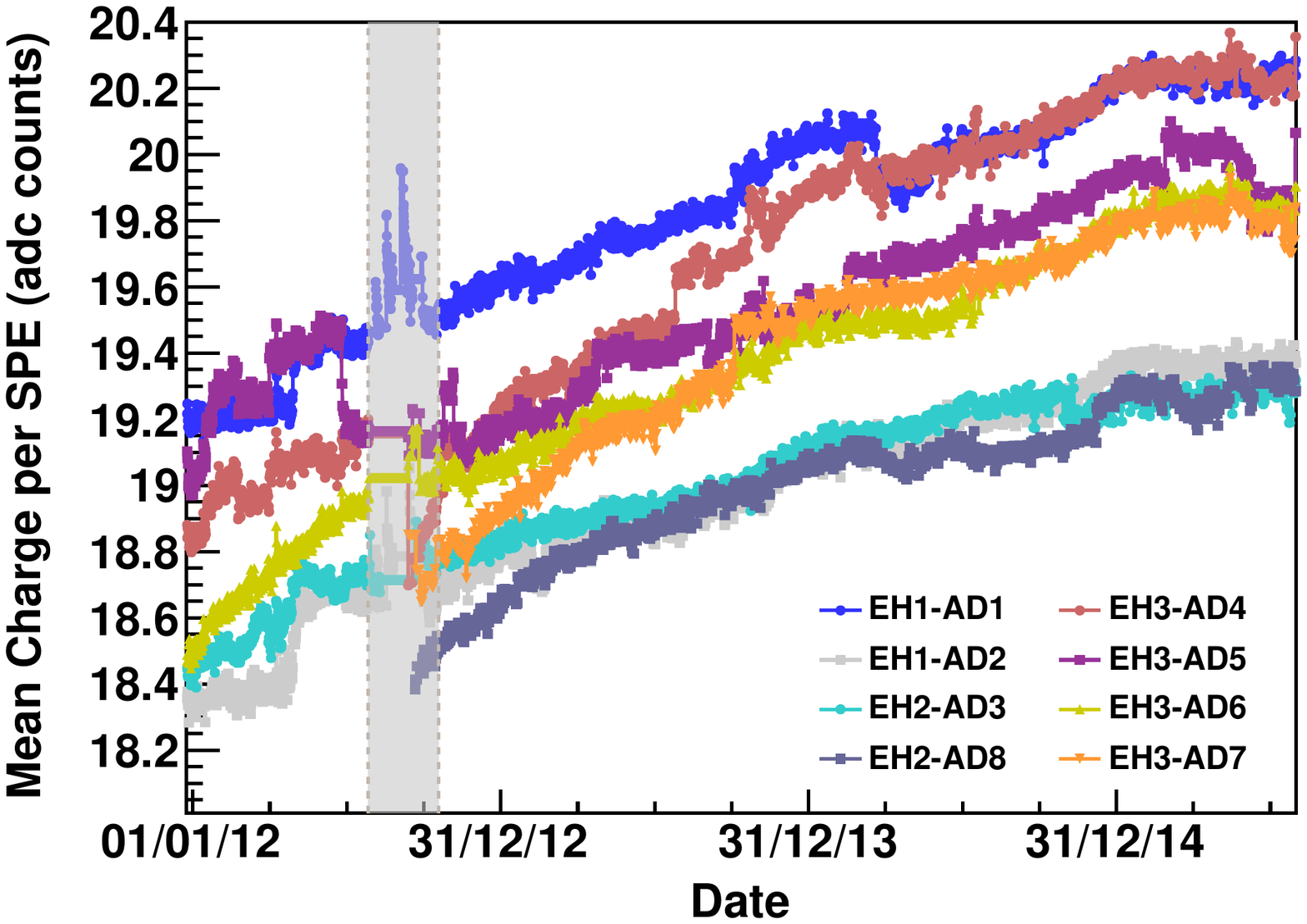}
\caption{\label{fig:ad_gains} Average
  number of ADC counts per single photoelectron, averaged over all PMT
  channels within each AD versus time. The shaded vertical band
  delineates the Summer 2012 shutdown period, during which the
  high-voltage mainframes were frequently power-cycled and data-taking
  was partially interrupted for installation activities.}
\end{figure}

%
The probability distribution of charge signals $Q$ from a PMT was
modeled using the convolution of a Poisson distribution with a
Gaussian function~\cite{nimPmtFit},
\begin{equation}\label{eq:spe_eq_fit}
S(Q) = \sum_n^\infty \frac{\mu^n e^{-\mu}}{n!} \frac{1}{\sigma_{\rm SPE} \sqrt{2n\pi}} \exp\left(- \frac{(Q-n\overline{Q}^{\rm SPE})^2}{2n\sigma_{\rm SPE}^2}   \right),
\end{equation}
where $\mu$ is the mean number of photoelectrons (PEs) collected by
the first dynode, and $\overline{Q}^{\rm SPE}$ and $\sigma_{\rm SPE}$
are the mean charge and resolution of the SPE distribution in units of
ADC counts.
The values of these three parameters which best described the observed
distribution $S$ were determined for each PMT\@.
Signals from PMT dark noise were predominantly single photoelectrons,
hence the sum was limited to $n\leq2$ without loss of precision.
Noise resulted in fluctuations of the distributions below 10~ADC
counts, and the results were more stable when this region was not used
to constrain the model.
Fig.~\ref{fig:spe_fit_example} shows an example SPE charge
distribution and corresponding model.
This procedure was applied to each PMT channel, and
Fig.~\ref{fig:ad_gains} shows the mean charge per SPE averaged over
all channels within each AD as a function of time.
The typical gain calibration constant was on the order of $19$~ADC
counts per SPE, although several-percent differences existed in the
average gain between ADs.
A slight upwards drift in gain was observed for all detectors.
This drift was partially but not completely attributable to changes in
temperature of the front-end electronics.
Jumps correlated among the ADs within the same hall were correlated
with power-cycling of the PMT high voltage mainframes.
An independent method of determining PMT channel gains, based on
weekly low-intensity calibration LED data samples, reproduced all of
these observed features.

\subsubsection{\label{sec:energy_active_pmts}Active PMT Correction}

An extensive data monitoring process was implemented in order to
ensure all data were of high quality.
In this process, suspect PMT channels were temporarily removed from
the analysis due to poor gain, high noise, or other features, as
identified by an automated algorithm run concurrently with data
collection.
The most common cause for disabling a PMT channel was temporary
instability in the supplied high voltage.
The total number of disabled channels at any one time typically
fluctuated around five, combined across all eight detectors, with
rarely more than one in a given detector.
The absence of a single channel biased the total number of detected
photoelectrons within an AD by an average of 1/192$\simeq$0.5\%.
Adjusting the observed number of photoelectrons using the simple
factor $f_{\rm act}(t) = N_{\rm total} / N_{\rm active}(t)$, where
$N_{\rm active}$ is the number of active PMT channels and $N_{\rm
  total}=192$, was found to sufficiently compensate for this bias.

\subsubsection{\label{sec:light_yield_calib}Light Yield Determination}

For a particle interaction of fixed energy, the mean number of
detected photoelectrons slightly varied between detectors, as well as
versus time within a single detector.
The mean number of observed PEs per MeV, $N^{\rm PE}(t)$, was
estimated with two independent and complementary methods:
(i) weekly $^{60}$Co deployments at the detector center
(calibration A), and
(ii) uniformly distributed spallation neutrons concurrent with
antineutrino data collection (calibration B).

\begin{figure}[h!]
  \includegraphics[width=.45\textwidth]{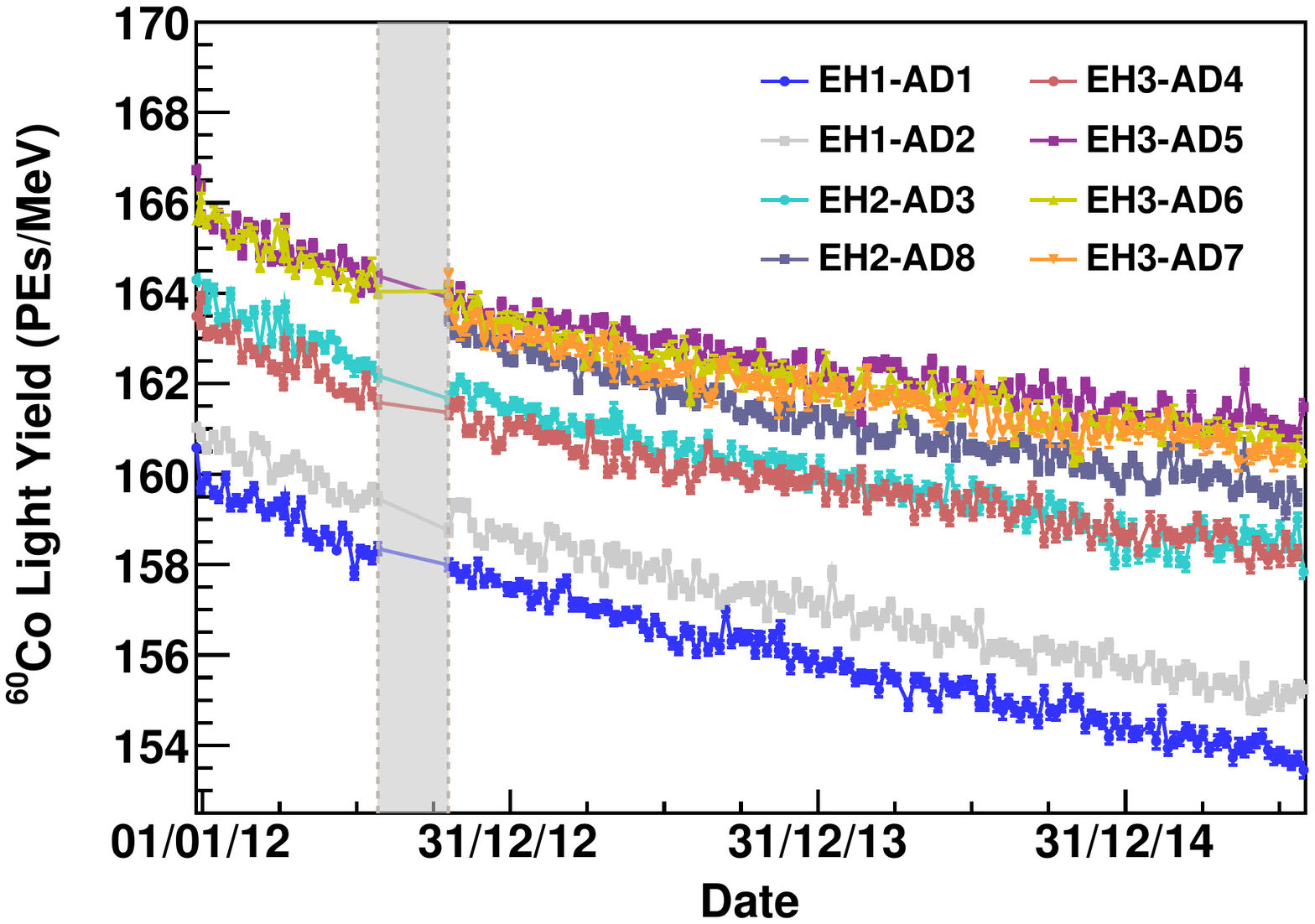}
  \includegraphics[width=.45\textwidth]{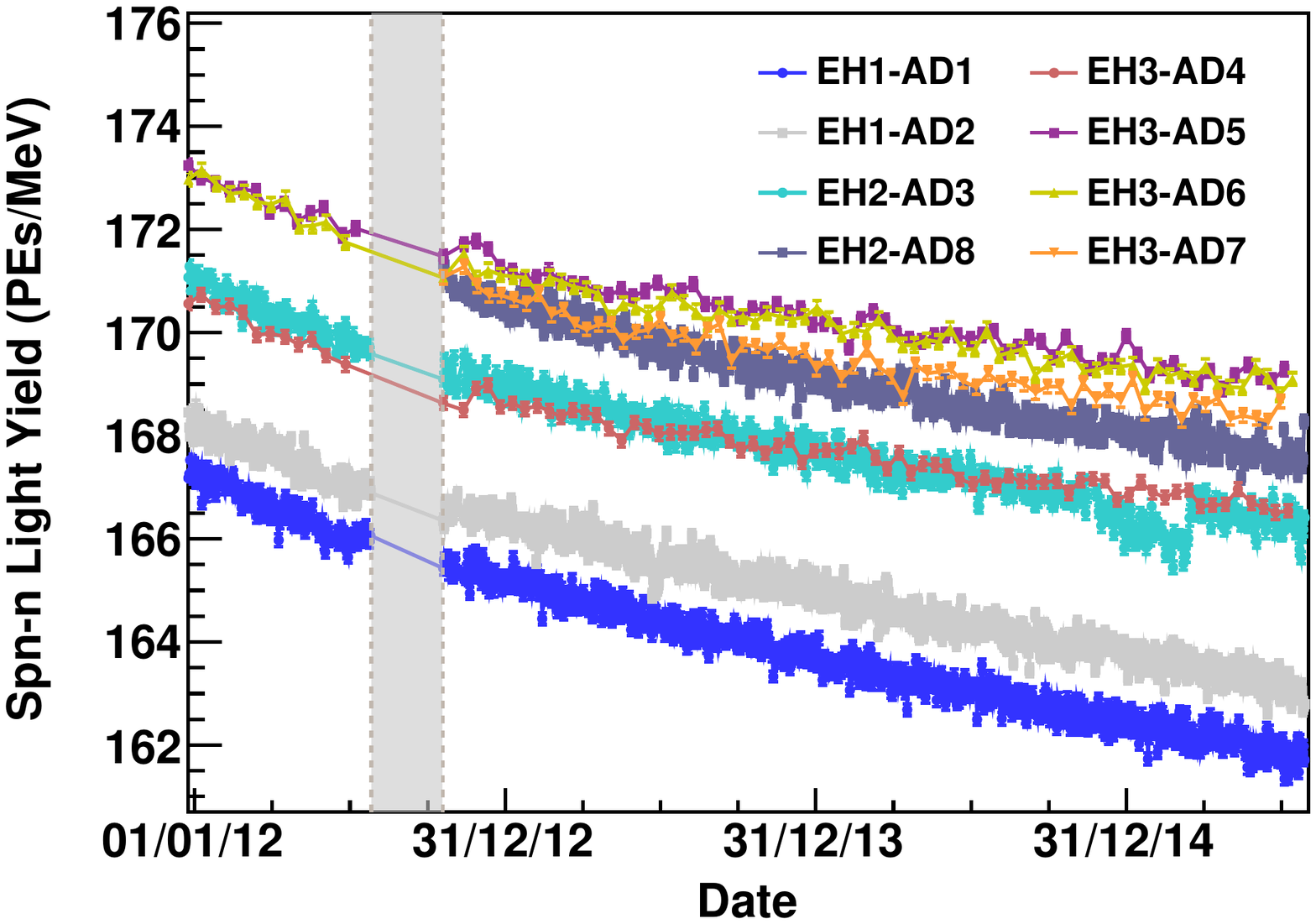}
\caption{\label{fig:energy_scale} {\em Top}: Observed light yield
  versus time, in units of observed photoelectrons (PEs) per MeV, as
  obtained using weekly deployments of a $^{60}$Co source at the
  center of each detector. {\em Bottom}: Observed light yield obtained
  using spallation neutron capture on Gd distributed throughout the
  target volume.  A consistent $\sim$1\% per year decline in the light
  yield was observed for all detectors.  The offset in light yield
  between these two calibration references was due to the nonlinear
  response of the detectors.  The vertical shaded band indicates the
  summer 2012 shutdown period, during which data-taking was partially
  interrupted for installation of the final two detectors.}
\end{figure}

The light yield was determined from the mean of a known gamma-ray peak
in the corresponding energy spectrum, either 2.506~MeV for $^{60}$Co
or the dual peaks of 7.95~MeV and 8.54~MeV for neutron capture on
Gd\@.
Escape of gamma-rays from the scintillator regions introduced a
low-energy tail, which was modeled using a Crystal Ball function for
each peak~\cite{cbfunction}.
The resulting energy scale constants obtained with both $^{60}$Co and
spallation neutrons can be seen in Fig.~\ref{fig:energy_scale}.
The observed light yield varied slightly with the particle energy and
type, primarily due to the intrinsic nonlinear light emission of the
scintillator (see Sec.~\ref{sec:abs_energy}).
Given that this variation was very similar for all detectors, it was
sufficient to choose the light yield at the 7.95~MeV peak of n-Gd
capture as a common convention.
Therefore the light yields obtained using the $^{60}$Co source were
scaled using the ratio of light yield relative to that from n-Gd
capture observed with an $^{241}$Am-$^{13}$C neutron source deployed
weekly at the detector centers.
A clear drift downwards of about 1\% to 1.5\% per year was seen in the
energy scale with both methods, the origin of which has not yet been
conclusively identified.
The drift was slightly more pronounced when measured with the
$^{60}$Co source at the center compared to the uniformly distributed
spallation neutrons, which suggested that the effect was related to a
slight degradation of light transmission in the liquid scintillator.
The drift resulted in a second-order time-dependent spatial
nonuniformity and a negligible degradation in energy resolution,
neither of which had a significant impact on the neutrino oscillation
analysis discussed here.
Likewise, it is not expected to compromise the operation of the
detectors in the near future.

\subsubsection{\label{sec:energy_nonuni}Spatial Nonuniformity}

The observed light yield varied with the position of a particle
interaction within a detector.
This spatial nonuniformity was attributed to the optical
characteristics of the detector, primarily from the geometric
acceptance of the PMTs\@.
It was similar for all detectors, and was reproduced using Monte-Carlo
simulation.
Correcting for this effect improved the energy resolution of each
detector, and improved the similarity of response among the detectors.

%
In order to correct for the spatial nonuniformity, a method for
determining the position of each particle interaction was needed.
Two independent reconstruction methods were developed.
Reconstruction A calculated a center-of-charge (COC) for each signal,
\begin{equation}
  \mathbf{r}_{\mathrm{COC}} = \left(\sum_i \mathbf{r}_i \frac{Q_{i}}{\overline{Q}^{\rm SPE}_{i}(t)}\right) / \left(\sum_i \frac{Q_{i}}{\overline{Q}^{\rm SPE}_{i}(t)}\right),
\end{equation}
where $r_i$ is the position of the $i$'th PMT.
The observed $\mathbf{r}_{\mathrm{COC}}$ was converted to the
estimated reconstructed position $\mathbf{r}_{\mathrm{rec}}$ in
cylindrical coordinates according to the relations
\begin{align}
r_{\mathrm{rec}} &= c_1 \times r_{\mathrm{COC}} - c_2 \times r^2_{\mathrm{COC}}, \nonumber\\
z_{\mathrm{rec}} &= (z_{\mathrm{COC}} - c_3 \times z^3_{\mathrm{COC}}) \times (c_4 - c_5 \times r_{\mathrm{COC}}), \nonumber\\
\phi_{\mathrm{rec}} &= \phi_{\mathrm{COC}}.
\end{align}
In this cylindrical coordinate system, the origin is the center of the
AD target region, $z$ gives the vertical distance from the origin,
while $r$ and $\phi$ define the position in the horizontal plane of
the detector.
A simulation, based on Geant4~\cite{Agostinelli:2002hh}, motivated the
functional form of this model with the five parameters $c_{i}$.
The values of $c_{i}$ were determined from data obtained by deploying
$^{60}$Co sources at known positions within the detectors.
The alternate method, reconstruction B, compared the distribution of
charge observed by the 192 PMTs with a library obtained from
simulation.
A library of 9600 charge-pattern templates was constructed by
simulating interactions in the detector on a grid with 20 bins in the
$r$ direction, 20 in the $z$ direction, and 24 in the $\phi$
direction.
The observed charge pattern was compared to a template using
\begin{equation}
\chi^2 = \sum_i \left( -2 \ln \frac{P(N^{\mathrm{obs}}_i;N^{\mathrm{temp}}_i(\mathbf{r}_{\mathrm{rec}}))}{P(N^{\mathrm{obs}}_i;N^{\mathrm{obs}}_i)} \right),
\end{equation}
where $P(n,\mu)$ is the Poisson probability of finding $n$
photoelectrons when the mean value is $\mu$, $N^{\mathrm{obs}}_i$ is
the observed number of photoelectrons in the $i$-th PMT, and
$N^{\mathrm{temp}}_i(r,z,\phi)$ is the expected number of
photoelectrons in the $i$-th PMT as predicted by the template.
The $\chi^2$ function was interpolated for $\mathbf{r}_{\mathrm{rec}}$
located between simulated templates.
The reconstructed position was determined as the
$\mathbf{r}_{\mathrm{rec}}$ that gave the minimum value of $\chi^2$.
The performance of both reconstruction methods was studied using
calibration data, and was found to be similar.
In particular, both methods estimated the position of signals from
$^{60}$Co sources with $<20$~cm of bias and with $<40$~cm resolution
within the GdLS and LS regions.
Both vertex reconstruction methods accounted for bad PMT channels by
not including them in the COC calculation (reconstruction A) or by
removing them from both the data and templates (reconstruction B)

\begin{figure}
\includegraphics[width=.45\textwidth]{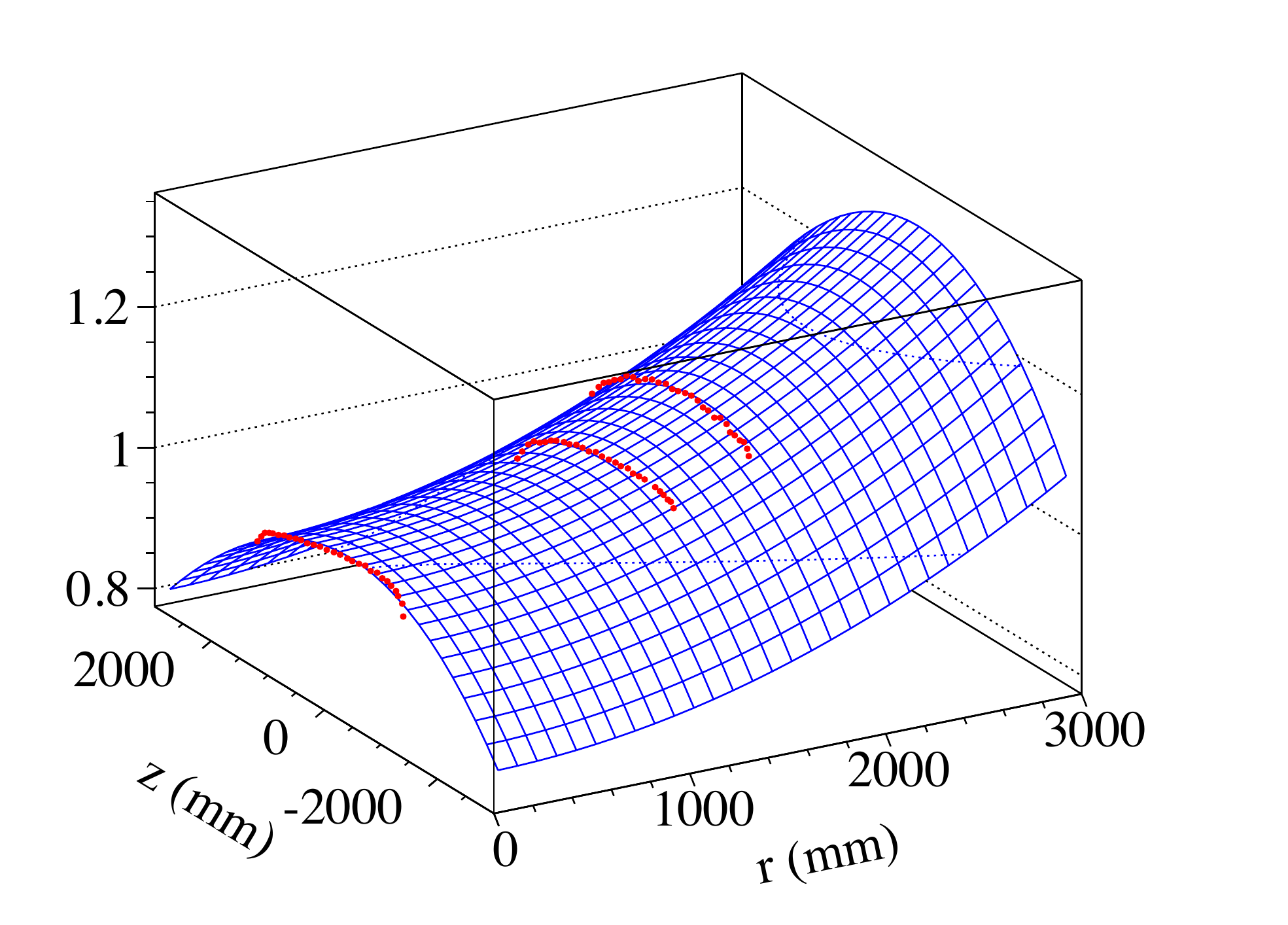}
\caption{\label{fig:Co60_nonunif} Variation in light yield versus
  vertical ($z$) and radial ($r$) position of an interaction within
  EH1-AD1 for reconstruction~A (blue surface)\@. The functional form
  of the variation $f_{\mathrm{b}}(z,r)$ was motivated by simulation,
  and constrained using the relative light yield of $^{60}$Co
  calibration data from ACU A (r=0~mm), ACU B (r=1350~mm) and ACU C
  (r=1772.5~mm) (red points)\@.  While the modeled correction was
  between -6\% to 17\% over the volume of a detector, only slight
  differences ($<$3.2\%) were observed between the eight detectors.}
\end{figure}

%
The reconstructed position was used to correct for the observed
variation in light yield versus interaction position in the detector.
The correction $f_{\mathrm{pos}}(\mathbf{r}_{\mathrm{rec}},t)$ was
decomposed into azimuthal, $z$-$r$, and $t$-$r$ variations,
\begin{equation}
  f_{\mathrm{pos}}(\mathbf{r}_{\mathrm{rec}},t) = \left[f_{\mathrm{a}}(\phi)\,f_{\mathrm{b}}(z,r)\,f_{\mathrm{c}}(t,r)\right]^{-1}.
\end{equation}
Only the radial component showed a significant variation versus time
$t$.
A $\sim$1\% dependence of the light yield with azimuthal angle $\phi$
was observed in all detectors, and was correlated with the orientation
of the PMTs relative to the local geomagnetic field.
The effect was modeled as
\begin{equation}
  f_{\mathrm{a}}(\phi) = 1 + \alpha^{\mathrm{a}}\sin(\phi - \phi_{0}),
\end{equation}
where the parameters $\alpha^{\mathrm{a}}$ and $\phi_0$ were
determined from the observed azimuthal variation in the light yield of
spallation neutron capture signals.
The variation in light yield versus $r$ and $z$ was more significant.
For this component, reconstruction~A used a parameterization motivated
by simulation,
\begin{equation}
  f_{\mathrm{b}}(z,r) = \left(\alpha^{\mathrm{b}}_0 r^2\right) \times \left( \alpha^{\mathrm{b}}_1 + \alpha^{\mathrm{b}}_2 z + \alpha^{\mathrm{b}}_3 z^2 + \alpha^{\mathrm{b}}_4 z^3 \right),
\end{equation}
where the parameters $\alpha_i^{\mathrm{b}}$ were determined using
$^{60}$Co sources located at known positions within each detector.
The variation modeled for EH1-AD1 is shown in
Figure~\ref{fig:Co60_nonunif}.

\begin{figure}
\includegraphics[width=.45\textwidth]{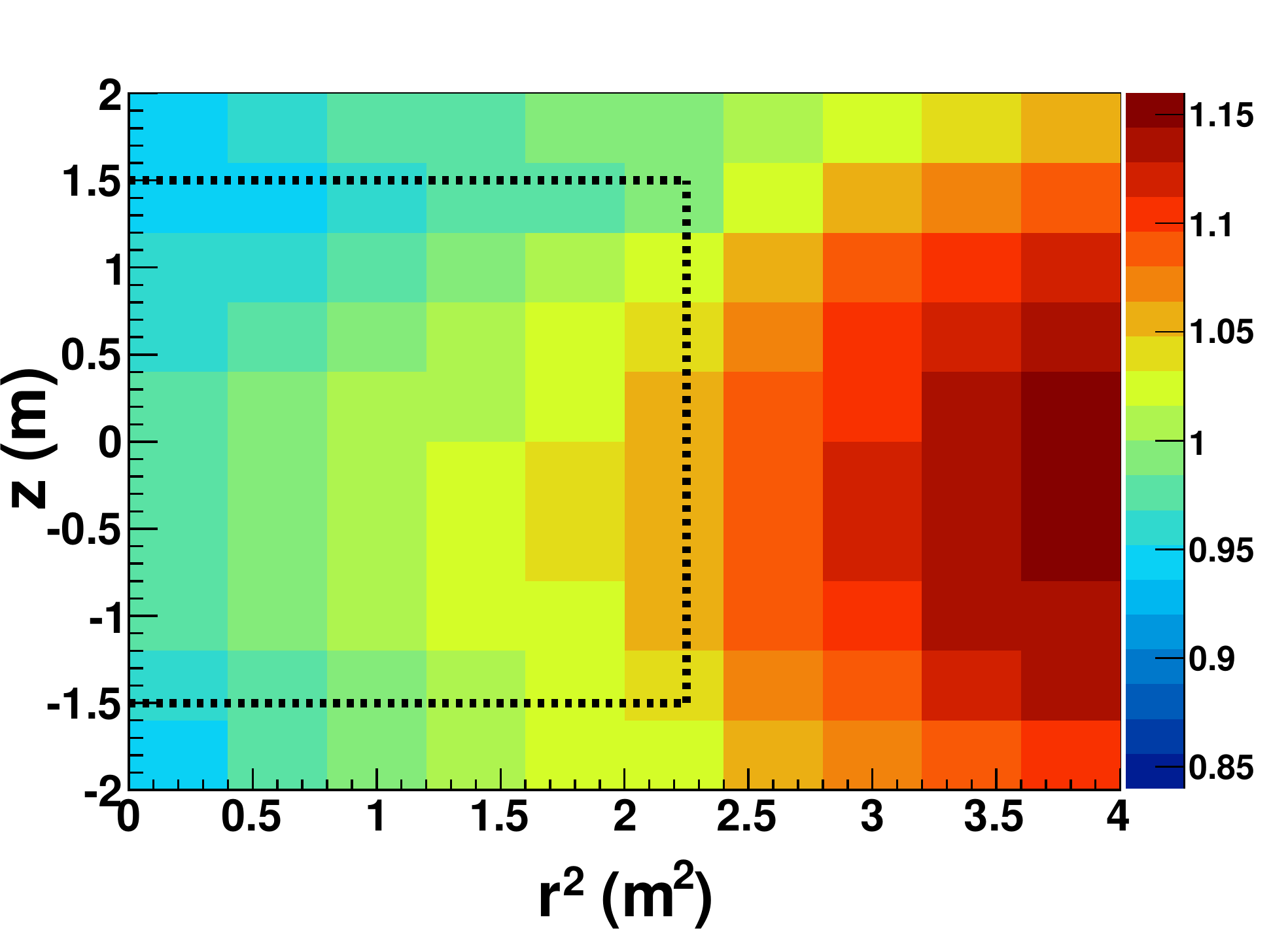}
\caption{\label{fig:spn-n_nonunif} Variation in light yield versus
  vertical ($z$) and radial ($r^2$) position of neutron capture
  interactions within EH1-AD1 for reconstruction~B\@.  Each pixel
  shows the ratio of the observed light yield over the average for the
  entire GdLS region. The dashed lines indicate the boundary between
  the GdLS and LS regions.  Neutrons which captured on Gd were used
  for the innermost GdLS region, while neutrons captured on hydrogen
  were used for the outermost LS region.  Only slight differences
  ($<$3.0\%) in the nonuniformity were observed between the eight
  detectors.}
\end{figure}

Reconstruction~B used signals from spallation neutron capture, divided
into 100 pixels in $z$ and $r^{2}$, to construct a nonuniformity
correction map.
For each pixel the ratio of the observed light yield over the average
light yield for the entire GdLS region was calculated, as illustrated
in Figure~\ref{fig:spn-n_nonunif} for the case of EH1-AD1\@.
The map was estimated using neutrons captured on Gd for the innermost
GdLS region, while neutrons captured on H were used for the outermost
LS region.
The average of the two were used for those pixels spanning the
boundary between the GdLS and LS.

\begin{figure}
\includegraphics[width=.45\textwidth]{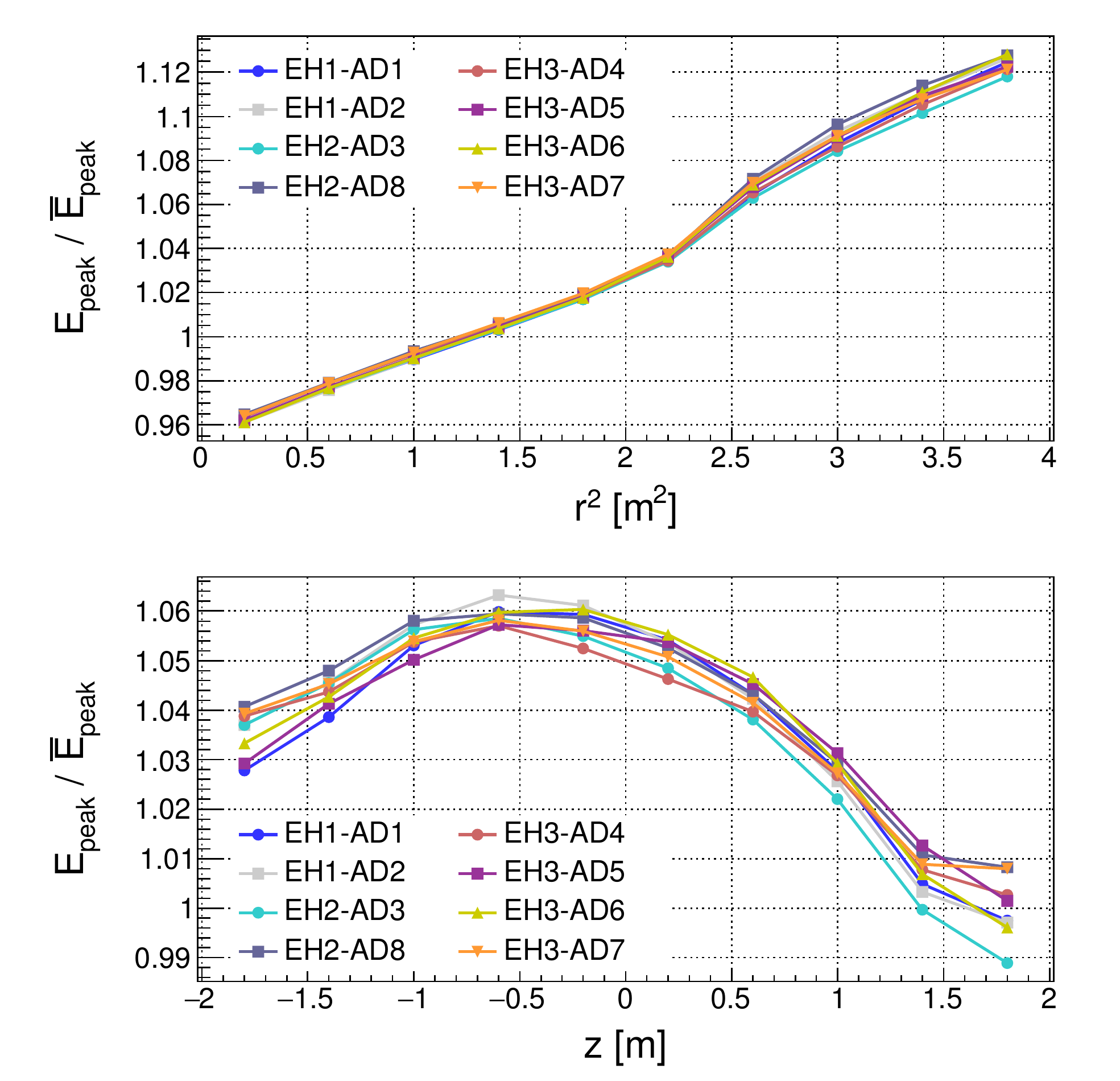}
\caption{\label{fig:spn-n_nonunif_alldets} Comparison of variation in
  light yield versus radial ({\em top}) and vertical ({\em bottom})
  position of neutron capture interactions for reconstruction~B\@.
  Each point shows the ratio of the observed light yield over the
  average for the entire the GdLS region.  The variation was primarily
  due to the optical acceptance versus position, and was reproduced by
  optical simulation.  Only slight differences ($<$3.0\%) in the
  nonuniformity were observed between the eight detectors.}
\end{figure}

Both techniques found consistent nonuniformities.
Variations on the order of 10\% and 17\% were observed in the vertical
and horizontal directions respectively across the volume formed by the
GdLS and LS regions.
Differences in nonuniformity of a few percent were observed for
signals at the extremities of the eight detectors, as shown in
Figure~\ref{fig:spn-n_nonunif_alldets}.

%
The spatial nonuniformity of the light yield showed a slight
variation with time.
Given that the average drift in light yield over time was accounted
for by $N^{\mathrm{PE}}(t)$, the time-dependent nonuniformity
correction only accounted for time-dependent drifts that differed
based on position in the detector.
The change of the nonuniformity over time was adequately modeled
using
\begin{equation}
  f_{\mathrm{c}}(t,r) = \left(\alpha^{\mathrm{c}}_0 + \alpha^{\mathrm{c}}_1 r^2 \right) t.
\end{equation}
The parameters $\alpha^{\mathrm{c}}_i$ were determined using data from $^{60}$Co
sources taken over the first 3~years of operation, while
reconstruction~B relied on signals from spallation neutron capture on
$^1$H\@.
The time-dependence was sufficiently similar between detectors, so
common values for the parameters were used for all eight detectors.
The correction was largest for signals near the edge of the LS region,
for which a $<$0.5\% shift in energy response per year was observed.

\subsubsection{\label{sec:perf}Calibration Performance}

\paragraph{\label{sec:energy_stabil}Energy stability}

\begin{figure}[h!]
\includegraphics[width=.45\textwidth]{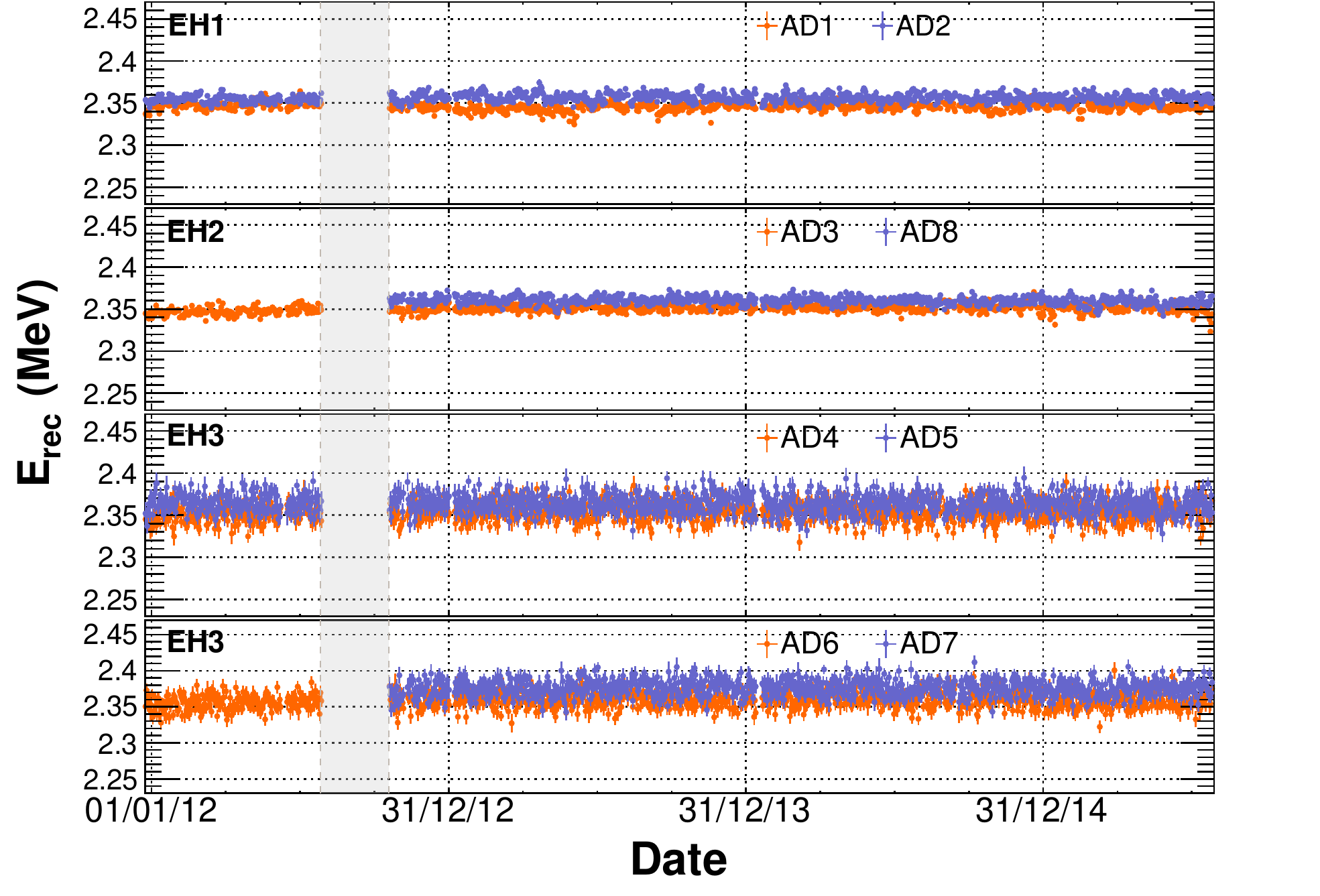}
\includegraphics[width=.45\textwidth]{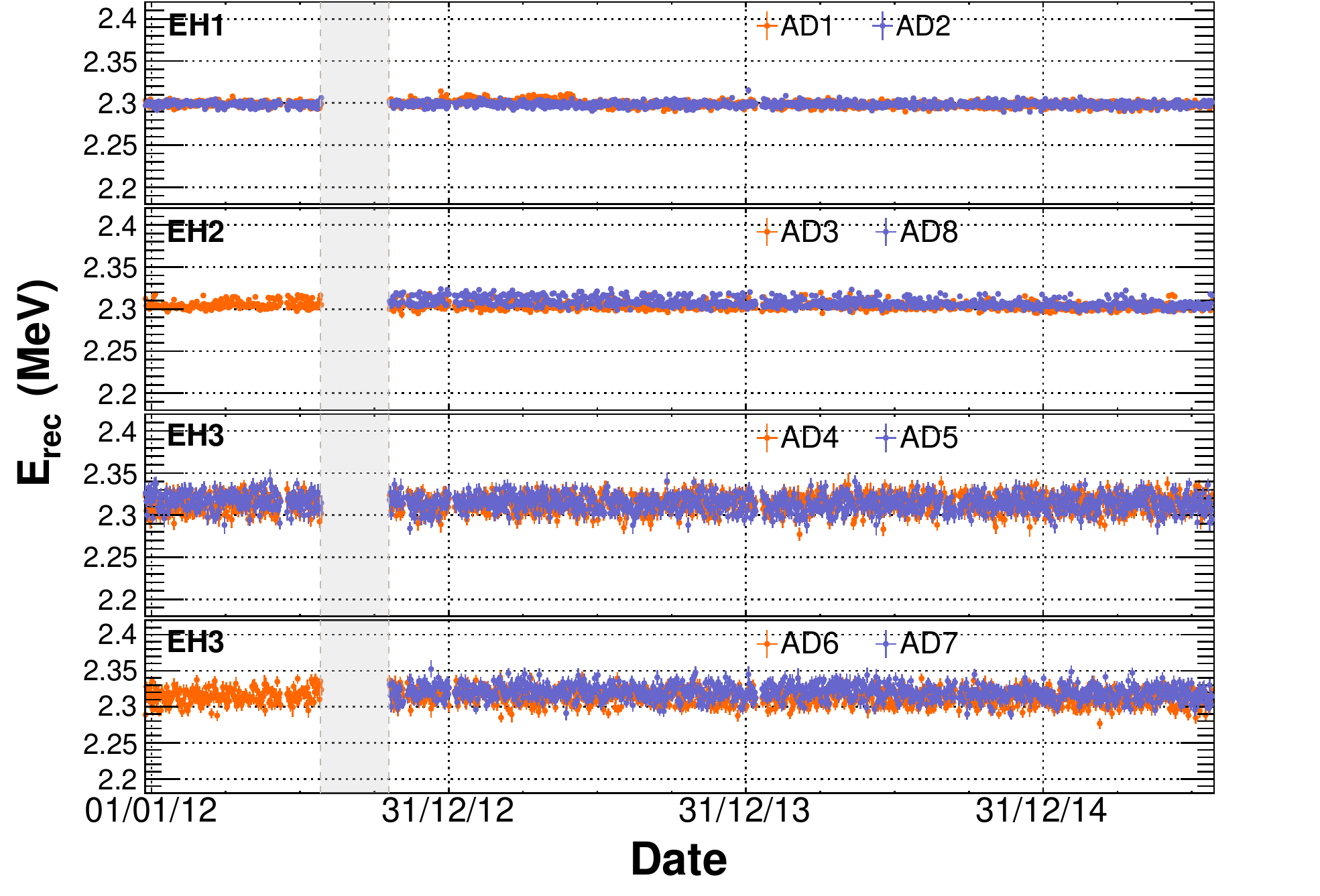}
\caption{{\em Top}: Stability of the mean reconstructed energy of
  signals from spallation neutron capture on H versus time, obtained
  from calibration using a $^{60}$Co source located at the detector
  center (reconstruction~A)\@.  {\em Bottom}: The same, but for the
  calibration method which relies on spallation neutrons
  (reconstruction~B)\@.  Each point corresponds to one day. The shaded
  vertical band represents the 2012 shutdown period.
  \label{fig:energy_stability}}
\end{figure}

As discussed in the preceding section, the charge calibration, channel
quality, AD light yield, and spatial nonuniformity calibration
corrections all varied with time.
The performance of the full energy calibration process was assessed by
examining the stability of the reconstructed energy over time for a
variety of calibration reference data.
Figure~\ref{fig:energy_stability} shows the reconstructed mean energy
of spallation neutrons captured on H for the entire period considered
in this analysis, for both the $^{60}$Co (Calibration~A) and
spallation neutron (Calibration~B) methods.
Both methods yielded an energy for n-H capture that was stable to
within 0.2\%, as determined from the RMS of the distributions in
Figure~\ref{fig:energy_stability}.

\paragraph{\label{sec:rel_energy} AD-to-AD differences}

Calibration was crucial to ensure that the reconstructed energy for
antineutrino interactions within all eight antineutrino detectors was
as identical as possible.
Small differences in the energy response between detectors could
distort the relative efficiency of the far versus near detectors,
biasing the measurement of $\theta_{13}$ as highlighted by
Eq.~\ref{eq:nearFarRatio}.
Slight distortions of the energy response between detectors could also
introduce an artificial distortion in the comparison of the
antineutrino energy spectra of the near and far detectors, degrading
the measurement of ${\Delta}m^2_{\mathrm{ee}}$.

AD-to-AD differences were estimated using 13 different calibration
references, both deployed and naturally occurring in the ADs.
Data from sources deployed on a weekly basis included $\gamma$-rays
from $^{68}$Ge and $^{60}$Co, and neutrons from $^{241}$Am-$^{13}$C\@.
Signals generated by natural radioactivity, including $\alpha$-decays
of $^{212}$Po, $^{214}$Po, $^{215}$Po and $^{219}$Po and $\gamma$-rays
from $^{40}$K and $^{208}$Tl, were also compared between detectors.
Neutrons from IBD and muon spallation that capture on H and Gd
provided additional calibration references.
These calibration reference data span the full time period used for
the oscillation analysis.

The spatial distribution of interactions differed among the
calibration references.
Some of the sources, e.g.\ neutron-capture on hydrogen and gammas from
external $^{40}$K and $^{208}$Tl decays, were concentrated at the edge
of the scintillating volumes close to the PMTs.
Selecting signals with reconstructed positions within the target
volume still resulted in distributions dominated by interactions
outside the target due to the limited precision of position
reconstruction.
Rejecting signals within 20~cm of the target boundary gave a
distribution sufficiently similar to that of antineutrinos.
This tightened selection had negligible impact on the estimated mean
energy for $\alpha$'s from $^{212}$Po, $^{215}$Po and $^{219}$Po
decays and $\gamma$-rays from neutrons capturing on Gd, which were
distributed nearly uniformly within the target volume.

\begin{figure}
  \centering
  \includegraphics[width=0.45\textwidth]{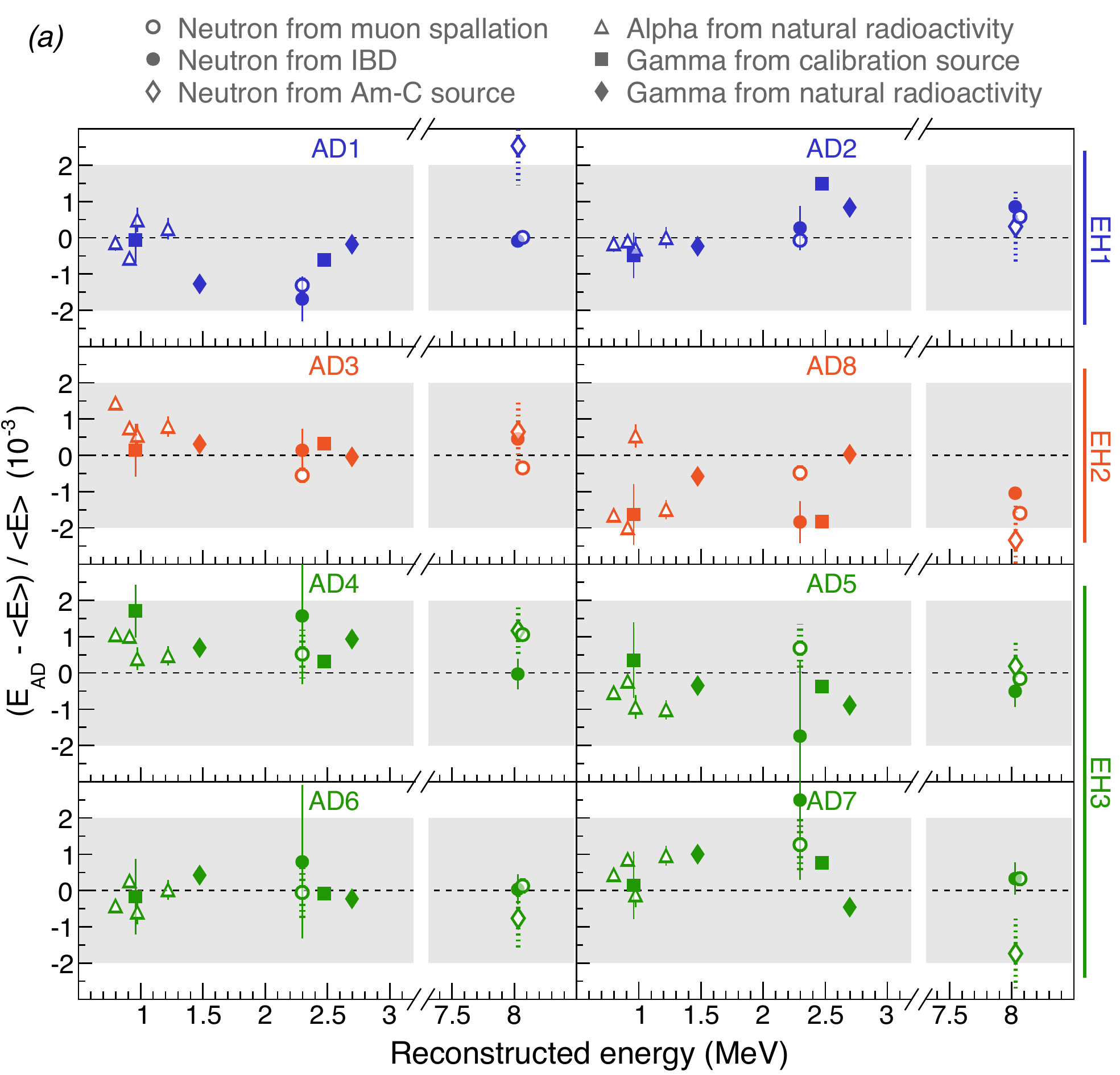}
  \vskip2ex
  \includegraphics[width=0.45\textwidth]{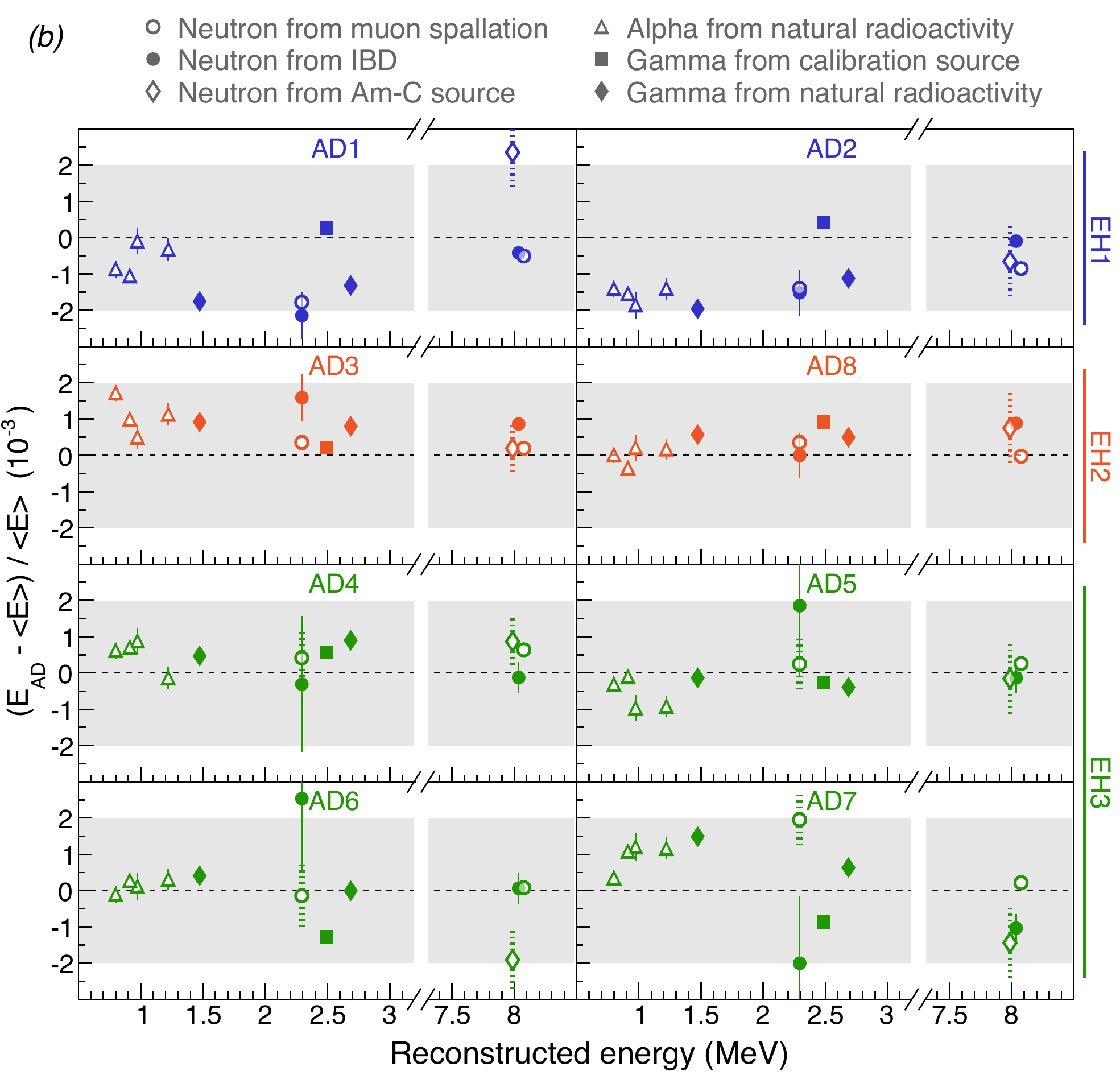}
  \caption{Comparison of the mean reconstructed energy between
    antineutrino detectors for a variety of calibration references
    using (a) calibration with $^{60}$Co sources (reconstruction A)
    and (b) calibration with spallation neutrons (reconstruction
    B). $E_{\mathrm{AD}}$ is the reconstructed energy determined for
    each AD, while $\langle E \rangle$ is the eight-detector average.
    The mean energy for each calibration reference was obtained from
    the corresponding peak in the energy spectrum of all regular data
    (natural alphas and gammas, neutrons from IBD and muon spallation)
    and all weekly calibration runs (gammas from $^{68}$Ge and
    $^{60}$Co sources, neutrons from Am-C sources) taken during the
    time period when all eight ADs were in operation.  An effective
    fiducial volume selection has been applied on distributed sources
    to suppress interactions outside the antineutrino target where
    AD-to-AD differences are larger.  Error bars are statistical only,
    and systematic variations between detectors for all calibration
    references were $<$0.2\% for both reconstruction methods.
  \label{fig:rel_energy}}
\end{figure}

Variations in the mean reconstructed energies of these calibration
references for all eight ADs are shown in Fig.~\ref{fig:rel_energy}.
These calibration references span the range of energies expected for
both the prompt and delayed signals from inverse beta decay.
Systematic variations were $<$0.2\% with typical deviations around
0.1\%, independent of the choice of the calibration and reconstruction
methods.
Therefore, a conservative 0.2\% systematic uncertainty in the
potential variation of the relative energy response between detectors
was used.

From 2012 until 2014, the levels of the liquids in the MO and LS
overflow tanks of EH3-AD4 shifted slowly relative to each
other~\cite{An:2015qga}.
The changes were consistent with a slow leak of 50~l of LS into the
42,800~l MO region of this detector over this two year period.
An increase in the absorbance of short-wavelength light in the MO and
an increase in the amount of light detected for muons which traversed
the MO verified that a leak had occurred.
Despite close examination of the data from this detector, no
significant deviation in performance was found.
Simulations which accounted for potential light production in the MO
supported this conclusion~\cite{An:2016bvr}.
As will be discussed, the rate of \nuebar{} interactions measured in
EH3-AD4 was consistent with the other three detectors in the far hall,
which confirmed that the leak did not adversely impact this detector.

\subsection{\label{sec:resolution}Energy resolution}

\begin{figure}
  \centering
  \includegraphics[width=0.45\textwidth]{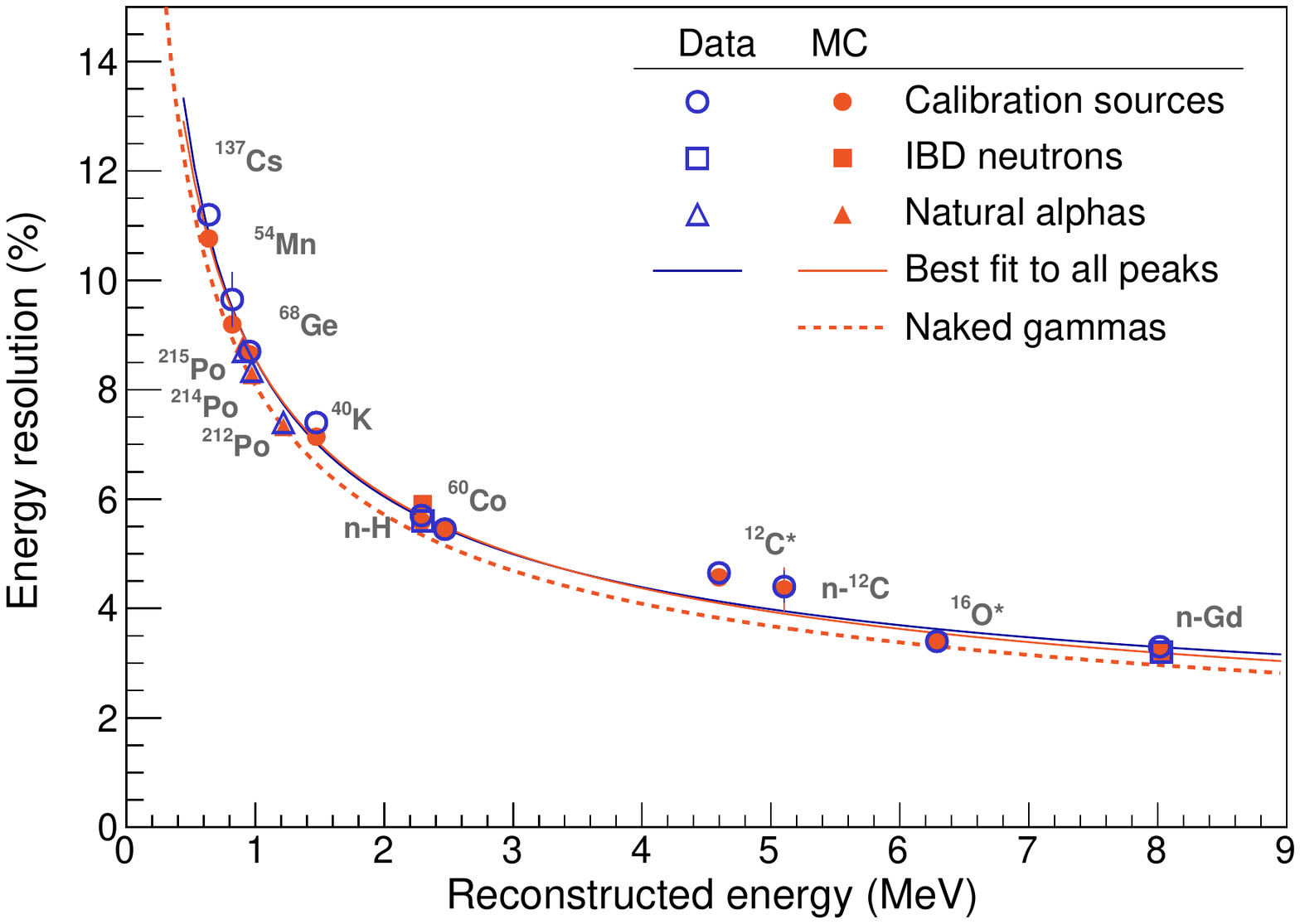}
  \caption{Energy resolutions measured for calibration sources,
    neutron captures following inverse beta decay, and natural
    $\alpha$ radioactivity within the scintillator (open blue
    markers).  The detector resolution was limited by the statistical
    uncertainty of photoelectron counting, as modeled using
    Eq.~\ref{eq:energy_resolution} (blue solid line).  The resolutions
    were consistent with Monte-Carlo simulation (solid orange
    markers).  The simulation predicted that the intrinsic detector
    resolution (dashed orange line) was slightly better than that
    estimated for the calibration sources (solid orange line), since
    the latter suffered from optical shadowing by the source
    encapsulation.  The energy resolutions for $\alpha$ particles
    emitted by natural radioactivity (open blue triangles) confirmed
    the intrinsic energy resolution.
    \label{fig:energy_resolution}}
\end{figure}

The energy resolution of the detectors was modeled using the
expression
\begin{equation}\label{eq:energy_resolution}
\frac{\sigma_E}{E_{\mathrm{rec}}} = \sqrt{a^2 + \frac{b^2}{E_{\mathrm{rec}}} + \frac{c^2}{E_{\mathrm{rec}}^2}}.
\end{equation}
The parameters $a$, $b$, and $c$ reflect the contributions to the
resolution from detector nonuniformity, photoelectron counting
statistics, and noise, respectively.
The resolutions observed for calibration sources, neutron captures
following inverse beta decay, and natural $\alpha$ radioactivity
within the scintillator are shown in
Figure~\ref{fig:energy_resolution}\@.
Modeling of the detector resolution using
Eq.~\ref{eq:energy_resolution} gave $a = 0.016$, $b =
0.081$~MeV$^{1/2}$, and $c = 0.026$~MeV\@.
The total resolution was dominated by photoelectron counting
statistics.
Simulation predicted that the intrinsic resolution was slightly better
than that observed for the calibration sources, since the latter
suffered from optical shadowing by the source encapsulation.
The intrinsic resolution was confirmed using natural $\alpha$
radioactivity within the scintillator.
No significant variations in detector energy resolution were observed
among the eight detectors.

\subsection{\label{sec:abs_energy}Absolute Antineutrino Energy}

%
Aside from the relative calibration of the energy response between
detectors, it was also necessary to calibrate the detectors in an
absolute sense.
In particular, interpretation of the distortion in the antineutrino
energy spectrum by oscillation required characterization of the
relationship between true \nuebar{} energy and the corresponding
reconstructed IBD positron energy.
While the uncertainty in absolute calibration had negligible impact on
the measurement of $\theta_{13}$, it influenced the estimate of the
neutrino mass-squared difference.
This can be seen as a direct consequence of the ratio of
${\Delta}m^2_{\mathrm{ee}}$ over $E_\nu$ in Eq.~\ref{eq:survProb_ee}.
%

%
To the lowest order, the kinematics of the IBD interaction implied
$E_{\mathrm{prompt}} \simeq E_{\nu} - 0.8$~MeV, where the prompt
positron energy included 1.022~MeV from annihilation.
The angular distribution of neutron recoil introduced a small
energy-dependent correction to the above relation, and negligibly
broadened the energy resolution.
The most significant bias arose from the nonlinear response of the
detector.
Nonlinearity of the reconstructed positron energy relative to the true
interaction energy originated from two sources: particle-dependent
nonlinear light yield of the scintillator, and charge-dependent
nonlinearity associated with the electronic readout of the PMT signal.
Positron interactions with the scintillator are predominantly identical
to electrons, except for their eventual annihilation.
Therefore, the visible energy for a positron was effectively modeled
as $E_{\mathrm{vis,}e^+} = E_{\mathrm{vis,}e^-} + 2\times
E_{\mathrm{vis,}\gamma}(0.511~\mathrm{MeV})$\@.

The scintillation light output for low-energy electrons was suppressed
due to ionization quenching.
A semi-empirical analytic approach based on Birks' law was used to
model this mechanism, expressed as the ratio of the quenched to true
electron energy $f_q$.
The energy-dependent contribution from Cherenkov light, predicted to
be at the level of a few percent relative to scintillation light,
induced an additional nonlinearity.
The average Cherenkov light emitted by an electron versus energy,
$f_{\mathrm{c}}$, was extracted from a Geant4-based simulation and
confirmed by an independent analytic calculation.
A scale factor $k_{\mathrm{c}}$, defined as the ratio of detected
Cherenkov to scintillation photons for 1~MeV electrons, accounted for
the difference in the magnitude and detection efficiency between these
two components of light.
The total detectable light from an electron in the scintillator, here
called the {\em visible} energy $E_{\mathrm{vis}}$, was therefore
related to the true kinetic energy $E_{\mathrm{true}}$ via
\begin{equation} \label{eq:electronNonlinearity}
\frac{E_{\mathrm{vis}}}{E_{\mathrm{true}}} = \beta_{\mathrm{vis}} \left[ f_{\mathrm{q}}(E_{\mathrm{true}},k_{\mathrm{B}}) + k_{\mathrm{c}} f_{\mathrm{c}}(E_{\mathrm{true}}) \right],
\end{equation}
where $k_{\mathrm{B}}$ is the Birks' constant for electrons and
$\beta_{\mathrm{vis}}$ is an arbitrary normalization.

%
A Geant4-based simulation was used to model the effective scintillator
response to $\gamma$-rays as a function of $k_B$, $k_{\mathrm{c}}$ and
the $\gamma$-ray energy using this electron model.
Monoenergetic $\gamma$-rays were tracked in the scintillator until all
energy was converted to scattered $e^-$'s and $e^+$'s.
The $E_{\mathrm{vis}}$ for each $\gamma$-ray was obtained from the sum
of the visible energy for each of these scattered electrons,
calculated according to Eq.~\ref{eq:electronNonlinearity}.

%
The charge nonlinearity induced by the electronics arose from a
complex interplay of the time distribution of detected light and the
response of the readout electronics, which could not be easily
calibrated at the single PMT channel level.
Instead, the resulting nonlinearity was modeled at the detector level.
A combination of measurements and modeling of the electronics response
suggested that the ratio of the reconstructed energy
$E_{\mathrm{rec}}$ to the visible energy $E_{\mathrm{vis}}$ could be
effectively parameterized using
\begin{equation} \label{eq:electronicsNonlinearity}
\frac{E_{\mathrm{rec}}}{E_{\mathrm{vis}}} = \beta_{\mathrm{rec}} \left[ 1 + \alpha \exp\left(-\frac{E_{\mathrm{vis}}}{\tau} \right)\right],
\end{equation}
where $\alpha$ determines the amplitude of the nonlinearity, $\tau$
sets the energy-dependence, and $\beta_{\mathrm{rec}}$ is an arbitrary
normalization.

\begin{figure}
  \centering
  \includegraphics[width=0.5\textwidth]{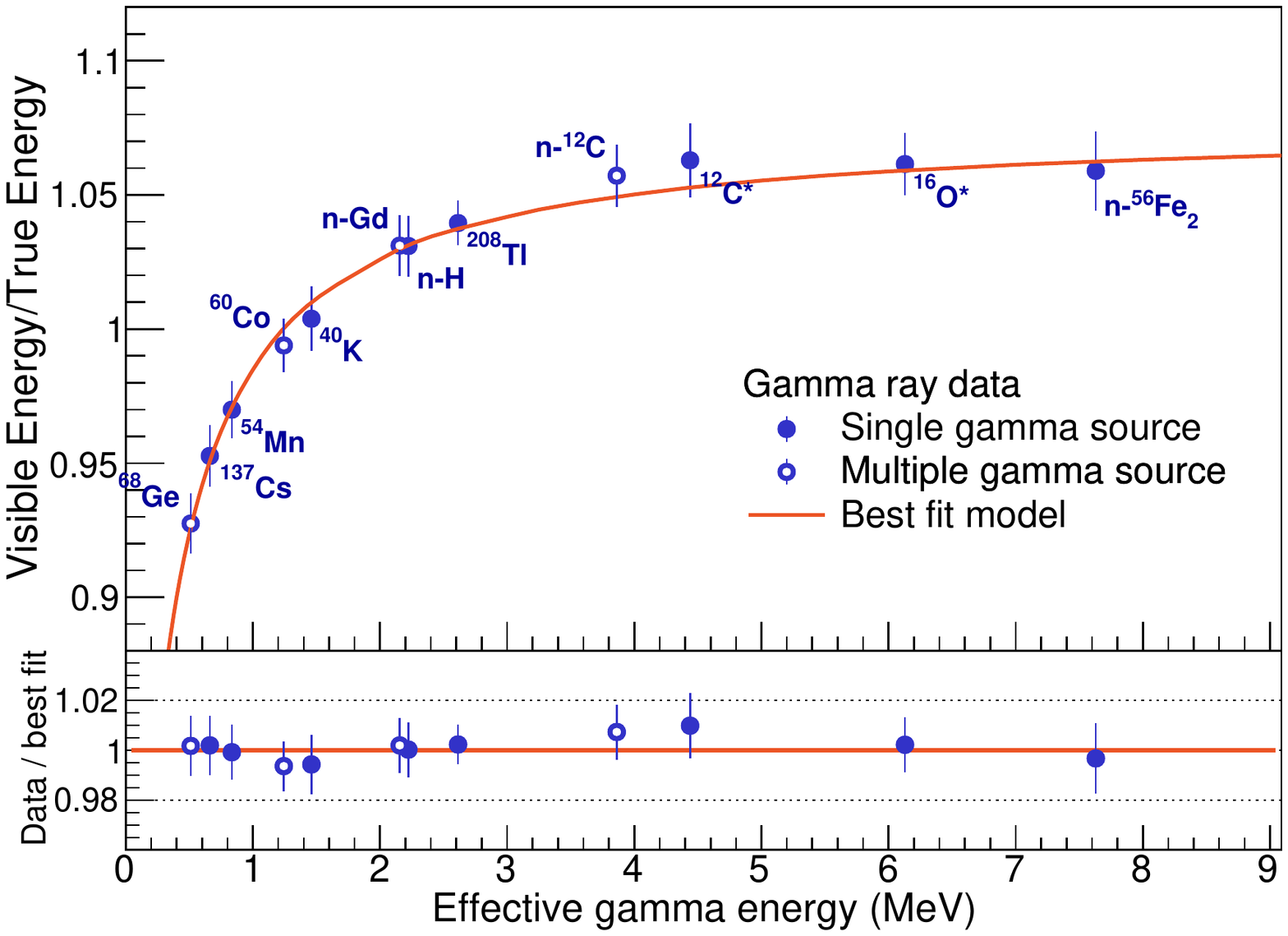}
  \caption{Ratio of observed energy to true energy for $\gamma$-ray
    reference data relative to the estimated model of the nonlinear
    scintillation and Cherenkov light emission.  For sources which
    consist of multiple $\gamma$-rays, the mean energy is used as an
    'effective' energy for the purposes of modeling the scintillator
    nonlinearity.  The best fit for a model of the scintillator
    nonlinearity is also shown (red solid line).  For clarity, the
    estimated nonlinearity contributed by the electronics has been
    removed from both the data and the model.}
  \label{fig:nl_nom_gamma_scint}
\end{figure}

\begin{figure}
  \centering
  \includegraphics[width=0.5\textwidth]{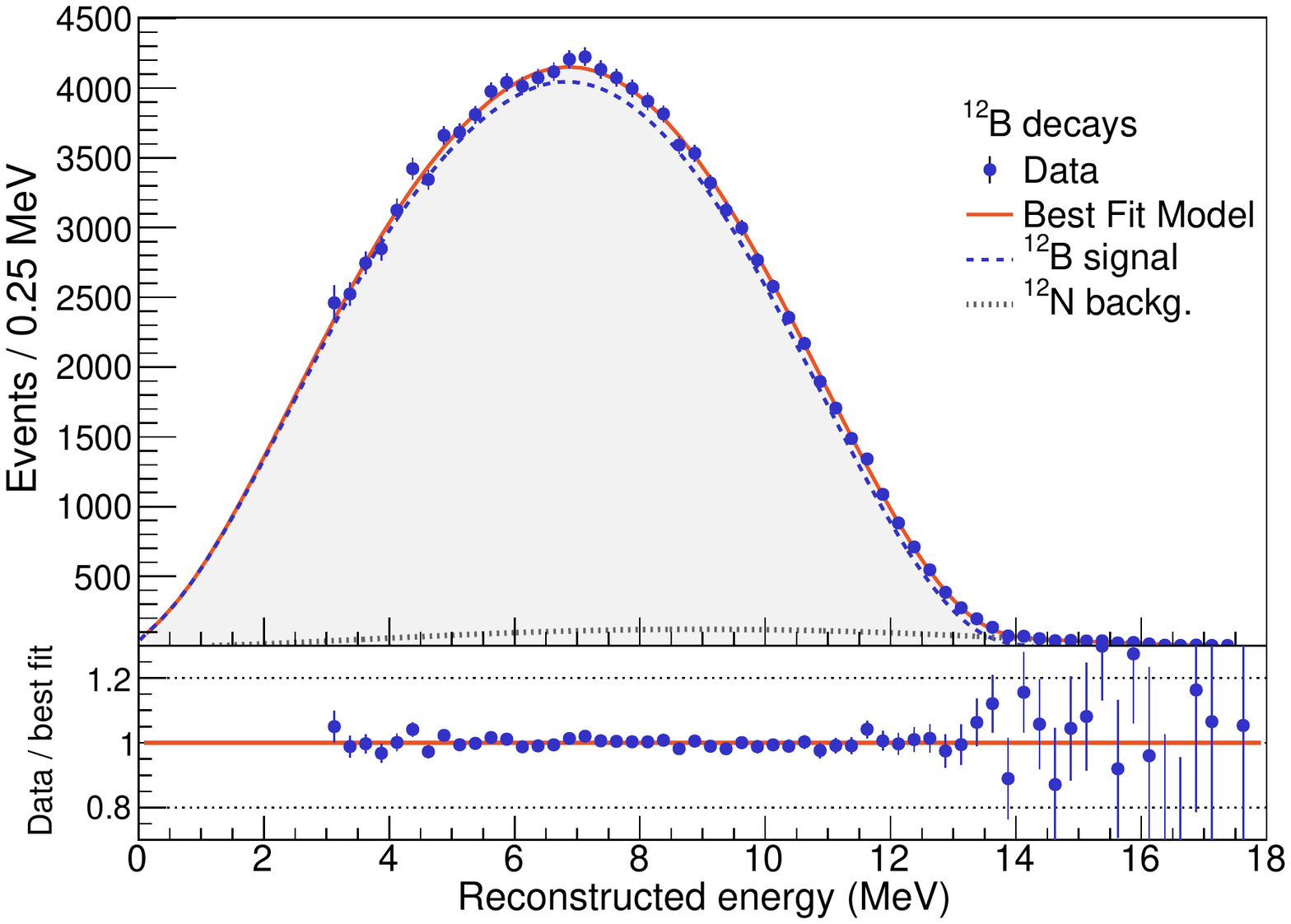}
  \caption{Comparison of the electron energy spectrum from $^{12}$B
    decay with simulation, including the estimated nonlinearity of the
    scintillator and electronics.}
  \label{fig:nl_nom_B12}
\end{figure}

%
The complete energy model, relating the reconstructed energy to true
particle kinetic energy, therefore contained four free parameters:
Birks' constant $k_{\mathrm{B}}$, the contribution from Cherenkov
radiation $k_{\mathrm{c}}$, and the two parameters $\alpha$ and $\tau$
of the electronics model.
The product of $\beta_{\mathrm{vis}}$ and $\beta_{\mathrm{rec}}$ were
defined such that $E_{\mathrm{rec}} = E_{\mathrm{true}}$ at the
reference energy of neutron capture on Gd.
There was no significant deviation in the nonlinearity between
detectors, so a common model was used for all eight ADs.
The parameter values were obtained from an unconstrained $\chi^2$ fit
to various AD calibration data, consisting of twelve gamma lines from
both deployed and naturally-occurring sources, as well as the
continuous electron spectrum from the decays of $^{12}$B produced by
muon interactions with the scintillator.
The study accounted for residual position-dependent variations in
light yield between the calibration references.
Figures~\ref{fig:nl_nom_gamma_scint} and~\ref{fig:nl_nom_B12} compare
the resulting energy response model with the calibration data.

\begin{figure}
  \includegraphics[width=\columnwidth]{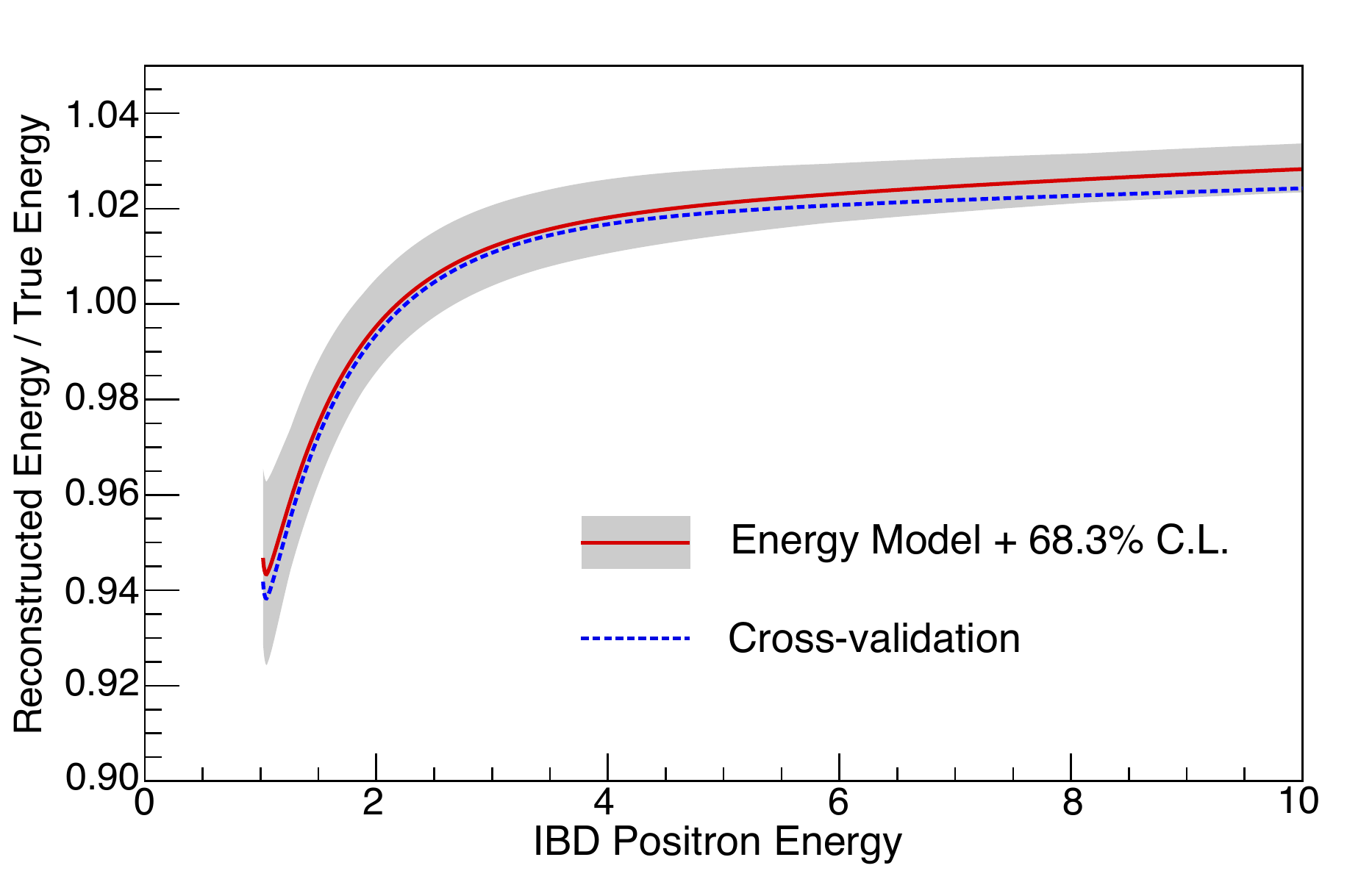}
  \caption{Estimated ratio of reconstructed over true energy,
    $E_{\mathrm{rec}}/E_{\mathrm{true}}$, for positron interactions
    totally contained within the antineutrino detector target (red
    solid line).  The two 0.511-MeV $\gamma$-rays from positron
    annihilation were included in the energy.  The resulting
    68\%~C.L. region (gray shaded band) constrains this ratio to
    better than 1\% over most of the energy range of interest.  An
    independent estimate, which relied on the $\beta$+$\gamma$ spectra
    from bismuth and thallium decay as well as the Michel electron
    spectrum endpoint, produced a consistent model (blue dotted line).
  \label{fig:ibdresponse}}
\end{figure}

The resulting estimate of the detector response to positrons is shown
in Figure~\ref{fig:ibdresponse}.
The depicted uncertainty band corresponds to variations of model
parameters consistent with the fitted calibration data within
68\%~C.L\@.
The $\chi^2$-based approach was used to constrain the energy response,
and resulted in sub-percent uncertainties of the absolute positron
energy scale above 2~MeV\@.
The best model gave the values
$k_{\mathrm{B}}=15\times10^{-3}$~cm~MeV$^{-1}$,
$k_{\mathrm{c}}=0.5\%$, $\alpha=0.078$, and $\tau=2.55$~MeV\@.
Strong correlations among the parameters resulted in large
uncertainties for each individually, although the combined model was
well-constrained as shown in Figure~\ref{fig:ibdresponse}.
Reproduction of the model from the best fit parameters was also
dependent on the specific configuration of low-energy electron
transport and quenching in Geant4.
Therefore, the best model is provided in a more convenient tabulated
form as Supplemental Material~\footnote{See the Supplemental Material
  at (provide link) for the following tabulated data: the observed
  prompt energy spectra for \nuebar{} inverse beta decay candidates
  and estimated backgrounds for each experimental hall, the
  distribution of ${\Delta}\chi^2$ versus $\sin^2(2\theta_{13})$ and
  ${\Delta}m^2_{\mathrm{ee}}$, and the components of the detector
  response model $P(E_{\mathrm{rec}};E_{\mathrm{true}})$.}\@.

This model of the detector response to positrons was validated using
independent calibration reference data.
These included the 53~MeV endpoint in the Michel electron spectrum of
muon decay, and the continuous $\beta$+$\gamma$ spectra from natural
bismuth and thallium decays.
The estimated model of the electronics nonlinearity was corroborated
by comparison with PMT data obtained using an independent waveform
digitizer system.
All measurements have been found to be consistent with the estimated
model within their respective uncertainties.
In turn, the estimated positron response model was stable within the
1-$\sigma$ uncertainty band under addition or removal of any single
calibration reference data set.
As a result of this extensive modeling of the positron energy
response, the final $\sim$1\% uncertainty is small when compared with
the overall uncertainty for ${\Delta}m^2_{\mathrm{ee}}$ found in the
study presented here.


\section{\label{sec:background} Antineutrino Signal and Backgrounds}

%
The relative far-versus-near measurement of neutrino oscillation, as
expressed in Eq.~\ref{eq:nearFarRatio}, motivated a particular
approach to \nuebar{} selection.
The selection criteria were not necessarily designed to maximize the
\nuebar{} detection efficiency and minimize backgrounds.
Instead, the criteria were chosen specifically to minimize relative
uncertainties in the comparison of signals observed among the eight
detectors.
The following section provides a detailed description of the \nuebar{}
selection criteria, assessment of the relative efficiencies between
detectors, and estimation of the residual backgrounds for the analysis
of neutrino oscillation.
To briefly summarize, a total of more than $2.5\times10^6$ \nuebar{}
candidate interactions were identified, with potential variation in
efficiency between detectors estimated at 0.13\%, while background
contamination was less than a few percent with an uncertainty of
$\lesssim$0.2\% in the sample.

%
Two independent methods and software were developed for selection of
the antineutrino candidates.
Here we refer to these two approaches as antineutrino selection A and
selection B\@.
These methods differed most significantly in their choice of energy
calibration and reconstruction: selection A used reconstruction A and
selection B used reconstruction B\@.
The two methods also differed slightly in their approach to background
rejection.
Both methods are discussed here, with their differences highlighted.
Table~\ref{tab:IBDselectionCuts} provides a side-by-side comparison of
the two selection methods.
\begin{table*}[!htb]
  \caption{Summary of two independent methods used in the selection of
    reactor \nuebar{} inverse beta decay interactions.  See text for
    details.}
  \label{tab:IBDselectionCuts}
  \begin{ruledtabular}
  \begin{tabular}{lcc}
    \textbf{Criterion} & \textbf{Selection A} & \textbf{Selection B} \\
    Calibration & $^{60}$Co and $^{241}$Am-$^{13}$C method & Spallation neutron method \\
    Reconstruction & Corrected center-of-charge & Charge template matching \\
    8-inch PMT light emission & Reject $f_{\mathrm{ID}}\geq0$ & Reject $f_{\mathrm{ID}}\geq0$ or $f_{\mathrm{PSD}}\geq0$ \\
    2-inch PMT light emission & \multicolumn{2}{c}{Reject $Q_{\mathrm{max}}$(2-inch PMTs) $ > 100$~p.e.} \\
    Prompt energy & \multicolumn{2}{c}{(0.7, 12.0)~MeV} \\
    Delayed energy & \multicolumn{2}{c}{(6, 12.0)~MeV} \\
    Prompt-delayed ${\Delta}t$ & \multicolumn{2}{c}{(1, 200) $\mu$s} \\
    Multiplicity veto \emph{(pre)} & No signal $>$0.7~MeV 200~$\mu$s before prompt & Only one signal (0.7, 12)~MeV 400~$\mu$s before delayed \\
    Multiplicity veto \emph{(post)} & No signal $>$0.7~MeV 200~$\mu$s after delayed & No signal (6, 12)~MeV 200~$\mu$s after delayed \\
    Water Shield muon veto & \multicolumn{2}{c}{Veto (-2, 600)~$\mu$s after NHIT $> 12$ in OWS or IWS} \\
    AD muon veto & Veto (0, 1)~ms after $>$20~MeV signal & Veto (0, 1.4)~ms after $>$3,000 p.e. ($\sim$18~MeV) signal\\
    AD shower veto & Veto (0, 1)~s after $>$2.5~GeV signal & Veto (0, 0.4)~s after $>$3$\times10^5$ p.e. ($\sim$1.8~GeV) signal \\
  \end{tabular}
  \end{ruledtabular}
\end{table*}

\begin{figure}[!ht]
  \includegraphics[width=0.5\textwidth]{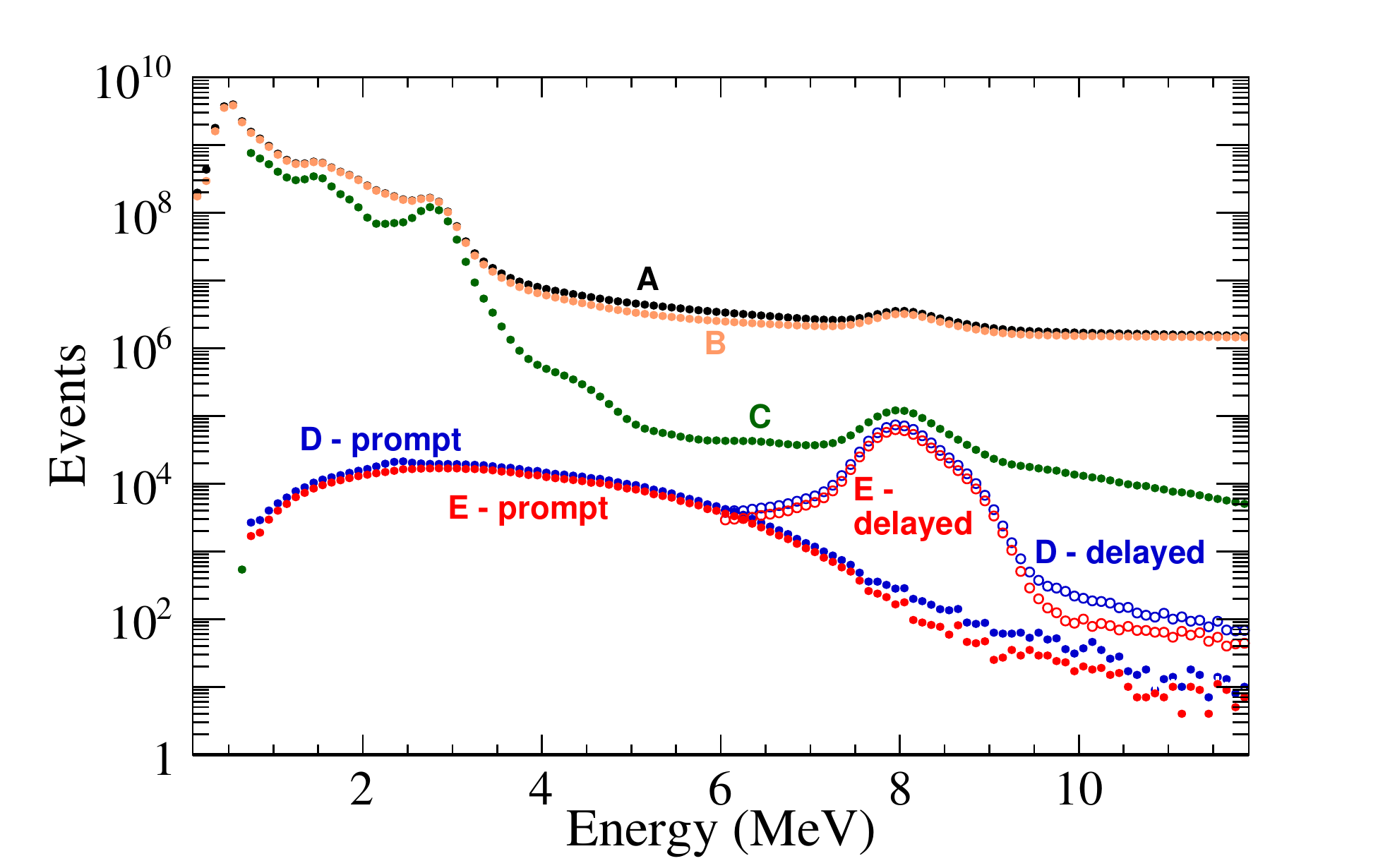}
  \caption{Reconstructed energy spectra of all signals in EH1-AD1 as
    successive cuts from selection A were applied: ({\em A}) all
    signals, ({\em B}) rejection of signals from PMT light emission,
    ({\em C}) Water Shield muon veto, ({\em D}) IBD signal pair
    selection: prompt energy, delayed energy, prompt-delayed time
    separation, and multiplicity veto, and ({\em E}) AD muon and
    shower vetos.  The spectra {\em E} are the final prompt and
    delayed energy spectra of the \nuebar{} candidates for EH1-AD1.
    \label{fig:ibd_spectra_vs_cut}}
\end{figure}

Fig.~\ref{fig:ibd_spectra_vs_cut} shows the reconstructed energy
spectra of all signals in EH1-AD1 as successive cuts from selection A
were applied to the data.
This figure provides a brief but helpful introduction to the selection
of \nuebar{} signals and rejection of backgrounds.
The spectrum for all signals ({\em A}) consisted of two prominent
components: radioactivity from natural sources below 3~MeV, and muons
generated in cosmic ray showers above 3~MeV\@.
A first step in the selection removed a minor instrumentation-related
background resulting from light emission by the PMTs, giving ({\em
  B})\@.
A veto following muon signals in the Water Shield efficiently rejected
muons and muon-induced neutrons, yielding ({\em C})\@.
The muon veto revealed an additional component of natural
radioactivity from 3 to 5~MeV ($^{208}$Tl decay within the
scintillator), as well as signals above 5~MeV from the $\beta$-decay
of unstable isotopes produced by muon-nuclear interactions in the
scintillator.
From these remaining signals, \nuebar{} inverse beta decay
interactions were identified by selecting pairs of signals consistent
with a positron, ({\em D-prompt}), followed soon after by the capture
of a neutron by Gd, ({\em D-delayed})\@.
As seen in the figure, the selection of prompt-delayed signal pairs
reduced the background by more than five orders of magnitude.
A veto following muon signals in the AD suppressed a small residual
muon-induced background, and gave the final prompt and delayed energy
spectra ({\em E}) of the \nuebar{} candidates.
Qualitatively similar results were found when the selection was
applied to the remaining seven antineutrino detectors.
The rest of this section describes the details of this selection,
including assessment of the residual background in the \nuebar{}
candidate sample.

\subsection{\label{subsec:eventseleff}Antineutrino Selection}

%
As discussed in Sec.~\ref{sec:intro}, antineutrino inverse beta decay
interactions provide a very characteristic pattern of two
time-correlated signals of specific energies.
The first, or prompt, signal is an $e^+$ which slows via ionization
and then annihilates in the scintillator, generating from 1 to 8~MeV
of visible energy.
The observed energy can be used to accurately estimate the original
energy of the incident \nuebar{}.
Each IBD interaction also produces a free neutron.
Carrying only O(10~keV) of kinetic energy, the neutron thermalizes in
$\sim$10~$\mu$s via collisions with nuclei in the scintillator.
For IBD interactions within the Daya Bay GdLS targets, the thermalized
neutron was subsequently captured by a nucleus with a characteristic
time constant of $\tau\simeq28$~$\mu$s.
The excited nucleus then relaxed via emission of $\gamma$-rays.
The interactions of these $\gamma$-rays within the scintillator
produced the second, or delayed, signal.
Within the Gd-loaded target region, $\sim$84\% of the captures
occurred on either $^{157}$Gd or $^{155}$Gd.
The relative probability for capture on these two Gd isotopes are
$\sim$81.5\% and $\sim$18.5\%, respectively.
The total $\gamma$-ray energy release per capture is 7.95~MeV for
$^{157}$Gd and 8.54~MeV for $^{155}$Gd.
These $\gamma$-rays were distinguished from natural radioactive
backgrounds, with energies predominantly below 5~MeV\@.

%
The remaining $\sim$16\% of neutrons captured almost entirely on
$^1$H, releasing a single 2.2-MeV $\gamma$-ray.
For capture on $^1$H, it was not possible to clearly discriminate
whether the \nuebar{} interacted within the target region or in the
gamma-catcher region.
Analysis of the n-$^1$H data therefore suffered from larger
uncertainties in the target volume and detector response, as well as a
much more significant background contamination.
Despite these obstacles, independent measurements of neutrino
oscillation have been obtained using these
interactions~\cite{An:2014ehw,An:2016bvr}, with results that were
consistent with the analysis of signals identified by n-Gd capture,
albeit with less precision.

%
IBD interactions followed by neutron capture on Gd are the focus of
the study presented here.
Antineutrino IBD interactions were selected by searching for pairs of
interactions separated by 1 to 200~$\mu$s, with a reconstructed prompt
energy, $E_{\mathrm{p}}$, between 0.7 and 12~MeV, and a reconstructed
delayed energy, $E_{\mathrm{d}}$, between 6 and 12~MeV\@.
All remaining selection criteria were designed for background
rejection, which will be discussed later in this section.

\begin{figure}[!ht]
  \includegraphics[width=0.45\textwidth]{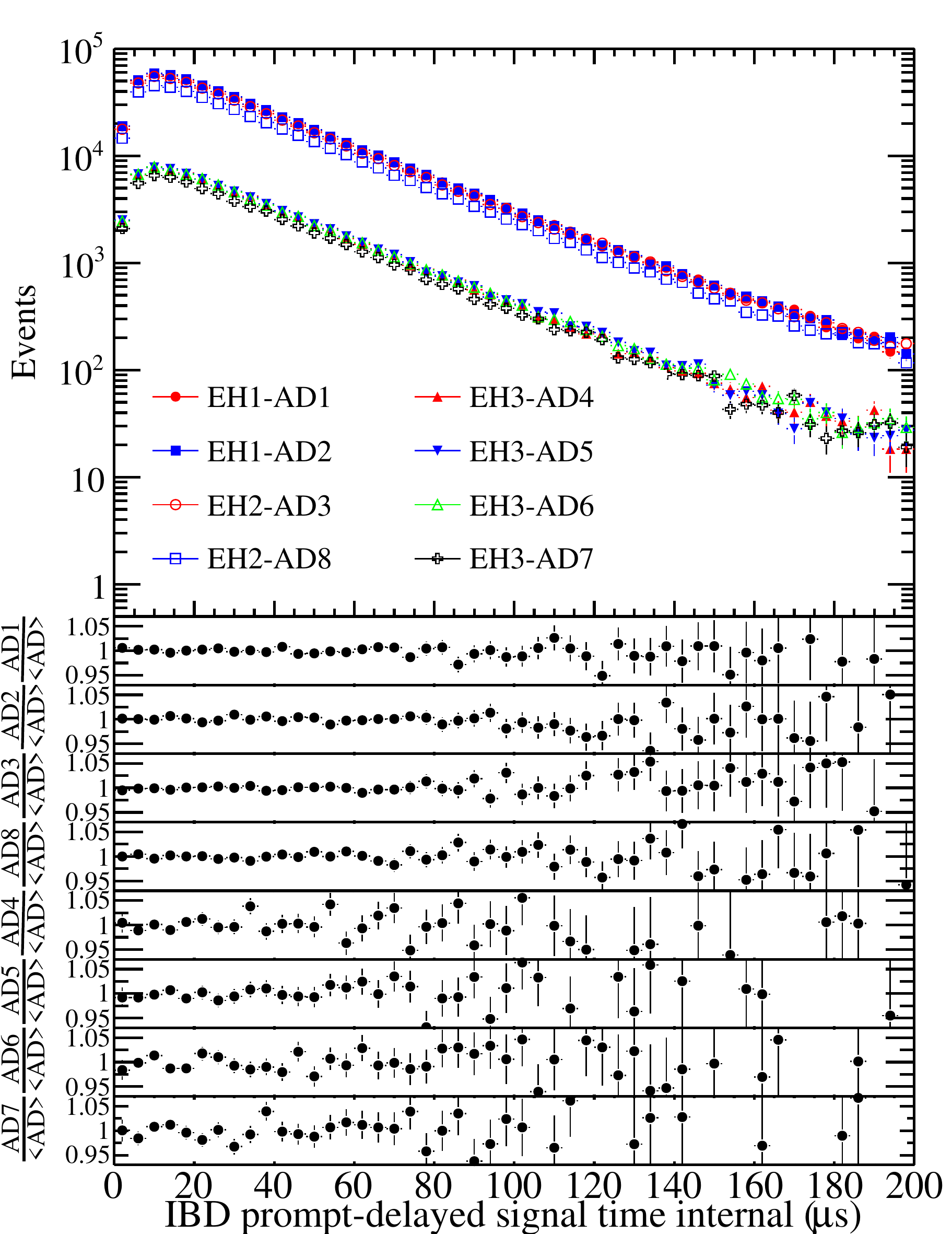}
  \caption{
    ({\em Top}) Distribution of the time separation between prompt and
    delayed \nuebar{} inverse beta decay interaction candidates for
    each detector.
    ({\em Bottom}) Ratio of the observed distribution to the
    normalized average for all eight detectors.  The consistency of
    these distributions constrained potential systematic variation in
    the fraction of IBD neutrons which captured on Gd among the eight
    detectors.
    \label{fig:ibd_dt}}
\end{figure}

%
Fig.~\ref{fig:ibd_dt} shows the temporal separation between prompt and
delayed signals of the IBD candidates, after applying all selection
criteria, of each detector for the entire data period used in this
study.
Delayed signals with ${\Delta}t>200$~$\mu$s were rejected since they
would not significantly improve signal statistics, yet they would
increase background contamination.
Signals with ${\Delta}t<1$~$\mu$s were rejected since they were
captured within a single triggered readout of the detector and were
therefore not easily discerned as separate interactions.
An absolute efficiency of (98.70$\pm$0.01)\% was estimated by
integrating the temporal distribution from 1~$\mu$s to 200~$\mu$s, and
was confirmed via simulation.
Potential variation in the efficiency of the prompt-delayed
${\Delta}t$ requirement between detectors was estimated to be 0.01\%,
by considering potential variation in the Gd concentration and
detector timing.

\begin{figure}[!htb]
\includegraphics[width=.5\textwidth]{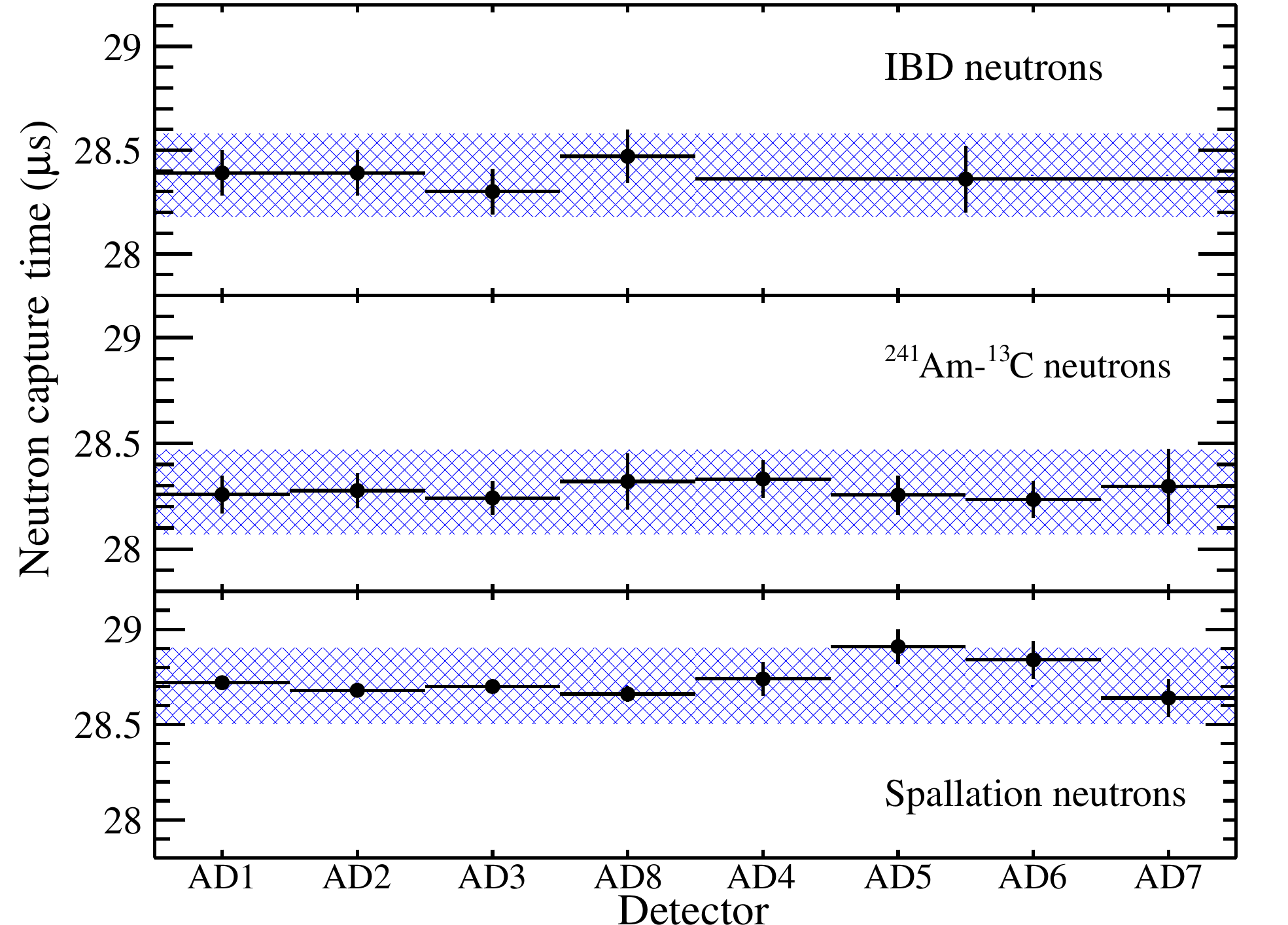}
\caption{Comparison of the capture time of neutrons on Gd measured in
  each detector, for neutrons from IBD, $^{241}$Am-$^{13}$C, and muon
  spallation. The IBD neutron data is combined for the four detectors
  in EH3 (AD4--AD7) in order to reduce the statistical
  uncertainty. The observed capture times vary by less than
  $\pm$0.2$\mu$s between detectors (blue band).  The spatial
  distributions of neutrons emitted by these sources differed,
  introducing slight offsets in their absolute capture times.
  \label{fig:s04_cap_time}}
\end{figure}

%
The fraction of IBD neutrons that were captured on Gd was estimated to
be (84.2$\pm$0.8)\%.
This was evaluated using the distributions of neutron capture time for
a variety of neutron sources, including muon spallation,
$^{241}$Am-$^{13}$C, $^{241}$Am-$^{9}$Be, $^{239}$Pu-$^{13}$C, and IBD
interactions, and confirmed by comparison with Monte-Carlo simulation.
More importantly, the similarity of the capture time distributions
between detectors for each of the neutron sources, as shown in
Fig.~\ref{fig:s04_cap_time}, constrained potential differences in Gd
capture fraction to $<$0.10\%~\cite{An:2016bvr}.
A difference in capture fraction could have resulted from differences
in the Gd concentration between the detectors, which was avoided by
using a common reservoir of GdLS to fill all eight detectors.

%
As discussed in Sec.~\ref{sec:experiment}, each detector was
individually triggered when the total number of channels over
threshold (NHIT) was $\geq$45 or the analog sum of all channels (ESUM)
was $\geq$65~PE\@.
This corresponds to a reconstructed energy threshold of
$\sim$0.4~MeV\@.
Comparisons of the measured rates and energy spectra of $^{68}$Ge
positron annihilation sources as a function of trigger threshold
demonstrated negligible inefficiency in detecting positrons for this
trigger threshold (see Fig.~9 of~\cite{DayaBay:2012aa}).
A combined study of both data and simulation showed that 0.19\% of the
prompt signals had $E_{\mathrm{p}}<0.7$~MeV, resulting from finite
detector resolution and $e^+$ energy loss in the IAV acrylic for
\nuebar{} interactions close to the target boundary.
Variation of the model of the detector response, in particular for
positrons which lose energy in the IAV, introduced a 0.10\%
uncertainty to this selection.
Therefore, the absolute efficiency of this selection was estimated to
be (99.81$\pm$0.10)\%.
Considering a conservative 2\% difference in energy scale at $E_p =
0.7$~MeV, a 0.01\% relative uncertainty between detectors was obtained.
Models of reactor \nuebar{} emission suggested no detectable signals
with $E_{\mathrm{p}}\gtrsim11$~MeV, which was consistent with the
observations presented here.
This analysis included signals with $E_{\mathrm{p}}$ up to 12~MeV,
introducing no additional inefficiency for reactor \nuebar{}
detection.
%

%
As shown, the selections based on prompt energy and time difference
had very limited potential for introducing differences in efficiency
between detectors.
This was not the case for the delayed energy selection.
In particular, escape of n-Gd $\gamma$-rays from the scintillating
regions of the detector introduced a low-energy tail to the peak at
8~MeV in the energy spectrum of delayed interactions.
The primary purpose of the gamma-catcher region surrounding each
antineutrino target was to significantly reduce the magnitude of this
spectral tail.
According to a Geant4-based simulation, the shape of this tail
depended on the $\gamma$-ray multiplicity and energy distribution from
neutron capture on Gd, which was not accurately known.
The observed energy spectra from n-Gd capture of neutrons from muon
spallation, $^{241}$Am-$^{13}$C sources, and IBD interactions were
used to constrain potential variations in the simulated n-Gd
$\gamma$-ray distributions.
Based on those simulated $\gamma$-ray distributions, an absolute
efficiency of (92.7$\pm$0.97)\% was determined for a selection of
$E_{\mathrm{d}}>6$~MeV\@.
The simulation determined $\sim$6\% of the selection inefficiency was
from signals with $E_{\mathrm{d}}$ between 3.6~MeV and 6~MeV\@, which
was strongly constrained by data.
Simulation predicted an additional $\sim$1\% inefficiency due to
signals with $E_{\mathrm{d}}<3.6$~MeV, but this estimate was poorly
constrained by data due to background from n-$^1$H capture in the
observed spectra.
Assuming a conservative 100\% uncertainty on this part of the
distribution resulted in a total uncertainty for the
$E_{\mathrm{d}}>6$~MeV selection of 0.97\%\@.
The same studies demonstrated that the selection of
$E_{\mathrm{d}}<12$~MeV was 100\% efficient.

%
Potential variation of the delayed energy selection efficiency between
detectors was of greater concern.
Three approaches all suggested a similar value for this systematic
uncertainty.
Direct $\chi^2$ comparisons of the observed n-Gd spectra between
detectors were consistent within the expected minor variations in
detector resolution and energy scale calibration.
A 0.2\% linear shift in energy scale between detectors, as discussed
in Sec.~\ref{sec:rel_energy}, corresponded to a 0.05\% shift in
efficiency, as evaluated from the observed data.
With slightly more rigor, correlations of position-dependent
variations in energy scale and n-Gd efficiency were considered.
Variations in energy scale between detectors were compared using data
from 16 regions of equal volume within the GdLS target.
Simulation of the n-Gd capture spectrum in each region determined the
relative impact of energy scale variations on the efficiency of each
region.
A volume-weighted average over the entire GdLS target estimated
potential variations in efficiency between detectors of 0.07\%.

An alternate technique, outlined in~\cite{An:2016bvr}, provided a more
general approach for assessing this uncertainty.
The efficiency for each detector $i$ is equivalent to the ratio of the
number of n-Gd signals above the 6~MeV threshold, $N^{\mathrm{d}}_i$,
to the total number of n-Gd signals, $N^{\mathrm{t}}_i$,
\begin{equation}
  \epsilon_i = \frac{N^{\mathrm{d}}_i}{N^{\mathrm{t}}_i}.
\end{equation}
It was not possible to directly measure $N^{\mathrm{t}}_i$, given that
backgrounds overwhelmed the observed spectra for
$E_{\mathrm{d}}<3.6$~MeV\@.
The number of signals above 3.6~MeV, $N^{\mathrm{a}}_i$, served as a
close approximation for $N^{\mathrm{t}}_i$, given that it contained
$\sim$99\% of all n-Gd capture signals.
Therefore, differences in the ratios of $N^{\mathrm{d}}_i$ to
$N^{\mathrm{a}}_i$ among the detectors were strongly correlated to
potential variations in efficiency $\epsilon_i$.
Explicitly, a linear model 
\begin{equation}
  \bar{N}^{\mathrm{d}}_i = a + b N^{\mathrm{a}}_i
\end{equation}
was fit to the distribution of $N^{\mathrm{d}}_i$ versus
$N^{\mathrm{a}}_i$ for the eight detectors.
The model estimate of $\bar{N}^{\mathrm{d}}_i$ for each detector
showed small deviations from the observed $N^{\mathrm{d}}_i$.
Variation in the efficiency of each detector from the model average,
$\delta\epsilon_i$/$\epsilon_i$, was then given by
\begin{equation}
  \frac{\delta\epsilon_i}{\epsilon_i} = \frac{{\delta}N^{\mathrm{d}}_i}{N^{\mathrm{d}}_i} = \frac{N^{\mathrm{d}}_i - \bar{N}^{\mathrm{d}}(N^{\mathrm{a}}_i)}{N^{\mathrm{d}}_i} = 1 - \frac{a + bN^{\mathrm{a}}_i}{N^{\mathrm{d}}_i}.
\end{equation}
Variations of 0.08\% were observed among the detectors, which was
adopted for the analysis here.

\begin{figure}[!ht]
\includegraphics[width=0.45\textwidth]{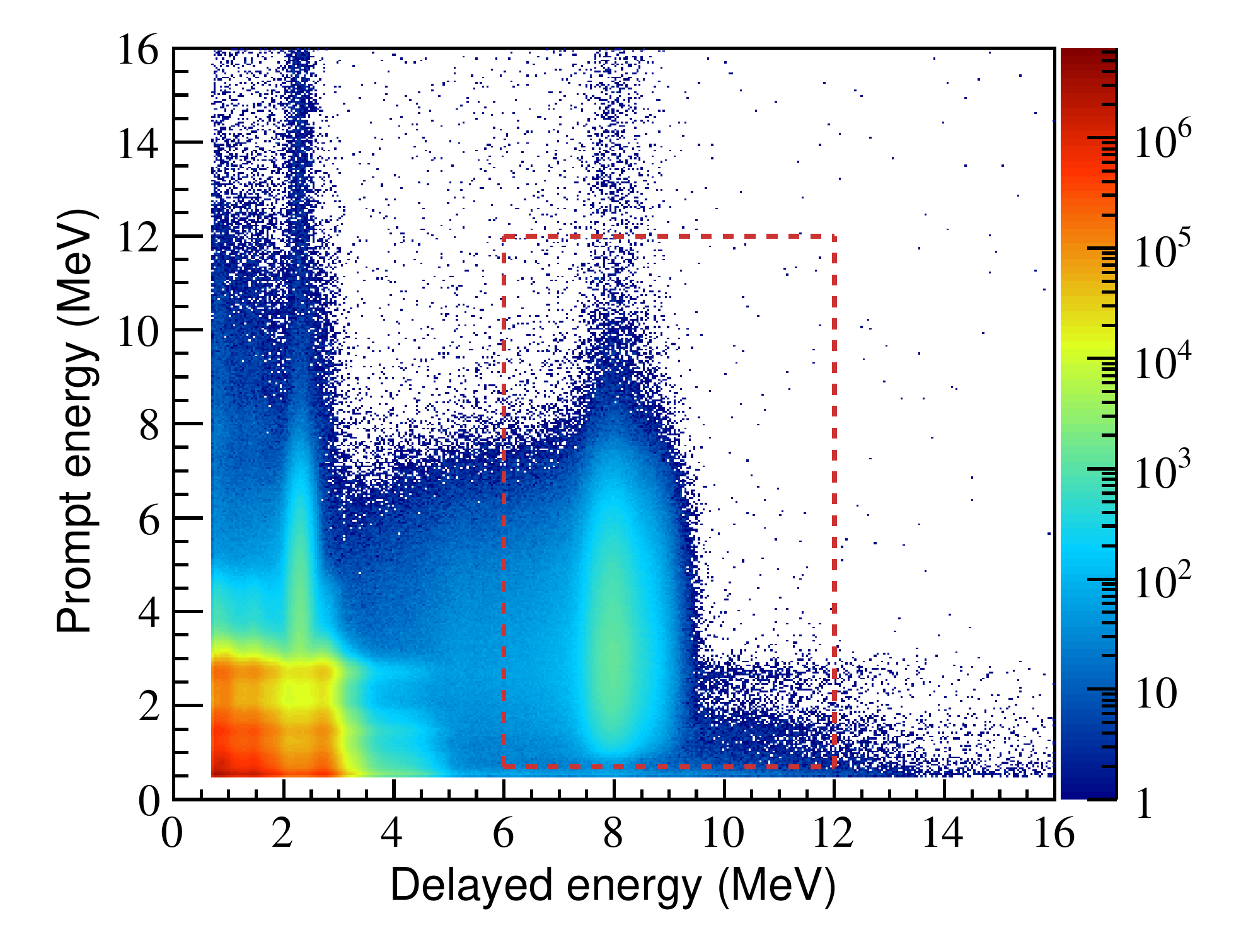}
\caption{The distribution of prompt versus delayed energy for signal
  pairs which satisfied the \nuebar{} inverse beta decay selection
  criteria.  Interactions with prompt energy of
  $0.7~\mathrm{MeV}<E_{\mathrm{p}}<12~\mathrm{MeV}$ and delayed energy
  of $6~\mathrm{MeV}<E_{\mathrm{d}}<12~\mathrm{MeV}$ were used for the
  neutrino oscillation analysis discussed here (red dashed box).  A
  few-percent contamination from accidental backgrounds (symmetric
  under interchange of prompt and delayed energy) and $^9$Li decay and
  fast neutron backgrounds (high prompt and $\sim$8~MeV delayed
  energy) are visible within the selected region.  Inverse beta decay
  interactions where the neutron was captured on $^1$H provided an
  additional signal region with $E_{\mathrm{d}}\sim2.2~\mathrm{MeV}$,
  albeit with much higher background.
  \label{fig:pd_cut}}
\end{figure}

%
Figure~\ref{fig:pd_cut} shows the distribution of prompt versus
delayed energy for all signal pairs which satisfied the \nuebar{}
selection criteria.
Only the $E_{\mathrm{d}}>6$~MeV selection truncated a significant
fraction of true signal events, which visibly span the boundary of the
selected region.

\subsection{\label{subsec:bkg_rejection}Background Rejection}

%
The vast majority of triggered signals in each detector were caused by
natural radioactive backgrounds, with less than one in $\sim$10$^5$
resulting from reactor \nuebar{} interactions.
This section describes the methods employed to reduce background
contamination to less than a few percent of the reactor \nuebar{}
sample used to measure neutrino oscillation.
%
Aside from natural radioactivity, a minor PMT instrumentation-related
background was discovered during detector assembly and commissioning.
Rejection of this instrumental background is also discussed here.
%
Selecting pairs of prompt plus delayed signals with the proper
energies and time separations rejected almost all backgrounds caused
by natural radioactivity.
%
Occasionally two such uncorrelated interactions would {\em
  accidentally} satisfy the antineutrino selection criteria.
Detailed studies of all uncorrelated interactions measured this
background contamination to be from 1\% to 2\% depending on
detector, with negligible uncertainty.
%
All remaining backgrounds were from physical processes that produce a
pair of correlated interactions that potentially mimic inverse beta
decay.
The majority of such correlated background were attributable to
atmospheric muons produced in cosmic ray showers.
Energetic muons penetrated the rock to reach the experimental halls,
and interacted with the detector or nearby environment.
%
Vetoing signals that occur during and soon after muon interactions
with the detector or muon systems effectively reduced the
contamination caused by these backgrounds to less than 0.5\%\@.
%
Lastly, three or more signals would occasionally occur close in time,
resulting in confusion as to which pair was most likely the result of
an antineutrino interaction.
To avoid this ambiguity, sets of signals with multiplicity $>$2 were
rejected.
The detailed characterization and mitigation of these backgrounds are
presented in the following sections.

\subsubsection{\label{subsubsec:eventsel}Instrumental Background}

%
Assessment of the PMTs during detector assembly revealed that some
PMTs emitted light.
Such emission is commonly called {\em flashing}, although the
mechanism of light emission can vary between PMT designs.
For the Daya Bay PMTs, direct imaging of the base using an
astronomical CCD camera pinpointed the light emission to an electrical
discharge occurring at the point of connection of the incoming HV
cable with the base circuit board.
Most of the PMTs were found to emit light at some level, with rates
and intensities which varied over time.
Once installed within an AD, the black radial shield prevented most of
the emitted light from entering the central AD volume.
A small fraction of the light propagated within the PMT, striking the
photocathode or passing into the scintillator region.
The emitted light produced background signals with reconstructed
energies up to $\sim$100~MeV, with a rate in the energy range of the
delayed IBD signals (6~MeV to 12~MeV) of approximately 0.7~Hz per
AD\@.

\begin{figure}[!ht]
\begin{center}
  \includegraphics[width=0.49\textwidth]{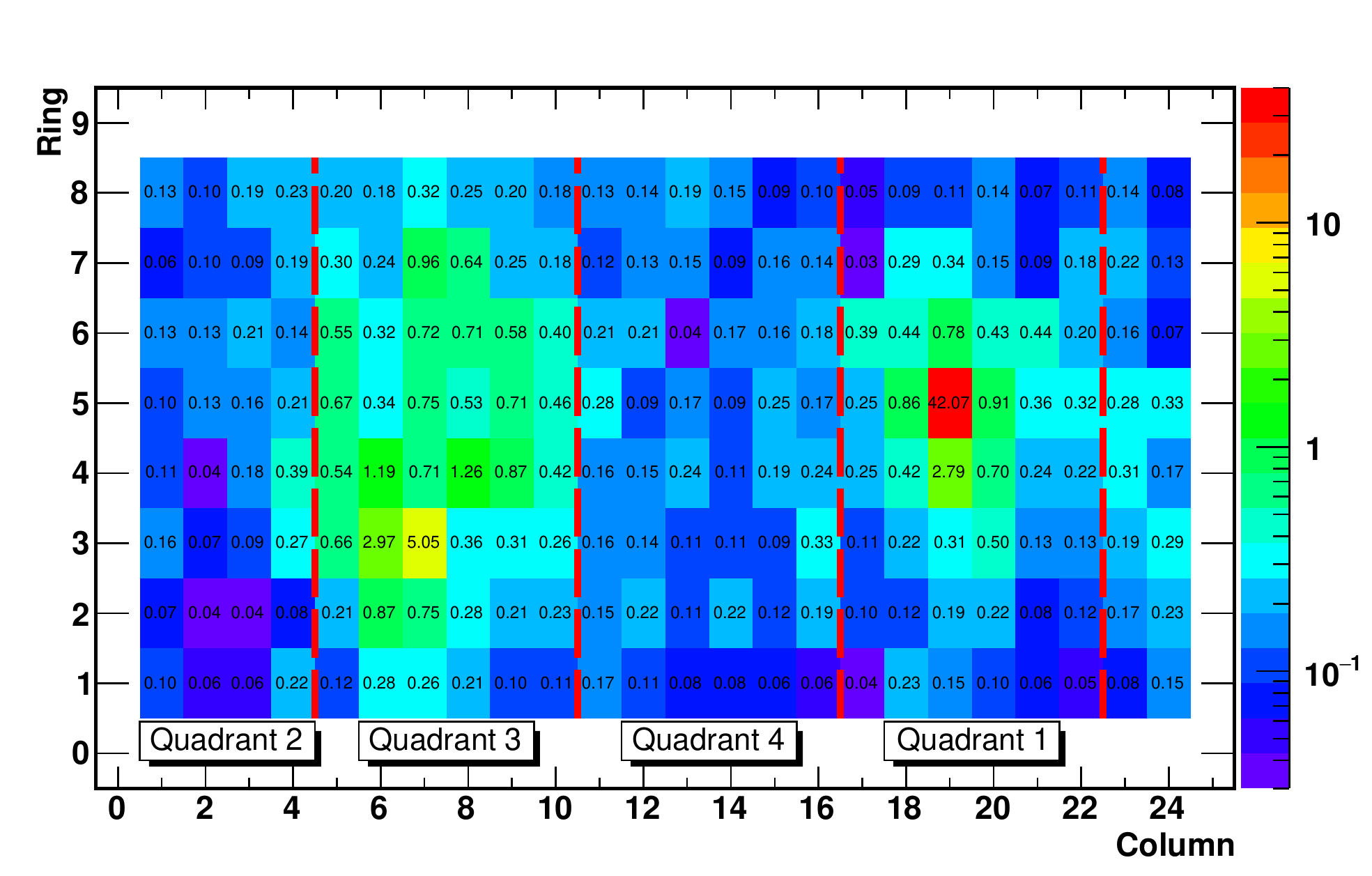}
  \caption{Typical charge distribution from PMT light emission.  The 8
    ring by 24 column cylindrical PMT array has been projected onto a
    plane.  The logarithmic color scale, as well as numbers, provides
    the percentage of the total signal charge observed by each PMT\@.
    In this example, light was emitted by the PMT at column 19 in ring
    5\@.  The discharging PMT and those PMTs directly opposite in
    Quadrant 3 observed a significant fraction of the total charge.
    The distinct charge pattern allowed efficient rejection of this
    instrumental background.
    \label{fig:pmt_flash}}
\end{center}
\end{figure}

Were they not removed, these false delayed signals would have
contributed a significant accidental background, comparable to the
observed antineutrino rate.
Fortunately, these signals had characteristic patterns in space and
time, easily distinguishable from genuine particle interactions within
the scintillator.
In particular, both the discharging PMT and those PMTs directly
opposite within the AD observed a large fraction of the charge.
Figure~\ref{fig:pmt_flash} shows a typical charge distribution from
PMT light emission.

%
A parameter was constructed to discriminate PMT light emission from
genuine particle interactions.
For every triggered signal, the single PMT which observed the greatest
charge, $Q_{\mathrm{max}}$, was identified as a potential light
emitter.
The spatial charge distribution of the AD signal was then quantified
using two variables, $f_{\mathrm{max}}$ and $f_{\mathrm{quad}}$\@.
$f_{\mathrm{max}}$ was the ratio of the maximum PMT charge over the
total observed charge, $Q_{\mathrm{max}}/Q_{\mathrm{total}}$.
The twenty-four columns of PMTs in one AD were grouped into four
quadrants such that the potential emitter was at the center of the
first quadrant.
The total charge observed in the $i$-th quadrant was defined as
$Q_{\mathrm{q}i}$.
$f_{\mathrm{quad}}$ was the ratio of the charge observed in the
opposite quadrant over the two adjacent quadrants,
$f_{\mathrm{quad}}=Q_{\mathrm{q3}}/(Q_{\mathrm{q2}}+Q_{\mathrm{q4}})$.
The discriminator, $f_{\mathrm{ID}}$, combined these two aspects of
the spatial distribution of light,
\begin{equation}
f_{\mathrm{ID}} = \log_{10}\left[f_{\mathrm{quad}}^2+\left(\frac{f_{\mathrm{max}}}{0.45}\right)^2\right].
\end{equation}

\begin{figure}[!ht]
  \includegraphics[width=0.49\textwidth]{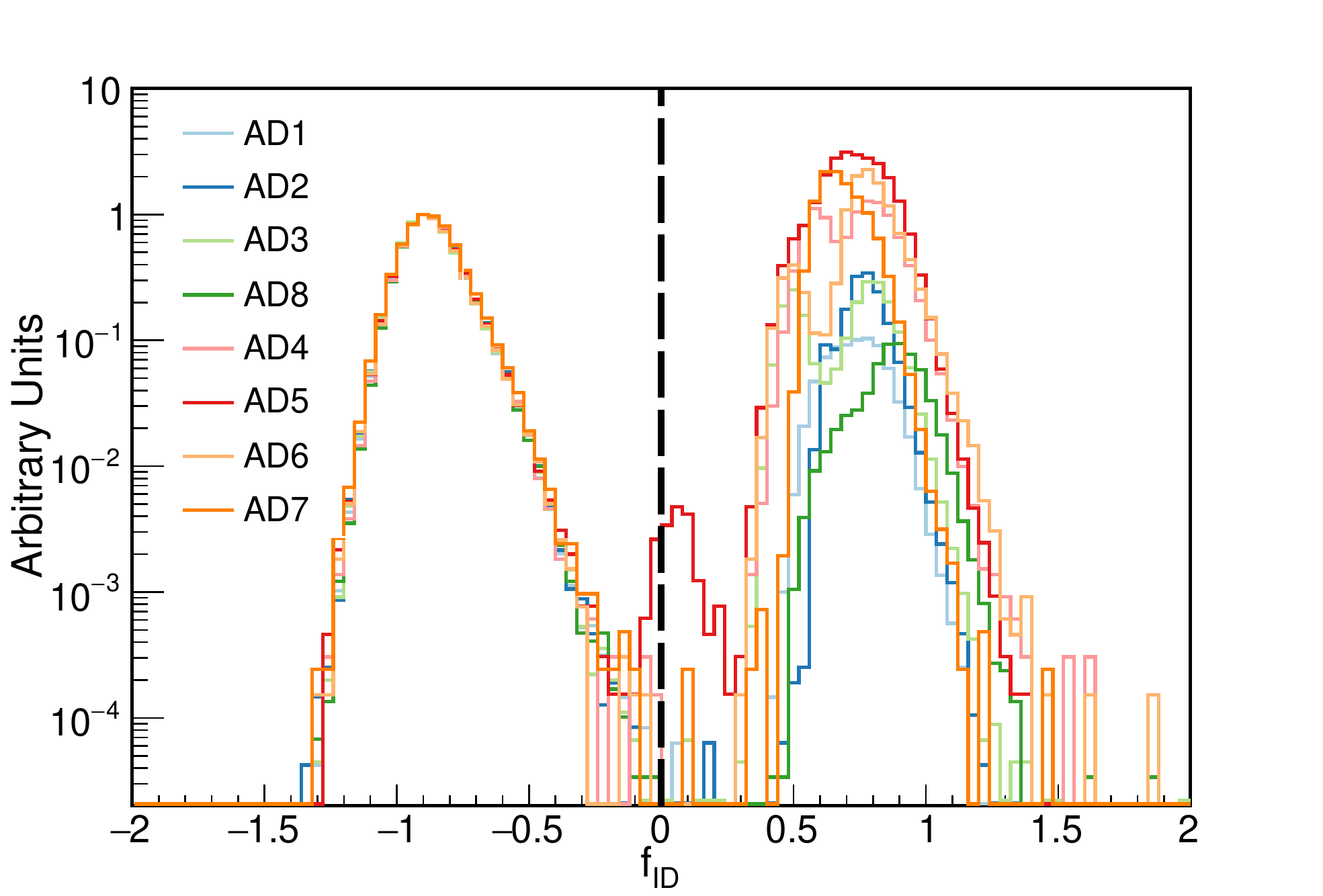}
  \caption{Distributions of the discriminator $f_{\mathrm{ID}}$ for
    all inverse beta decay delayed signal candidates from the eight
    detectors.  The distributions are normalized to demonstrate the
    consistency of the distributions among the ADs for genuine
    antineutrino candidates ($f_{\mathrm{ID}}<0$), while the
    background distributions depended on the characteristics of the
    light-emitting PMTs in each AD ($f_{\mathrm{ID}}\geq0$)\@.
    \label{fig:flasherFID}}
\end{figure}

Figure~\ref{fig:flasherFID} shows the normalized distributions of this
discriminator for the delayed signals of the antineutrino candidates,
including the background from PMT light emission.
The discriminator had a consistent distribution for genuine IBD
candidates ($f_{\mathrm{ID}}<0$) among the eight ADs, while the
signals from light emission ($f_{\mathrm{ID}}\geq0$) varied as a
result of the particular characteristics of the light-emitting PMTs in
each AD\@.
Light emission by a PMT located in the bottom ring of AD5 produced the
small peak near $f_{\mathrm{ID}}=0$ in Fig.~\ref{fig:flasherFID}.

%
PMT light emission also generally exhibited a broader distribution of
relative times between the PMT signals.
An additional discriminator, $f_{\mathrm{t1}}$, was defined as the
ratio of the number of PMT signals in the first 200~ns of the signal,
relative to the total number of PMT signals in the first 400~ns.
The variable $f_{\mathrm{t2}}$ was the same except for only using the
number of PMT signals in the first 150~ns in the numerator.
A time-based discriminator $f_{\mathrm{PSD}}$ combined these two
variables using an optimized weighting,
\begin{equation}
f_{\mathrm{PSD}} = \log_{10}\left[4\cdot(1-f_{\mathrm{t1}})^2+1.8\cdot(1-f_{\mathrm{t2}})^2\right].
\end{equation}
Signals with $f_{\mathrm{PSD}}\geq0$ were identified as PMT light
emission and rejected from further studies.
This discriminator had the added benefit of rejecting pile-up, in
which two independent particle interactions occurred within a single
triggered readout of the detector.

%
Each AD also included six 2-inch PMTs which were located at the top
and bottom of the detector adjacent to the paths for ACU calibration
deployment.
These PMTs were used to monitor the scintillator characteristics, and
were not employed for triggering or energy and position
reconstruction.
Light emission by these PMTs was easily rejected whenever the charge
observed by any one of them exceeded 100~PE\@.

%
After applying the $f_{\mathrm{ID}}$ and $f_{\mathrm{PSD}}$
discriminators, negligible background from PMT light emission remained
in the antineutrino candidate sample.
A study of high-purity samples of particle interactions showed that
very few were incorrectly rejected by these discriminators, and an
efficiency of 99.98\% was determined for antineutrino signals.
Negligible uncertainties of 0.01\% correlated among the detectors and
0.01\% uncorrelated among the detectors were also determined from this
study.
Any residual light emission was incorporated into the assessment of
uncorrelated backgrounds, as discussed in the next section.

\subsubsection{\label{subsubsec:bkg_uncorr}Uncorrelated Backgrounds}

%
Two uncorrelated signals occasionally satisfied the antineutrino
selection criteria, giving rise to backgrounds that are commonly
referred to as {\em accidentals}.
The rate, energy spectrum, and other characteristics of these
backgrounds were precisely modeled from studies of individual
uncorrelated signals.
Each day, only a few of the $\sim$10$^7$ uncorrelated signals were
estimated to form a pair which satisfied the antineutrino selection.
As a result, the residual accidental backgrounds in the final sample
of antineutrino candidates were reliably determined to be only
$\sim$1\% in the near detectors and $\sim$2\% in the far detectors,
with negligible systematic uncertainty.
The detailed assessment of this background is discussed here.

\begin{figure}
  \includegraphics[width=\columnwidth]{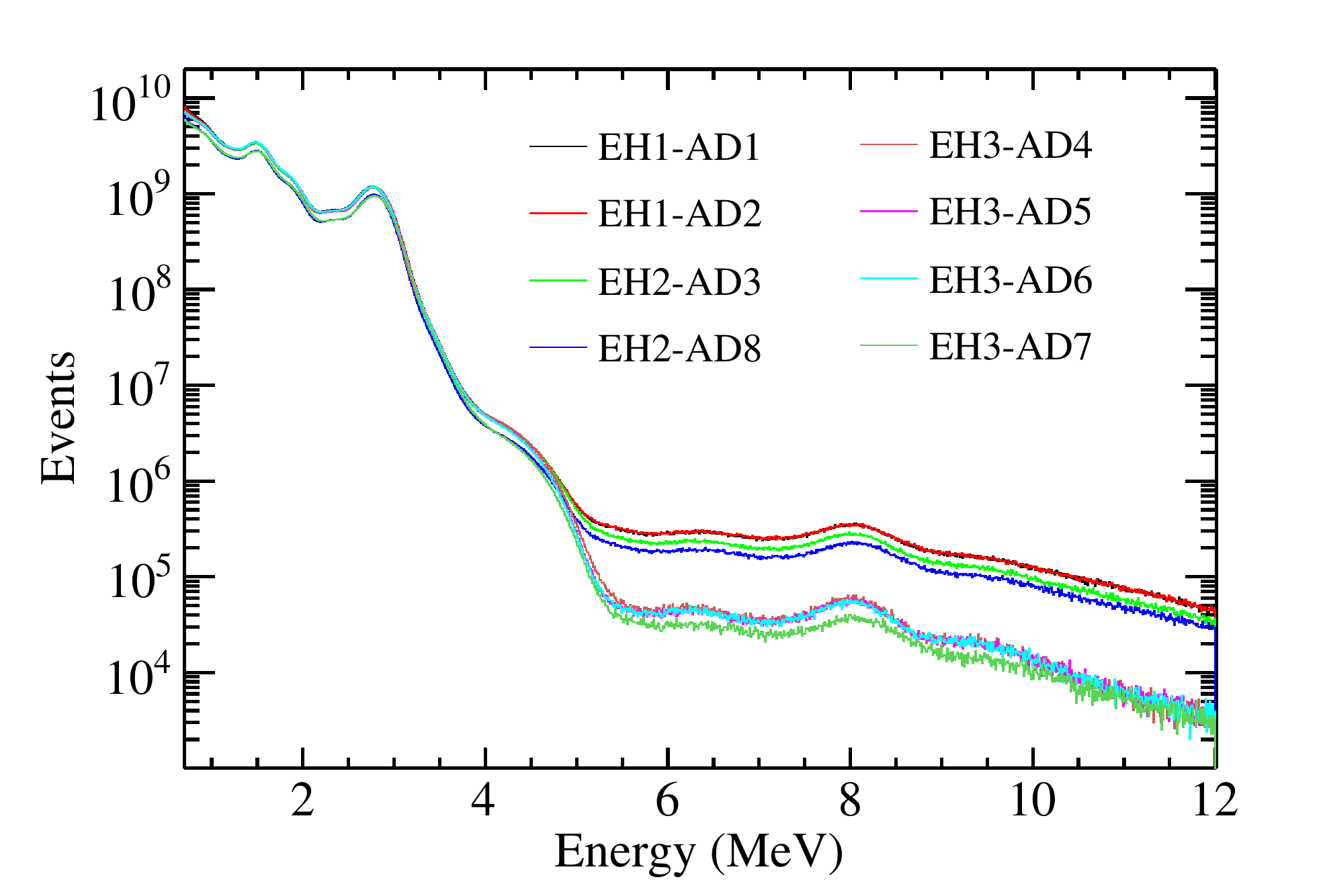}
  \caption{The reconstructed energy spectra of isolated uncorrelated
    signals for all eight detectors.  The prompt-like signal rate was
    dominated by natural radioactivity with energies below 3~MeV\@.
    The delayed-like signal rate was dominated by beta decay of
    muon-induced unstable isotopes, mainly $^{12}$B, and $\gamma$-rays
    from capture of $^{241}$Am-$^{13}$C calibration source neutrons on
    materials at the top of each detector. The energy spectrum of the
    accidental background prompt signals in each AD were equivalent to
    these distributions.
    \label{fig:singles-spectra}}
\end{figure}

%
An uncorrelated signal was identified as prompt-like if it satisfied
the prompt energy selection 0.7~MeV~$< E_{\mathrm{rec}} <$~12~MeV\@.
Correspondingly, it was also identified as delayed-like if it
satisfied 6~MeV~$< E_{\mathrm{rec}} <$~12~MeV\@.
The majority of prompt-like uncorrelated signals were from natural
radioactivity in the detector components and the surrounding
environment, and had $E_{\mathrm{rec}} < 3$~MeV\@.
Delayed-like uncorrelated signals were primarily from two sources.
Muon-nuclear interactions produced unstable nuclei within the
detectors, which would subsequently beta decay.
$^{12}$B was by far the most prominent of such isotopes, although
others such as $^{12}$N were also produced.
The second were high-energy $\gamma$-rays produced by the capture of
neutrons emitted by the $^{241}$Am-$^{13}$C calibration sources
located in the ACUs on the AD lid.
A small fraction of these $\gamma$-rays reached the scintillator
volume and produced delayed-like signals.
Figure~\ref{fig:singles-spectra} shows the reconstructed energy
spectrum for all isolated prompt-like signals for all eight detectors.
Note that uncorrelated signals which occurred during time periods
vetoed by the muon or multiplicity selections were appropriately
excluded from the studied samples.
Delayed-like signals were the subset with $E_{\mathrm{rec}}>6$~MeV\@.
The prompt signal energy spectra of the accidental background for each
detector were estimated directly from these distributions.

%
The accidental background was modeled using combinations of these
uncorrelated prompt-like and delayed-like signals.
The accidental background rate for selection A was
\begin{equation}\label{eq:accRateA}
  R^{\mathrm{A}}_{\mathrm{acc}} = R_{\mathrm{d}}\,\left(1-e^{-R_{\mathrm{p}}{\Delta}t}\right)\,e^{-2R_{\mathrm{p}}{\Delta}t},
\end{equation}
where $R_{\mathrm{p}}$ was the measured rate of prompt-like signals,
$R_{\mathrm{d}}$ was the measured rate of delayed-like signals, and
${\Delta}t \cong 200$~$\mu$s was the length of the time window for
selection of antineutrino candidate pairs.
The factor in parentheses was the probability for an uncorrelated
prompt-like signal to fall in the selected time window preceding a
delayed-like signal.
The final term accounted for the efficiency of the multiplicity veto,
which rejected accidental backgrounds when a second prompt-like signal
occurred either before the prompt or after the delayed signal.
Given that all other antineutrino selection criteria (e.g.\ light
emission rejection, muon veto, etc.) were applied when selecting
uncorrelated signals for the calculation of $R_{\mathrm{p}}$ and
$R_{\mathrm{d}}$, the estimated background rate
$R^{\mathrm{A}}_{\mathrm{acc}}$ automatically included these selection
efficiencies.

%
The slight difference in the multiplicity requirement for selection B,
as outlined in Table~\ref{tab:IBDselectionCuts}, resulted in an
alternate expression for the rate of accidental backgrounds.
The accidental rate for selection B was
\begin{equation}\label{eq:accRateB}
  R^{\mathrm{B}}_{\mathrm{acc}} = R_{\mathrm{d}}\,R_{\mathrm{p}}\,{\Delta}t\,e^{-2R_{\mathrm{p}}{\Delta}t}\,e^{-R_{\mathrm{d}}{\Delta}t}.
\end{equation}
%

\begin{figure}
  \includegraphics[width=\columnwidth]{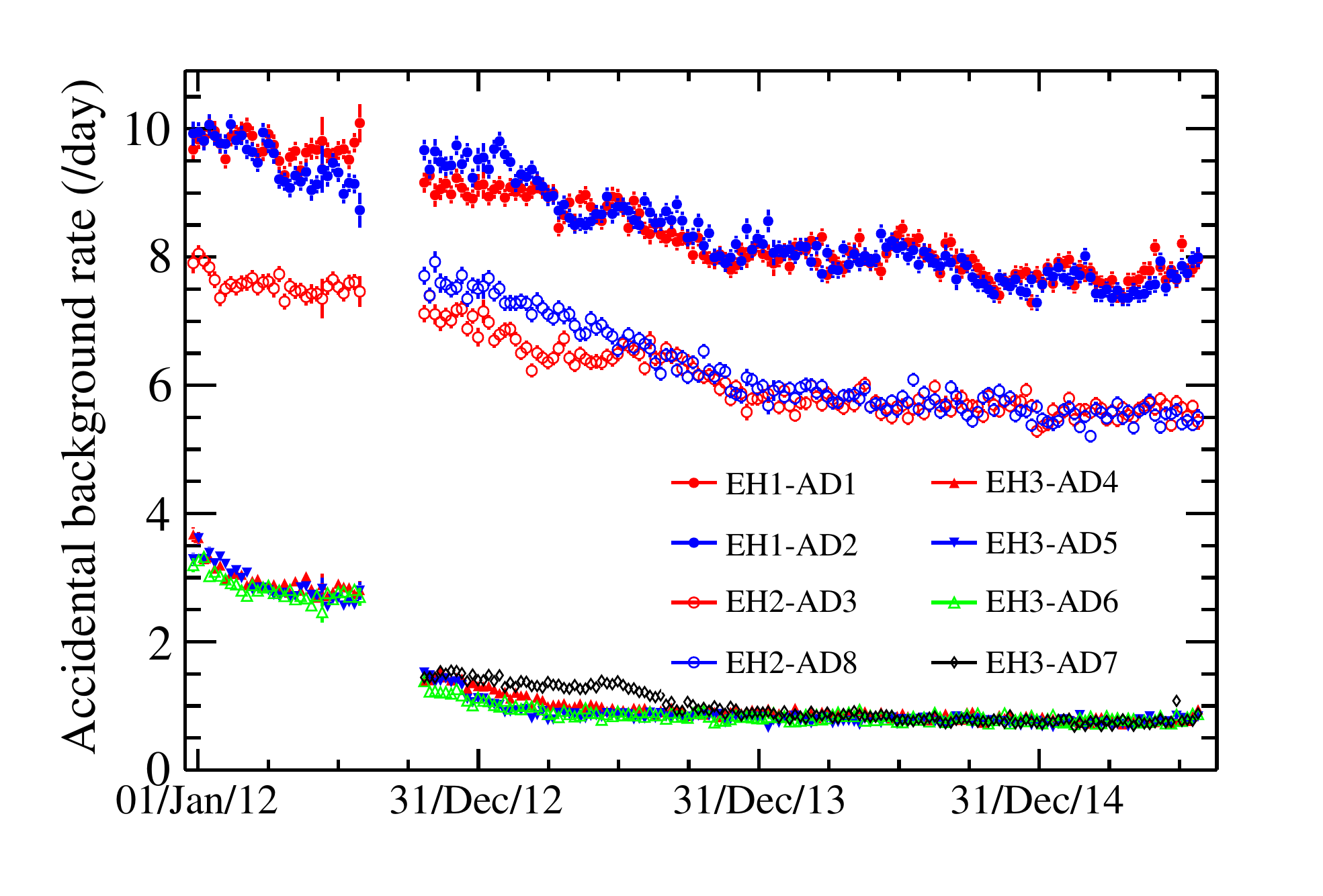}
  \caption{The accidental background rate for selection A as a
    function of time for each antineutrino detector, as calculated
    from measurements of the rates of uncorrelated signals which
    satisfy the prompt \nuebar{} signal selection, $R_{\mathrm{p}}$,
    and delayed \nuebar{} signal selection, $R_{\mathrm{d}}$.  The
    accidental rate primarily depended on experimental hall depth due
    to the relative rate of unstable isotope production by muons.  The
    decline of the accidental background versus time was due to
    combination of a decrease in the natural radioactivity following
    detector installation and a reduction of the neutron emission by
    the $^{241}$Am-$^{13}$C calibration sources.  $R_{\mathrm{p}}$ was
    sensitive to small changes in electrical noise, which resulted in
    the observed instability in the accidental background rate.
    Installation of the final two detectors in the summer of 2012 is
    evident as a gap during this time.  Removal of two of the three
    $^{241}$Am-$^{13}$C sources in each far detector during the
    installation period reduced the accidental background rate by
    $\sim$50\%.
    \label{fig:accidental-rate-vs-time}}
\end{figure}

%
The rates of uncorrelated signals varied over time for each detector.
For the first few months following commissioning of each detector, the
prompt-like signal rate $R_{\mathrm{p}}$ showed an initial decline of
$\sim$20\% attributed to removal of natural radioactivity during
purification of the water in the Water Shields.
The rates eventually stabilized in the range of 55~Hz to 60~Hz for all
eight detectors.
As can be seen in Figure~\ref{fig:singles-spectra}, $R_{\mathrm{p}}$
was dominated by signals which occurred just above the 0.7~MeV
selection threshold.
Small changes in the electrical noise environment within each
experimental hall would result in slight changes in the efficiency of
this threshold.
These shifts are consequently visible as slight instability in the
rate of accidental background versus time.
The delayed-like signal rate $R_{\mathrm{d}}$ showed significant
differences between experimental halls; a decrease in the muon flux
versus hall depth resulted in a corresponding reduction in $^{12}$B
production.
The neutron production by the $^{241}$Am-$^{13}$C sources declined
over time, which reduced the delayed-like rate of high-energy
$\gamma$-rays from this source.
During installation of the final two detectors in the summer of 2012,
the $^{241}$Am-$^{13}$C sources were removed from ACU-B and ACU-C for
all detectors in EH3.
While reduction of the $^{241}$Am-$^{13}$C correlated background was
the primary purpose, as will be described in the next section, it had
the added benefit of cutting the delayed-like signal rate in half for
the far detectors.
In summary, each detector initially had about 1000 (EH1), 800 (EH2),
and 250 (EH3) delayed-like signals per day, but this has declined by
$\sim$20\% for the near detectors and $\sim$65\% for the far
detectors.
Since the rates of uncorrelated signals varied with time, so did the
accidental background.
Consequently, the data was divided into short intervals in time and
the accidental background was independently estimated for each.
For selection A, these periods corresponded to every four hours, while
for selection B this was once every day.
Figure~\ref{fig:accidental-rate-vs-time} shows the estimated rate of
accidental background for each detector as a function of time.

\begin{figure}
  \includegraphics[width=\columnwidth]{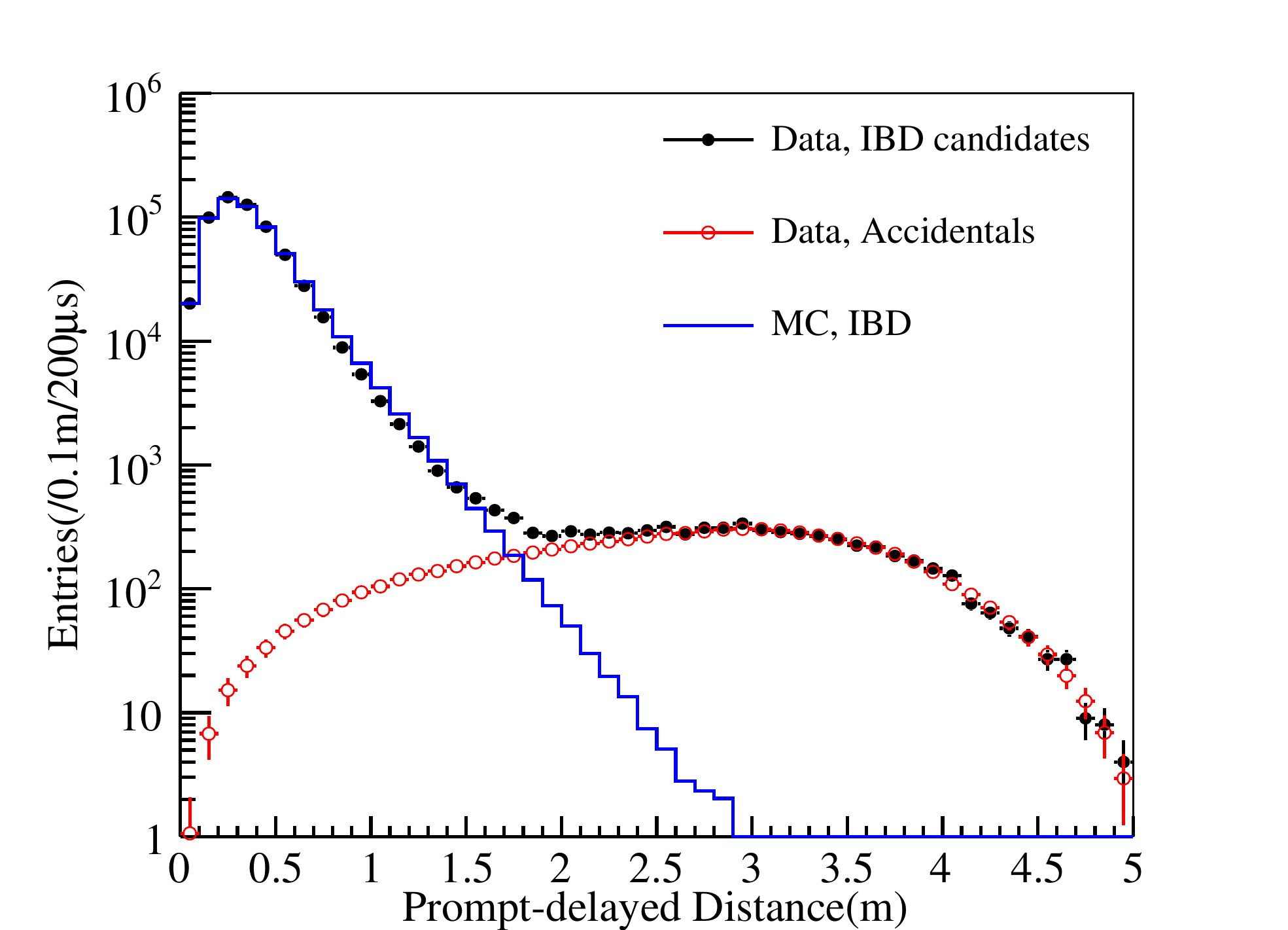}
  \caption{Distribution of distances between the reconstructed
    positions of the prompt and delayed signals of the antineutrino
    inverse beta decay candidate signals from all detectors (black
    points).  The positions were highly correlated for true \nuebar{}
    interactions, as demonstrated by Monte-Carlo simulation (blue
    line).  Repeating the selection of \nuebar{} interactions, but
    with the time window for selection of the delayed signals offset
    by 1~ms to 20~ms, enhanced the uncorrelated accidental background
    (red points) and suppressed correlated signals.  The consistency
    of the distributions confirmed the estimate of contamination by
    accidental backgrounds, which varied from 1\% to 2\% depending on
    detector.
    \label{fig:accidental-Distance}}
\end{figure}

%
The accidental rates determined according to Eqs.~\ref{eq:accRateA}
and~\ref{eq:accRateB} were cross-checked by using an {\em
  offset-window} method.
In this approach the antineutrino selection was repeated, but with a
time offset $t^{\mathrm{off}}$ introduced between prompt and delayed
signal pairs (i.e. $1~\mu\mathrm{s} + t^{\mathrm{off}} < {\Delta}t <
200~\mu\mathrm{s} + t^{\mathrm{off}}$).
A minimum offset of 1~ms suppressed correlated signals such as IBD,
fast neutrons, and $^9$Li, thereby providing a sample enriched in
accidental background.
Repeating the process with $t^{\mathrm{off}}$ from 1~ms to 20~ms in
200~$\mu$s steps increased the statistical precision of this method.
Figure~\ref{fig:accidental-Distance} shows the distribution of
distances between the reconstructed positions of the prompt and
delayed signals, ${\Delta}r_{\mathrm{p,d}}$, for the \nuebar{}
candidates and for the accidental background determined using this
method.
Genuine correlated signals favor small ${\Delta}r_{\mathrm{p,d}}$, as
shown by simulation, while accidentals dominate for
${\Delta}r_{\mathrm{p,d}}>2$~m\@.
Both the shape and normalization of the accidental background
distribution agreed with that of the \nuebar{} candidate sample for
${\Delta}r_{\mathrm{p,d}}>2$~m, confirming the estimate of this
background.
The rate determined from the offset-window method was also consistent
with those obtained using Eqs.~\ref{eq:accRateA}
and~\ref{eq:accRateB}, albeit less precise.

%
Uncertainty in the estimation of the accidental background was
negligible.
The largest statistical contribution was from the uncertainty in
$R_{\mathrm{d}}$, which was 0.2\% in the near detectors and 0.4\% in
the far detectors.
Variation in the methods used to select only isolated prompt-like
signals resulted in 0.3\% variation in $R_{\mathrm{p}}$ in the most
extreme cases.
In the end, the total uncertainty was dominated by the 1\% precision
of the cross-check of the accidental rate using the offset-window
method, which was used as a conservative estimate of potential
systematic uncertainty.
Given that the estimated accidental background contamination amounted
to only 1\% to 2\% of the antineutrino candidates, accidentals
contributed $\lesssim$0.01\% uncertainty to the observed rate of
antineutrinos.

\subsubsection{\label{subsubsec:bkg_corr}Correlated Backgrounds}

%
The remaining backgrounds were from physical processes which produced
correlated pairs of prompt and delayed signals, capable of mimicking
\nuebar{} inverse beta decay.
Six potential sources of correlated backgrounds were identified:
\begin{itemize}
\item Muons: Muons produced in cosmic ray showers would fragment
  stable nuclei in or near the detectors, creating free neutrons and
  unstable nuclei.  The signals from the initial muon interaction,
  subsequent neutron captures, or beta decays of unstable nuclei could
  potentially form a pair which satisfied the antineutrino selection.
  Muons were easily identified by their large scintillation light
  production or by Cherenkov light in the water shield.  Vetoing
  concurrent or subsequent signals rendered this background
  negligible, at the cost of 2\% of the far-hall and 14\%--18\% of the
  near-hall livetimes.
\item Fast neutrons: Muon interactions in the environment near the
  detector generated energetic, or {\em fast}, neutrons.  A nuclear
  collision of a fast neutron within the scintillator could mimic a
  prompt signal, while the subsequent neutron capture was identical to
  a true IBD delayed signal.  The contamination from this background
  was $\lesssim$0.1\%.
\item $\beta$-n decays: Muon interactions occasionally produced the
  rare unstable isotopes $^9$Li and $^8$He, which $\beta$-decay with a
  chance of simultaneously emitting a neutron.  The muon veto reduced
  contamination from this background to 0.3\%--0.4\% depending on
  experimental hall, with a $\sim$50\% systematic uncertainty.
\item $^{241}$Am-$^{13}$C neutron sources: During detector operation
  it was found that neutrons from the $^{241}$Am-$^{13}$C calibration
  sources within the ACUs occasionally introduced several
  $\gamma$-rays, correlated in time, into the detector.  Contamination
  from this backgrounds was $\lesssim$0.1\%.
\item ($\alpha$,n) interactions: $\alpha$'s emitted by natural
  radioactivity within the detector could eject neutrons from stable
  nuclei, with $^{13}$C($\alpha$,n)$^{16}$O being the most prevalent
  interaction.  Protons scattered by the neutron or $^{16}$O$^*$
  de-excitation $\gamma$-rays could mimic a prompt signal, while the
  eventual neutron capture provided a delayed signal.  Contamination
  from this background was $\lesssim$0.07\%.
\item High-multiplicity signals: The pile-up of correlated \nuebar{}
  signal pairs with uncorrelated radioactive backgrounds resulted in
  three or more signals within the correlation time for IBD
  candidates.  While not a background per se, such sets of signals
  with multiplicity $\geq$3 resulted in ambiguity in determination of
  the actual prompt and delayed \nuebar{} signals.  Complete rejection
  of these high-multiplicity combinations resolved this ambiguity, but
  introduced a $\sim$2.5\% loss of \nuebar{} efficiency.
\end{itemize}
The characteristics of each of these backgrounds, the methods used to
mitigate them, and their residual contamination in the \nuebar{}
candidates will be discussed in the following sections.
The mitigation methods, e.g.\ the muon and high-multiplicity vetoes,
and their associated impact on \nuebar{} detection efficiency, are
also presented here.
Of the correlated backgrounds, only $\beta$-n decays contributed
enough residual background contamination as well as systematic
uncertainty to significantly impact the oscillation measurement.

\paragraph{\label{prgph:bkg_cosmicrays}Muon Background and Veto}

%
Minimization of the muon background was a primary reason behind the
selection of the Daya Bay site for the measurement of neutrino
oscillation.
The mountains directly adjacent to the nuclear power facility provided
ample shielding for each experimental hall, as listed in
Table~\ref{tab:site_survey}\@.
The attenuated muon fluxes for EH1, EH2, and EH3 were found to be
1.16, 0.86, and 0.054~Hz/m$^2$ respectively, defined according to a
spherical acceptance~\cite{Dayabay:2014vka}.

%
Muons which traversed the antineutrino detectors deposited an average
of $\sim$0.6~GeV in the scintillator, and were easily discriminated
from antineutrino signals.
Selection A identified these {\em AD muons} as any signal with
reconstructed energy $>$20~MeV, while selection B used any signal with
$>$3,000~PE ($\gtrsim$18~MeV).
Any delayed-like candidate signal following an AD muon signal within a
veto time window of $t^{\mathrm{veto}}_{\mathrm{AD}\mu}$ was rejected.
This veto was applied independently for each antineutrino detector.
For selection A, $t^{\mathrm{veto}}_{\mathrm{AD}\mu} = 1$~ms, while
selection B used a longer 1.4~ms veto to avoid correlations between
the muon and multiplicity veto efficiencies.
The length of the veto was dictated by the timescale for neutron
capture in the antineutrino detectors.
Although the veto may seem rather long in comparison with the
$\sim$28~$\mu$s neutron capture time in the GdLS target, some neutrons
lingered in the LS region and slowly diffused into the GdLS\@.
Given a mean neutron capture time in the LS region of
$\sim$200~$\mu$s, the veto corresponded to an $\sim\exp(-5)$ (or
$\sim\exp(-7)$, for selection B) suppression for such neutrons.
%

%
Both of the selections identified {\em Water Shield muons} as any
signal in either the IWS or OWS in which more than 12~PMTs were above
threshold.
Neutrons generated by muon interactions in the water shield also had
the potential to reach the GdLS target, although with a much lower
probability.
A shorter veto time of $t^{\mathrm{veto}}_{\mathrm{WS}\mu} =
600$~$\mu$s was sufficient to reject these neutrons.
As discussed in Sec.~\ref{sec:experiment}, the IWS, OWS, and AD were
all independently triggered.
Small differences in detector latency resulted in some AD signals
arriving in advance of simultaneous signals in the water shield.
Therefore, any delayed-like AD signal which occurred in the 2~$\mu$s
preceding a WS muon was also rejected.
This veto was applied to all ADs within an experimental hall for both
selections A and B\@.
%

%
A minority of muons produced a significantly higher proportion of
neutrons and unstable nuclei in the ADs.
Such interactions were associated with scintillation light production
in excess of that expected for minimum-ionizing muons, and were
assumed to be associated with muon-induced particle showers.
Selection A identified these {\em AD showers} as any signal with
reconstructed energy $>$2.5~GeV, while selection B used any signal
with $>$30,000~PE ($\gtrsim$1.8~GeV).
The veto time following these signals,
$t^{\mathrm{veto}}_{\mathrm{shower}}$, was significantly longer; 1~s
for selection~A and 0.4~s for selection~B\@.
This significantly longer veto was necessitated by the 178.3~ms
half-life of $^9$Li, as will be discussed.
This veto was applied independently to each detector for both
selections A and B\@.

%
The combination of these muon veto criteria resulted in negligible
background contamination from muons, with the exception of fast
neutrons and $\beta$-n decays discussed in the following sections.

%
The veto reduced the livetime for \nuebar{} detection, which was
quantified as an effective contribution $\epsilon_{\mu}$ to the
antineutrino selection efficiency for each detector.
This efficiency was directly measured from the data using
\begin{equation}\label{eq:mu_eff}
  \epsilon_{\mu} = \left(\sum_it^i_{\mathrm{s}}\right)/t_{\text{DAQ}},
\end{equation}
where $t^{i}_{\mathrm{s}}$ were the individual segments $i$ of
livetime between each vetoed period in a detector.
The total DAQ livetime, $t_{\mathrm{DAQ}}$, was the time between the
first and last signal in the data period, accounting for gaps in the
data due to downtime and periods of poor data quality.
The resulting efficiency was $\sim$82\%, $\sim$86\%, and $\sim$98\%
for the detectors in EH1, EH2, and EH3, respectively.
The dominant uncertainty in this calculation was from jitter in the
recorded time of each signal, but was found to be negligible.

\paragraph{\label{prgph:bkg_fastn}Fast Neutron Background}

%
While muon interactions within the detectors or water shield were
efficiently identified, interactions with the cavern rock surrounding
the experimental hall were missed.
Neutrons produced in these interactions could reach the detectors
without producing a detectable signal in the water shield.
In order to attenuate these neutrons, the thickness of the water
shield surrounding the ADs was at least 2.5~m in all directions.
As a result, only the most energetic, or {\em fast}, neutrons had the
potential to penetrate all the way to the GdLS target.
A fast neutron could stop in the scintillator target through an
energetic collision with a nucleus, primarily $^1$H\@.
Ionization of the scintillator by a recoiling proton could mimic a
prompt \nuebar{} interaction, while the slowed neutron could capture
and provide a delayed signal.

%
The broad and smooth energy spectrum of recoils from fast neutrons
resulted in a reconstructed prompt spectrum that was approximately
flat up to energies of $\sim$100~MeV\@.
Fast neutrons were the dominant correlated signal at these high
energies.
Correlated signals with prompt energy greater than the selection for
\nuebar{} interactions, $E_{\mathrm{p}}>$12~MeV, were used to directly
measure this background.
A smooth extrapolation of the background into the \nuebar{} prompt
energy range, 0.7~MeV$< E_{\mathrm{p}}<$12~MeV, provided an initial
estimate of the contamination of this background in the \nuebar{}
sample.

\begin{figure}
  \includegraphics[width=\columnwidth]{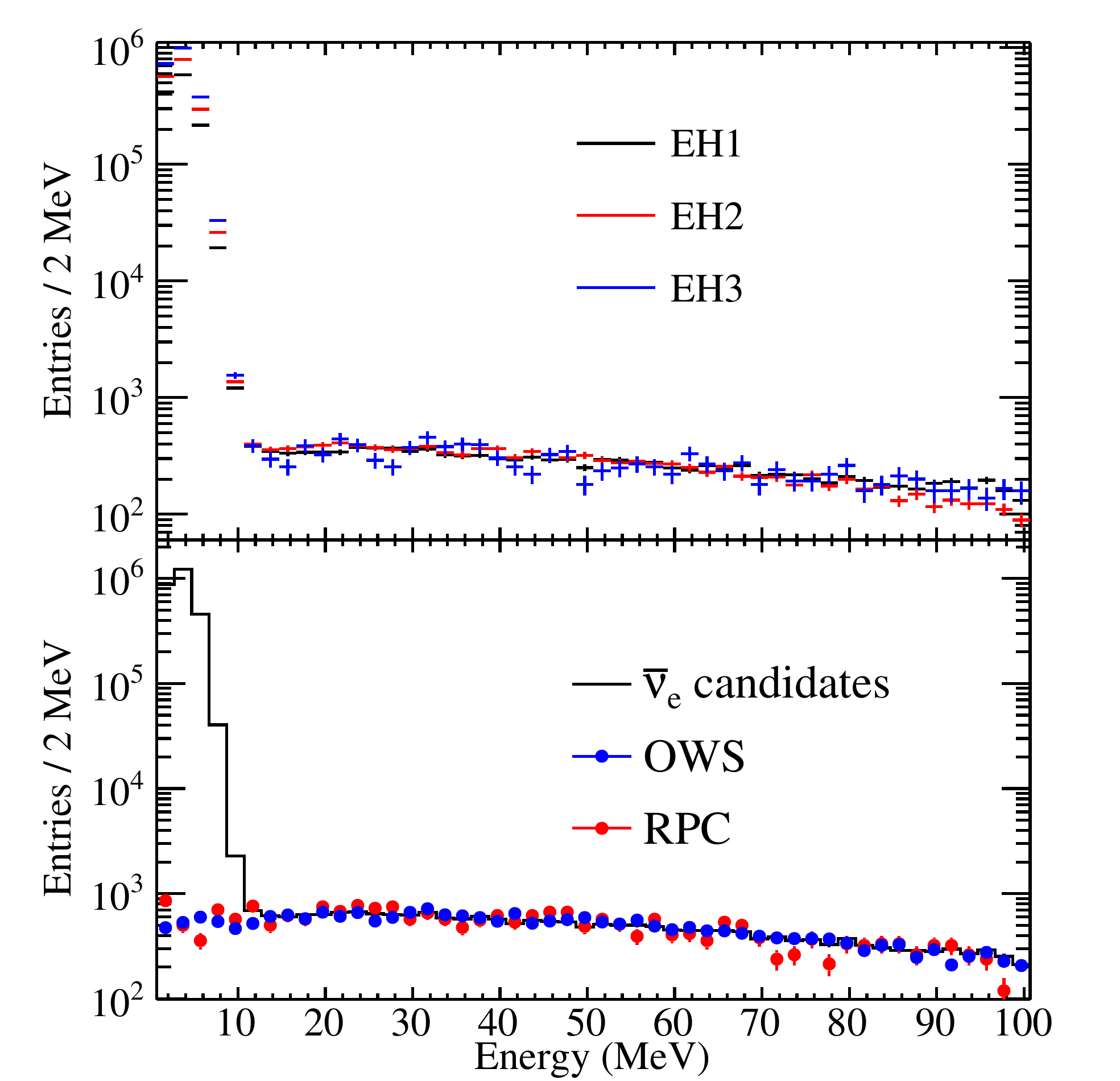}
  \caption{({\em Top}) An expanded selection of \nuebar{} candidate
    signals with prompt energies from 0.7~MeV to 100~MeV revealed a
    broad continuum of background, attributed to energetic, or {\em
      fast}, neutrons.  Actual \nuebar{} signals were visible as the
    dominant signals below $\sim$12~MeV\@.  The energy spectra of the
    fast neutron background above 12~MeV were similar in all three
    experimental halls (EH1: black, EH2: red, EH3: blue).  ({\em
      Bottom}) The sum of the \nuebar{} candidate spectra from all
    three halls (black line) was compared to a high-purity fast
    neutron sample obtained by selecting those \nuebar{} candidate
    signals which had been vetoed in the 200~$\mu$s after an OWS or an
    RPC, but not an IWS, muon signal (OWS: blue, RPC: red).  The
    energy spectra of the vetoed samples were consistent with the
    background observed above 12~MeV in the \nuebar{} candidate
    sample.  The normalization of the vetoed samples were adjusted to
    match the \nuebar{} candidates above 12~MeV\@.  The fast neutron
    contamination in the \nuebar{} candidate sample below 12~MeV was
    estimated using the vetoed sample.
    \label{fig:fastn_spectrum}}
\end{figure}

%
Although simulation supported the validity of a linear extrapolation
of this background into the \nuebar{} signal region, a method which
solely relied on data was more robust.
The energy spectrum of prompt recoils was directly measured using
\nuebar{}-like correlated signals coincident with muons.
In particular, a sample of {\em boundary muons} were selected when a
muon was identified only in the OWS or the RPC, but not in the IWS\@.
Figure~\ref{fig:fastn_spectrum} compares the observed prompt energy
spectra for the standard antineutrino candidate sample for selection A
but with the prompt signal extended up to 100~MeV\@.
The spectra for candidates whose delayed signal was within 200~$\mu$s
after a boundary muon are also shown.
The energy spectra above 12~MeV were consistent for the two samples,
and the prompt recoil spectrum in the range of 0.7~MeV to 12~MeV is
clearly visible for the boundary muon sample.
The spectra from only the OWS or the RPC boundary muon samples are
also consistent with each other.

%
The residual fast-neutron contamination rate $R^i_{\mathrm{fn}}$ in
the \nuebar{} sample for each detector $i$ was estimated from the
observed rate of \nuebar{}-like candidates with high prompt energy,
12~MeV $\le E_{\mathrm{p}}\le$ 99~MeV, $R^i_{\mathrm{he}}$.
The fast neutron sample identified, or {\it tagged}, using boundary
muons was used to estimate the proportion of the background in the low
energy region of interest, 0.7~MeV$< E_{\mathrm{p}} <$12~MeV\@.
The tagged spectra were consistent between detectors, so the entire
tagged sample from all eight detectors was combined to improve the
statistical accuracy of this technique.
From the tagged fast neutron rate at low, $R^{\mu}_{\mathrm{fn}}$, and
high, $R^{\mu}_{\mathrm{he}}$, prompt energies, the actual fast
neutron background rate was determined as
\begin{equation}
  R^i_{\mathrm{fn}} = R^i_{\mathrm{he}} \frac{R^{\mu}_{\mathrm{fn}}}{R^{\mu}_{\mathrm{he}}}.
\end{equation}
This method found the fast neutron contamination in the \nuebar{}
candidate sample to be less than $\lesssim$0.1\% for all detectors.
The contamination varied by a negligible amount, $\lesssim$0.03\%,
assuming a range of variations in spectra between the tagged and
untagged fast neutron samples.

\paragraph{\label{prgph:bkg_betan}$\beta$-n Decay Background}

%
An unstable nuclide which $\beta$-decayed with the simultaneous
emission of a neutron generated a correlated pair of signals nearly
identical to those from \nuebar{} inverse beta decay.
The production of $\beta$-n nuclides by muon spallation in organic
liquid scintillator has been measured~\cite{Abe:2009aa}.
$^9$Li, with a lifetime of $\tau=257.2$~ms and a maximum $\beta$
energy of 13.6~MeV, and $^8$He, with a lifetime of $\tau=171.8$~ms and
maximum $\beta$ energy of 10.7~MeV, are the most prominent $\beta$-n
nuclides.
A FLUKA-based simulation suggested that the dominant production method
was the fragmentation of $^{12}$C by $\pi^{-}$ in muon-induced
hadronic showers, with a relative yield of 10:1 for these two
$\beta$-n nuclides~\cite{Abe:2009aa}.

%
The natural site-dependent variation in $\beta$-n production presented
the most direct route for a potential bias in the far versus near
detector measurement of neutrino oscillation.
As will be shown, uncertainty in the $\beta$-n background contributed
the most significant systematic uncertainty in the oscillation
measurement.

%
$\beta$-n background was discriminated from \nuebar{} interactions by
association with a preceding muon signal in the same detector.
This association was complicated by the high muon rates in the
detectors relative to the lifetime of these nuclides, rendering
signal-by-signal discrimination impossible.
Still, the distribution of times between each \nuebar{} candidate and
the most recent AD muon signal could be used to estimate the $\beta$-n
production rate~\cite{Wen:2006hx}, allowing statistical estimation of
this background.
The distribution of total correlated signal rate $R_{\mathrm{total}}$
versus time since the most recent muon signal was expressed
\begin{equation}\label{eq:Li9_time_fit}
  R_{\mathrm{total}} = R_{\mathrm{IBD}}R_{\mu}e^{-R_{\mu}t} + \sum_i^{\mathrm{^9Li,^8He,^{12}B}} R_i\lambda_i e^{-\lambda_it},
\end{equation}
where $\lambda_{i}=R_{\mu}+\frac{n_i}{\tau_{i}}$ was the effective
time constant for nuclide $i$ with lifetime $\tau_{i}$ and rate
$R_{i}$, accounting for the impact of the muon rate $R_{\mu}$\@.
The factor $n_i$ is 1 for $^9$Li and $^8$He.
A term for $^{12}$B with $n_i$ equal 2 was included to accommodate a
potential increase in the background following muons, where two
$^{12}$B decays mimicked a prompt and delayed \nuebar{} signal.
%

\begin{figure}[h!]
\includegraphics[width=.5\textwidth]{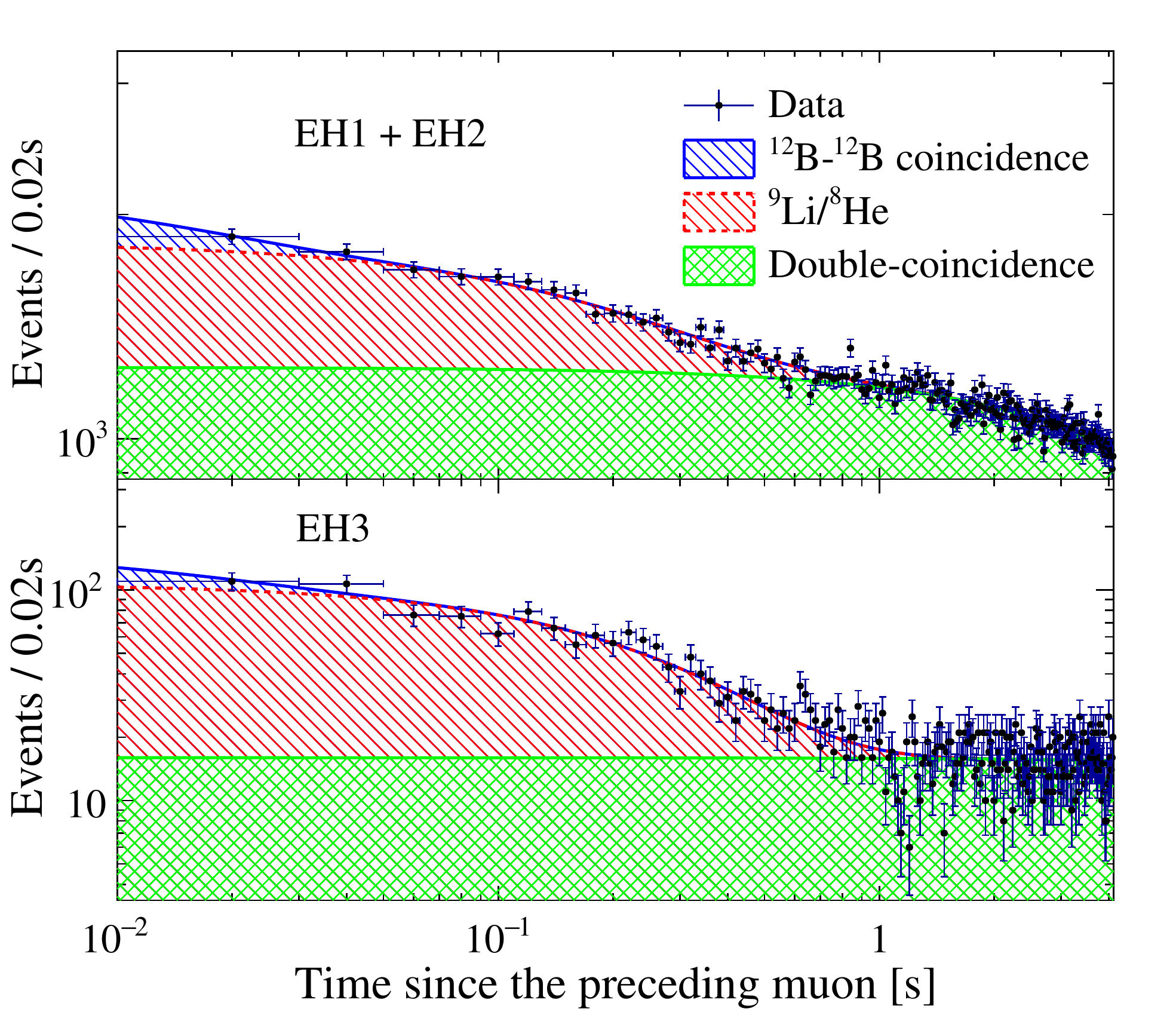}
\caption{The distribution of \nuebar{} inverse beta decay candidates
  versus time since the most recent muon-induced particle shower
  ($E_{rec} > 2.5$~GeV) in the same detector (blue points).  A fit to
  this distribution statistically distinguishes the muon-uncorrelated
  \nuebar{} and accidentals (green hatched region), from the
  muon-correlated background from $\beta$-n decay (red lined region).
  A minor contribution from accidental background, where the delayed
  signal was from $^{12}$B $\beta$ decay, is barely visible in the
  first bin (blue lined region).  A more restrictive \nuebar{}
  selection of 3.5~MeV~$< E_{\mathrm{p}} <$~12~MeV, and 1~$\mu$s~$<
  {\Delta}t <$~100~$\mu$s suppressed the accidental contribution,
  providing a more precise estimate of the $\beta$-n background.
\label{fig:time_since_last_muon}}
\end{figure}

%
A fit to the distribution of $R_{\mathrm{total}}$ for all \nuebar{}
candidates, not including the muon veto, unfortunately does not
provide a precise estimate of the $\beta$-n background.
As reported previously~\cite{Abe:2009aa}, unstable isotope production
was generally associated with muons which yield significantly higher
scintillation light and was attributed to energetic particle showers.
Limiting the time distribution to the most recent {\em AD shower}, as
defined earlier, enhanced the signature of $\beta$-n decays relative
to \nuebar{} signals.
Also, accidental background in the distribution was suppressed by
further limiting the sample to correlated signals with 3.5~MeV~$<
E_{\mathrm{p}} <$~12~MeV, and 1~$\mu$s~$< {\Delta}t <$~100~$\mu$s\@.
From the modeled spectra of $^9$Li and $^8$He, this selection reduced
the $\beta$-n acceptance to 67\%\@.
Figure~\ref{fig:time_since_last_muon} shows the resulting distribution
for selection A, with the contribution from $^{9}$Li and $^{8}$He
clearly visible.

%
The observed $\beta$-n decay constant preferred $^{9}$Li over
$^{8}$He, consistent with the observation by the KamLAND experiment.
Varying the $^8$He contribution from 0 to 15\% resulted in only a 4\%
change in the estimated $\beta$-n rate.
The veto following AD showers, 1~s for selection~A and 0.4~s for
selection~B, removed the majority of this background.
The residual background contamination following showering muons was
estimated at $<$0.01\% from integration of the tail of the fitted
$\beta$-n time distribution beyond the veto window.

%
Although $\beta$-n production was primarily associated with
muon-induced particle showers, a fraction of this background was also
produced by those muons with no associated shower.
While the $\beta$-n yield was expected to be much lower for an
individual non-showering muon, this was compensated by the much higher
non-showering muon rate.
The high muon rate resulted in a time-since-muon distribution where the
time constants for the $\beta$-n background and \nuebar{} signal are
nearly degenerate.
Therefore, an alternate technique was required.
Given that $\beta$-n production required nuclear fragmentation, the
presence of free neutrons was correlated with unstable nuclide
production~\cite{Li:2015lxa}.
A neutron-tagged muon sample was identified as those muons which were
followed within 20~$\mu$s to 200~$\mu$s by a potential neutron capture
signal, identified using the loose criteria
1.8~MeV$<$$E_{rec}$$<$12~MeV\@.
Table~\ref{tab:betan_rates} summarizes the number of $\beta$-n decays
for all muons as well as for the the neutron-tagged subset, as
estimated using the time distribution to the preceding muon
(e.g. Fig.~\ref{fig:time_since_last_muon})\@.
\begin{table}[!htb]
  \caption{The estimated number of $\beta$-n decays in the \nuebar{}
  candidate sample from each experimental hall, based on the
  characteristics of the preceding muon.  The values were determined
  from modeling the distribution of time since the preceding muon, as
  shown in Figure~\ref{fig:time_since_last_muon}.  Modeling the
  distribution for all muons with
  0.02~GeV$<$$E^{\mu}_{\mathrm{rec}}$$<$1.8~GeV was inconclusive.
  Instead, the values given in parentheses were estimated using the
  neutron-tagged sample, with the assumption that the efficiency of
  the neutron tagging of $\beta$-n production was independent of
  $E^{\mu}_{\mathrm{rec}}$.
  \label{tab:betan_rates}}
\begin{ruledtabular}
\begin{tabular}{crrr}
Hall & \multicolumn{3}{c}{Muon energy deposition, $E^\mu_{\mathrm{rec}}$ [GeV] } \\
     & 0.02--1.8 & 1.8--2.5 & $>$2.5 \\
\hline
     & \multicolumn{3}{c}{$\beta$-n decays (all muons)} \\
EH1  & (2106$\pm$193) & 1169$\pm$185 & 4086$\pm$177 \\
EH2  & (1282$\pm$165) &  879$\pm$162 & 3065$\pm$154 \\
EH3  &  (276$\pm$36)  &  167$\pm$29  & 1142$\pm$43 \\
\hline
     & \multicolumn{3}{c}{$\beta$-n decays (neutron-tagged muons)} \\
EH1  & 1847$\pm$169 & 818$\pm$63 & 3793$\pm$97 \\
EH2  & 1087$\pm$140 & 614$\pm$47 & 2730$\pm$81 \\
EH3  &  245$\pm$32  & 120$\pm$14 &  994$\pm$34 \\
\end{tabular}
\end{ruledtabular}
\end{table}

More than half of the $\beta$-n decays followed the small number of
muons with associated particle showers, identified using
$E^{\mu}_{\mathrm{rec}}>$2.5~GeV\@.
Muons with 1.8~GeV$<$$E^{\mu}_{\mathrm{rec}}$$<$2.5~GeV contributed
$<$20\% of the estimated $\beta$-n decays.
The estimates for muons with lower reconstructed energy,
0.02~GeV$<$$E^{\mu}_{\mathrm{rec}}$$<$1.8~GeV, were inconclusive due to
degeneracy of the $\beta$-n and \nuebar{} time constants in the
distribution relative to those muons.
Instead, $\beta$-n production by these muons was estimated using the
neutron-tagged muon sample.
For muons with $E^{\mu}_{\mathrm{rec}}$$>$1.8~GeV, $\sim$86\% of
$\beta$-n production was found to follow neutron-tagged muons.
The neutron-tagging efficiency for $\beta$-n production was assumed to
be the same for those muons with $E^{\mu}_{\mathrm{rec}}$$<$1.8~GeV\@.
A limited variation in the observed tagging efficiency versus
$E^{\mu}_{\mathrm{rec}}$ was consistent with this assumption, with a
$\sim$40\% systematic uncertainty.

%
The total $\beta$-n background in the final \nuebar{} sample was
determined using the values given in Table~\ref{tab:betan_rates},
corrected for the selection efficiency and sample livetime.
The showering muon veto efficiently rejected the contribution from
muons with $E^{\mu}_{\mathrm{rec}}$$>$2.5~GeV\@.
The contamination was estimated at (0.37$\pm$0.16)\%,
(0.29$\pm$0.13)\%, and (0.20$\pm$0.08)\% per detector in EH1, EH2, and
EH3 respectively.

%
An independent analysis of the $\beta$-n background was done for
selection~B\@.
Although the shower muon threshold for selection~B was more stringent,
the veto time was less so.
Analysis of the time distribution to all muons in the past 5~s,
instead of only the most recent, resulted in a flat distribution for
muon-uncorrelated \nuebar{} and accidental signals.
Discrimination of the muon-correlated $\beta$-n component from this
flat distribution was easier in principle, but came at the cost of
increased statistical uncertainty from the larger number of muons
considered in the study.
Despite these alternate choices, this analysis found a similar
background contamination of (0.41$\pm$0.14)\%, (0.32$\pm$0.12)\%, and
(0.30$\pm$0.09)\% per detector in EH1, EH2, and EH3 respectively.

\begin{figure}[h!]
\includegraphics[width=.5\textwidth]{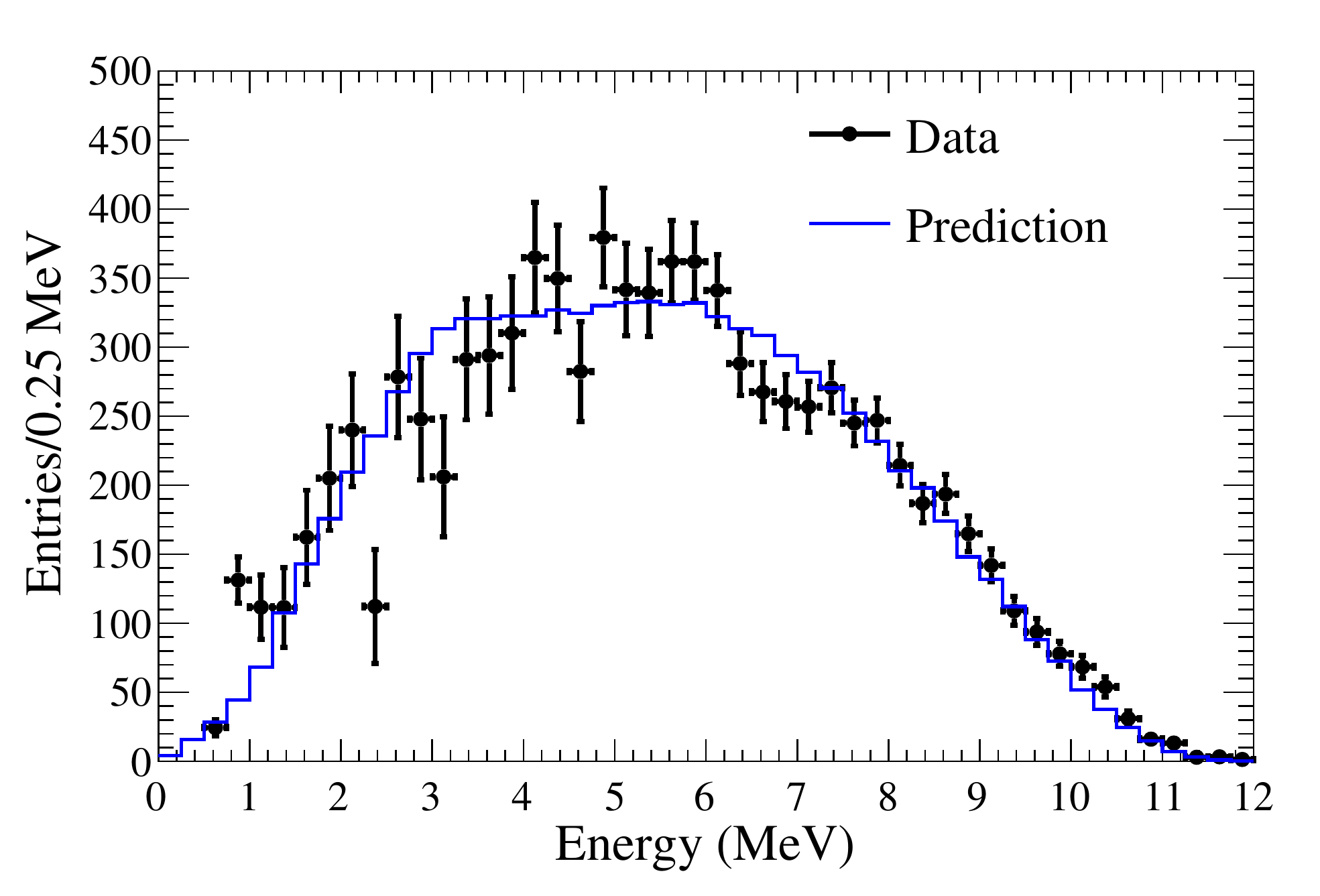}
\caption{A comparison of the observed (black points) and predicted
  (blue line) prompt reconstructed energy spectrum from $^{9}$Li
  $\beta$-n decay.  The observed spectrum was obtained from those
  \nuebar{} candidate interactions which followed muon-induced
  particle showers, with the expected background from \nuebar{} and
  accidentals subtracted.  The prediction was estimated considering
  the known $\beta$-n decay branches and contribution from the nuclear
  break-up to a neutron and two $\alpha$'s.
  \label{fig:li9_spec}}
\end{figure}

%
In addition to measuring the $\beta$-n background rate,
characterization of the energy spectrum was also necessary for the
spectral analysis of neutrino oscillation.
The prompt energy spectrum was determined from the $\beta$-n sample
following muon-induced particle showers.
An off-time window from 10~ms to 1010~ms after the shower was used to
measure and then subtract the \nuebar{} and accidental contribution to
the spectrum.
Figure~\ref{fig:li9_spec} shows the measured prompt energy spectrum.
The poorer statistical precision at lower prompt signal energies was a
consequence of the subtraction of the \nuebar{} and accidental
signals.
A prediction of the prompt energy spectrum was calculated from the
known $\beta$-n decay of $^{9}$Li.
After decay, the nucleus fragments into an energetic neutron and two
$\alpha$-particles.
An empirical model was constructed to match the observed branching
$\beta$-n decay fractions~\cite{Tilley:2004zz}, and observed neutron
and $\alpha$ energy spectra~\cite{Nyman1990189,Bjornstad:1981ad}.
Although their impact on the spectrum was minor, the contribution of
the neutron and $\alpha$'s to the observed energy were included based
on the detector response model previously discussed.
The spectrum was consistent with sole production of $^9$Li, as also
suggested by the fit to the lifetime.

\paragraph{\label{prgph:bkg_amc}$^{241}$Am-$^{13}$C Neutron Source Background}

%
The three automated calibration units (ACUs) present on the lid of
each antineutrino detector each contained a low-intensity
$^{241}$Am-$^{13}$C neutron source which was used to assess the
response of the detector to neutron capture~\cite{Liu:2015cra}.
When not in use for calibration, these $\sim$0.7~Hz neutron sources
were withdrawn into each ACU\@.
Concerns about correlated backgrounds motivated the choice and design
of the neutron source.
For example, $^{252}$Cf was avoided due to the emission of multiple
neutrons and associated $\gamma$-rays.
In the chosen design, $\alpha$'s from an $^{241}$Am source were
degraded using a thin gold foil.
With their energy reduced, they were below the threshold for
$\gamma$-ray production upon interaction with $^{13}$C.
Design simulations had demonstrated that there was a negligible
probability that a neutron emitted by the withdrawn source would reach
the GdLS target and introduce background.

%
Despite these efforts, correlated signals were produced by the
following rare mechanism.
First, a neutron collided inelastically with an Fe, Cr, Mn, or Ni
nucleus, present in the stainless steel of the detector, which emitted
prompt $\gamma$-rays.
The neutron was subsequently captured by a nucleus in the stainless
steel, or by Gd present in the GdLS overflow tank, generating delayed
$\gamma$-rays.
When an energetic $\gamma$-ray from each of these prompt and delayed
interactions happened to penetrate into and interact in the
scintillator region of the detector, it could mimic a \nuebar{}
interaction.

%
Initial observation of the background from the $^{241}$Am-$^{13}$C
sources came from an excess of uncorrelated $\gamma$-rays signals
above the delayed signal threshold of $E_{\mathrm{rec}}>6$~MeV\@.
This $\gamma$-ray excess had reconstructed positions primarily in the
upper half of each detector~\cite{CPC}, with a rate and energy
spectrum in agreement with simulation.
As discussed in Sec.~\ref{subsubsec:bkg_uncorr}, these uncorrelated
$\gamma$-rays enhanced the rate of accidental backgrounds.
Monte-Carlo studies suggested the corresponding correlated background,
discussed previously.
The detailed modeling of this background is discussed
in~\cite{Gu:2015inc}.

\begin{figure}[!htb]
  \includegraphics[width=0.45\textwidth]{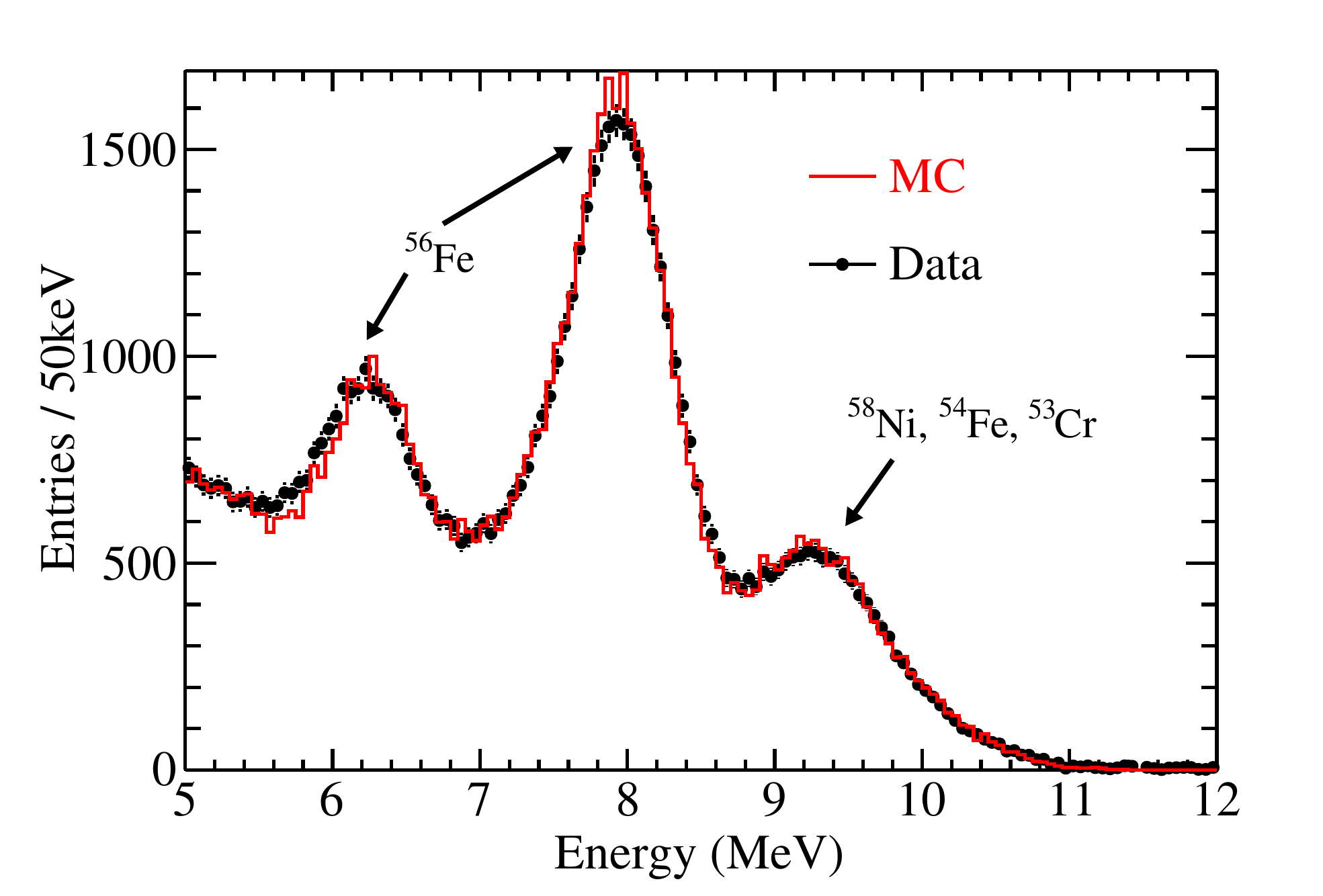}
  \caption{Uncorrelated individual signals with
    $E_{\mathrm{rec}}>5$~MeV generated by a $\sim$59~Hz
    $^{241}$Am-$^{13}$C neutron source temporarily installed on the
    lid of EH3-AD5 (black points).  Peaks in the spectra are
    attributable to neutron capture on Fe, Cr, or Ni nuclei present in
    stainless steel of the lid, where an energetic capture
    $\gamma$-ray subsequently penetrates to the scintillating region
    of the detector.  The measurement was used to benchmark a
    Monte-Carlo simulation of the uncorrelated and correlated
    background induced by this source (red solid line), and thereby
    estimate the correlated background induced by the regular weak
    $\sim$0.7~Hz $^{241}$Am-$^{13}$C sources present in the automated
    calibration units (ACUs), also on the detector lid.
    \label{fig:s04_amc_strong_single}}
\end{figure}

\begin{figure}[!htb]
  \includegraphics[width=0.45\textwidth]{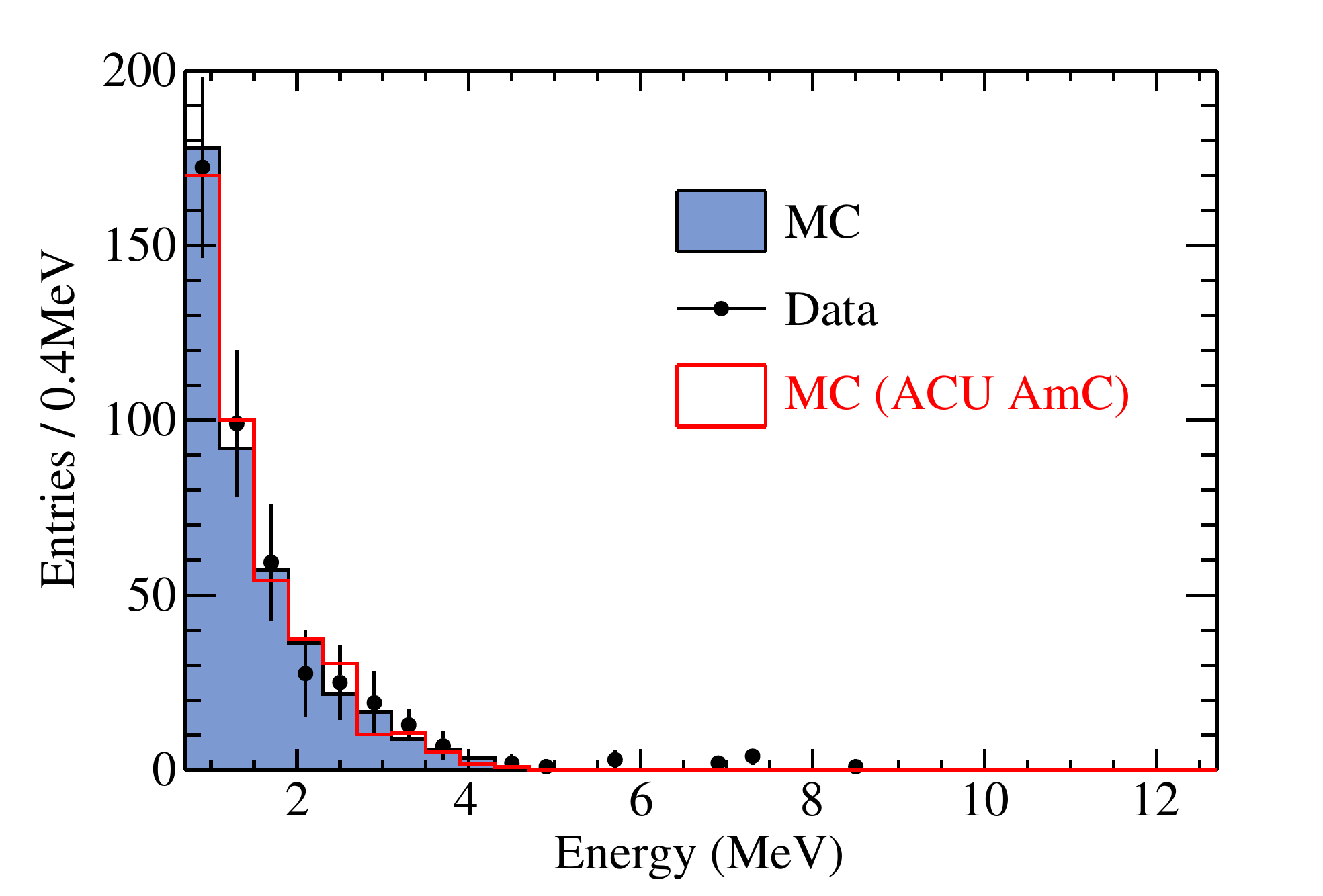}
  \caption{The prompt reconstructed energy spectrum from correlated
    \nuebar{}-like signal pairs induced by a $\sim$59~Hz
    $^{241}$Am-$^{13}$C neutron source temporarily installed on the
    lid of EH3-AD5 (black points).  Inelastic collisions of a neutron
    with nuclei present in stainless steel produced $\gamma$-rays
    which occasionally penetrated to the scintillating region of the
    detector, generating signals with a steeply-falling energy
    spectrum.  A benchmarked Monte-Carlo simulation predicted a
    consistent spectrum (solid blue region).  The prompt energy
    spectrum for correlated background from the regular $\sim$0.7~Hz
    $^{241}$Am-$^{13}$C sources present in the automated calibration
    units (ACUs) was estimated to be identical, within the statistical
    precision of the simulation (red line).
    \label{fig:s04_amc_strong_prompt}}
\end{figure}

%
Simulation of the correlated background was uncertain since it
depended on accurate knowledge of many factors:
the initial neutron energy spectrum, the $\gamma$-ray spectra of
inelastic nuclear collisions, the eventual location and isotope of the
neutron capture, and the penetration plus degradation of the
$\gamma$-rays which reach the scintillator all impacted the modeling
of this background.
Therefore, an empirical approach was used to assess this background.
A more intense $\sim$59~Hz $^{241}$Am-$^{13}$C source was prepared and
installed on the lid of EH3-AD5 for ten days during the summer of
2012.
An increase in the rate of uncorrelated single interactions with
$E_{\mathrm{rec}}>6$~MeV was measured in the upper half of the
detector.
An extra 613$\pm$64 correlated backgrounds were observed in AD5
relative to the neighboring detector AD4, after accounting for the
enhanced accidental background in the former.
For the intense $^{241}$Am-$^{13}$C source, the ratio of the rates of
correlated to uncorrelated signals,
\begin{equation}
  f^{\mathrm{int}} = \frac{R^{\mathrm{int}}_{\mathrm{corr}}}{R^{\mathrm{int}}_{\mathrm{uncorr}}},
\end{equation}
was $(1.5\pm0.3)\times10^{-3}$\@.
The simulation of this background was benchmarked against the observed
signal rates, energy spectra, and distribution of reconstructed
positions for both correlated and uncorrelated signals.
Figure~\ref{fig:s04_amc_strong_single} shows the energy spectrum of
uncorrelated $^{241}$Am-$^{13}$C signals with $E_{\mathrm{rec}}>5$~MeV
measured using the intense source.
Simulation produced a consistent spectrum, with peaks attributed to
prominent $\gamma$-rays from neutron capture on stainless steel.
For correlated signal pairs, the $\gamma$-ray spectrum from prompt
inelastic neutron collisions on nuclei in stainless steel is shown in
Figure~\ref{fig:s04_amc_strong_prompt}.
The simulation predicted a consistent energy spectrum for prompt
signals from both the intense and regular $^{241}$Am-$^{13}$C sources.
The prompt spectrum shape was effectively modeled using $S(E) =
\exp(-E/E_o)$ with $E_o = 0.8\pm0.1$~MeV\@.

%
Simulation of the regular $^{241}$Am-$^{13}$C sources predicted fewer
correlated signals for an equivalent number of uncorrelated signals,
given that they were located further from the scintillator.
This offset was captured by a simulated double-ratio of the relative
correlated-to-uncorrelated rates for the two sources,
\begin{equation}
  \eta = \frac{f^{\mathrm{reg}}}{f^{\mathrm{int}}},
\end{equation}
where simulation predicted $\eta$ to be 0.8$\pm$0.2.
The differences in rates for data and simulation determined the
systematic uncertainty in $\eta$.

%
The rate of correlated background from the regular $^{241}$Am-$^{13}$C
sources, $R^{\mathrm{reg}}_{\mathrm{corr}}$, was estimated as
\begin{equation}
  R^{\mathrm{reg}}_{\mathrm{corr}} = \eta\,f^{\mathrm{int}}\,R^{\mathrm{reg}}_{\mathrm{uncorr}},
\end{equation}
where $R^{\mathrm{reg}}_{\mathrm{uncorr}}$ was the excess rate of
uncorrelated signals with $E_{\mathrm{rec}}>6$~MeV attributed to the
regular $^{241}$Am-$^{13}$C sources.
This method found a daily background rate of $\sim$0.27$\pm$0.12 per
near detector and $\sim$0.22$\pm$0.10 per far detector during the
operation of the first six detectors.
%
%
While this background was negligible compared to the \nuebar{} rate in
the near detectors, the rate in the far detectors was almost equal to
that of the $\beta$-n background.
To mitigate this background, the $^{241}$Am-$^{13}$C sources present
in ACU-B and ACU-C were removed from each of the far site detectors
during the installation of the final two detectors in 2012.
Removal of these sources reduced the correlated background by a factor
of $\sim$3.
The remaining sources each showed a $\sim$50\% decline in neutron
emission rate during the first two years of operation.
This decline was attributed to scintillator infiltration into the
sources, which reduced the efficiency of ($\alpha$,n) neutron
production.
Combined, the two effects reduced the background rate by a factor of
$\sim$6 for far detector data collected after the summer of 2012.
For the entire data period presented here, the $^{241}$Am-$^{13}$C
background contamination in the \nuebar{} sample was 0.03\% for the
near detectors and 0.09\% for the far detectors.
A 45\% uncertainty in the correlated background rate covered the range
of variations between simulation and measurement obtained when
comparing the regular and intense sources.

\paragraph{\label{prgph:bkg_alphan}$\alpha$-n Interaction Background}

%
The last minor correlated background resulted from ($\alpha$,n)
nuclear interactions.
In these interactions, an $\alpha$-particle, produced by natural
radioactivity in the scintillator, fused with a nucleus in the
detector materials, resulting in the emission of a neutron.
$^{13}$C($\alpha$,n)$^{16}$O was the dominant process of this type, as
determined from the composition of organic scintillator and the known
cross-sections for ($\alpha$,n) interactions.
Details of modeling the $^{13}$C($\alpha$,n)$^{16}$O background are
presented in~\cite{AlphaN}.

%
Capture of the neutron emitted by $^{13}$C($\alpha$,n)$^{16}$O
interactions is identical to the delayed signal from \nuebar{}
interactions.
The false prompt signal was not as obvious, and resulted from three
different potential processes.
In the first process, the neutron was ejected with sufficient energy
such that successive collisions with protons generated enough
scintillation light to mimic a prompt signal with
$E_{\mathrm{rec}}>0.7$~MeV\@.
For the second process, the energetic neutron collided inelastically
with $^{12}$C, leaving this nucleus in the first excited state,
$^{12}$C$^*$(4.4~MeV)\@.
The nucleus would immediately de-excite via emission of a 4.4-MeV
$\gamma$-ray, producing a prompt signal.
In the third process, $\alpha$-particles with energy greater than
$\sim$5~MeV would preferentially leave the $^{16}$O in an excited
state.
The $^{16}$O$^*$ would de-excite via emission of a $\gamma$-ray or an
electron-positron pair.

\begin{figure}[!htb]
  \includegraphics[width=.5\textwidth]{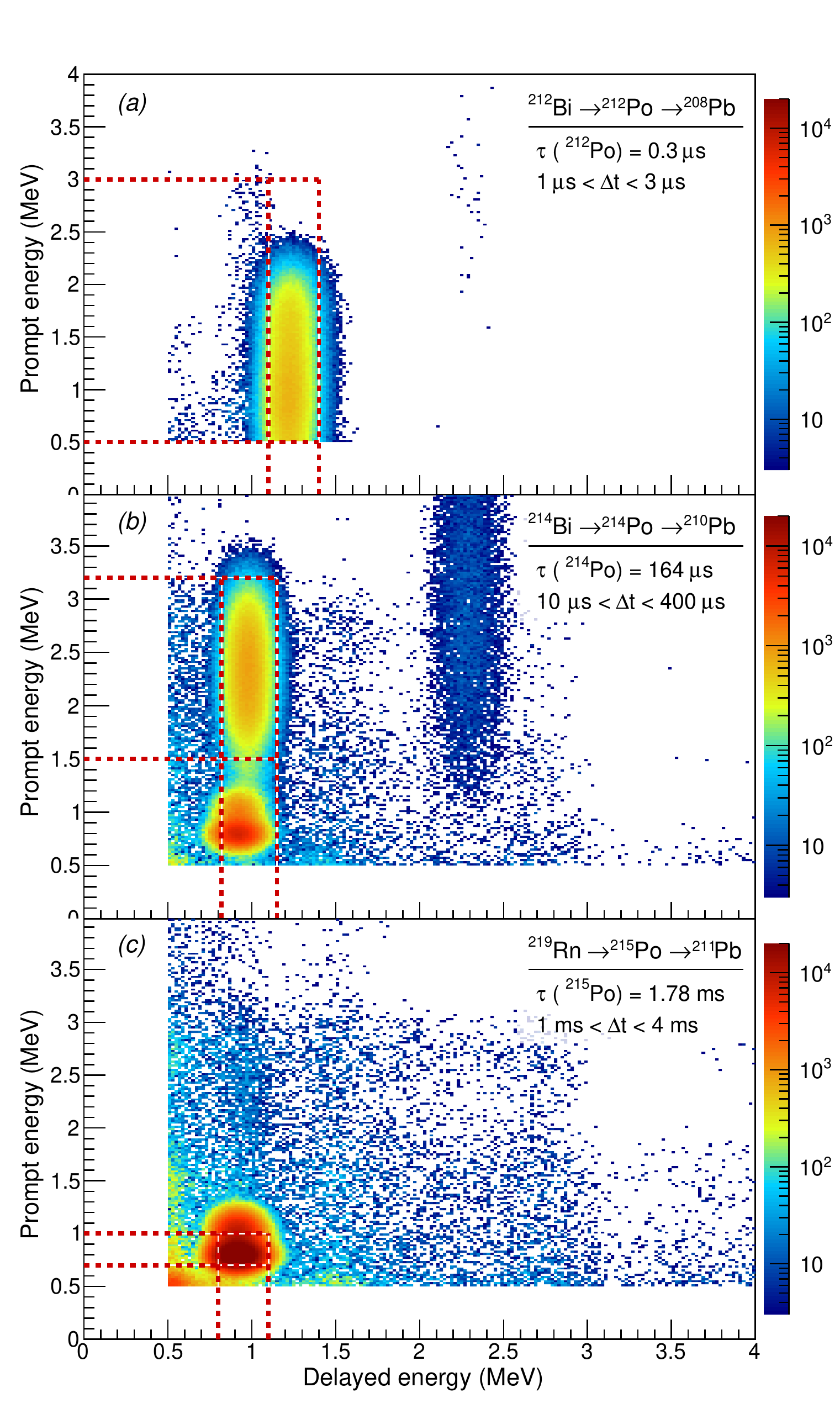}
  \caption{The distributions of low-energy prompt versus delayed
    reconstructed energy for the time intervals (a) 1~$\mu$s to
    3~$\mu$s, (b) 10~$\mu$s to 400~$\mu$s, and (c) 1~ms to 4~ms for
    all detectors combined.  The distributions revealed
    time-correlated signal pairs from actinide contamination in the
    GdLS regions of each detector.  While these interactions were not
    a background for \nuebar{} detection, the resulting estimates of
    actinide $\alpha$-particle activity constrained potential
    background from subsequent $^{13}$C($\alpha$,n)$^{16}$O
    interactions.
    \label{fig:alphaRate}}
\end{figure}

%
The rate and energy of natural alpha activity was determined for each
detector.
Within the GdLS, the dominant alpha sources are the $^{238}$U,
$^{232}$Th, and $^{227}$Ac actinide decay chains.
With chemical properties similar to Gd, these nuclides were introduced
at trace levels during the scintillator doping process.
Each actinide chain contains a polonium cascade decay, which generates
a time-correlated pair of $\beta$-$\alpha$ or $\alpha$-$\alpha$
interactions.
The half-life of the delayed polonium $\alpha$ decay for each actinide
chain is 164.3 $\mu$s, 0.3 $\mu$s, and 1.781 ms, respectively.
The scintillator quenched the alpha from polonium decay, as discussed
in Sec~\ref{sec:abs_energy}, producing a delayed signal with
$E_{\mathrm{rec}} \approx 1$~MeV\@.
Given the low-energy of the delayed signal, it did not mimic a
\nuebar{} interaction.
Instead, the time-correlated signals were used to determine the
$\alpha$ production rate for each of these actinide decay chains.
Figure~\ref{fig:alphaRate} shows the correlated prompt-delayed energy
distributions for various time intervals corresponding to these Po
cascade decays.
A time interval of 1 to 3~$\mu$s revealed $^{212}$Bi-$^{212}$Po decays
from the $^{232}$Th decay chain.
An interval from 10 to 400 $\mu$s revealed $^{214}$Bi-$^{214}$Po
decays from the $^{238}$U decay chain.
A 1 to 4~ms interval showed $^{219}$Rn-$^{215}$Po decays from the
$^{227}$Ac decay chain.
Assuming each chain was in equilibrium with the observed Po cascades,
the average decay rates of 0.009, 0.16 and 0.2~Bq were found for the
$^{238}$U, $^{232}$Th, and $^{227}$Ac chains within the GdLS region
for the first two years of data.
The measured rates were consistent among all eight detectors.
A 40\% decrease of the $^{212}$Bi-$^{212}$Po rate and 9\% decrease of
the $^{219}$Rn-$^{215}$Po rate were observed, consistent with the
half-lives of $^{228}$Th~(1.9~yr) and $^{227}$Ac~(21.8~yr).
%

%
Monoenergetic 5.3~MeV $\alpha$-particles were also emitted by
$^{210}$Po, a long-lived daughter of $^{222}$Rn.
These $\alpha$-particles were visible as a peak at $E_{\mathrm{rec}}
\simeq 0.5$~MeV in the energy spectrum of uncorrelated signals in each
detector.
The amplitude of these peaks determined a $^{210}$Po $\alpha$ rate of
4~Hz for AD3-AD7, while larger rates of 8 to 10~Hz were found for AD1,
AD2, and AD8.
The distribution of reconstructed positions for these signals suggests
that the $^{210}$Po background was concentrated on the wall of the
inner acrylic vessel.
This observation agreed with previous experiments which reported
$^{210}$Po contamination on detector surfaces, with a variable amount
of contamination dependent on the history of material exposure to
$^{222}$Rn~\cite{Wojcik:2013}.

%
A Geant4-based simulation was used to model the probability for
$^{13}$C($\alpha$,n)$^{16}$O interactions in the detector.
Each $\alpha$-particle was attenuated via interactions with the
detector material, with the probability for an
$^{13}$C($\alpha$,n)$^{16}$O interaction determined using the JENDL
tabulated cross-sections~\cite{JENDL:2011}.
Table~\ref{tab:neutronYield} summarizes the estimated probability for
a $^{13}$C($\alpha$,n)$^{16}$O interaction to occur, per initial
$\alpha$ decay for the various natural $\alpha$ sources.
Probabilities of interaction to the $^{16}$O ground state only, to the
excited states, and the sum of the two are provided.
The uncertainty in the simulated probability was determined from
comparison with an alternate simulation which relied on EXFOR tabulated
cross-sections~\cite{Otuka2014272} and a SRIM simulation of alpha
attenuation~\cite{SRIM}.
\begin{table}[!htb]
  \caption{The simulated probability for a $^{13}$C($\alpha$,n)$^{16}$O
  interaction to occur, per equilibrium decay of the natural actinide
  contaminants in the detectors.  The separate probability for the
  interaction to occur to the $^{16}$O ground state and the excited
  states, as well as the sum of the two, are given.  The uncertainty
  in the probability $\sigma_{\mathrm{tot}}$ was determined from a
  comparison of simulation techniques.
  \label{tab:neutronYield}}
\begin{ruledtabular}
\begin{tabular}{crrrr}
 $\alpha$ Source & $ P_{\mathrm{gnd}}$ [10$^{-8}$] & $P_{\mathrm{exc}}$ [10$^{-8}$] & $P_{\mathrm{tot}}$ [10$^{-8}$] & $\sigma_{\mathrm{tot}}$ [\%] \\
\hline
 $^{210}$Po  &  5.26 &   0.49 &   5.75  &  7.2\% \rule{0pt}{2.6ex}\\
 $^{238}$U   & 43.40  & 29.60 &  73.00  & 16.9\% \\
 $^{232}$Th  & 44.90  & 49.20 &  94.10  & 27.7\% \\
 $^{227}$Ac  & 47.20  & 61.80 & 109.00  & 25.9\% \\
\end{tabular}
\end{ruledtabular}
\end{table}

\begin{figure}[!htb]
  \includegraphics[width=.5\textwidth]{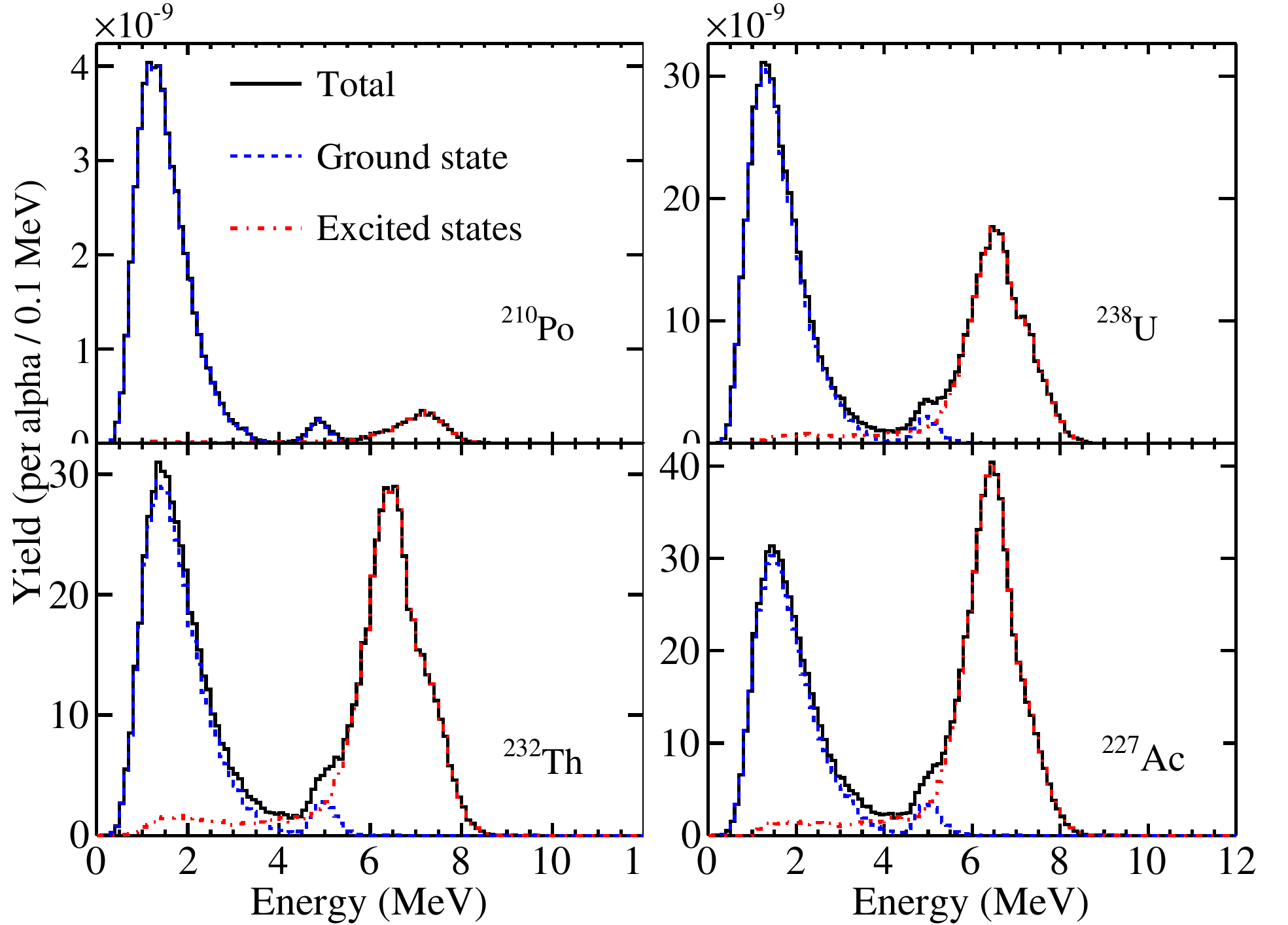}
  \caption{Simulation of energy spectra of the prompt signals for the
    $^{13}$C($\alpha$,n)$^{16}$O background, for each of the primary
    sources of natural $\alpha$ activity in the detectors.  Energetic
    neutrons produced a broad peak from proton recoils below 4~MeV,
    and a small peak near 5~MeV from inelastic scattering on
    $^{12}$C\@.  Higher-energy $\alpha$-particles increasingly
    populated the excited states of $^{16}$O, resulting in a broad
    peak from 5~MeV to 8~MeV from $^{16}$O$^*$ de-excitation.
  \label{fig:s04_fig_alphan_spec4iso}}
\end{figure}

%
The same simulation was also used to estimate the reconstructed energy
spectra for prompt signals from $^{13}$C($\alpha$,n)$^{16}$O
interactions, as presented in
Figure~\ref{fig:s04_fig_alphan_spec4iso}.
Energetic neutrons produced a broad peak from proton recoils below
4~MeV, and a small peak near 5~MeV from inelastic scattering on
$^{12}$C\@.
Higher-energy $\alpha$-particles increasingly populated the excited
states of $^{16}$O, resulting in a broad peak from 5~MeV to 8~MeV from
$^{16}$O$^*$ de-excitation.
Uncertainties in the spectra, although substantial, were safely
ignored due to the insignificant contribution of this background to
the \nuebar{} sample.
%
%
The rate of correlated ($\alpha$,n) background was estimated from the
product of the measured $\alpha$ activity, the simulated probabilities
of $^{13}$C($\alpha$,n)$^{16}$O interaction, and the modeled
efficiency for the corresponding signals to satisfy the \nuebar{}
signal selection.
The contamination of this background in the \nuebar{} candidate sample
was found to be negligible, 0.07\% for the far detectors and 0.01\% for
the near detectors.
A conservative $\sim$50\% total uncertainty in the rate of this
background resulted from a combination of uncertainties in Po cascade
selection, assumptions of actinide chain equilibrium, and simulation
of $^{13}$C($\alpha$,n)$^{16}$O interactions.

\paragraph{\label{prgph:mult_bkg}High-Multiplicity Background}

%
In the search for a pair of prompt and delayed signals consistent with
\nuebar{} inverse beta decay, occasionally three or more signals were
found within the relevant time window.
Such signal sets, referred to as having multiplicity $\geq$3, were
generally from the pile-up of actual \nuebar{} interactions with
additional uncorrelated signals from natural radioactivity.
Classification of each signal as a potential prompt or delayed
interaction was ambiguous in these cases.
Rejection of signal sets with multiplicity $\geq$3 removed all
concerns about this ambiguity, while resulting in an insignificant
loss of efficiency.
Two different techniques were implemented to reject these ambiguous
signals.

%
For selection A, a \nuebar{} candidate was rejected if one or more
additional signals with reconstructed energy $>$0.7~MeV occurred within
200~$\mu$s before the prompt signal, or within 200~$\mu$s following
the delayed signal.
%
The requirement for selection~B differed slightly.
Each delayed signal was required to have one and only one signal with
energy $>$0.7~MeV in the preceding 400~$\mu$s.
It was also required to have no other signal which satisfied the
delayed energy selection 6~MeV$< E_{\mathrm{rec}} < $12~MeV in the
200~$\mu$s following the delayed signal.
This approach had an efficiency which was independent of the time
between the prompt and delayed signals, and also avoided unwanted
correlations in the estimation of the muon and multiplicity veto
efficiencies.

%
Both methods effectively resolved signal ambiguities, while
maintaining a signal acceptance efficiency of $\sim$97.5\%.
The efficiency was estimated in a fashion similar to that used for the
accidental background, by calculating the probability for an
uncorrelated signal to randomly occur in close time proximity to a
true \nuebar{} interaction.
For selection A, the efficiency was
\begin{equation}
  \epsilon^{\mathrm{A}}_{\mathrm{m}} = e^{-R_{\mathrm{p}}{\Delta}t}\,\epsilon_{\mathrm{mid}}\,\epsilon_{\mathrm{post}},
\end{equation}
where $R_{\mathrm{p}}$ was the rate of prompt-like signals defined in
the discussion of the accidental background, and ${\Delta}t$ was
200~$\mu$s.
The probability for an uncorrelated signal to fall between the true
prompt and delayed signals, $\epsilon_{\mathrm{mid}}$, varied with the
time between these signals.
On average it was
\begin{equation}
  \epsilon_{\mathrm{mid}} = \int^{{\Delta}t}_0 e^{-R_{\mathrm{p}}t}\,f(t)\,dt \cong 1 - R_{\mathrm{p}}\overline{t}_{\mathrm{c}},
\end{equation}
where $f(t)$ was the distribution of times between prompt and delayed
signals within 200~$\mu$s, and $\overline{t}_{\mathrm{c}}$ was the
mean of this distribution.
At lowest order, the probability for an uncorrelated signal to occur
after the delayed signal, $\epsilon_{\mathrm{post}}$, was simply
$\exp(-R_{\mathrm{p}}{\Delta}t)$.
The probability that the muon veto truncated the time window after the
candidate signal, reducing the probability for detecting uncorrelated
signals after the \nuebar{} interaction, was not negligible.
A correction for this time-truncation was incorporated into
$\epsilon_{\mathrm{post}}$.
As a result, for a time segment $t^{i}_{\mathrm{s}}$ between vetoed
periods,
\begin{equation}
  \epsilon^{i}_{\mathrm{post}} =
  \begin{cases}
    \left(1-\frac{{\Delta}t}{t^{i}_{\mathrm{s}}}\right)e^{-R_{\mathrm{p}}{\Delta}t} + \frac{1-e^{-R_{\mathrm{p}}{\Delta}t}}{R_{\mathrm{p}}t^{i}_{\mathrm{s}}}, & \text{for } t^{i}_{\mathrm{s}} \geq {\Delta}t, \\
      \frac{1-e^{-R_{\mathrm{p}}{\Delta}t}}{R_{\mathrm{p}}t^{i}_{\mathrm{s}}}, & \text{for } t^{i}_{\mathrm{s}} < {\Delta}t.
  \end{cases}
\end{equation}
%
%
Given the correlation of the multiplicity selection and muon veto, the
combined \nuebar{} selection efficiency was calculated from a
time-weighted average of the multiplicity efficiency over the time
segments $t^{i}_{\mathrm{s}}$ between muon-vetoed periods,
\begin{equation}
  \epsilon_\mu\overline{\epsilon}^{\mathrm{A}}_{\mathrm{m}} = \left(\sum_i\epsilon^i_mt^{i}_{\mathrm{s}}\right)/t_{\mathrm{DAQ}},
\end{equation}
where $t_{\mathrm{DAQ}}$ was the total DAQ livetime before application
of the muon veto.
The average efficiency of the multiplicity veto could be determined
from comparison of
$\epsilon_\mu\overline{\epsilon}^{\mathrm{A}}_{\mathrm{m}}$ with
$\epsilon_\mu$ calculated using Eq.~\ref{eq:mu_eff}.
Uncertainty in $f(t)$ resulted in a systematic uncertainty of 0.02\%
in $\overline{\epsilon}^{\mathrm{A}}_{\mathrm{m}}$, correlated between
detectors.
Similarity of the observed $f(t)$ between detectors constrained
potential uncorrelated variations in efficiency to $<$0.01\%.

%
Estimation of the multiplicity efficiency for selection B was trivial
by design.
The efficiency was calculated from the expression
\begin{equation}
  \epsilon^{\mathrm{B}}_{\mathrm{m}} = e^{-2R_{\mathrm{p}}{\Delta}t}\,e^{-R_{\mathrm{d}}{\Delta}t}.
\end{equation}
The probability of simultaneous multiplicity and muon vetos was
reduced to a negligible level, given the very low rate of delayed-like
signals.
Uncertainty in $\epsilon^{\mathrm{B}}_{\mathrm{m}}$ was insignificant.

\subsection{\label{subsec:selec_summary}Summary of Antineutrino Selection}

\subsubsection{\label{subsubsec:eff_summary}Detection Efficiencies}

%
Table~\ref{tab:s04_tab_effsys} summarizes the efficiencies for
detection of \nuebar{} inverse beta decay in the GdLS target of each
detector.
The combined efficiency was estimated to be 80.6\%.
Neutrons which did not capture on Gd, as well as those n-Gd captures
which failed to produce signals with $E_{\mathrm{rec}} > 6$~MeV, had
the greatest impact on the efficiency.
The number of target protons and multiplicity cut efficiencies varied
slightly between detectors, and their precisely measured differences
are provided in Table~\ref{tab:site_survey} and
Table~\ref{tab:s04_table_ibdsummary}.
\begin{table}[!hbp]
  \caption{A summary of the estimated efficiencies and their relative
    uncertainties for detection of \nuebar{} inverse beta decay in the
    GdLS target region of each antineutrino detector.  The values are
    provided for selection A, but differ negligibly for selection~B\@.
    The number of target protons and multiplicity cut efficiencies
    varied slightly between detectors, and their precisely measured
    differences are provided in Table~\ref{tab:site_survey} and
    Table~\ref{tab:s04_table_ibdsummary}.  The estimated uncertainties
    are divided into a correlated component, which was common for all
    detectors, and an uncorrelated component, which captured potential
    variations in efficiency between detectors.  This latter component
    was relevant for the measurement of neutrino oscillation.
    \label{tab:s04_tab_effsys}}
  \begin{ruledtabular}
  \begin{tabular}{lrrr}
    & Efficiency & Correlated & Uncorrelated \\
    \hline
    Target protons      &  -       & 0.92\%    & 0.03\% \\
    Flasher cut         &  99.98\% & 0.01\%    & 0.01\% \\
    Delayed energy cut  &  92.7\%  & 0.97\%    & 0.08\% \\
    Prompt energy cut   &  99.8\%  & 0.10\%    & 0.01\% \\
    Multiplicity cut    &          & 0.02\%    & 0.01\% \\
    Capture time cut    &  98.7\%  & 0.12\%    & 0.01\% \\
    Gd capture fraction &  84.2\%  & 0.95\%    & 0.10\% \\
    Spill-in            & 104.9\%  & 1.00\%    & 0.02\% \\
    Livetime            &  -       & 0.002\%   & 0.01\% \\
    \hline
    Combined            &  80.6\%  & 1.93\%    & 0.13\% \\
  \end{tabular}
  \end{ruledtabular}
\end{table}

%
Uncertainties in the detection efficiencies were divided into
correlated components, which were common for all detectors, and
uncorrelated components, which captured potential variations in
efficiency between detectors.
The total correlated relative uncertainty in efficiency was estimated
to be 1.93\%.
Spill-in neutrons, generated by \nuebar{} interactions outside the
GdLS target but which diffused into the target and captured on Gd,
caused the largest correlated uncertainty in detection efficiency.
Absolute uncertainties in the fraction of neutrons in the GdLS target
which captured on Gd, as well as in the fraction of n-Gd captures
which produced signals with $E_{\mathrm{rec}}<6$~MeV, were also
significant.
A detailed assessment of correlated uncertainties is given
in~\cite{An:2016srz}.

The absolute efficiencies and their correlated uncertainties canceled
when comparing the ratio of signals in the far versus near detectors,
as presented in Eq.~\ref{eq:nearFarRatio}.
Therefore, only the uncorrelated uncertainties were relevant for the
far versus near detector measurement of neutrino oscillation.
Variations in efficiency between detectors were estimated to be
0.13\%.
The most significant variation in efficiency came from potential
differences in the fraction of neutrons which captured on Gd, as
constrained by comparisons of the capture time distributions between
detectors.

%
The Daya Bay experiment was designed to minimize potential variations
in efficiency between detectors, but the actual detectors exceeded the
design goal.
Given this achievement, it was important to have an independent method
to verify the low 0.13\% estimate of uncorrelated variation between
detector efficiency.
Comparison of the \nuebar{} rates observed in detectors located
side-by-side within the same experimental hall provided a direct test
of the estimated variations between detector efficiency.
The results of these tests are discussed in the following section.

\begin{table*}[!htb]
    \caption{Summary of the \nuebar{} inverse beta decay interaction
      candidate sample. Results obtained using both selection~A and
      selection~B are provided for comparison (Selection criteria are
      given in Table\ref{tab:IBDselectionCuts}).  The number of
      signals selected as \nuebar{} inverse beta decay candidates, DAQ
      livetime, efficiency lost to vetoes, and estimated background
      rates are listed for each antineutrino detector.  The
      background-subtracted rates of \nuebar{} interactions for each
      detector were estimated from these quantities. All rates are
      corrected for the detector-dependent loss of live time from the
      muon and multiplicity vetoes, expressed as the efficiencies
      $\epsilon_{\mu}$ and $\epsilon_{\mathrm{m}}$\@.  Slight
      differences in the number of protons in each target region,
      ${\Delta}N_{\mathrm{p}}$, relative to AD1 in EH1 are also
      provided.
    \label{tab:s04_table_ibdsummary}}
    \begin{ruledtabular}
      {\footnotesize
    \begin{tabular}{ccccccccc}
  & \multicolumn{2}{c}{EH1}&\multicolumn{2}{c}{EH2}&\multicolumn{4}{c}{EH3} \\
  & AD1  & AD2  & AD3 & AD8 & AD4 & AD5 & AD6 & AD7 \\
\hline
${\Delta}N_{\mathrm{p}}$ [\%] & $0.00\pm0.03$ & $0.13\pm0.03$ & $-0.25\pm0.03$ & $0.02\pm0.03$ & $-0.12\pm0.03$ & $0.24\pm0.03$ & $-0.25\pm0.03$ & $-0.05\pm0.03$ \\
\hline
& \multicolumn{8}{c}{Selection A} \\
\hline
\nuebar{} candidates & 597616	& 606349 & 567196 & 466013 & 80479 & 80742 & 80067 & 66862 \\
DAQ live time [days] & 1117.178 & 1117.178 & 1114.337 & 924.933 & 1106.915 & 1106.915 & 1106.915 & 917.417 \\
$\epsilon_{\mu}$ & 0.8255 & 0.8221	&0.8573	 &0.8571	 &0.9824	&0.9823	&0.9821	&0.9826 \\
$\overline{\epsilon}_{\mathrm{m}}$ & 0.9744	&0.9747	&0.9757	&0.9757	 &0.9759	 &0.9758	&0.9756	&0.9758 \\
Accidentals [day$^{-1}$] & $8.46\pm0.09$ & $8.46\pm0.09$ & $6.29\pm0.06$ & $6.18\pm0.06$ & $1.27\pm0.01$ & $1.19\pm0.01$ & $1.20\pm0.01$ & $0.98\pm0.01$ \\
Fast neutron [AD$^{-1}$ day$^{-1}$] & \multicolumn{2}{c}{$0.79\pm0.10$} & \multicolumn{2}{c}{$0.57\pm0.07$} & \multicolumn{4}{c}{$0.05\pm0.01$} \\
$^9$Li,$^8$He [AD$^{-1}$ day$^{-1}$] & \multicolumn{2}{c}{$2.46\pm1.06$} & \multicolumn{2}{c}{$1.72\pm0.77$} & \multicolumn{4}{c}{$0.15\pm0.06$} \\
$^{241}$Am-$^{13}$C, 6-AD [day$^{-1}$] & $0.27\pm0.12$ & $0.25\pm0.11$ & $0.28\pm0.13$ & & $0.22\pm0.10$ & $0.21\pm0.10$ & $0.21\pm0.10$  \\
$^{241}$Am-$^{13}$C, 8-AD [day$^{-1}$] & $0.15\pm0.07$ & $0.16\pm0.07$ & $0.13\pm0.06$ & $0.15\pm0.07$ & $0.04\pm0.02$ & $0.03\pm0.02$ & $0.03\pm0.02$ & $0.05\pm0.02$ \\
$^{13}$C($\alpha$, n)$^{16}$O [day$^{-1}$] & $0.08\pm0.04$ & $0.07\pm0.04$ & $0.05\pm0.03$ & $0.07\pm0.04$ & $0.05\pm0.03$ & $0.05\pm0.03$ & $0.05\pm0.03$ & $0.05\pm0.03$ \\
\hline
\nuebar{} rate, $R_{\overline{\nu}}$ [day$^{-1}$] & $653.03\pm1.37$ & $665.42\pm1.38$ & $599.71\pm1.12$	& $593.82\pm1.18$ & $74.25\pm0.28$ & $74.60\pm0.28$ & $73.98\pm0.28$	& $74.73\pm0.30$ \\ 
\hline
& \multicolumn{8}{c}{Selection B} \\
\hline
\nuebar{} candidates & 594737 & 603092 & 562681 & 462129 & 80508 & 80769 & 80112 & 67018 \\
DAQ live time [days] & 1117.162 & 1117.162 & 1114.334 & 924.930 & 1106.898 & 1106.898 & 1106.898 & 917.401 \\
$\epsilon_{\mu}$ & 0.8210 & 0.8178 & 0.8502 & 0.8496 & 0.9824 & 0.9821 & 0.9820 & 0.9825 \\
$\epsilon_{\mathrm{m}}$ & 0.9768 & 0.9773 & 0.9782 & 0.9781 & 0.9783 & 0.9783 & 0.9781 & 0.9784 \\
Accidentals [day$^{-1}$] & $7.99\pm0.01$ & $7.88\pm0.01$ & $5.94\pm0.01$ & $5.81\pm0.01$ & $1.20\pm0.00$ & $1.13\pm0.00$ & $1.14\pm0.00$ & $0.92\pm0.00$ \\
Fast neutron [AD$^{-1}$ day$^{-1}$] & \multicolumn{2}{c}{$0.84\pm0.08$} & \multicolumn{2}{c}{$0.64\pm0.06$} & \multicolumn{4}{c}{$0.05\pm0.01$} \\
$^9$Li,$^8$He [AD$^{-1}$ day$^{-1}$] & \multicolumn{2}{c}{$2.71\pm0.90$} & \multicolumn{2}{c}{$1.91\pm0.73$} & \multicolumn{4}{c}{$0.22\pm0.07$} \\
$^{241}$Am-$^{13}$C, 6-AD [day$^{-1}$] & $0.26\pm0.12$ & $0.25\pm0.11$ & $0.28\pm0.12$ & & $0.22\pm0.10$ & $0.21\pm0.09$ & $0.21\pm0.09$  \\
$^{241}$Am-$^{13}$C, 8-AD [day$^{-1}$] & $0.15\pm0.07$ & $0.15\pm0.07$ & $0.13\pm0.06$ & $0.15\pm0.07$ & $0.04\pm0.02$ & $0.03\pm0.02$ & $0.04\pm0.02$ & $0.05\pm0.02$ \\
$^{13}$C($\alpha$, n)$^{16}$O [day$^{-1}$] & $0.08\pm0.04$ & $0.07\pm0.04$ & $0.05\pm0.03$ & $0.07\pm0.04$ & $0.05\pm0.03$ & $0.05\pm0.03$ & $0.05\pm0.03$ & $0.05\pm0.03$ \\ \hline
\nuebar{} rate, $R_{\overline{\nu}}$ [day$^{-1}$] & $651.99\pm1.25$ & $663.74\pm1.26$ & $598.47\pm1.09$ & $592.67\pm1.15$ & $74.08\pm0.28$ & $74.43\pm0.28$ & $73.83\pm0.28$ & $74.70\pm0.30$ \\
    \end{tabular}
    }
 \end{ruledtabular}
\end{table*}

\subsubsection{\label{subsubsec:obs_antinus}Final Antineutrino Sample}

%
Table~\ref{tab:s04_table_ibdsummary} summarizes the antineutrino
candidate data sample.
Refer to Table~\ref{tab:IBDselectionCuts} for an overview of the
selection criteria.
More than $2.5\times10^{6}$ \nuebar{} inverse beta decay interactions
were identified using the eight detectors.
Backgrounds were estimated to contribute 1.8\% to the sample from EH1,
1.5\% to EH2, and 2.0\% to EH3, primarily from uncorrelated
accidentals.
Uncertainty in the background was 0.2\%, 0.1\%, and 0.1\% for the
three experimental halls, and was dominated by the contribution from
$\beta$-n decay of $^{9}$Li and $^{8}$He.
For each detector $i$, the observed rate of \nuebar{} interactions was
\begin{equation}
  R^i_{\overline{\nu}} = \frac{N^i_{\mathrm{cand}}}{t^i_{\mathrm{DAQ}}\,\epsilon^i_\mu\,\epsilon^i_{\mathrm{m}}} - R^i_{\mathrm{bkg}},
\end{equation}
where $N_{\mathrm{cand}}$ was the number of \nuebar{} candidates
selected, $t_{\mathrm{live}}$ was the DAQ operation live time,
$\epsilon_\mu$ was the reduced signal efficiency from live time
rejected by the muon veto, $\epsilon_{\mathrm{m}}$ was the same for
the multiplicity veto, and $R_{\mathrm{bkg}}$ was the total background
rate.
These rates have not been corrected for the absolute selection
efficiencies shown in Table~\ref{tab:s04_tab_effsys}, nor for the
slight differences in the number of protons, ${\Delta}N_{\mathrm{p}}$,
in the target region of each detector.

\begin{figure}[!htb]
\includegraphics[width=.5\textwidth]{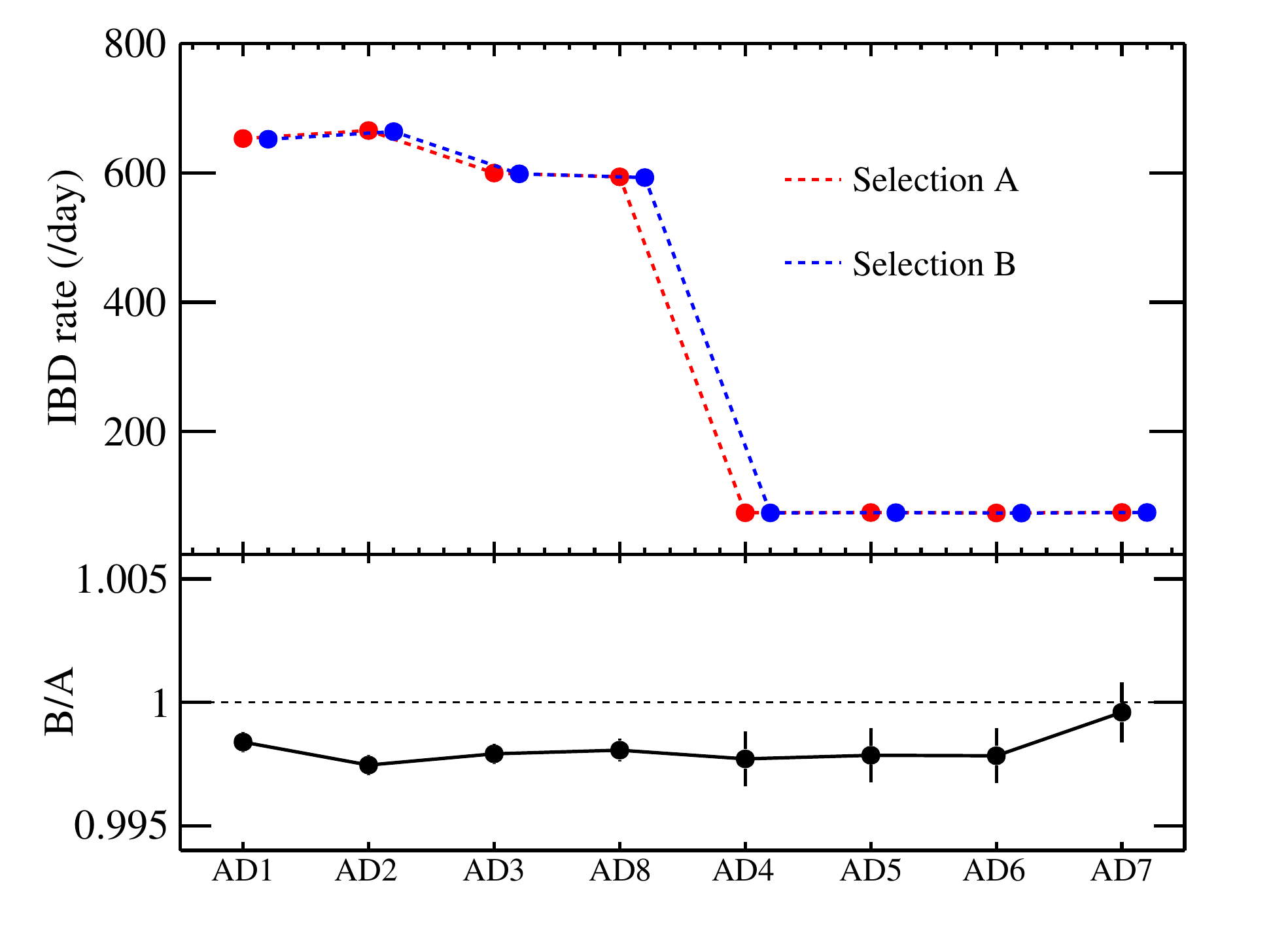}
\caption{({\em Top}) The rates of \nuebar{} inverse beta decay
  observed in each antineutrino detector, for both selection~A and
  selection~B\@.  ({\em Bottom}) The rates obtained by selection B
  were $\sim$0.2\% lower than those of selection A, demonstrating a
  difference in absolute selection efficiency within expectations.
  The ratio of the far versus near detector rates are consistent for
  the two selections, considering the statistical uncertainty from
  candidate signals uncommon between the two samples.
  \label{fig:s04_ibd_comp}}
\end{figure}

%
Figure~\ref{fig:s04_ibd_comp} compares the \nuebar{} rates obtained
using selection~A with those from selection~B\@.
The consistency of the two results served as an independent
cross-check of the \nuebar{} selection process.
10\% of the candidates differed between the two selections.
For the near detectors, differences in the muon vetoes caused most of
the discrepant candidate signals.
For the far detectors, discrepancies were primarily a result of the
different multiplicity veto criteria.

\begin{figure*}[!htbp]
  \includegraphics[width=0.49\textwidth]{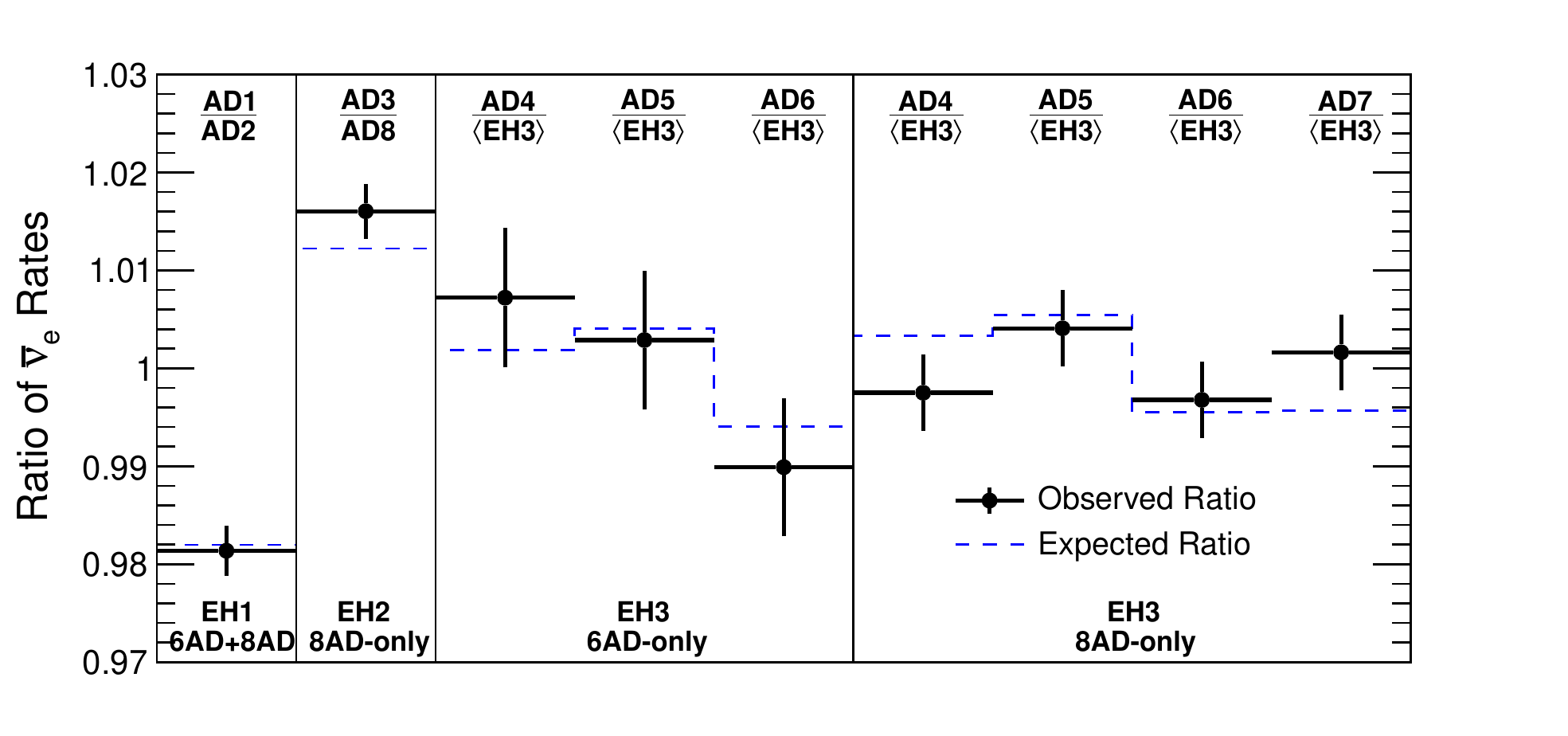}
  \includegraphics[width=0.49\textwidth]{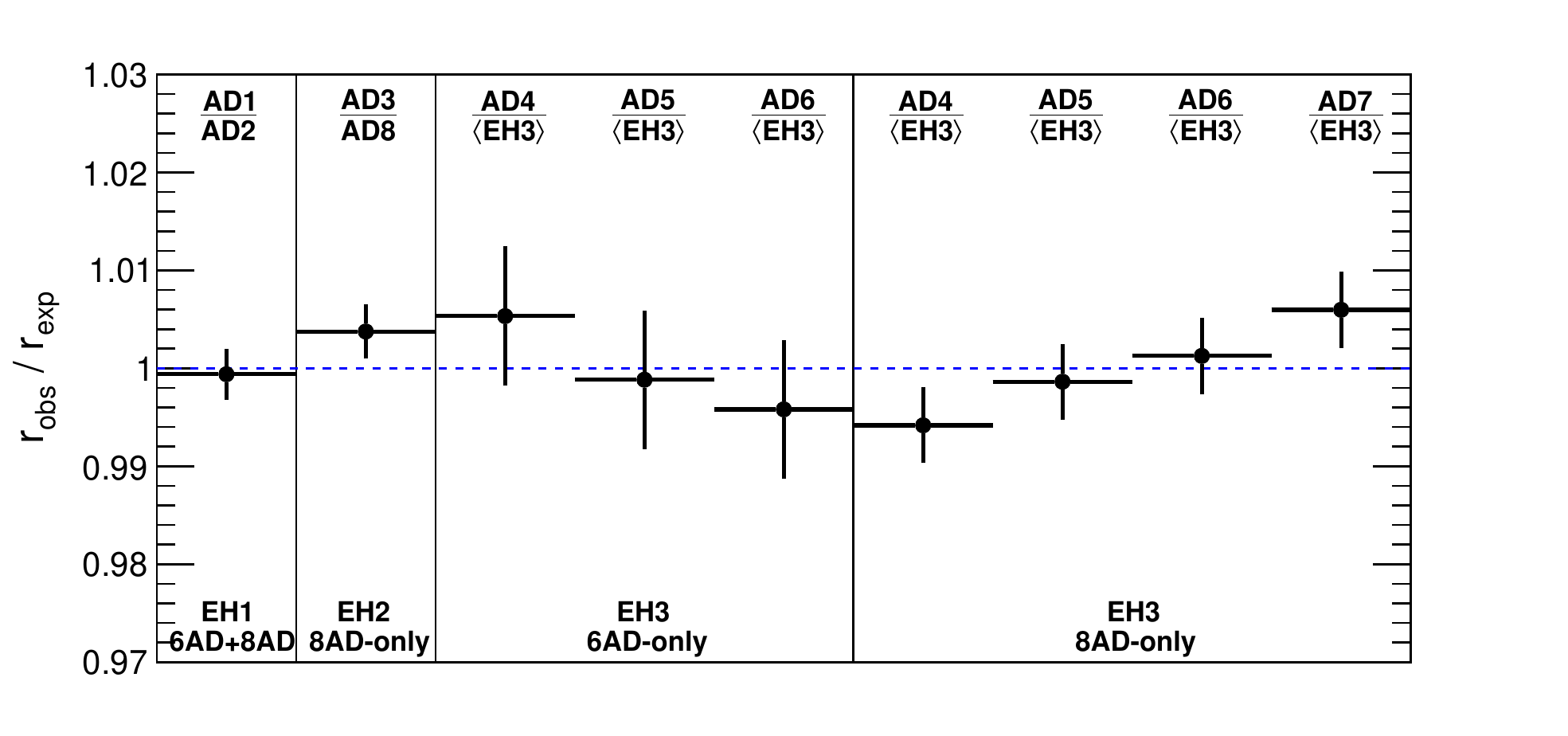}
  \caption{({\em Left}) Ratios of the \nuebar{} interaction rates
    observed by detectors within the same experimental hall,
    $r_{\mathrm{obs}}$ (black points).  The ratio of the rate in AD1
    to that of AD2 is shown for EH1 for the entire data period.  For
    the period following installation of AD8 in EH2, the ratio of AD3
    to AD8 is provided.  For EH3, the ratio of the rate in each
    detector relative to the site average is given separately for the
    period before and after installation of AD7.  Slight differences
    in the distances of each detector relative to each reactor as well
    as small variations in detector target mass
    ${\Delta}N_{\mathrm{p}}$ predicted minor deviations in these
    ratios, $r_{\mathrm{exp}}$ (blue dashed line).  The uncertainties
    are dominated by statistics and the estimated 0.13\% variation in
    efficiency between detectors.  The consistency of the side-by-side
    detector rates confirmed the stringent limits on this latter
    systematic uncertainty.  ({\em Right}) The same figure, expressed
    as the double ratio $r_{\mathrm{obs}}/r_{\mathrm{exp}}$.
    \label{fig:side_by_side_rates}}
\end{figure*}

%
The rates observed by detectors located side-by-side within the same
experimental hall were used to independently assess potential
variations in \nuebar{} efficiency between detectors.
Figure~\ref{fig:side_by_side_rates} compares the observed rates for
detectors within the same hall.
Slight differences in distances from the reactors and
${\Delta}N_{\mathrm{p}}$ predict $\lesssim$1\% deviations between
detectors.
Comparisons for the detectors in EH2 and EH3 are shown separately for
the 6-AD and 8-AD operation periods.
The consistency of the detected rates, relative to the slight
differences in predictions, provided independent confirmation of the
estimated 0.13\% variation in efficiency between detectors.

\begin{figure}[htbp]
  \includegraphics[width=3in]{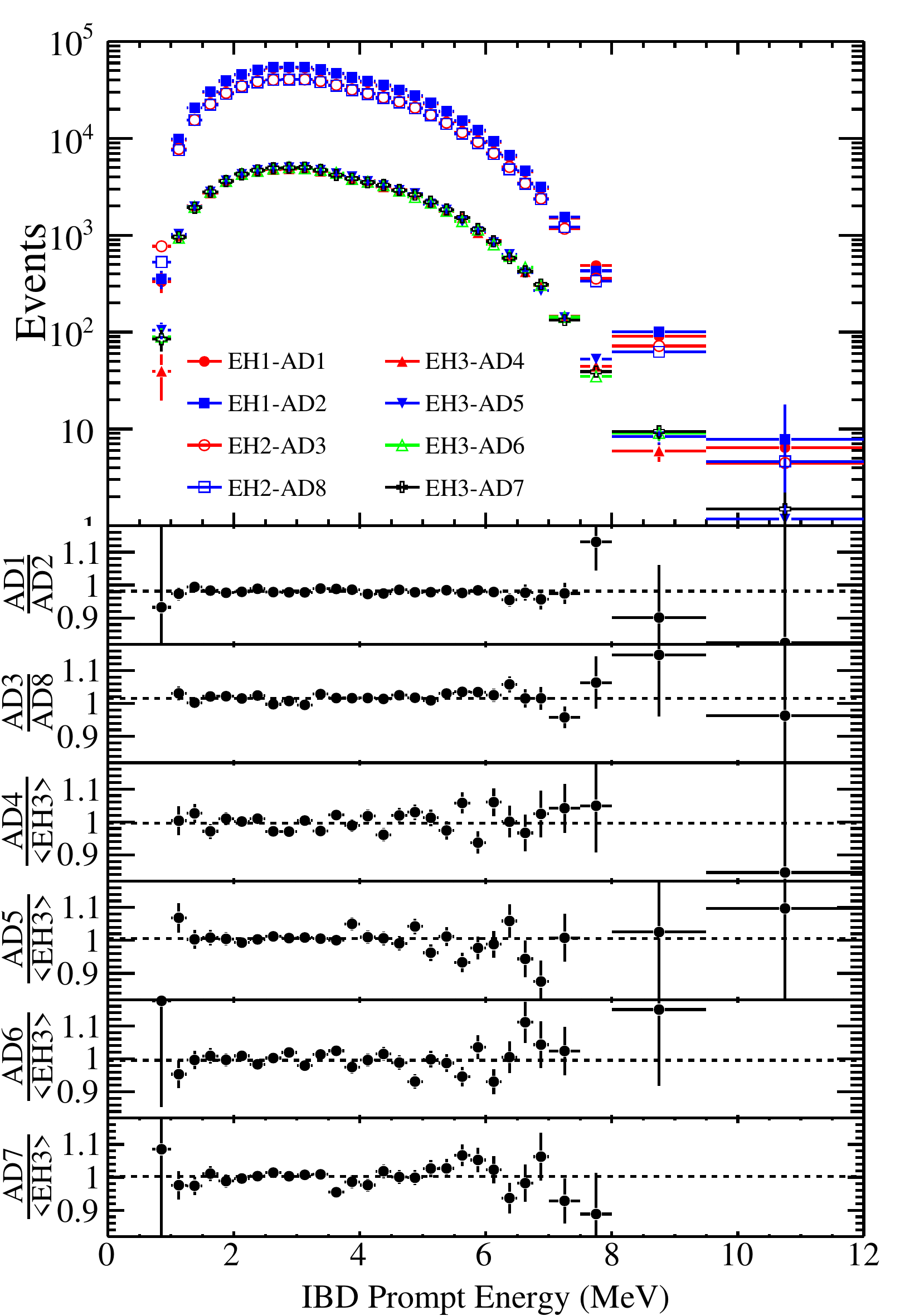}
  \caption{({\em Top}) Distributions of the reconstructed prompt
    energy for the selected \nuebar{} candidates in each of the eight
    detectors.  ({\em Bottom}) The ratio of the distributions for
    detectors within the same experimental hall showed no significant
    deviations between detectors.
    \label{fig:ibd_cand_Ep}}
\end{figure}

\begin{figure}[htbp]
  \includegraphics[width=3in]{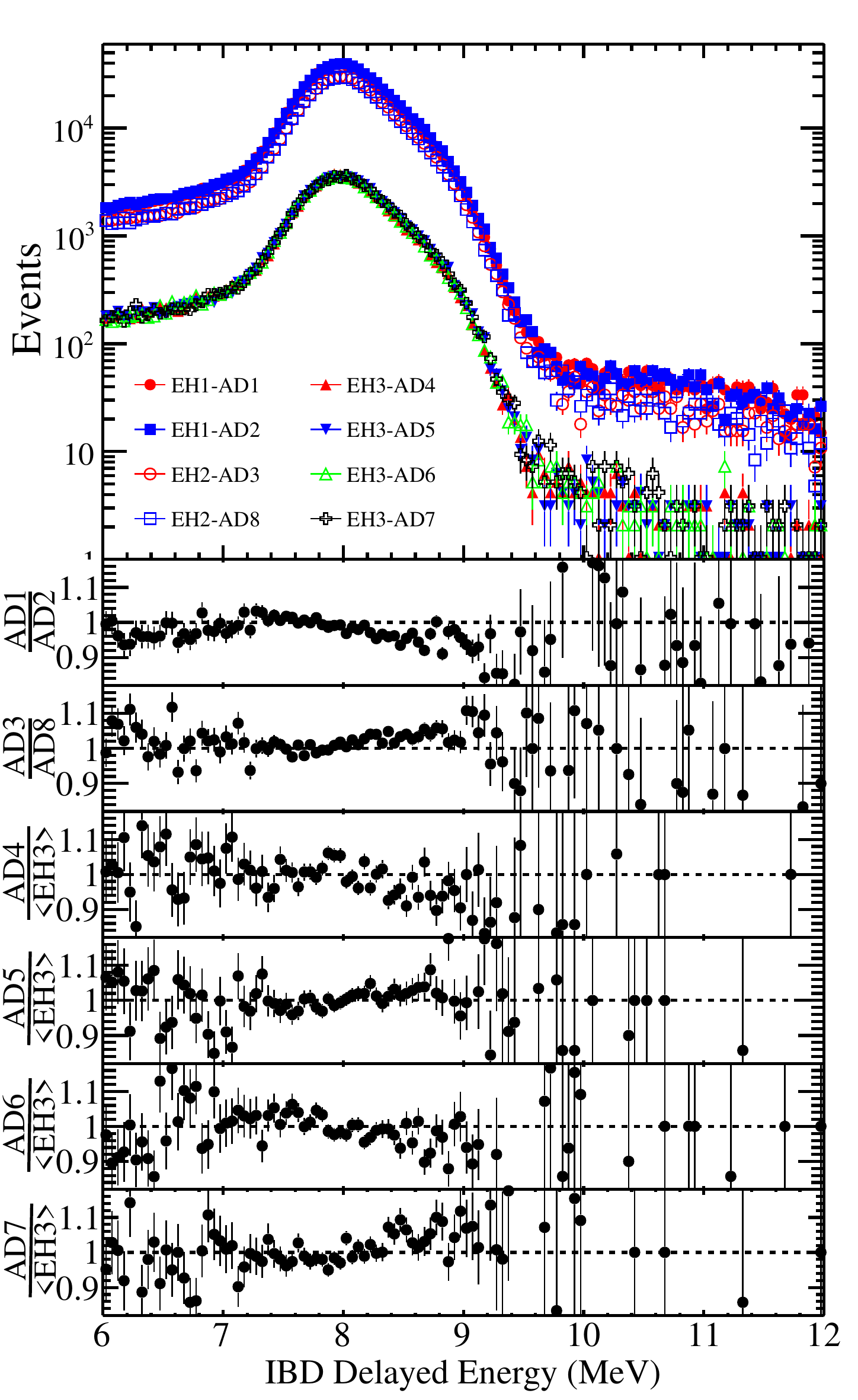}
  \caption{({\em Top}) Distributions of the reconstructed delayed
    energy for the selected \nuebar{} candidates in each of the eight
    detectors.  ({\em Bottom}) The ratio of the distributions for
    detectors within the same experimental hall showed no significant
    deviations between detectors; the slight slopes in the ratios are
    consistent with the $\lesssim$0.2\% relative differences in the
    energy scale among the detectors.
    \label{fig:ibd_cand_Ed}}
\end{figure}

%
Potential variation in performance between detectors was assessed by
comparing the capture time, prompt energy, and delayed energy
distributions for the selected \nuebar{} candidates for side-by-side
detectors.
Figures~\ref{fig:ibd_dt},~\ref{fig:ibd_cand_Ep},
and~\ref{fig:ibd_cand_Ed} compare these three distributions for all
eight detectors.
For spectral comparisons, simple ratios of the distributions for AD2
to AD1 and AD8 to AD3 are shown for the near detectors.
For the far hall, the distribution for each AD was divided by the site
average.
No significant deviations in the distributions for detectors within
the same experimental hall were found.

\section{\label{sec:osc_analysis} Oscillation Analysis}

%
The \nuebar{} inverse beta decay interactions observed with the eight
detectors were used to measure the oscillation of neutrino flavor.
Comparison of the rates observed in the three experimental halls
revealed \nuebar{} disappearance between the near and far detectors.
The amplitude of the disappearance determined the value of the mixing
angle $\theta_{13}$.
The energy-dependence of \nuebar{} disappearance distorted the energy
spectra of the prompt positrons observed in the far detectors relative
to the near detectors.
The neutrino mass-squared difference ${\Delta}m^2_{\mathrm{32}}$ was
measured from this spectral distortion.
The details of the measurement of neutrino oscillation are presented
in this section.

\subsection{\label{subsec:nue_disapp}Antineutrino Disappearance}

%
Before discussing detailed statistical methods, it is useful to
examine a simple ratio of the signal observed by the far detectors
versus the near detectors.
After correcting for differences in detector exposure, any deficit in
this ratio would be evidence of \nuebar{} disappearance.
This example demonstrates the robustness of the observation of
\nuebar{} disappearance due to $\theta_{13}$, independent of the
statistical models that will be presented.

The estimated number of \nuebar{} interactions at the far site was
determined from the sum of the background-subtracted \nuebar{} rate
$R^{i}_{\overline{\nu}}$ times the livetime $t^i_{\mathrm{live}}$ of
each far detector $i$,
\begin{equation}\label{eq:site_nuebar_rate}
  \widetilde{N}^{\mathrm{far}}_{\mathrm{obs}} = \sum_{i}^{\mathrm{Far ADs}}\,R^{i}_{\overline{\nu}}\,t^{i}_{\mathrm{live}}.
\end{equation}
The \nuebar{} rates and livetimes were taken from
Table~\ref{tab:s04_table_ibdsummary}.
The estimated numbers for each near site,
$\widetilde{N}^{\mathrm{EH1}}_{\mathrm{obs}}$ and
$\widetilde{N}^{\mathrm{EH2}}_{\mathrm{obs}}$, are defined in the same
fashion as given in Eq.~\ref{eq:site_nuebar_rate}.
Assuming no oscillation, the \nuebar{} signal at the far site was
predicted by a suitable weighted combination of the signals observed
at the two near sites,
\begin{equation}
  \widetilde{N}^{\mathrm{far}}_{\mathrm{exp}} = w^{\mathrm{EH1}}\,\widetilde{N}^{\mathrm{EH1}}_{\mathrm{obs}} + w^{\mathrm{EH2}}\,\widetilde{N}^{\mathrm{EH2}}_{\mathrm{obs}}.
\end{equation}
The weights $w^{\mathrm{EH1}}$ and $w^{\mathrm{EH2}}$ are defined such
that $\widetilde{N}^{\mathrm{far}}_{\mathrm{exp}}$ consisted of
\nuebar{}'s from the Daya Bay and Ling Ao reactors in roughly the same
proportion as determined for the far site.

\begin{table}[!hbp]
  \caption{Estimated relative contribution of the two Daya Bay
    reactors and four Ling Ao reactors to the \nuebar{} signal in each
    of the experimental halls for the combined six detector and eight
    detector data periods.  The values are normalized relative to the
    total estimated signal in EH3.
  \label{tab:hall_rate_breakdown}}
  \begin{ruledtabular}
  \begin{tabular}{lccc}
    & \multicolumn{3}{c}{Relative \nuebar{} Signal, $f^{k}_{l}$} \\ 
    Reactors & EH1    & EH2    & EH3 \\
    \hline
    Daya Bay & 3.5022 & 0.2338 & 0.2423 \\   
    Ling Ao  & 0.9255 & 3.4333 & 0.7577 \\
  \end{tabular}
  \end{ruledtabular}
\end{table}

Table~\ref{tab:hall_rate_breakdown} gives the estimated relative
\nuebar{} signal contribution $f^k_l$ of each set of reactors $l$ to
the detectors in each hall $k$, including hall-dependent differences
in livetime, efficiency, and the number of target protons.
Variations in the contributions were primarily due to the distances of
each reactor to each detector, as given in
Table~\ref{tab:site_survey}.
Differences in the fluxes from each reactor were a minor effect given
the common average thermal power of the six reactors during this data
period, as listed in Table~\ref{tab:site_survey}.
Minor differences in the target mass of each detector, ${\Delta}N_{p}$
in Table~\ref{tab:s04_table_ibdsummary}, were included in this
calculation.
The weights that sample the flux of the two sets of reactors in equal
proportion to the far detectors are simply
\begin{align}
  w^{\mathrm{EH1}} =& \frac{f^{\mathrm{far}}_{\mathrm{D}}f^{\mathrm{EH2}}_{\mathrm{L}} - f^{\mathrm{far}}_{\mathrm{L}}f^{\mathrm{EH2}}_{\mathrm{D}}}{f^{\mathrm{EH1}}_{\mathrm{D}}f^{\mathrm{EH2}}_{\mathrm{L}} - f^{\mathrm{EH1}}_{\mathrm{L}}f^{\mathrm{EH2}}_{\mathrm{D}}},  \\
  w^{\mathrm{EH2}} =& \frac{f^{\mathrm{far}}_{\mathrm{L}}f^{\mathrm{EH1}}_{\mathrm{D}} - f^{\mathrm{far}}_{\mathrm{D}}f^{\mathrm{EH1}}_{\mathrm{L}}}{f^{\mathrm{EH2}}_{\mathrm{L}}f^{\mathrm{EH1}}_{\mathrm{D}} - f^{\mathrm{EH2}}_{\mathrm{D}}f^{\mathrm{EH1}}_{\mathrm{L}}}.
\end{align}
Using the relative signal contributions provided in
Table~\ref{tab:hall_rate_breakdown}, $w^{\mathrm{EH1}}=0.05545$ and
$w^{\mathrm{EH2}}=0.2057$.
This result was independent of the specific model of reactor \nuebar{}
emission.
In fact, an equivalent result can be obtained using the average
thermal power of each reactor, given in Table~\ref{tab:site_survey},
as a proxy for the relative \nuebar{} flux.

From this simple analytic method, a clear \nuebar{} rate deficit was
observed in the far detectors,
\begin{equation}
  R = \frac{\widetilde{N}^{\mathrm{far}}_{\mathrm{obs}}}{\widetilde{N}^{\mathrm{far}}_{\mathrm{exp}}} = 0.949 \pm 0.002 (\mathrm{stat.}) \pm 0.002 (\mathrm{syst.}),
\end{equation}
for the combined six detector and eight detector data periods.
The statistical uncertainty was primarily determined from the observed
signal rate in the far detectors.
Uncertainty from the reactor \nuebar{} flux was almost completely
canceled in this ratio.
An uncorrelated 0.9\% uncertainty in the estimated flux of each
reactor resulted in $\lesssim$0.1\% uncertainty in $R$, as
demonstrated by simple error propagation.
The $\beta$-n background and relative variations in detector
efficiency contributed the most significant systematic uncertainties.

%
The value of $\theta_{13}$ can be estimated from the observed rate
deficit using a simple calculation.
The rate deficit can be expressed as
\begin{equation}
  R = \frac{N^{\mathrm{far}}_{\mathrm{pred}}}{w^{\mathrm{EH1}}N^{\mathrm{EH1}}_{\mathrm{pred}} + w^{\mathrm{EH2}}N^{\mathrm{EH2}}_{\mathrm{pred}}},
\end{equation}
where $N^{k}_{\mathrm{pred}}$ is a prediction for the number of
\nuebar{} interactions in each hall $k$ as a function of
$\theta_{13}$.
The \nuebar{} signal can be rewritten as the product of the \nuebar{}
signal assuming no oscillation and the mean survival probability
$\bar{P}^{k}_{\mathrm{sur}}$,
\begin{equation}
  N^{k}_{\mathrm{pred}} = N^{k}_{\mathrm{no\,osc}}\,\bar{P}^{k}_{\mathrm{sur}}. 
\end{equation}
$\bar{P}^{k}_{\mathrm{sur}}$ is approximately linear in
$\sin^22\theta_{13}$, as shown by Eq.~\ref{eq:survProb_31}.
For short-baseline reactor measurements,
\begin{equation}
  \bar{P}^{k}_{\mathrm{sur}} = 1 - \eta^{k}\,\sin^22\theta_{13} + O(10^{-5}),
\end{equation}
given that $\theta_{13}$ is small.
The terms $\eta^{k}$ were determined to be 0.180, 0.206, and 0.789 for
EH1, EH2, and EH3 respectively.
%
For this calculation, values for $\sin^{2}2\theta_{12}$ and
${\Delta}m^2_{21}$ were taken from~\cite{PDG2015}.
A value of ${\Delta}m^2_{\mathrm{32}}=2.43\times10^{-3}$~eV$^2$ was
also assumed, based on the error-weighted average of measurements by
the T2K and MINOS experiments~\cite{PDG2015}.
Expressing $R$ as in terms of $\sin^22\theta_{13}$ gives
\begin{equation}\label{eq:oscRatioVsTheta13}
  R = \frac{1 - \eta^{\mathrm{far}}\,\sin^22\theta_{13}}{1 - \eta^{\mathrm{near}}\,\sin^22\theta_{13}},
\end{equation}
where
\begin{equation}
  \eta^{\mathrm{near}} = w^{\mathrm{EH1}}\beta^{\mathrm{EH1}}\eta^{\mathrm{EH1}} + w^{\mathrm{EH2}}\beta^{\mathrm{EH2}}\eta^{\mathrm{EH2}},
\end{equation}
and
\begin{equation}
  \beta^{k} = \frac{N^{k}_{\mathrm{no\,osc}}}{N^{\mathrm{far}}_{\mathrm{no\,osc}}} = \frac{(f^{k}_{\mathrm{D}}+f^{k}_{\mathrm{L}})}{(f^{\mathrm{far}}_{\mathrm{D}}+f^{\mathrm{far}}_{\mathrm{L}})}.
\end{equation}
Eq.~\ref{eq:oscRatioVsTheta13} is interpreted as a numerator which
gives the mean \nuebar{} survival probability at the far site, and a
denominator which accounts for oscillation present in the near site
measurements.
Consequently,
\begin{align}\label{eq:analytic_theta13_result}
  \sin^{2}2\theta_{13} &= \frac{1 - R}{\eta^{\mathrm{far}} - R\,\eta^{\mathrm{near}}}, \\
  &= 0.085 \pm 0.003 (\mathrm{stat.}) \pm 0.003 (\mathrm{syst.})
\end{align}
was obtained from this simple calculation.

\subsection{\label{subsec:stat_meth}Statistical Methods}
%
Standard frequentist statistical techniques were applied to the
measurement, providing
(i) best estimates of $\theta_{13}$ and ${\Delta}m^2_{\mathrm{32}}$,
(ii) confidence intervals for these parameters,
and (iii) a goodness-of-fit test for the observations relative to the
three-flavor neutrino oscillation model.
Five independent statistical calculations were performed, with each
relying on complementary approaches.
The conceptual details, approach to modeling of systematic
uncertainties, and validation of the calculations are discussed in
this section.

%
All methods defined a $\chi^2$ expression for comparison of the
observation to prediction.
The observation consisted of the reconstructed energy of the prompt
$e^+$ interaction candidates in each detector.
In each case, $N^{\mathrm{obs}}_{ik}$ was the observed number of
candidates in the $k$-th bin of the prompt energy spectrum from the
$i$-th detector, while $N^{\mathrm{exp}}_{ik}$ was the prediction.
The observed and expected counts per energy bin for all detectors can
be expressed as the vectors $\bm{N}^{\mathrm{obs}}$ and
$\bm{N}^{\mathrm{exp}}$\@.
The prediction $\bm{N}^{\mathrm{exp}}$ was a function of the neutrino
oscillation parameters $\theta_{13}$ and
${\Delta}m^2_{\mathrm{32}}$\@.

%
Definitions of the $\chi^2$ statistics differed primarily based on how
additional model parameters and systematic uncertainties were
incorporated into the calculation.
In the {\em profile} approach, additional systematic parameters were
incorporated into the prediction.
The values of these nuisance parameters were profiled; that is, their
value was allowed to vary during minimization of the $\chi^2$\@.
These nuisance parameters were described using the vector $\bm{\nu}$,
following the notation of Sec.~38 of~\cite{PDG2015}.
A systematic term, $\chi^2_{\mathrm{syst}}$, was added to the original
$\chi^2$ test statistic, now labeled $\chi^2_{\mathrm{stat}}$, to
obtain
\begin{equation}\label{eq:chi_nuis}
  \begin{split}
  \chi^2_{\mathrm{prof}}(\theta_{13},{\Delta}m^2_{\mathrm{32}},\bm{\nu}) & = \\
  & \chi^2_{\mathrm{stat}}(\theta_{13},{\Delta}m^2_{\mathrm{32}},\bm{\nu}) + \chi^2_{\mathrm{syst}}(\bm{\nu}).
  \end{split}
\end{equation}
The term $\chi^2_{\mathrm{syst}}$ penalized the total
$\chi^2_{\mathrm{prof}}$ based on deviations of the systematic
nuisance parameters from their expected values.
For the analyses discussed here, either a Poisson maximum likelihood
estimator (see Eq.~38.16 of~\cite{PDG2015}) or a standard Pearson
$\chi^2$ were used for $\chi^2_{\mathrm{stat}}$.

In an alternate {\em covariance} approach, the impact of systematic
variations are integrated into the calculation,
\begin{equation}\label{eq:chi_covar}
  \begin{split}
    \chi^2_{\mathrm{cov}}(\theta_{13},{\Delta}m^2_{\mathrm{32}}) & = \\
    & \left(\bm{N}^{\mathrm{obs}} - \bm{N}^{\mathrm{exp}}\right)^{T}V^{-1}\left(\bm{N}^{\mathrm{obs}} - \bm{N}^{\mathrm{exp}}\right).
  \end{split}
\end{equation}
The covariance matrix $V$ includes both statistical and systematic
components,
\begin{equation}
  V = V_{\mathrm{stat}} + V_{\mathrm{syst}}, 
\end{equation}
where $V_{\mathrm{syst}}$ accounted for the correlated variation
between different energy bins and detectors within the expected
deviations of the systematic parameters.
This approach is mathematically equivalent to the profiling approach
for the simple case of systematic uncertainties which are linear in
character, and has the added benefit of significantly faster
calculation.
Dependence of the covariance matrix $V$ on the parameters, in
particular $\theta_{13}$ and ${\Delta}m^2_{\mathrm{32}}$, was also
included.
Between these two extremes are hybrid calculations where some
systematic parameters are profiled while others are modeled in the
covariance.

\subsubsection{\label{subsec:syst_model}Modeling Systematic Uncertainties}
%
Table~\ref{tab:syst_model} summarizes the systematic components
incorporated into the prediction of $\bm{N}^{\mathrm{exp}}$.
%
\begin{table*}[htb]
  \caption{Summary of systematic uncertainties considered in the
    analysis of \nuebar{} oscillation.
    \label{tab:syst_model}}
 \begin{ruledtabular}
  \newcommand{\ParBoxWidth}{5cm}
  \begin{tabular}{llll}
    \multicolumn{2}{l}{Source} & Uncertainty & Correlation\\
    \hline
    \multicolumn{2}{l}{\bf{Reactor antineutrino flux}} \\
                               & Actinide fission fractions  & 5\%
                               & Correlation between isotopes from Ref.~\cite{Djurcic:2008ny}, \\
                               &&& correlated among all reactors\\
                               & Average energy per fission  & Uncertainties from Ref.~\cite{bib:fr_ma} 
                               & Correlated among all reactors \\
                               & \nuebar{} flux per actinide fission
                               & \parbox[t]{\ParBoxWidth}{\raggedright Uncorrelated uncertainties from Huber+Mueller
                                                                       Model~\cite{HuberAnomaly, Mueller}} 
                               & Correlated  among all reactors\\
                               & Non-equilibrium \nuebar{} emission
                               & 30\% of predicted contribution
                               & Uncorrelated among all reactors\\
                               & Spent nuclear fuel
                               & 100\% of predicted contribution
                               & Uncorrelated among all reactors\\
                               & Reactor power & 0.5\% 
                               & Uncorrelated  among all reactors\\
    \multicolumn{2}{l}{\bf{Detector response model}} \\
                               & Absolute energy scale (non-linearity)   
                               & $<$1\%, constrained by calibration data
                               & Correlated among all detectors\\
                               & Relative energy scale  & 0.2\%
                               & Uncorrelated among all detectors\\
                               & Detection efficiency
                               & 0.13\%.
                                 See Tab.~\ref{tab:s04_table_ibdsummary}
                                 for the breakdown.
                               & Uncorrelated among all detectors,\\
                               &&& partial correlation with 
                                   relative energy scale \\
                               & \parbox[b]{\ParBoxWidth}{\raggedright Detection efficiency and relative 
                                                                       energy scale correlation coefficient} 
                               & 0.54 & -- \\
                               & IAV thickness  
                               & \parbox[t]{\ParBoxWidth}{\raggedright 4\% (0.1\%) of signal for energies below (above) 1.25 MeV} 
                               & Uncorrelated among all detectors\\
                               & Energy resolution   & Negligible
                               & Correlated among all detectors \\
    \multicolumn{2}{l}{\bf{Background prediction}} & ({\em Uncert.'s for EH1, EH2, EH3 given separately.})\\
                               & Accidental rate
                               & 1\% of predicted contribution
                               & Uncorrelated among all detectors\\
                               & Accidental spectral shape   
                               & Negligible & --\\

                               & $^9$Li, $^8$He rate
                               & 44\% of predicted contribution
                               & Correlated among same-site detectors \\ 
                               & $^9$Li fraction   
                               & Negligible
                               & Correlated among all detectors \\
                               & $^9$Li, $^8$He spectral shape   
                               & Negligible
                               & -- \\
                               & Fast neutron rate
                               & 13\%, 13\%, and 17\% of predicted contribution
                               & Correlated among same-site detectors \\
                               & Fast neutron spectral shape
                               & Negligible & -- \\ 

                               & $^{241}$Am-$^{13}$C rate
                               & 45\% of predicted contribution
                               & Correlated among all detectors \\
                               & $^{241}$Am-$^{13}$C spectral shape
                               & Negligible & -- \\ 

                               & ($\alpha$,n) background rate
                               & 50\% of predicted contribution
                               & Uncorrelated among all detectors \\
                               & ($\alpha$,n) spectral shape
                               & Negligible & --\\


  \end{tabular}
 \end{ruledtabular}
\end{table*}

\paragraph{\label{prgh:three_flav_syst}Three-flavor parameters}
%
There was a minor impact on the \nuebar{} survival probability at the
far detectors due to the value of the solar and long-baseline reactor
oscillation parameters, as shown in Eq.~\ref{eq:survProb_ee}.
We adopted the best estimates of $\sin^{2}2\theta_{12} =
0.846\pm0.021$ and ${\Delta}m^2_{21} =
(7.53\pm0.18)\times10^{-5}$~eV$^2$ according to~\cite{PDG2015}.

\paragraph{\label{prgh:reactor_flux_syst}Reactor \nuebar{} Flux}
%
The far versus near detector measurement of oscillation was designed
to be largely insensitive to the model of reactor \nuebar{} emission.
Still, a nominal prediction of the \nuebar{} emission was used to
assess the residual uncertainty not canceled by the far versus near
measurement.
A brief summary of the model and uncertainties are presented here,
while a detailed description is given in~\cite{An:2016srz}.

The emission or \nuebar{} from each reactor was estimated as
\begin{align}
 \frac{d^2R_\nu (E_\nu,t)}{dE_\nu dt} &= \frac{W_\text{th}(t)}{\langle e(t)\rangle}\sum^{\mathrm{isotope}}_i f_i(t) S_i(E_\nu) c_{\text{ne},i} (E_\nu,t) \nonumber\\
& +S_\text{SNF}(E_\nu,t),
\label{eq:reactor_flux_model}
\end{align}
with the following description for each term:
\begin{itemize}
\item $W_{\mathrm{th}}(t)$: The thermal power of the reactor core as a
  function of time $t$\@.  These data were obtained at hourly
  intervals through collaboration with the reactor company. A
  systematic uncertainty of 0.5\% was attributed to these
  data~\cite{Cao:2011gb, Djurcic:2008ny}, and assumed to be
  uncorrelated between reactors.
\item $f_i(t)$: The fraction of nuclear fissions attributed to the
  parent isotope $i$ in each reactor.  The four parent isotopes of
  relevance were $^{235}$U, $^{238}$U, $^{239}$Pu, and $^{241}$Pu.
  These data were also obtained through agreement with the reactor
  company, and validated using an independent simulation of the
  reactors.  An uncertainty of 5\% was assumed, uncorrelated between
  reactors, but with a correlation among isotopes taken
  from~\cite{Djurcic:2008ny}.
\item $\langle e(t)\rangle = \sum_j f_j(t)e_j$: The mean thermal
  energy released per fission.  The energy released $e_j$ per fission
  of parent $j$ were taken from~\cite{bib:fr_ma}.  Uncertainty was
  $<$0.2\% and correlated between reactors.
\item $S_i(E_\nu)$: The estimated \nuebar{} emission versus energy per
  fission of parent isotope $i$.  The predictions by
  Huber~\cite{HuberAnomaly} for $^{235}$U, $^{239}$Pu, and $^{241}$Pu,
  and the prediction for $^{238}$U by Mueller~\cite{Mueller}
  were used.  Uncertainties as described in these references were
  adopted, and taken as correlated between reactors.
\item $c_{\mathrm{ne},i}$: A sub-percent correction in the emitted
  \nuebar{} flux attributed to a non-equilibrium population of fission
  daughters, as described in~\cite{Kopeikin:2012zz}.  This correction
  introduced 0.15\% uncertainty in the predicted number of \nuebar{}
  interactions from one reactor.
\item $S_\text{SNF}(E_\nu,t)$: A sub-percent contribution to the
  \nuebar{} flux from spent nuclear fuel present in the cooling pool
  adjacent to each reactor core.  An uncertainty of 100\%,
  uncorrelated between reactors, was used for this term.  This
  correction introduced 0.38\% uncertainty in the predicted number of
  \nuebar{} interactions from one reactor.
\end{itemize}

%
The predictions of reactor \nuebar{} emission according to these
models have shown a $\sim$6\% rate excess as well as a deviation in
prompt positron energy spectra when compared with
observations~\cite{An:2015nua}.
Relaxing the model uncertainties in normalization and spectral shape
allowed the near detector measurements to accurately constrain the
intrinsic reactor \nuebar{} flux and spectrum.
A minor reactor-related residual uncertainty in the oscillation
measurement was primarily due to the uncorrelated uncertainty in
reactor thermal power and fission fractions.

\begin{table*}[!htb]
    \caption[Summary of Statistical Methods]{Summary of the
      characteristics of the five independent statistical methods
      (labeled A, B, C, D, and E) used to compare the \nuebar{}
      observation with the predictions of the three-flavor model of
      neutrino oscillation.
  \label{tab:stat_meth_summary}}
\begin{minipage}[c]{\textwidth}
\begin{ruledtabular}
\begin{tabular}{ccl}
  Model Component & Method & Description \\
  \hline
  \multirow{4}{*}{Reactor \nuebar{} flux} & A & Analytic prediction based on near detector observation \\
  & B & Unconstrained absolute \nuebar{} spectrum, bin-to-bin uncorrelated. \\
  & C & Huber-Mueller model~\cite{HuberAnomaly,Mueller}, with inflated uncertainty \\
  & D & Unconstrained absolute \nuebar{} spectrum, with piece-wise continuous analytic model \\
  & E & Unconstrained absolute \nuebar{} spectrum, bin-to-bin uncorrelated. \\
  \hline  
  \multirow{4}{*}{Detector response} & A & Analytic model with Geant4-based correction for energy loss in the IAV acrylic \\
  & B & Full Geant4-based detector response, tuned to reproduce the observed energy nonlinearity \\
  & C & Analytic model with Geant4-based correction for energy loss in the IAV acrylic \\
  & D & Analytic model with Geant4-based correction for energy loss in the IAV acrylic \\
  & E & Analytic model including potential correlations between model components \\
  \hline
  \multirow{8}{*}{Systematic Modeling} & A & Pure $\chi^2$ covariance approach for all systematics \\
  & B & Full profiling of systematics via nuisance parameters, with corresponding $\chi^2$ penalty \\
  & C & Hybrid: Full profiling of systematics, with a covariance based penalty used to reduce the dimension of \\
  &   & the reactor model systematics. \\
  & Da & Full profiling of systematics, with penalties for all but reactor spectra coefficients. \\
  & Db & Hybrid: $\chi^2$ covariance approach for all systematics, except for reactor spectra coefficients profiled \\
  &    & with no penalty. \\
  & E & Full profiling of systematics via nuisance parameters, with corresponding $\chi^2$ penalty \\
\end{tabular}
\end{ruledtabular}
\end{minipage}
\end{table*}

\paragraph{\label{prgh:det_syst}Detector Response}
%
Detector-related systematic uncertainties were presented in
Sec.~\ref{sec:background}.
These included:
(i) a 0.2\% uncertainty in energy scale, uncorrelated between
detectors,
(ii) a 0.13\% uncertainty in efficiency between detectors, of which
0.07\% was contributed by the above-mentioned energy scale
uncertainty,
%
(iii) an absolute relation between observed positron energy and true
\nuebar{} energy, constrained according to the model presented in
Sec.~\ref{sec:abs_energy} and assumed to be common for all detectors,
(iv) a slight distortion of the prompt positron spectrum caused by
energy loss in the acrylic of the IAV, 4\% below 1.25~MeV and 0.1\%
above, estimated using simulation and assumed correlated between
detectors.
Of these systematic effects, only the relative energy scale and
efficiency variations were significant to the oscillation measurement.

While it has become common to rely on Monte-Carlo techniques to model
the response of detectors to particle interactions, the specific case
of \nuebar{} interactions in large scintillator detectors was well
suited to an analytic approach.
In this manner, detector systematic parameters were easily
incorporated directly into the analytic response model, without
incurring the computational cost of repeated Monte-Carlo simulations.
Appendix~\ref{sec:signal_prediction} provides a description of the
analytic approach to modeling the expected signal.

\paragraph{\label{prgh:bkgd_syst}Residual Backgrounds}
%
Uncertainties in residual background rates were discussed in
Sec.~\ref{subsec:bkg_rejection}.
Rate uncertainties were assumed to be uncorrelated between detectors
for the accidental background, $^{241}$Am-$^{13}$C and ($\alpha$,n) backgrounds.
The fast neutron and $\beta$-n rate uncertainties were assumed
uncorrelated between experimental halls.
%
%
Uncertainty in the energy spectra of all the backgrounds were considered negligible.
%
Of these, only the $\beta$-n background rate uncertainty was
significant to the oscillation measurement.

%
\begin{figure*}[!htb]
 \includegraphics[width=0.45\textwidth]{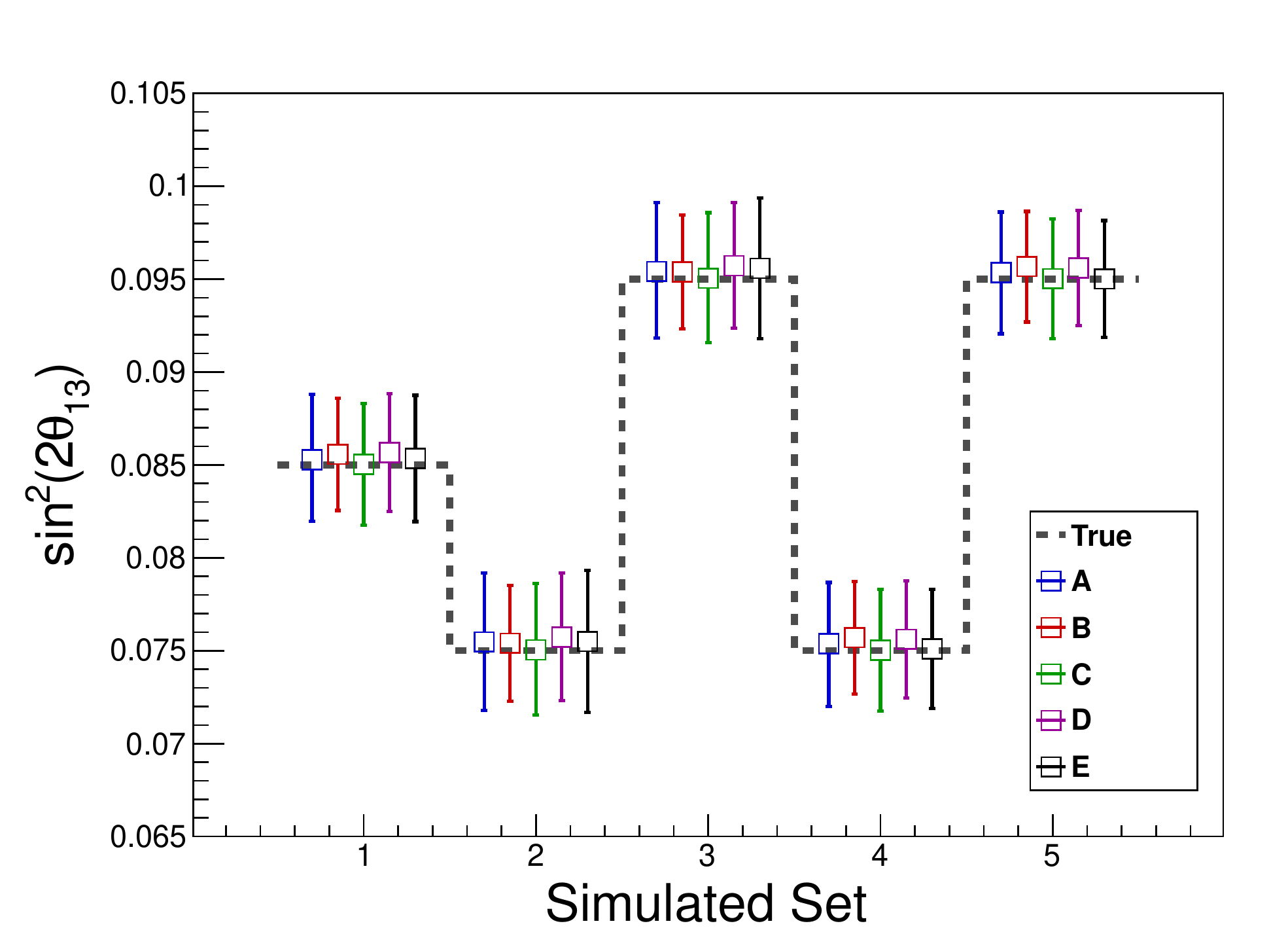}
 \includegraphics[width=0.45\textwidth]{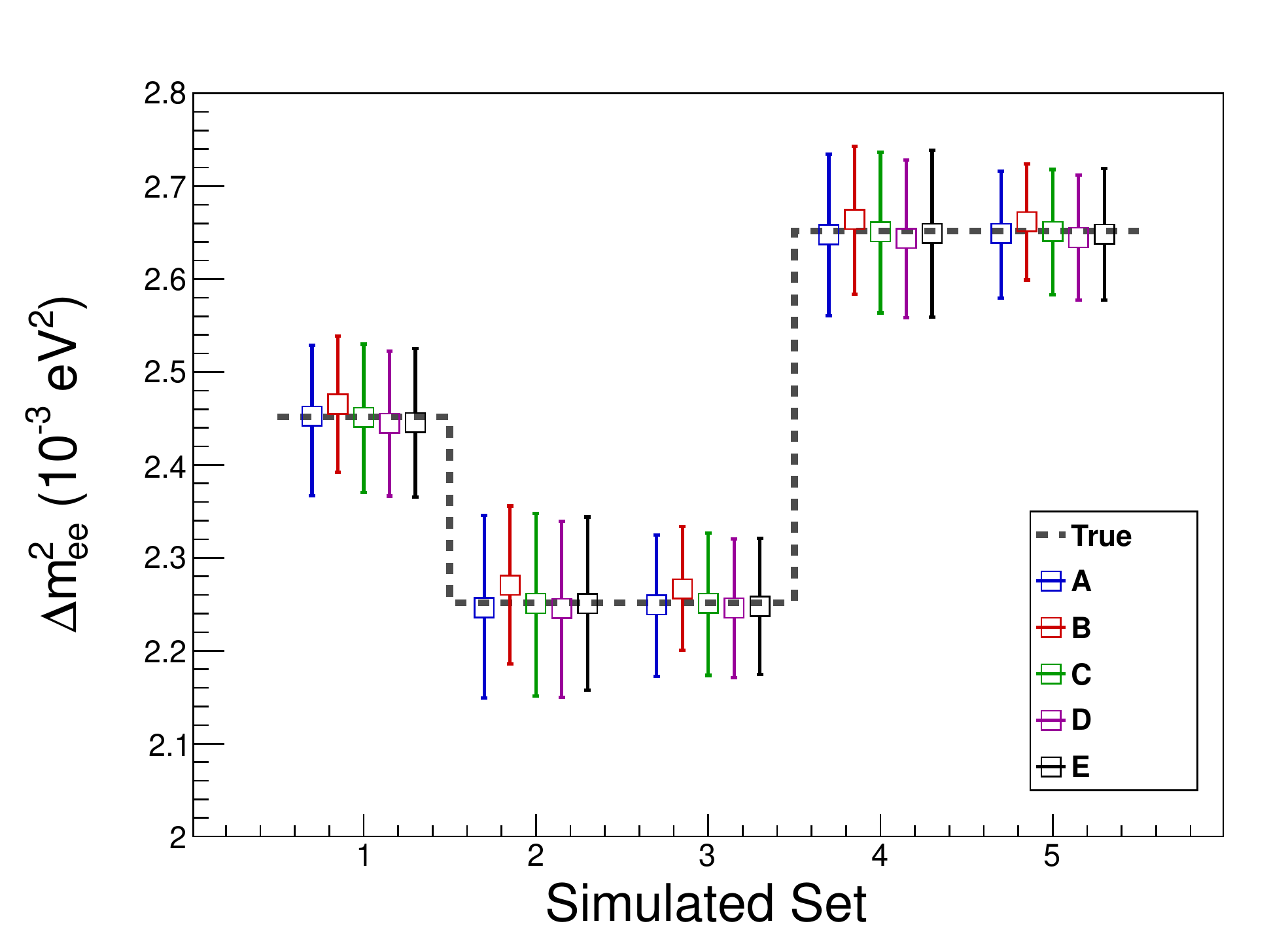}
 \caption{Results of a validation study of the independent statistical
   methods (A through E) used to compare observation with the
   three-flavor model of neutrino oscillation.  An independent program
   was used to generate fake observations under the assumption of a
   range of true values for the oscillation parameters $\theta_{13}$
   and $\left|{\Delta}m^2_{\mathrm{ee}}\right|$ (grey dashed lines).
   All methods demonstrated a consistent unbiased estimation of the
   true input parameters (colored points) for $\theta_{13}$ (left
   panel) and $\left|{\Delta}m^2_{\mathrm{ee}}\right|$ (right panel).
   The final estimated total uncertainties, given by the error bars on
   each point, were also consistent between methods.
   \label{fig:toymc_crosschecks}}
\end{figure*}

\subsubsection{\label{subsec:stat_models}Description of Models}
%
As mentioned, five independent statistical calculations were
performed.
The details of each are presented in this section.

%
Method A was designed, in the spirit of Eq~\ref{eq:nearFarRatio}, to
directly compare the near and far measurements with minimal dependence
on models of reactor \nuebar{} emission.
This method is identical to the method described in Ref.~\cite{An:2015rpe}.

The data from detectors at the same site were combined into a single
observed spectrum for each hall.
The predicted signal in the $i$-th bin for the far-hall energy
spectrum, $N^{\mathrm{far,exp}}_i$, was estimated as
\begin{equation}
  \label{eq:lbnl_prediction}
  N^{\mathrm{far,exp}}_i = w_{i}\left(\theta_{13},\left|{\Delta}m^2_{\mathrm{ee}}\right|\right) (N^{\mathrm{near,obs}}_i - B^{\mathrm{near}}_i) + B^{\mathrm{far}}_i,
\end{equation}
where $N^{\mathrm{near,obs}}_i$ was the signal observed in one of the
near halls, and $B^{\mathrm{near}}_i$ and $B^{\mathrm{far}}_i$ were
the estimated background contamination in each hall.
The weights $w_{i}$ captured the expected ratio of far signal to the
near signal versus observed energy, including the dependence on
oscillation via $\theta_{13}$ and
$\left|{\Delta}m^2_{\mathrm{ee}}\right|$.
These weights were calculated analytically, and were shown to be
effectively independent of the specific model of \nuebar{} emission.
The spectra for the six detector and eight detector data periods were
kept distinct in $\bm{N}^{\mathrm{obs}}$.

Method A used a covariance approach for $\chi^2$ calculation, as given
by Eq.~\ref{eq:chi_covar}.
The statistical component of the covariance matrix was estimated
analytically, while the systematic component was estimated by Monte
Carlo calculations which included the discussed systematic effects.

%
Method B relied on the more traditional approach of predicting the
signal in all the detectors from the reactor flux model
(Eq.~\ref{eq:reactor_flux_model}) convolved with the detector response
determined from a Geant4-based detector simulation.
This method was used for the analysis presented in
Ref.~\cite{DYB2012}.
All systematic uncertainties were profiled, as reflected in
Eq.~\ref{eq:chi_nuis}.

To accommodate the discrepancies between the reactor flux model and
observation, the normalization of each bin in the energy spectrum was
allowed to deviate from the prediction.
Nuisance parameters were used to implement these variations, which
were uncorrelated between energy bins, but identical for all
detectors.
With no systematic penalty applied to these additional terms in the
$\chi^2$, the near detectors effectively constrained the predicted
spectrum.

The detector response model was determined by a full Geant4-based
simulation, instead of the semi-analytic model used in the other four
methods.
This detector response model provided a combined estimate of the
various detector effects, including energy non-linearity, energy loss
in the IAV acrylic, and energy resolution.
The simulated energy non-linearity was adjusted so that it reproduced
the non-linearity observed with data.

%
Method C also relied on a traditional $\chi^2$ comparison of
observation with reactor model prediction, and was used for the
analysis presented in~\cite{DBPRL2014}.
The antineutrino interaction rate and spectrum at each experimental
hall was constrained based on the
Huber-Mueller~\cite{HuberAnomaly,Mueller} model.
To avoid excess tension between the reactor model and observation, the
uncertainty in the normalization of each energy bin in the model was
inflated to have at least 4\% uncertainty~\cite{Hayes:2013wra,
  Hayes:2015yka}.
The overall normalization of the reactor flux was also allowed to
freely float.
The detector response model was determined analytically.
A profile $\chi^2$ was used to account for systematic uncertainties.
To reduce the number of nuisance parameters, the various parameters in
the reactor model were condensed to a single parameter per energy bin.
The penalty on these nuisance parameters were calculated using a
covariance matrix constructed from the expected variation in the
Huber-Mueller model with inflated uncertainties.

%
Method D also used a traditional $\chi^2$ comparison of observation
with prediction based on a reactor flux model, but developed a unique
approach to accommodate deviation between the flux model and
observation.
The \nuebar{} energy spectrum $S_i(E_\nu)$ of each fission parent
isotope $i$ was modeled as a parameterized piece-wise smooth function,
\begin{equation}
 S_{ij}(E_\nu) = n_j k_{ij}\text{e}^{-b_{ij}(E_\nu-E_\nu^j)}, \quad E_\nu \in 
\left(E_\nu^j,E_\nu^{j+1}\right),
\end{equation}
where $k_{ij}$ is the \nuebar{} intensity in the $j$-th energy bin.
The parameters $n_j$ allowed the combined spectral shape to vary in
order to match observation, while the coefficients $b_{ij}$ ensured
that the spectrum for each isotope remained continuous.

Two versions were developed for method D in order to facilitate
comparisons.
In method Da, all systematic uncertainties were accommodated using nuisance
parameters.
In method Db, all parameters were accommodated using a covariance
term, except for the reactor flux spectral parameters $n_{j}$.
For both methods, the parameters $n_{j}$ were profiled with no
systematic penalty.
The developers of method D also extended the calculation to flexibly
accommodate many of the techniques seen in the other methods.
As such, method D served as a valuable tool for understanding subtle
differences between the various methods.

%
Method E was similar in design to method B, except in the approach to
modeling the detector response.
An analytic model was used, but special attention was given to
potential correlations between the various components of the detector
response.
The nonlinear scintillation light emission was separately calculated
for those interactions where a fraction of visible energy was lost via
energy deposition in the IAV\@.
Expected correlations between the nonuniformity of light collection
versus position and the dependence of the electronics nonlinearity
with observed light were also included.
The agreement between this more rigorous approach and the other
analytic predictions confirmed that these potential correlations were
negligible for the current analysis.
This method was also used for an independent measurement of neutrino
oscillation using \nuebar{} interactions followed by neutron capture
on hydrogen.
Further details of the method and the n-H measurement are given
in~\cite{An:2016bvr}.

%
The main features of the five statistical methods are summarized in
Tab.~\ref{tab:stat_meth_summary}.
The consistency of the measurements obtained using these complementary
treatments demonstrated the robustness of the final result.

\subsubsection{\label{subsec:stat_valid}Validation of Methods}
%
Before application of these statistical methods to actual data, each
was tested using simulation.
An independent program was developed to generate simulated
observations, including no statistical or systematic fluctuations.
All statistical methods were then tested with the simulated samples.
The resulting estimated parameter values and total uncertainties were
highly consistent with the true input parameters, as demonstrated in
Figure~\ref{fig:toymc_crosschecks}.

\subsection{\label{subsec:ana_results}Analysis Results}
%
For the final results presented here, statistical method Da was
applied to the data sample provided by selection~A\@.
All results, including parameter estimation, confidence intervals, and
goodness-of-fit, were consistently reproduced using the alternate
statistical methods as well as when applied to selection~B\@.

\subsubsection{\label{subsec:rate_result}Rate-only Analysis}
\begin{figure}[!ht]
  \includegraphics[width=0.9\columnwidth]{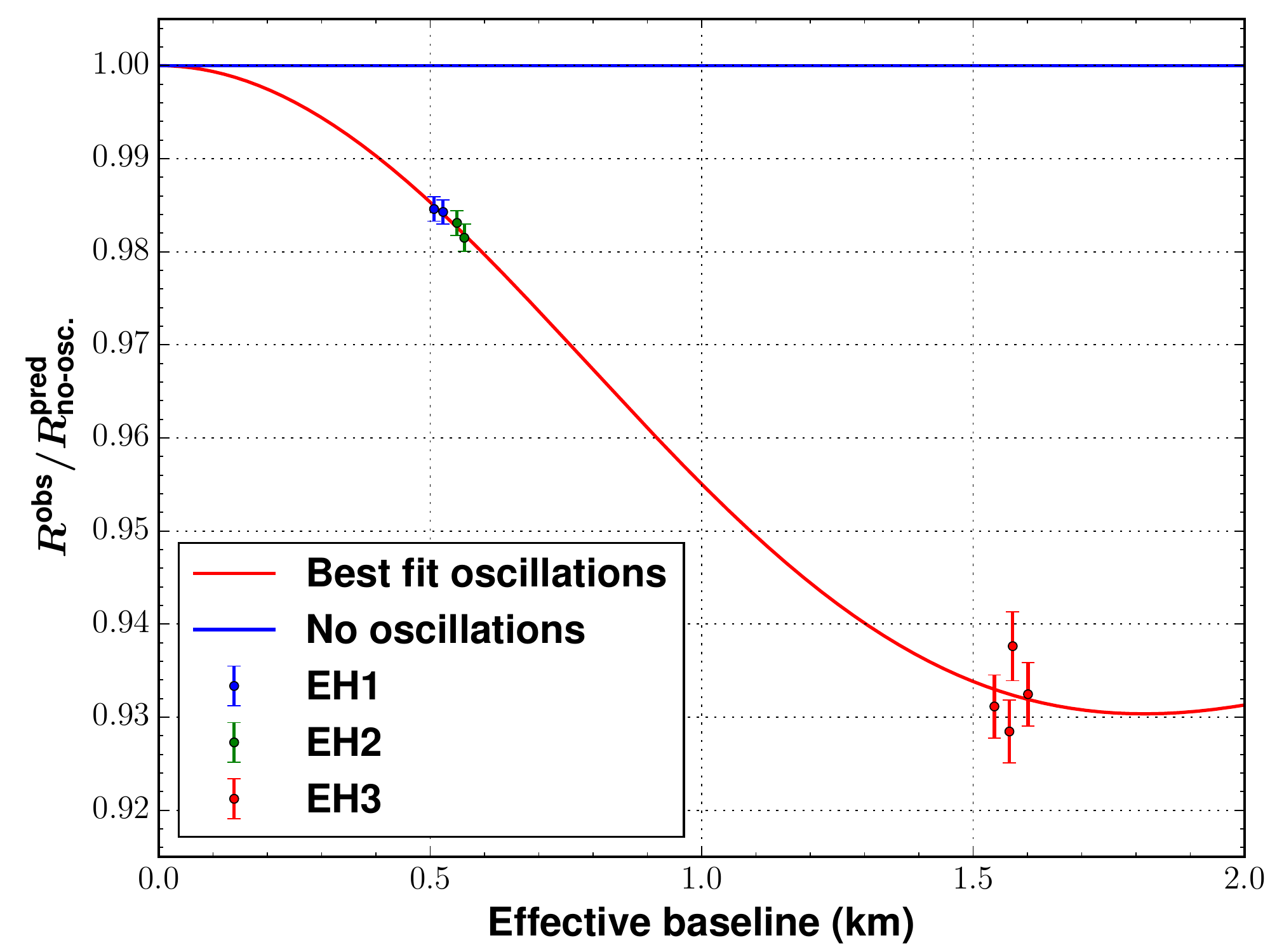}
  \caption{Ratio of the detected \nuebar{} signal to that expected
    assuming no oscillation versus the effective baseline for all
    eight antineutrino detectors.  Oscillation due to $\theta_{13}$
    introduced a deficit in the far detectors relative to the near
    detectors, and the best-fit three-flavor oscillation model from
    the rate-only analysis is shown (red line).  Extrapolation of the
    model to a baseline of zero determined the absolute normalization
    of the reactor \nuebar{} flux, $R^{\mathrm{pred}}(L=0)$.  The
    points representing the near (far) detectors are displaced by $\pm
    6$ m ($\pm 30$ m) for clarity.
  \label{fig:rate_vs_baseline}}
\end{figure}

%
A rate-only statistical analysis was obtained by simplifying the
$\chi^2$ expression to consider only the rates observed in the eight
detectors.
In this case, an external estimate of $\left|{\Delta}m^2_{32}\right|$
was required as input.
The existing value from accelerator muon neutrino disappearance,
$(2.43\pm0.07) \times 10^{-3}$~eV$^2$, was used as
input~\cite{PDG2015}.
The rate-only measurement found
\begin{equation}
 \begin{aligned}
 \sin^22\theta_{13} & = 0.0850 \pm 0.0030\stat \pm 0.0028\syst,\\
 \frac{\chi^2}{\mathrm{NDF}} & = \frac{5.07}{8-2}=0.85,
 \end{aligned}
\label{eq:results_rate_only}
\end{equation}
consistent with the simple analytic estimate given in
Eq.~\ref{eq:analytic_theta13_result}.
Figure~\ref{fig:rate_vs_baseline} illustrates the ratio of the
observed \nuebar{} signal over the no-oscillation prediction versus
the effective baseline for all eight detectors.
For this figure, the effective baseline $L_{\mathrm{eff}}$ between a
given detector and the six reactors was defined as the smallest
positive solution of
\begin{align}
  &\int \sin^2(\Delta m^2_\text{ee} L_\text{eff}/E_\nu)\frac{dN_0}{dE_\nu}dE_\nu \nonumber 
\\
&= \sum_j^\text{reactors} \int\sin^2(\Delta m^2_\text{ee} L_{j}/E_\nu)\frac{dN^{j}_0}{dE_\nu}dE_\nu,
\label{eq:effective_baseline}
\end{align}
where $dN^{j}_0/dE_\nu$ is the expected signal in a given detector due
to reactor $j$ assuming no neutrino oscillations, $dN_0/dE_\nu =
\sum_j dN^{j}_0/dE_\nu$, and $L_j$ is the distance between the
detector and the reactor.
The deficit in the rate observed in the far detectors relative to that
of the near detectors was consistent with \nuebar{} disappearance due
to oscillation.
The absolute normalization of the reactor \nuebar{} flux was
determined from the data, ensuring a ratio of one at a baseline of
zero.
%
%

\subsubsection{\label{subsec:spectral_result}Spectral Analysis}
\begin{figure}[!ht]
\includegraphics[width=0.85\columnwidth]{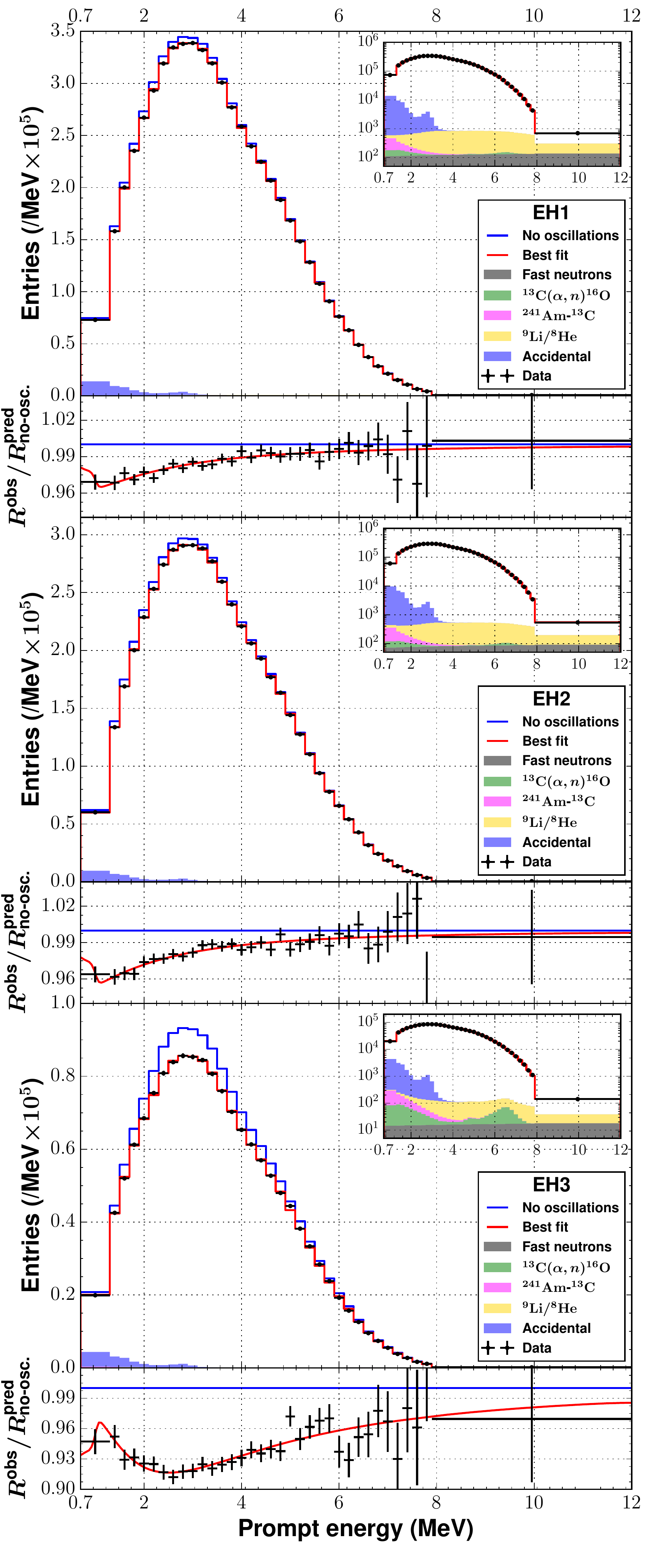}
\caption{Reconstructed positron energy spectra for the \nuebar{}
  candidate interactions (black points).  The spectra of the detectors
  in each experimental hall are combined: EH1 (top), EH2 (middle), and
  EH3 (bottom).  The measurements are compared with the prediction
  assuming no oscillation (blue line) and the best-fit three-flavor
  neutrino oscillation model (red line).  The inset in semi-logarithmic
  scale shows the backgrounds.  The ratio of the background-subtracted
  spectra to prediction assuming no oscillation is shown in the panel
  beneath each energy spectrum.
  \label{fig:shape_eh123}}
\end{figure}

%
Figure~\ref{fig:shape_eh123} shows the reconstructed positron energy
spectra for each experimental hall.
Each spectrum is compared to the prediction assuming no oscillation,
as well as for the three-flavor neutrino oscillation model in best
agreement with the observation.
The distortion of the energy spectrum at the far hall relative to near
halls was consistent with oscillation, and allowed measurement of
$\left|{\Delta}m^2_{\mathrm{ee}}\right|$.
Detailed spectral data are provided as Supplemental
Material~\cite{Note1}\@.
%

%
The parameters of the three-flavor model in best agreement with the
observed rate and energy spectra were
\small
\begin{equation}\label{eq:results_spectral_full}
 \begin{aligned}
   \sin^2 2 \theta_{13}                   & = \\
   & \phantom{-}0.0841\pm0.0027\stat\pm0.0019\syst, \\
   \left|{\Delta}m^2_{\mathrm{ee}}\right|  & = \\
   & \phantom{-}\left[2.50\pm0.06\stat\pm0.06\syst\right]\times 10^{-3}\,\mathrm{eV}^2,    \\
   \Delta m^2_{\mathrm{32}} (\mathrm{NH}) & = \\
   & \phantom{-}\left[2.45\pm0.06\stat\pm0.06\syst\right]\times 10^{-3}\,\mathrm{eV}^2,      \\
   \Delta m^2_{\mathrm{32}} (\mathrm{IH})  & = \\
   & \left[-2.56\pm0.06\stat\pm0.06\syst\right]\times 10^{-3}\,\mathrm{eV}^2, \\
  \frac{\chi^2}{\mathrm{NDF}}           & = \frac{234.7}{280-17} = 0.89.
\end{aligned}
\end{equation}
\normalsize
The ${\Delta}m^2_{\mathrm{32}}$ values were obtained using the full
three-flavor expression from Eq.~\ref{eq:survProb_31}, under the
assumptions of normal (NH) and inverted (IH) mass hierarchy.
$\left|{\Delta}m^2_{\mathrm{ee}}\right|$ was obtained from comparison
of the observation with the effective oscillation model given in
Eq.~\ref{eq:survProb_ee}.
The offset between the values of ${\Delta}m^2_{\mathrm{ee}}$ and
${\Delta}m^2_{\mathrm{32}}$ was identical to an analytic
estimate~\cite{Nunokawa:2005nx}.

\begin{figure}[!htb]
\includegraphics[width=0.9\columnwidth]{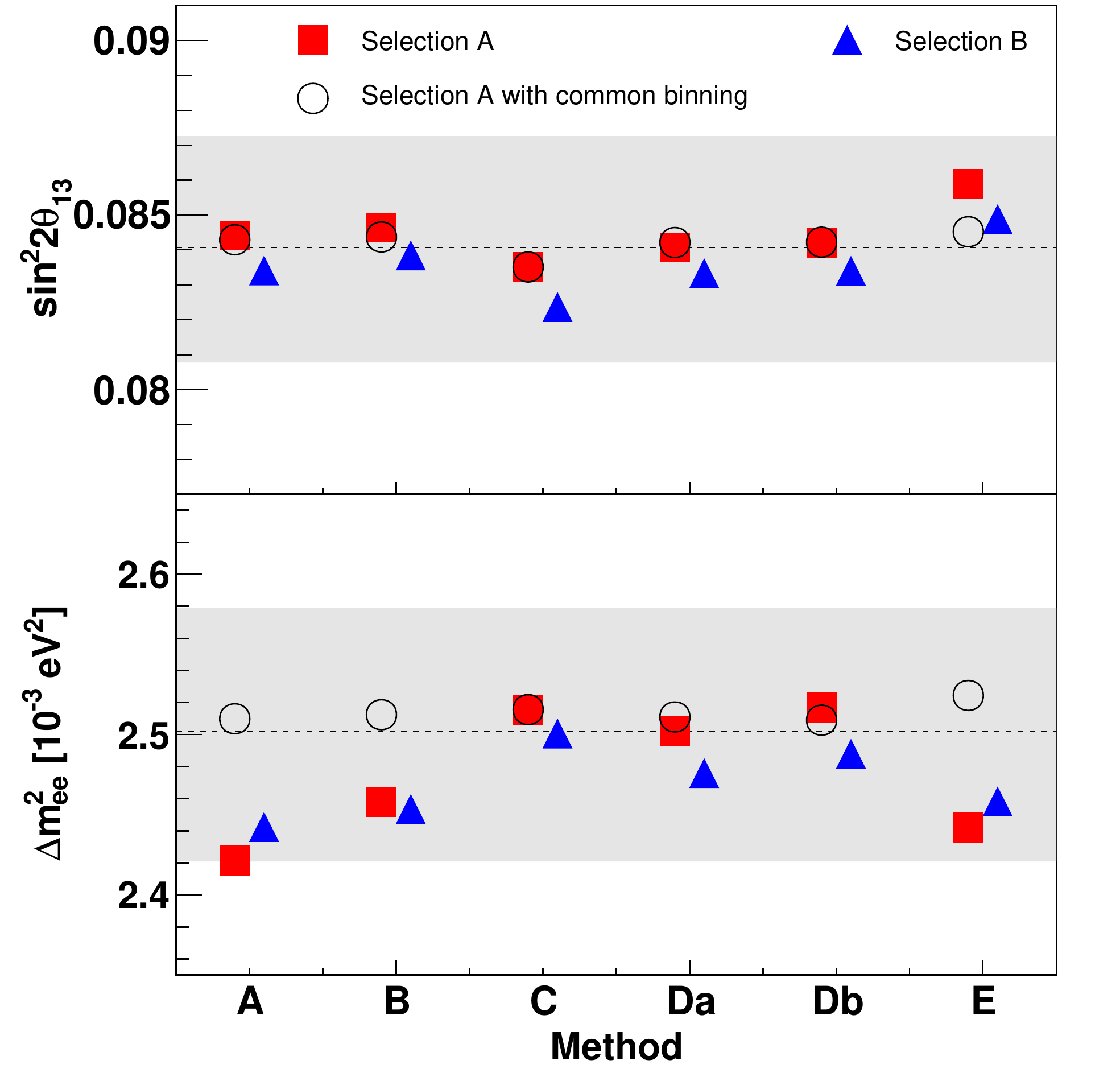}
\caption{A comparison of the estimated values of $\sin^22\theta_{13}$
  (top) and $\Delta m^2_{\mathrm{ee}}$ (bottom) obtained using various
  combinations of the two selected \nuebar{} samples, statistical
  methods, and reactor \nuebar{} flux models.  The horizontal dashed
  lines show the best estimate of each parameter, while the gray
  regions show the $\pm1\sigma$ confidence interval from the reference
  results (selection~A, method~D, and the Huber-Mueller reactor flux
  model).  The $\lesssim$1$\sigma$ offsets for methods A, B, and E
  were due to their choice of binning of the prompt energy spectrum,
  particularly below 1.3~MeV\@.  When all methods used the same
  binning as method~C, consistent results were obtained (open
  circles).  See the text for details.
  \label{fig:oscillation_result_comparison}}
\end{figure}

Figure~\ref{fig:oscillation_result_comparison} compares these
estimates to those obtained using the other statistical methods, as
well as for the alternate sample obtained using selection~B\@.
The slight shift in the estimated value of $\sin^22\theta_{13}$ for
selection~B was consistent with statistical uncertainty from those
candidate signals uncommon between the two selections.
The offsets in the estimated value of $\Delta m^2_{\mathrm{ee}}$ for
the methods A, B and E were predominantly caused by the choice of
binning of the prompt energy spectrum below 1.3~MeV\@.
These three methods divided the low-energy data among multiple bins,
while methods C and D combined the data from 0.7 to 1.3~MeV into a
single bin.
The estimated values for all methods were consistent to
$\lesssim$0.1$\sigma$ when data below 1.3~MeV was combined into a
single bin.
Finely-binning the region below 1.3~MeV was also found to sizeably
worsen the goodness-of-fit.
For example, the $\chi^2$ of method D increased by $\sim$43
(${\Delta}$NDF=16) when the spectrum below 1.3~MeV was binned
identically to method A, and the shift of the estimated value of
$\Delta m^2_{\mathrm{ee}}$ observed by method A was reproduced.
Alternatively, increasing the systematic uncertainty of the
finely-binned energy spectrum below 1.3~MeV also resolved the
discrepancies.
These observations indicated that the combined modeling of the large
systematics at low energies, including relative energy scale
differences and energy loss in the IAV, was insufficient for the case
of a finely-binned low-energy spectrum.
In contrast, the results had negligible dependence on the choice of
binning above 1.3~MeV\@.
Variations were $\lesssim$0.1$\sigma$ for ${\Delta}m^2_{\mathrm{ee}}$,
while those for $\sin^22\theta_{13}$ were even smaller.
%

\begin{figure}[!ht]
\includegraphics[width=\columnwidth]{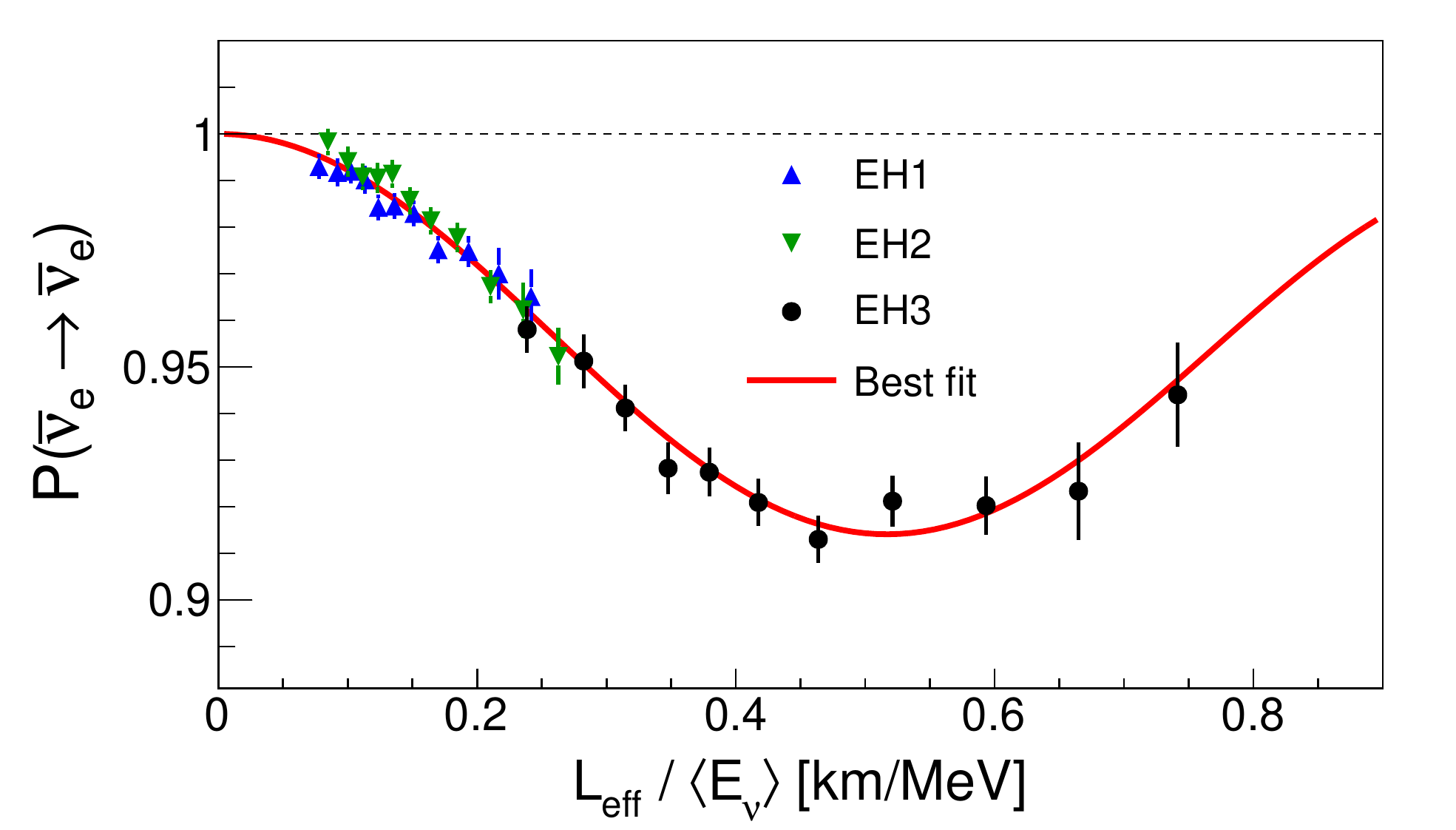}
\caption{Measured reactor \nuebar{} spectral distortion, displayed as
  the oscillation survival probability versus
  $L_{\mathrm{eff}}/E_{\nu}$.  The effective propagation distance
  $L_{\mathrm{eff}}$ was estimated for each hall based on the
  distribution of reactors contributing to the signal
  (see~Eq.~\ref{eq:effective_baseline}).  The average true \nuebar{}
  energy $\langle E_{\nu} \rangle$ was determined for each bin in the
  observed prompt positron spectrum based on the model of the detector
  response.  The \nuebar{} survival probability was given by the
  observed signal in each bin divided by the prediction assuming no
  oscillation.  The measurement sampled \nuebar{} survival over almost
  one full cycle, demonstrating distinct evidence in support of
  neutrino flavor oscillation.
  \label{fig:loe}}
\end{figure}

%
For illustrative purposes, the spectral distortion shown in
Fig.~\ref{fig:shape_eh123} can be displayed as the \nuebar{} survival
probability versus $L$/$E_{\nu}$.
The probability of \nuebar{} disappearance for each bin in the prompt
positron energy spectrum was given by the observed signal divided by
the prediction assuming no oscillation, after subtraction of
background.
The prediction includes corrections to the absolute reactor \nuebar{}
flux as constrained by the observation.
An average \nuebar{} energy $\langle E_{\nu} \rangle$ was estimated
for each bin in the prompt positron spectra from the model of the
detector response previously discussed.
Given that it was not possible to determine the reactor-of-origin for
each \nuebar{} interaction, an effective baseline $L_{\mathrm{eff}}$
was determined for each experimental hall, according to
Eq.~\ref{eq:effective_baseline}.
Figure~\ref{fig:loe} shows the observed \nuebar{} survival probability
as a function of effective baseline $L_{\mathrm{eff}}$ divided by the
average antineutrino energy $\langle E_{\nu} \rangle$.
Almost one full oscillation cycle was sampled, given the range of
$L$/$E_{\nu}$ values which were measured.
The data from all three experimental halls were consistent with the
three-flavor oscillation hypothesis.

\begin{figure}[!htb]
\includegraphics[width=\columnwidth]{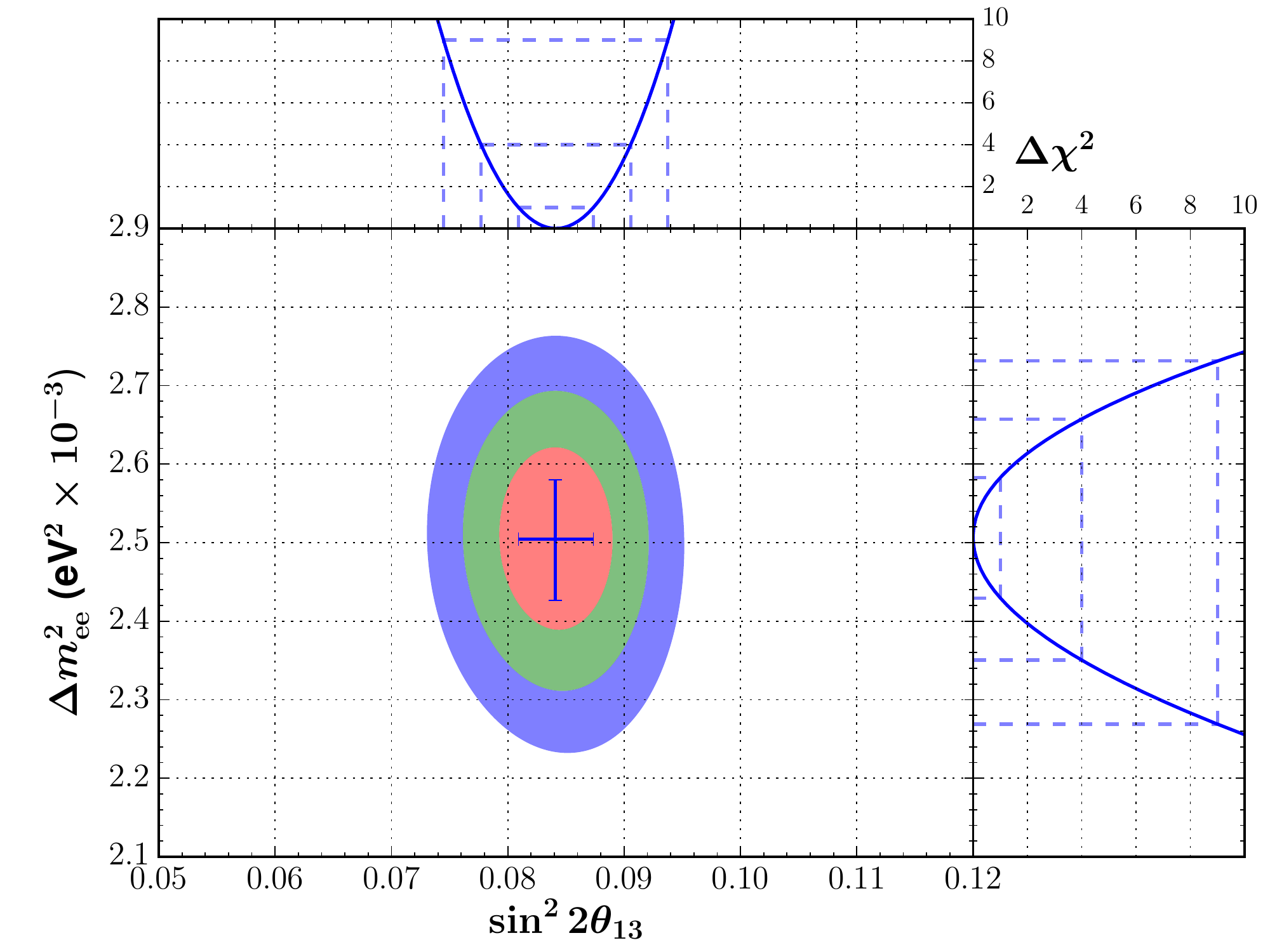}
\caption{Confidence regions of $\sin^22\theta_{13}$ and
  $\left|{\Delta}m^2_{\mathrm{ee}}\right|$ from a combined analysis of
  the prompt positron spectra and rates.  The 1$\sigma$, 2$\sigma$,
  and 3$\sigma$ 2-D confidence regions are estimated using
  ${\Delta}\chi^2$ values of 2.30 (red), 6.18 (green), and 11.83
  (blue) relative to the best fit.  The upper panel provides the 1-D
  ${\Delta}\chi^2$ for $\sin^22\theta_{13}$ obtained by profiling
  $\left|{\Delta}m^2_{\mathrm{ee}}\right|$ (blue line), and the dashed
  lines mark the corresponding 1$\sigma$, 2$\sigma$, and 3$\sigma$
  intervals.  The right panel is the same, but for
  $\left|{\Delta}m^2_{\mathrm{ee}}\right|$, with $\sin^22\theta_{13}$
  profiled. The point marks the best estimates, and the error bars
  display their 1-D 1$\sigma$ confidence intervals.
\label{fig:oscillation_allowed_region}}
\end{figure}

%
The confidence regions for ${\Delta}m^2_{\mathrm{ee}}$ versus $\sin^2
2\theta_{13}$ are shown in
Figure~\ref{fig:oscillation_allowed_region}.
The confidence regions were obtained using the change of the $\chi^2$
value relative to that of the best fit, $\Delta\chi^2 =
\chi^2-\chi^2_{\mathrm{min}}$, as a function of $\sin^22\theta_{13}$
and $\left|{\Delta}m^2_{\mathrm{ee}}\right|$.
All other model parameters were profiled during the determination of
the value of $\Delta\chi^2$.
The confidence regions are defined as $\Delta\chi^2$ less than 2.30
(68.27\%~C.L.), 6.18 (95.45\%~C.L.), and 11.83 (99.73\%~C.L.).
The 1-D distribution of $\Delta\chi^2$ are also provided for each
individual parameter, where the alternate parameter has been profiled.
A table of $\Delta\chi^2$ values as a function of $\sin^22\theta_{13}$
and $\left|{\Delta}m^2_{\mathrm{ee}}\right|$ is provided as
Supplemental Material~\cite{Note1}\@.

The precision of this measurement of $\theta_{13}$ was limited by
statistics, although systematic uncertainty from differences of the
\nuebar{} efficiency between detectors and predicted reactor flux also
contributed significantly.
For $\left|{\Delta}m^2_{\mathrm{ee}}\right|$, statistical and
systematic uncertainties were approximately equal in size.
The largest systematic uncertainty arose from potential variation in
the energy calibration of the far versus near detectors, which was
well characterized using multiple redundant low-energy radioactive
sources.
Systematic uncertainty from \nuebar{} interactions in the IAV also
contributed.

\begin{figure}[!htb]
  \centering
 \includegraphics[width=\columnwidth]{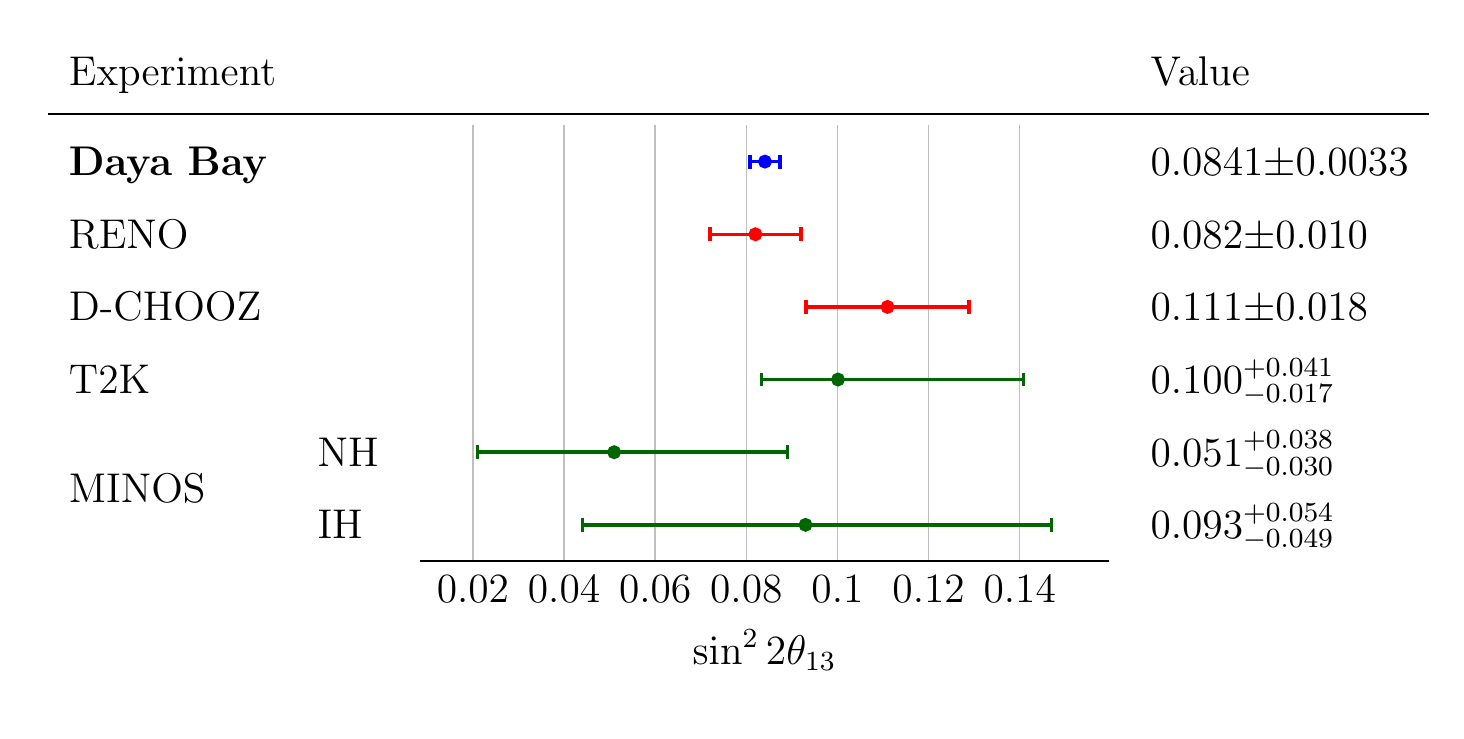}
 \caption{Comparison of measurements of $\sin^22\theta_{13}$: this
   measurement (blue point); RENO~\cite{RENO:2015ksa} and
   Double-CHOOZ~\cite{Ishitsuka:Moriond2016} (red points);
   T2K~\cite{Iwamoto:ICHEP2016} and MINOS~\cite{Adamson:2013ue} (green
   points).  The T2K and MINOS values were deduced from
   $2\sin^2\theta_{23}\sin^22\theta_{13}$, and are presented for the
   two cases of the normal (upper) and inverted (lower) mass
   hierarchy.  The MINOS measurement assumed $\sin^2\theta_{23} =
   0.5$, $\delta_{\mathrm{CP}} = 0$, while the T2K measurement
   marginalized over these unknown parameters.
 \label{fig:s06_theta13_global}}
\end{figure}

\begin{figure}[!ht]
 \includegraphics[width=\columnwidth]{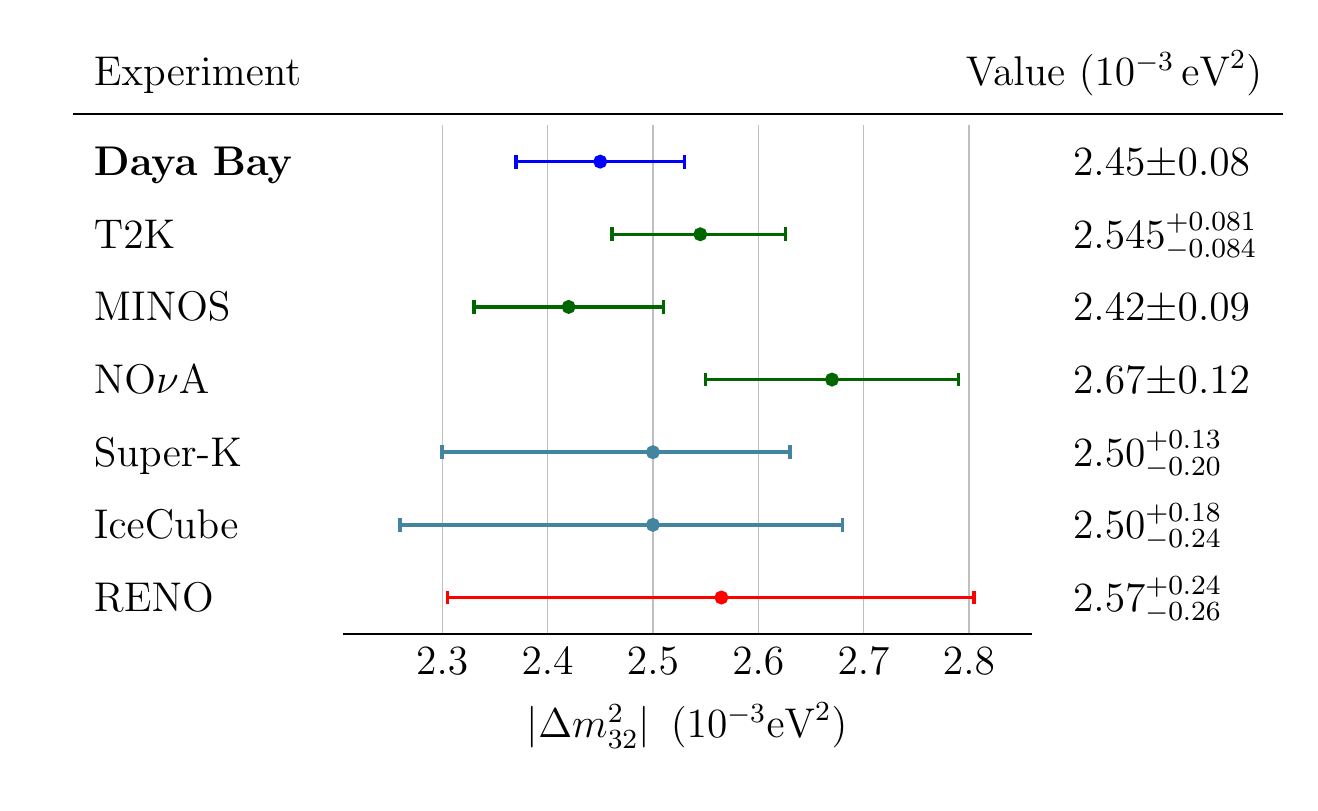}
 \caption{Comparison of measurements of ${\Delta}m^2_{32}$: this
   measurement (blue point); RENO~\cite{RENO:2015ksa} (red point);
   T2K~\cite{Iwamoto:ICHEP2016}, MINOS~\cite{Evans:Neutrino2016}, and
   NO$\nu$A~\cite{Vahle:Neutrino2016} (green points);
   Super-Kamiokande~\cite{Moriyama:Neutrino2016} and
   IceCube~\cite{Koskinen:Neutrino2016} (cyan points).  All values are
   given for the case of the normal mass hierarchy; the comparison for
   the inverted ordering was qualitatively similar.
   \label{fig:s06_dm32_global}}
\end{figure}

%
Figure~\ref{fig:s06_theta13_global} compares the estimate of
$\sin^22\theta_{13}$ with those values obtained by other experiments,
while Figure~\ref{fig:s06_dm32_global} provides a similar comparison
for measurements of ${\Delta}m^2_{32}$.
The measurements relied on a variety of $\nu$ observations:
\begin{itemize}
\item the disappearance of MeV-energy reactor \nuebar{}'s over
  $\sim$km distances,
\item the disappearance of $\nu_\mu$ produced by particle accelerators
  with mean energies of $\sim$600~MeV~\cite{Abe:2014ugx},
  $\sim$3~GeV~\cite{Adamson:2014vgd}, and
  $\sim$2~GeV~\cite{Adamson:2016xxw} which had propagated
  distances of $\sim$295~km, $\sim$735~km, and $\sim$810~km
  respectively,
\item the appearance of $\nu_{\mathrm{e}}$ in those same neutrino
  beams, and
\item the disappearance of $\nu_\mu$ produced by particle interactions
  in the upper atmosphere~\cite{Wendell:2014dka,Aartsen:2014yll}, with
  energies $>$1~GeV and baselines up to the diameter of the Earth.
\end{itemize}
The consistency of the values of ${\Delta}m^2_{32}$ measured via these
various techniques firmly establishes the three-flavor model of
neutrino mass and mixing.


\section{\label{sec:conclusions} Summary} 

From Dec.~4,~2011 to Jul.~28,~2015, the Daya~Bay experiment measured
the rate and energy spectrum of electron antineutrinos emitted by the
six 2.9~GW$_{\mathrm{th}}$ reactors of the Daya~Bay and Ling~Ao
nuclear power facilities.
Combining 217 days of data collected using six antineutrino detectors
with 1013 days of data using eight detectors, a total of
$2.5\times10^6$ \nuebar{} inverse beta decay interactions were
observed.
The unprecedented statistics of this sample allowed the most precise
measurement of \nuebar{} disappearance to date.
A relative comparison of the rates and positron energy spectra of the
detectors located far ($\sim$1500-1950~m) relative to those near the
reactors ($\sim$350-600~m) gave $\sin^22\theta_{13}=\TRS{}$ and the
effective neutrino mass-squared difference of
$\left|{\Delta}m^2_{\mathrm{ee}}\right|=\DMEE{}$\@.
This is equivalent to ${\Delta}m^2_{32}=\DMNH{}$ assuming the normal
mass hierarchy, or ${\Delta}m^2_{32}=\DMIH{}$ assuming the inverse
hierarchy.
The consistency with ${\Delta}m^2_{32}$ measured using $\sim$GeV
accelerator and atmospheric $\nu_{\mu}$ disappearance strongly
supports the three-flavor model of neutrino oscillation.

\begin{acknowledgments}

Daya Bay is supported in part by the Ministry of Science and Technology of China, the U.S. Department of Energy, the Chinese Academy of Sciences, the CAS Center for Excellence in Particle Physics, the National Natural Science Foundation of China, the Guangdong provincial government, the Shenzhen municipal government, the China Guangdong Nuclear Power Group, Key Laboratory of Particle and Radiation Imaging (Tsinghua University), the Ministry of Education, Key Laboratory of Particle Physics and Particle Irradiation (Shandong University), the Ministry of Education, Shanghai Laboratory for Particle Physics and Cosmology, the Research Grants Council of the Hong Kong Special Administrative Region of China, the University Development Fund of The University of Hong Kong, the MOE program for Research of Excellence at National Taiwan University, National Chiao-Tung University, and NSC fund support from Taiwan, the U.S. National Science Foundation, the Alfred~P.~Sloan Foundation, the Ministry of Education, Youth, and Sports of the Czech Republic, the Joint Institute of Nuclear Research in Dubna, Russia, the CNFC-RFBR joint research program, the National Commission of Scientific and Technological Research of Chile, and the Tsinghua University Initiative Scientific Research Program. We acknowledge Yellow River Engineering Consulting Co., Ltd., and China Railway 15th Bureau Group Co., Ltd., for building the underground laboratory. We are grateful for the ongoing cooperation from the China General Nuclear Power Group and China Light and Power Company.

\end{acknowledgments}

\appendix
\section{\label{sec:signal_prediction}Prediction of the \nuebar{} Signal}

A method to calculate the expected rate and reconstructed positron
energy spectrum from \nuebar{} inverse beta decay interactions in the
Daya Bay detectors is summarized in this appendix.
The total number of signals in the reconstructed energy interval
$\{E^{k}_{\mathrm{rec}},E^{k+1}_{\mathrm{rec}}\}$ of the
prompt energy spectrum for detector $i$ is given by
\begin{equation}
  N^{\mathrm{exp}}_{ik} = N^{\mathrm{IBD}}_{ik} + N^{\mathrm{bkg}}_{ik}, 
\end{equation}  
where $N^{\mathrm{IBD}}_{ik}$ are from \nuebar{} inverse beta decay
positrons and $N^{\mathrm{bkg}}_{ik}$ are the contributions from
backgrounds.
The background spectra are displayed in Fig.~\ref{fig:shape_eh123},
while their rates are summarized in
Table~\ref{tab:s04_table_ibdsummary}.
The IBD signal is given by
\begin{equation}
N^{\mathrm{IBD}}_{ik} = \int_{E^{k}_{\mathrm{rec}}}^{E^{k+1}_{\mathrm{rec}}}dE_{\mathrm{rec}}\int_{t_{\mathrm{DAQ}}} dt\;\frac{d^2N_i}{dE_{\mathrm{rec}}dt} \;\varepsilon_{i}(t),
\label{eq:app:expectation_reco} 
\end{equation}
where $d^2N_i(E_{\mathrm{rec}},t)/dE_{\mathrm{rec}}dt$ is the expected
signal number density as a function of time and reconstructed energy.
The integral includes the efficiency of detector $i$,
$\varepsilon_{i}(t)$, which accounts for the slight variations in
detector livetime and veto efficiency versus time.
Eq.~\ref{eq:app:expectation_reco} was designed for use in the combined
analysis of the spectrum and rate, but was also applied to the
rate-only analysis by using only a single energy interval per
detector.

Given the true IBD positron energy, including the energy from
annihilation,
\begin{equation}
  E_{\mathrm{true}} = E_{e} + m_e,
\end{equation}
the expected signal number density can be obtained by a convolution of
the true signal number density with the estimated detector response,
\begin{equation}
\frac{d^2N_i}{dE_{\mathrm{rec}}dt} = \int_{2m_e}^\infty dE_{\mathrm{true}}\;\frac{d^2N_i}{dE_{\mathrm{true}}dt} \;P(E_{\mathrm{rec}};E_{\mathrm{true}}).
\label{eq:app_expectation_reco_density}
\end{equation}
The estimated detector response,
$P(E_{\mathrm{rec}};E_{\mathrm{true}})$, describes the conditional
probability of obtaining a reconstructed energy $E_{\mathrm{rec}}$
given a true energy of $E_{\mathrm{true}}$.

The expected number density of IBD signals per interval of true
positron energy,
\begin{equation}
\frac{d^2N_i}{dE_{\mathrm{true}} dt} = N_i^p \int_{-1}^{1}d\cos\theta_e\frac{d\sigma}{d\cos\theta_e}\frac{d^2N^\nu_i}{dE_\nu dt} \frac{dE_\nu}{dE_{\mathrm{true}}}.
\label{eq:app_expectation_true_density}
\end{equation}
is the product of the number density of antineutrinos,
$d^2N^\nu_i/dE_{\nu}dt$, the number of protons in the detector,
$N_i^p$, and the \nuebar{}-proton IBD interaction cross section,
$d\sigma/d\cos\theta_e$.
The emission angle of the positron, $\theta_e$, was not resolved by
the detectors, and so it is integrated in this calculation.

The IBD cross-section as a function of positron scattering angle and
\nuebar{} energy, $d\sigma/d\cos\theta_e$, was taken
from~\cite{Vogel:1999zy}.
This tree-level calculation was performed up to first order in
$1/m_p$, and accounted for recoil, weak magnetism, and inner radiative
corrections.

\begin{figure}[htb]
  \includegraphics[width=0.45\textwidth]{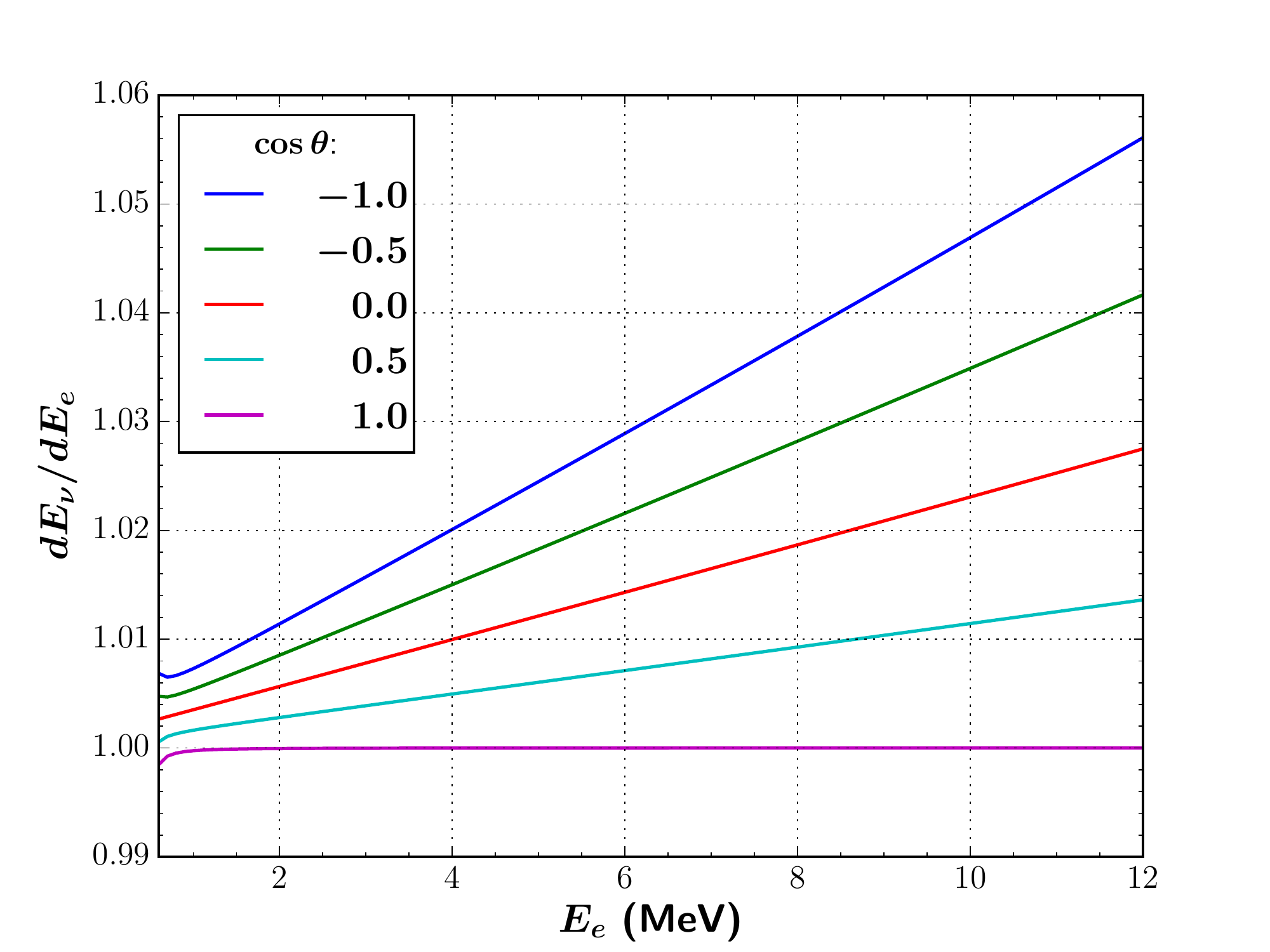}
  \caption{The Jacobian $dE_\nu/dE_{\mathrm{true}}$ as a
    function of positron energy $E_e$ for a range of values of
    the positron emission angle $\cos\theta_e$.
  \label{fig:s06_csc_jacobian}}
\end{figure}

The dependence of the positron's energy $E_e$ and scattering angle
$\theta_e$ on antineutrino energy $E_\nu$ is
\begin{equation}
E_\nu = E_\nu(E_e,\cos\theta_e) = \frac{E_e + \widetilde{\Delta}}{1-\frac{E_e}{m_p}(1-\beta_e\cos\theta_e)},
\label{eq:s06_osc_analysis_Enu_vs_Ee_cos} 
\end{equation}
where $\widetilde{\Delta} = (m_n^2-m_p^2-m_e^2)/2m_p \approx m_n-m_p$
and $\beta_e$ is the positron velocity.
The corresponding Jacobian is,
\begin{equation}
\begin{aligned}
\frac{dE_\nu}{dE_{\mathrm{true}}} = \frac{1+\frac{E_\nu}{m_p}(1-\beta_e^{-1}\cos\theta_e)}{1-\frac{E_e}{m_p}(1-\beta_e\cos\theta_e)},
\end{aligned}
\label{eq:app_jacobian} 
\end{equation}
and is shown in Fig.~\ref{fig:s06_csc_jacobian} as a function of
positron energy $E_e$.

The number density of reactor \nuebar{} passing through the detector
is estimated as
\begin{equation}
 \frac{d^2N^\nu_i(E_\nu,t)}{dE_\nu dt} = \sum_j \frac{d^2R^j_\nu (E_\nu,t)}{dE_\nu dt} \frac{P_{\mathrm{sur}}(E_\nu,L_{ij})}{4\pi L^2_{ij}},
\end{equation}
where $d^2R^j_\nu (E_\nu,t)/dE_\nu dt$ is given by
Eq.~\ref{eq:reactor_flux_model}, $L_{ij}$ is the distance between
centers of reactor core $j$ and detector $i$, and
$P_{\mathrm{sur}}(E_\nu,L)$ is the \nuebar{} survival probability
given by Eq.~\ref{eq:survProb_31}.

The detector response, $P(E_{\mathrm{rec}};E_{\mathrm{true}})$,
accounted for the detector and reconstruction effects discussed in
Sec.~\ref{sec:calibration}.
Two approaches were used to estimate this relationship.
The first method decomposed the detector response into three prominent
components:
\begin{itemize}
  \item $P(E_{\mathrm{dep}};E_{\mathrm{true}})$: The probability to
    have an energy $E_{\mathrm{dep}}$ deposited in the scintillator
    for a given true positron energy $E_{\mathrm{true}}$\@.  The two
    energies are usually identical, except for $\sim$1\% of positrons
    which lost a fraction of their energy in the inner acrylic vessel.
    This component, commonly referred to as the IAV effect, was
    modeled via a Geant4-based simulation.
  \item $P(\bar{E}_{\mathrm{rec}};E_{\mathrm{dep}})$: The probability
    to obtain a mean reconstructed energy $\bar{E}_{\mathrm{rec}}$
    given a given deposited energy $E_{\mathrm{dep}}$\@.  This
    component accounts for the nonlinear light emission of the
    scintillator and the nonlinear response of the PMT charge
    measurement, and is given by the function shown in
    Fig.~\ref{fig:ibdresponse}.
  \item $P(E_{\mathrm{rec}};\bar{E}_{\mathrm{rec}})$: The probability
    to obtain a reconstructed energy $E_{\mathrm{rec}}$ for a given
    expected mean reconstructed energy $\bar{E}_{\mathrm{rec}}$.  This
    term accounts for the detector resolution, and was modeled as a
    Gaussian distribution about $\bar{E}_{\mathrm{rec}}$ with a
    standard deviation given by Eq.~\ref{eq:energy_resolution}.
\end{itemize}
The total detector response is simply the consecutive convolution of
these three distributions.
The distributions $P(E_{\mathrm{dep}};E_{\mathrm{true}})$ and
$P(\bar{E}_{\mathrm{rec}};E_{\mathrm{dep}})$, as well as the combined
distribution $P(E_{\mathrm{rec}};E_{\mathrm{true}})$, are provided as
Supplemental Material~\cite{Note1}\@.

The second method used full Geant4-based simulation to construct the
detector response as a single unified distribution
$P(E_{\mathrm{rec}};E_{\mathrm{true}})$.
The simulation was adjusted to reproduce the observed calibration
data, and naturally included the potential interplay between the
different components of the detector response.
In principle, this technique allowed for more accurately modeling of
the detector response near $E_{\mathrm{rec}} \simeq 1$~MeV where the
effect of the IAV was most significant.
The two methods yielded consistent predictions for the observed prompt
energy spectrum.

\bibliographystyle{apsrev4-1}
\bibliography{DayaBay}

\begin{thebibliography}{79}%
\makeatletter
\providecommand \@ifxundefined [1]{%
 \@ifx{#1\undefined}
}%
\providecommand \@ifnum [1]{%
 \ifnum #1\expandafter \@firstoftwo
 \else \expandafter \@secondoftwo
 \fi
}%
\providecommand \@ifx [1]{%
 \ifx #1\expandafter \@firstoftwo
 \else \expandafter \@secondoftwo
 \fi
}%
\providecommand \natexlab [1]{#1}%
\providecommand \enquote  [1]{``#1''}%
\providecommand \bibnamefont  [1]{#1}%
\providecommand \bibfnamefont [1]{#1}%
\providecommand \citenamefont [1]{#1}%
\providecommand \href@noop [0]{\@secondoftwo}%
\providecommand \href [0]{\begingroup \@sanitize@url \@href}%
\providecommand \@href[1]{\@@startlink{#1}\@@href}%
\providecommand \@@href[1]{\endgroup#1\@@endlink}%
\providecommand \@sanitize@url [0]{\catcode `\\12\catcode `\$12\catcode
  `\&12\catcode `\#12\catcode `\^12\catcode `\_12\catcode `\%12\relax}%
\providecommand \@@startlink[1]{}%
\providecommand \@@endlink[0]{}%
\providecommand \url  [0]{\begingroup\@sanitize@url \@url }%
\providecommand \@url [1]{\endgroup\@href {#1}{\urlprefix }}%
\providecommand \urlprefix  [0]{URL }%
\providecommand \Eprint [0]{\href }%
\providecommand \doibase [0]{http://dx.doi.org/}%
\providecommand \selectlanguage [0]{\@gobble}%
\providecommand \bibinfo  [0]{\@secondoftwo}%
\providecommand \bibfield  [0]{\@secondoftwo}%
\providecommand \translation [1]{[#1]}%
\providecommand \BibitemOpen [0]{}%
\providecommand \bibitemStop [0]{}%
\providecommand \bibitemNoStop [0]{.\EOS\space}%
\providecommand \EOS [0]{\spacefactor3000\relax}%
\providecommand \BibitemShut  [1]{\csname bibitem#1\endcsname}%
\let\auto@bib@innerbib\@empty
\bibitem [{\citenamefont {Cleveland}\ \emph {et~al.}(1998)\citenamefont
  {Cleveland}, \citenamefont {Daily}, \citenamefont {Davis}, \citenamefont
  {Distel}, \citenamefont {Lande}, \citenamefont {Lee}, \citenamefont
  {Wildenhain},\ and\ \citenamefont {Ullman}}]{Cleveland:1998nv}%
  \BibitemOpen
  \bibfield  {author} {\bibinfo {author} {\bibfnamefont {B.~T.}\ \bibnamefont
  {Cleveland}}, \bibinfo {author} {\bibfnamefont {T.}~\bibnamefont {Daily}},
  \bibinfo {author} {\bibfnamefont {R.}~\bibnamefont {Davis}, \bibfnamefont
  {Jr.}}, \bibinfo {author} {\bibfnamefont {J.~R.}\ \bibnamefont {Distel}},
  \bibinfo {author} {\bibfnamefont {K.}~\bibnamefont {Lande}}, \bibinfo
  {author} {\bibfnamefont {C.~K.}\ \bibnamefont {Lee}}, \bibinfo {author}
  {\bibfnamefont {P.~S.}\ \bibnamefont {Wildenhain}}, \ and\ \bibinfo {author}
  {\bibfnamefont {J.}~\bibnamefont {Ullman}},\ }\href {\doibase 10.1086/305343}
  {\bibfield  {journal} {\bibinfo  {journal} {Astrophys. J.}\ }\textbf
  {\bibinfo {volume} {496}},\ \bibinfo {pages} {505} (\bibinfo {year}
  {1998})}\BibitemShut {NoStop}%
\bibitem [{\citenamefont {Kaether}\ \emph {et~al.}(2010)\citenamefont
  {Kaether}, \citenamefont {Hampel}, \citenamefont {Heusser}, \citenamefont
  {Kiko},\ and\ \citenamefont {Kirsten}}]{Kaether:2010ag}%
  \BibitemOpen
  \bibfield  {author} {\bibinfo {author} {\bibfnamefont {F.}~\bibnamefont
  {Kaether}}, \bibinfo {author} {\bibfnamefont {W.}~\bibnamefont {Hampel}},
  \bibinfo {author} {\bibfnamefont {G.}~\bibnamefont {Heusser}}, \bibinfo
  {author} {\bibfnamefont {J.}~\bibnamefont {Kiko}}, \ and\ \bibinfo {author}
  {\bibfnamefont {T.}~\bibnamefont {Kirsten}},\ }\href {\doibase
  10.1016/j.physletb.2010.01.030} {\bibfield  {journal} {\bibinfo  {journal}
  {Phys. Lett.}\ }\textbf {\bibinfo {volume} {B685}},\ \bibinfo {pages} {47}
  (\bibinfo {year} {2010})},\ \Eprint {http://arxiv.org/abs/1001.2731}
  {arXiv:1001.2731 [hep-ex]} \BibitemShut {NoStop}%
\bibitem [{\citenamefont {Abdurashitov}\ \emph {et~al.}(2009)\citenamefont
  {Abdurashitov} \emph {et~al.}}]{Abdurashitov:2009tn}%
  \BibitemOpen
  \bibfield  {author} {\bibinfo {author} {\bibfnamefont {J.~N.}\ \bibnamefont
  {Abdurashitov}} \emph {et~al.} (\bibinfo {collaboration} {SAGE
  Collaboration}),\ }\href {\doibase 10.1103/PhysRevC.80.015807} {\bibfield
  {journal} {\bibinfo  {journal} {Phys. Rev.}\ }\textbf {\bibinfo {volume}
  {C80}},\ \bibinfo {pages} {015807} (\bibinfo {year} {2009})},\ \Eprint
  {http://arxiv.org/abs/0901.2200} {arXiv:0901.2200 [nucl-ex]} \BibitemShut
  {NoStop}%
\bibitem [{\citenamefont {Hirata}\ \emph {et~al.}(1990)\citenamefont {Hirata}
  \emph {et~al.}}]{Hirata:1990xa}%
  \BibitemOpen
  \bibfield  {author} {\bibinfo {author} {\bibfnamefont {K.~S.}\ \bibnamefont
  {Hirata}} \emph {et~al.} (\bibinfo {collaboration} {Kamiokande-II
  Collaboration}),\ }\href {\doibase 10.1103/PhysRevLett.65.1297} {\bibfield
  {journal} {\bibinfo  {journal} {Phys. Rev. Lett.}\ }\textbf {\bibinfo
  {volume} {65}},\ \bibinfo {pages} {1297} (\bibinfo {year}
  {1990})}\BibitemShut {NoStop}%
\bibitem [{\citenamefont {Fukuda}\ \emph
  {et~al.}(1998{\natexlab{a}})\citenamefont {Fukuda} \emph
  {et~al.}}]{Fukuda:1998fd}%
  \BibitemOpen
  \bibfield  {author} {\bibinfo {author} {\bibfnamefont {Y.}~\bibnamefont
  {Fukuda}} \emph {et~al.} (\bibinfo {collaboration} {Super-Kamiokande
  Collaboration}),\ }\href {\doibase 10.1103/PhysRevLett.81.1158} {\bibfield
  {journal} {\bibinfo  {journal} {Phys. Rev. Lett.}\ }\textbf {\bibinfo
  {volume} {81}},\ \bibinfo {pages} {1158} (\bibinfo {year}
  {1998}{\natexlab{a}})},\ \bibinfo {note} {[Erratum: Phys. Rev.
  Lett.81,4279(1998)]},\ \Eprint {http://arxiv.org/abs/hep-ex/9805021}
  {arXiv:hep-ex/9805021 [hep-ex]} \BibitemShut {NoStop}%
\bibitem [{\citenamefont {Fukuda}\ \emph
  {et~al.}(1998{\natexlab{b}})\citenamefont {Fukuda} \emph
  {et~al.}}]{Fukuda:1998mi}%
  \BibitemOpen
  \bibfield  {author} {\bibinfo {author} {\bibfnamefont {Y.}~\bibnamefont
  {Fukuda}} \emph {et~al.} (\bibinfo {collaboration} {Super-Kamiokande
  Collaboration}),\ }\href {\doibase 10.1103/PhysRevLett.81.1562} {\bibfield
  {journal} {\bibinfo  {journal} {Phys. Rev. Lett.}\ }\textbf {\bibinfo
  {volume} {81}},\ \bibinfo {pages} {1562} (\bibinfo {year}
  {1998}{\natexlab{b}})}\BibitemShut {NoStop}%
\bibitem [{\citenamefont {Ahn}\ \emph {et~al.}(2003)\citenamefont {Ahn} \emph
  {et~al.}}]{Ahn:2002up}%
  \BibitemOpen
  \bibfield  {author} {\bibinfo {author} {\bibfnamefont {M.~H.}\ \bibnamefont
  {Ahn}} \emph {et~al.} (\bibinfo {collaboration} {K2K Collaboration}),\ }\href
  {\doibase 10.1103/PhysRevLett.90.041801} {\bibfield  {journal} {\bibinfo
  {journal} {Phys. Rev. Lett.}\ }\textbf {\bibinfo {volume} {90}},\ \bibinfo
  {pages} {041801} (\bibinfo {year} {2003})}\BibitemShut {NoStop}%
\bibitem [{\citenamefont {Ahmad}\ \emph {et~al.}(2001)\citenamefont {Ahmad}
  \emph {et~al.}}]{Ahmad:2001an}%
  \BibitemOpen
  \bibfield  {author} {\bibinfo {author} {\bibfnamefont {Q.~R.}\ \bibnamefont
  {Ahmad}} \emph {et~al.} (\bibinfo {collaboration} {SNO Collaboration}),\
  }\href {\doibase 10.1103/PhysRevLett.87.071301} {\bibfield  {journal}
  {\bibinfo  {journal} {Phys. Rev. Lett.}\ }\textbf {\bibinfo {volume} {87}},\
  \bibinfo {pages} {071301} (\bibinfo {year} {2001})}\BibitemShut {NoStop}%
\bibitem [{\citenamefont {Ahmad}\ \emph {et~al.}(2002)\citenamefont {Ahmad}
  \emph {et~al.}}]{Ahmad:2002jz}%
  \BibitemOpen
  \bibfield  {author} {\bibinfo {author} {\bibfnamefont {Q.~R.}\ \bibnamefont
  {Ahmad}} \emph {et~al.} (\bibinfo {collaboration} {SNO Collaboration}),\
  }\href {\doibase 10.1103/PhysRevLett.89.011301} {\bibfield  {journal}
  {\bibinfo  {journal} {Phys. Rev. Lett.}\ }\textbf {\bibinfo {volume} {89}},\
  \bibinfo {pages} {011301} (\bibinfo {year} {2002})}\BibitemShut {NoStop}%
\bibitem [{\citenamefont {Abe}\ \emph {et~al.}(2008)\citenamefont {Abe} \emph
  {et~al.}}]{Abe:2008aa}%
  \BibitemOpen
  \bibfield  {author} {\bibinfo {author} {\bibfnamefont {S.}~\bibnamefont
  {Abe}} \emph {et~al.} (\bibinfo {collaboration} {KamLAND Collaboration}),\
  }\href {\doibase 10.1103/PhysRevLett.100.221803} {\bibfield  {journal}
  {\bibinfo  {journal} {Phys. Rev. Lett.}\ }\textbf {\bibinfo {volume} {100}},\
  \bibinfo {pages} {221803} (\bibinfo {year} {2008})}\BibitemShut {NoStop}%
\bibitem [{\citenamefont {de~Gouvea}\ \emph {et~al.}(2013)\citenamefont
  {de~Gouvea} \emph {et~al.}}]{deGouvea:2013onf}%
  \BibitemOpen
  \bibfield  {author} {\bibinfo {author} {\bibfnamefont {A.}~\bibnamefont
  {de~Gouvea}} \emph {et~al.} (\bibinfo {collaboration} {Intensity Frontier
  Neutrino Working Group}),\ }\href
  {https://inspirehep.net/record/1260555/files/arXiv:1310.4340.pdf} {}
  (\bibinfo {year} {2013}),\ \Eprint {http://arxiv.org/abs/1310.4340}
  {arXiv:1310.4340 [hep-ex]} \BibitemShut {NoStop}%
\bibitem [{\citenamefont {de~Gouvea}(2004)}]{DeGouvea:2005gd}%
  \BibitemOpen
  \bibfield  {author} {\bibinfo {author} {\bibfnamefont {A.}~\bibnamefont
  {de~Gouvea}},\ }\href {\doibase 10.1142/S0217732304016032} {\bibfield
  {journal} {\bibinfo  {journal} {Mod. Phys. Lett.}\ }\textbf {\bibinfo
  {volume} {A19}},\ \bibinfo {pages} {2799} (\bibinfo {year} {2004})},\ \Eprint
  {http://arxiv.org/abs/hep-ph/0503086} {arXiv:hep-ph/0503086 [hep-ph]}
  \BibitemShut {NoStop}%
\bibitem [{\citenamefont {Mohapatra}\ \emph {et~al.}(2007)\citenamefont
  {Mohapatra} \emph {et~al.}}]{Mohapatra:2005wg}%
  \BibitemOpen
  \bibfield  {author} {\bibinfo {author} {\bibfnamefont {R.~N.}\ \bibnamefont
  {Mohapatra}} \emph {et~al.},\ }\href {\doibase 10.1088/0034-4885/70/11/R02}
  {\bibfield  {journal} {\bibinfo  {journal} {Rept. Prog. Phys.}\ }\textbf
  {\bibinfo {volume} {70}},\ \bibinfo {pages} {1757} (\bibinfo {year}
  {2007})},\ \Eprint {http://arxiv.org/abs/hep-ph/0510213}
  {arXiv:hep-ph/0510213 [hep-ph]} \BibitemShut {NoStop}%
\bibitem [{\citenamefont {Olive}\ \emph {et~al.}(2014)\citenamefont {Olive}
  \emph {et~al.}}]{PDG2015}%
  \BibitemOpen
  \bibfield  {author} {\bibinfo {author} {\bibfnamefont {K.~A.}\ \bibnamefont
  {Olive}} \emph {et~al.} (\bibinfo {collaboration} {Particle Data Group}),\
  }\href {http://pdg.lbl.gov} {\bibfield  {journal} {\bibinfo  {journal}
  {Chin.Phys.}\ }\textbf {\bibinfo {volume} {C38}},\ \bibinfo {pages} {090001}
  (\bibinfo {year} {2014})},\ \bibinfo {note} {and 2015 update}\BibitemShut
  {NoStop}%
\bibitem [{\citenamefont {Apollonio}\ \emph {et~al.}(2003)\citenamefont
  {Apollonio} \emph {et~al.}}]{Apollonio:2002gd}%
  \BibitemOpen
  \bibfield  {author} {\bibinfo {author} {\bibfnamefont {M.}~\bibnamefont
  {Apollonio}} \emph {et~al.} (\bibinfo {collaboration} {CHOOZ
  Collaboration}),\ }\href {\doibase 10.1140/epjc/s2002-01127-9} {\bibfield
  {journal} {\bibinfo  {journal} {Eur. Phys. J.}\ }\textbf {\bibinfo {volume}
  {C27}},\ \bibinfo {pages} {331} (\bibinfo {year} {2003})}\BibitemShut
  {NoStop}%
\bibitem [{\citenamefont {Boehm}\ \emph {et~al.}(2001)\citenamefont {Boehm}
  \emph {et~al.}}]{Boehm:2001ik}%
  \BibitemOpen
  \bibfield  {author} {\bibinfo {author} {\bibfnamefont {F.}~\bibnamefont
  {Boehm}} \emph {et~al.},\ }\href {\doibase 10.1103/PhysRevD.64.112001}
  {\bibfield  {journal} {\bibinfo  {journal} {Phys. Rev.}\ }\textbf {\bibinfo
  {volume} {D64}},\ \bibinfo {pages} {112001} (\bibinfo {year} {2001})},\
  \Eprint {http://arxiv.org/abs/hep-ex/0107009} {arXiv:hep-ex/0107009 [hep-ex]}
  \BibitemShut {NoStop}%
\bibitem [{\citenamefont {Vogel}\ and\ \citenamefont
  {Beacom}(1999)}]{Vogel:1999zy}%
  \BibitemOpen
  \bibfield  {author} {\bibinfo {author} {\bibfnamefont {P.}~\bibnamefont
  {Vogel}}\ and\ \bibinfo {author} {\bibfnamefont {J.~F.}\ \bibnamefont
  {Beacom}},\ }\href {\doibase 10.1103/PhysRevD.60.053003} {\bibfield
  {journal} {\bibinfo  {journal} {Phys. Rev.}\ }\textbf {\bibinfo {volume}
  {D60}},\ \bibinfo {pages} {053003} (\bibinfo {year} {1999})},\ \Eprint
  {http://arxiv.org/abs/hep-ph/9903554} {arXiv:hep-ph/9903554 [hep-ph]}
  \BibitemShut {NoStop}%
\bibitem [{\citenamefont {Minakata}\ \emph {et~al.}(2006)\citenamefont
  {Minakata}, \citenamefont {Nunokawa}, \citenamefont {Parke},\ and\
  \citenamefont {Zukanovich~Funchal}}]{Minakata:2006gq}%
  \BibitemOpen
  \bibfield  {author} {\bibinfo {author} {\bibfnamefont {H.}~\bibnamefont
  {Minakata}}, \bibinfo {author} {\bibfnamefont {H.}~\bibnamefont {Nunokawa}},
  \bibinfo {author} {\bibfnamefont {S.~J.}\ \bibnamefont {Parke}}, \ and\
  \bibinfo {author} {\bibfnamefont {R.}~\bibnamefont {Zukanovich~Funchal}},\
  }\href {\doibase 10.1103/PhysRevD.74.053008} {\bibfield  {journal} {\bibinfo
  {journal} {Phys.Rev.}\ }\textbf {\bibinfo {volume} {D74}},\ \bibinfo {pages}
  {053008} (\bibinfo {year} {2006})}\BibitemShut {NoStop}%
\bibitem [{\citenamefont {Mikaelyan}\ and\ \citenamefont
  {Sinev}(1999)}]{Mikaelyan:1998yg}%
  \BibitemOpen
  \bibfield  {author} {\bibinfo {author} {\bibfnamefont {L.}~\bibnamefont
  {Mikaelyan}}\ and\ \bibinfo {author} {\bibfnamefont {V.}~\bibnamefont
  {Sinev}},\ }\href@noop {} {\bibfield  {journal} {\bibinfo  {journal} {Phys.
  Atom. Nucl.}\ }\textbf {\bibinfo {volume} {62}},\ \bibinfo {pages} {2008}
  (\bibinfo {year} {1999})},\ \bibinfo {note} {[Yad. Fiz.62,2177(1999)]},\
  \Eprint {http://arxiv.org/abs/hep-ph/9811228} {arXiv:hep-ph/9811228 [hep-ph]}
  \BibitemShut {NoStop}%
\bibitem [{\citenamefont {An}\ \emph {et~al.}(2007)\citenamefont {An} \emph
  {et~al.}}]{DYBProposal}%
  \BibitemOpen
  \bibfield  {author} {\bibinfo {author} {\bibfnamefont {F.~P.}\ \bibnamefont
  {An}} \emph {et~al.} (\bibinfo {collaboration} {Daya Bay Collaboration}),\
  }\href@noop {} {} (\bibinfo {year} {2007}),\ \Eprint
  {http://arxiv.org/abs/hep-ex/0701029} {arXiv:hep-ex/0701029} \BibitemShut
  {NoStop}%
\bibitem [{\citenamefont {Ahn}\ \emph {et~al.}(2010)\citenamefont {Ahn} \emph
  {et~al.}}]{Ahn:2010vy}%
  \BibitemOpen
  \bibfield  {author} {\bibinfo {author} {\bibfnamefont {J.~K.}\ \bibnamefont
  {Ahn}} \emph {et~al.} (\bibinfo {collaboration} {RENO Collaboration}),\
  }\href@noop {} {} (\bibinfo {year} {2010}),\ \Eprint
  {http://arxiv.org/abs/1003.1391} {arXiv:1003.1391 [hep-ex]} \BibitemShut
  {NoStop}%
\bibitem [{\citenamefont {Ardellier}\ \emph {et~al.}(2006)\citenamefont
  {Ardellier} \emph {et~al.}}]{DCProposal}%
  \BibitemOpen
  \bibfield  {author} {\bibinfo {author} {\bibfnamefont {F.}~\bibnamefont
  {Ardellier}} \emph {et~al.} (\bibinfo {collaboration} {Double Chooz
  Collaboration}),\ }\href@noop {} {} (\bibinfo {year} {2006}),\ \Eprint
  {http://arxiv.org/abs/hep-ex/0606025} {arXiv:hep-ex/0606025} \BibitemShut
  {NoStop}%
\bibitem [{\citenamefont {An}\ \emph {et~al.}(2012{\natexlab{a}})\citenamefont
  {An} \emph {et~al.}}]{DYB2012}%
  \BibitemOpen
  \bibfield  {author} {\bibinfo {author} {\bibfnamefont {F.~P.}\ \bibnamefont
  {An}} \emph {et~al.} (\bibinfo {collaboration} {Daya Bay Collaboration}),\
  }\href@noop {} {\bibfield  {journal} {\bibinfo  {journal} {Phys. Rev. Lett.}\
  }\textbf {\bibinfo {volume} {108}},\ \bibinfo {pages} {171803} (\bibinfo
  {year} {2012}{\natexlab{a}})}\BibitemShut {NoStop}%
\bibitem [{\citenamefont {Ahn}\ \emph {et~al.}(2012)\citenamefont {Ahn} \emph
  {et~al.}}]{RENO2012}%
  \BibitemOpen
  \bibfield  {author} {\bibinfo {author} {\bibfnamefont {J.}~\bibnamefont
  {Ahn}} \emph {et~al.} (\bibinfo {collaboration} {RENO Collaboration}),\
  }\href@noop {} {\bibfield  {journal} {\bibinfo  {journal} {Phys. Rev. Lett.}\
  }\textbf {\bibinfo {volume} {108}},\ \bibinfo {pages} {191802} (\bibinfo
  {year} {2012})}\BibitemShut {NoStop}%
\bibitem [{\citenamefont {Abe}(2012)}]{DC2012}%
  \BibitemOpen
  \bibfield  {author} {\bibinfo {author} {\bibfnamefont {Y.}~\bibnamefont
  {Abe}} (\bibinfo {collaboration} {Double Chooz Collaboration}),\ }\href@noop
  {} {\bibfield  {journal} {\bibinfo  {journal} {Phys. Rev. Lett.}\ }\textbf
  {\bibinfo {volume} {108}},\ \bibinfo {pages} {131801} (\bibinfo {year}
  {2012})}\BibitemShut {NoStop}%
\bibitem [{\citenamefont {Abe}\ \emph {et~al.}(2015)\citenamefont {Abe} \emph
  {et~al.}}]{Abe:2015awa}%
  \BibitemOpen
  \bibfield  {author} {\bibinfo {author} {\bibfnamefont {K.}~\bibnamefont
  {Abe}} \emph {et~al.} (\bibinfo {collaboration} {T2K Collaboration}),\ }\href
  {\doibase 10.1103/PhysRevD.91.072010} {\bibfield  {journal} {\bibinfo
  {journal} {Phys. Rev.}\ }\textbf {\bibinfo {volume} {D91}},\ \bibinfo {pages}
  {072010} (\bibinfo {year} {2015})},\ \Eprint
  {http://arxiv.org/abs/1502.01550} {arXiv:1502.01550 [hep-ex]} \BibitemShut
  {NoStop}%
\bibitem [{\citenamefont {Adamson}\ \emph
  {et~al.}(2016{\natexlab{a}})\citenamefont {Adamson} \emph
  {et~al.}}]{Adamson:2016tbq}%
  \BibitemOpen
  \bibfield  {author} {\bibinfo {author} {\bibfnamefont {P.}~\bibnamefont
  {Adamson}} \emph {et~al.} (\bibinfo {collaboration} {NOvA Collaboration}),\
  }\href {\doibase 10.1103/PhysRevLett.116.151806} {\bibfield  {journal}
  {\bibinfo  {journal} {Phys. Rev. Lett.}\ }\textbf {\bibinfo {volume} {116}},\
  \bibinfo {pages} {151806} (\bibinfo {year} {2016}{\natexlab{a}})},\ \Eprint
  {http://arxiv.org/abs/1601.05022} {arXiv:1601.05022 [hep-ex]} \BibitemShut
  {NoStop}%
\bibitem [{\citenamefont {An}\ \emph {et~al.}(2015{\natexlab{a}})\citenamefont
  {An} \emph {et~al.}}]{An:2015rpe}%
  \BibitemOpen
  \bibfield  {author} {\bibinfo {author} {\bibfnamefont {F.~P.}\ \bibnamefont
  {An}} \emph {et~al.} (\bibinfo {collaboration} {Daya Bay Collaboration}),\
  }\href {\doibase 10.1103/PhysRevLett.115.111802} {\bibfield  {journal}
  {\bibinfo  {journal} {Phys. Rev. Lett.}\ }\textbf {\bibinfo {volume} {115}},\
  \bibinfo {pages} {111802} (\bibinfo {year} {2015}{\natexlab{a}})},\ \Eprint
  {http://arxiv.org/abs/1505.03456} {arXiv:1505.03456 [hep-ex]} \BibitemShut
  {NoStop}%
\bibitem [{\citenamefont {An}\ \emph {et~al.}(2012{\natexlab{b}})\citenamefont
  {An} \emph {et~al.}}]{DayaBay:2012aa}%
  \BibitemOpen
  \bibfield  {author} {\bibinfo {author} {\bibfnamefont {F.~P.}\ \bibnamefont
  {An}} \emph {et~al.} (\bibinfo {collaboration} {Daya Bay Collaboration}),\
  }\href {\doibase 10.1016/j.nima.2012.05.030} {\bibfield  {journal} {\bibinfo
  {journal} {Nucl. Instrum. Meth.}\ }\textbf {\bibinfo {volume} {A685}},\
  \bibinfo {pages} {78} (\bibinfo {year} {2012}{\natexlab{b}})},\ \Eprint
  {http://arxiv.org/abs/1202.6181} {arXiv:1202.6181 [physics.ins-det]}
  \BibitemShut {NoStop}%
\bibitem [{\citenamefont {An}\ \emph {et~al.}(2016{\natexlab{a}})\citenamefont
  {An} \emph {et~al.}}]{An:2015qga}%
  \BibitemOpen
  \bibfield  {author} {\bibinfo {author} {\bibfnamefont {F.~P.}\ \bibnamefont
  {An}} \emph {et~al.} (\bibinfo {collaboration} {Daya Bay Collaboration}),\
  }\href {\doibase 10.1016/j.nima.2015.11.144} {\bibfield  {journal} {\bibinfo
  {journal} {Nucl. Instrum. Meth.}\ }\textbf {\bibinfo {volume} {A811}},\
  \bibinfo {pages} {133} (\bibinfo {year} {2016}{\natexlab{a}})},\ \Eprint
  {http://arxiv.org/abs/1508.03943} {arXiv:1508.03943 [physics.ins-det]}
  \BibitemShut {NoStop}%
\bibitem [{\citenamefont {Beriguete}\ \emph {et~al.}(2014)\citenamefont
  {Beriguete} \emph {et~al.}}]{Beriguete:2014gua}%
  \BibitemOpen
  \bibfield  {author} {\bibinfo {author} {\bibfnamefont {W.}~\bibnamefont
  {Beriguete}} \emph {et~al.},\ }\href {\doibase 10.1016/j.nima.2014.05.119}
  {\bibfield  {journal} {\bibinfo  {journal} {Nucl. Instrum. Meth.}\ }\textbf
  {\bibinfo {volume} {A763}},\ \bibinfo {pages} {82} (\bibinfo {year}
  {2014})},\ \Eprint {http://arxiv.org/abs/1402.6694} {arXiv:1402.6694
  [physics.ins-det]} \BibitemShut {NoStop}%
\bibitem [{\citenamefont {Liu}\ \emph {et~al.}(2014)\citenamefont {Liu},
  \citenamefont {Cai}, \citenamefont {Carr}, \citenamefont {Dwyer},
  \citenamefont {Gu}, \citenamefont {Li}, \citenamefont {Qian}, \citenamefont
  {McKeown}, \citenamefont {Tsang}, \citenamefont {Wang} \emph
  {et~al.}}]{Liu:2013ava}%
  \BibitemOpen
  \bibfield  {author} {\bibinfo {author} {\bibfnamefont {J.~L.}\ \bibnamefont
  {Liu}}, \bibinfo {author} {\bibfnamefont {B.}~\bibnamefont {Cai}}, \bibinfo
  {author} {\bibfnamefont {R.}~\bibnamefont {Carr}}, \bibinfo {author}
  {\bibfnamefont {D.~A.}\ \bibnamefont {Dwyer}}, \bibinfo {author}
  {\bibfnamefont {W.~Q.}\ \bibnamefont {Gu}}, \bibinfo {author} {\bibfnamefont
  {G.~S.}\ \bibnamefont {Li}}, \bibinfo {author} {\bibfnamefont
  {X.}~\bibnamefont {Qian}}, \bibinfo {author} {\bibfnamefont {R.~D.}\
  \bibnamefont {McKeown}}, \bibinfo {author} {\bibfnamefont {R.~H.~M.}\
  \bibnamefont {Tsang}}, \bibinfo {author} {\bibfnamefont {W.}~\bibnamefont
  {Wang}},  \emph {et~al.},\ }\href {\doibase 10.1016/j.nima.2014.02.049}
  {\bibfield  {journal} {\bibinfo  {journal} {Nucl. Instr. Meth. A}\ }\textbf
  {\bibinfo {volume} {750}},\ \bibinfo {pages} {19} (\bibinfo {year}
  {2014})}\BibitemShut {NoStop}%
\bibitem [{\citenamefont {An}\ \emph {et~al.}(2015{\natexlab{b}})\citenamefont
  {An} \emph {et~al.}}]{Dayabay:2014vka}%
  \BibitemOpen
  \bibfield  {author} {\bibinfo {author} {\bibfnamefont {F.~P.}\ \bibnamefont
  {An}} \emph {et~al.} (\bibinfo {collaboration} {Daya Bay Collaboration}),\
  }\href {\doibase 10.1016/j.nima.2014.09.070} {\bibfield  {journal} {\bibinfo
  {journal} {Nucl. Instr. Meth. A}\ }\textbf {\bibinfo {volume} {773}},\
  \bibinfo {pages} {8} (\bibinfo {year} {2015}{\natexlab{b}})}\BibitemShut
  {NoStop}%
\bibitem [{\citenamefont {Knoll}(2010)}]{Knoll:2010xta}%
  \BibitemOpen
  \bibfield  {author} {\bibinfo {author} {\bibfnamefont {G.}~\bibnamefont
  {Knoll}},\ }\href
  {http://www-spires.fnal.gov/spires/find/books/www?cl=QC787.C6K56::2010}
  {\emph {\bibinfo {title} {{Radiation Detection and Measurement (4th ed.)}}}}\
  (\bibinfo  {publisher} {John Wiley},\ \bibinfo {address} {Hoboken, NJ},\
  \bibinfo {year} {2010})\BibitemShut {NoStop}%
\bibitem [{\citenamefont {Bellamy}\ \emph {et~al.}(1993)\citenamefont {Bellamy}
  \emph {et~al.}}]{nimPmtFit}%
  \BibitemOpen
  \bibfield  {author} {\bibinfo {author} {\bibfnamefont {E.~H.}\ \bibnamefont
  {Bellamy}} \emph {et~al.},\ }\href@noop {} {\bibfield  {journal} {\bibinfo
  {journal} {Nucl. Instr. Meth. A}\ }\textbf {\bibinfo {volume} {339}},\
  \bibinfo {pages} {468} (\bibinfo {year} {1993})}\BibitemShut {NoStop}%
\bibitem [{cbf()}]{cbfunction}%
  \BibitemOpen
  \href@noop {} {}\bibinfo {note} {T.~Skwarnicki, Ph.D Thesis, DESY
  F31-86-02(1986), Appendix E; M.J.~Oreglia, Ph.D Thesis, SLAC-236(1980),
  Appendix D; J.~E.~Gaiser, Ph.D Thesis, SLAC-255(1982), Appendix
  F.}\BibitemShut {Stop}%
\bibitem [{\citenamefont {Agostinelli}\ \emph {et~al.}(2003)\citenamefont
  {Agostinelli} \emph {et~al.}}]{Agostinelli:2002hh}%
  \BibitemOpen
  \bibfield  {author} {\bibinfo {author} {\bibfnamefont {S.}~\bibnamefont
  {Agostinelli}} \emph {et~al.} (\bibinfo {collaboration} {GEANT4
  Collaboration}),\ }\href {\doibase 10.1016/S0168-9002(03)01368-8} {\bibfield
  {journal} {\bibinfo  {journal} {Nucl. Instrum. Meth.}\ }\textbf {\bibinfo
  {volume} {A506}},\ \bibinfo {pages} {250} (\bibinfo {year}
  {2003})}\BibitemShut {NoStop}%
\bibitem [{\citenamefont {An}\ \emph {et~al.}(2016{\natexlab{b}})\citenamefont
  {An} \emph {et~al.}}]{An:2016bvr}%
  \BibitemOpen
  \bibfield  {author} {\bibinfo {author} {\bibfnamefont {F.~P.}\ \bibnamefont
  {An}} \emph {et~al.} (\bibinfo {collaboration} {Daya Bay Collaboration}),\
  }\href {\doibase 10.1103/PhysRevD.93.072011} {\bibfield  {journal} {\bibinfo
  {journal} {Phys. Rev.}\ }\textbf {\bibinfo {volume} {D93}},\ \bibinfo {pages}
  {072011} (\bibinfo {year} {2016}{\natexlab{b}})},\ \Eprint
  {http://arxiv.org/abs/1603.03549} {arXiv:1603.03549 [hep-ex]} \BibitemShut
  {NoStop}%
\bibitem [{Note1()}]{Note1}%
  \BibitemOpen
  \bibinfo {note} {See the Supplemental Material at (provide link) for the
  following tabulated data: the observed prompt energy spectra for $\protect
  \overline {\nu }_{e}${} inverse beta decay candidates and estimated
  backgrounds for each experimental hall, the distribution of ${\Delta }\chi
  ^2$ versus $\protect \qopname \relax o{sin}^2(2\theta _{13})$ and ${\Delta
  }m^2_{\protect \mathrm {ee}}$, and the components of the detector response
  model $P(E_{\protect \mathrm {rec}};E_{\protect \mathrm
  {true}})$.}\BibitemShut {Stop}%
\bibitem [{\citenamefont {An}\ \emph {et~al.}(2014{\natexlab{a}})\citenamefont
  {An} \emph {et~al.}}]{An:2014ehw}%
  \BibitemOpen
  \bibfield  {author} {\bibinfo {author} {\bibfnamefont {F.~P.}\ \bibnamefont
  {An}} \emph {et~al.} (\bibinfo {collaboration} {Daya Bay Collaboration}),\
  }\href {\doibase 10.1103/PhysRevD.90.071101} {\bibfield  {journal} {\bibinfo
  {journal} {Phys. Rev. D}\ }\textbf {\bibinfo {volume} {90}},\ \bibinfo
  {pages} {071101} (\bibinfo {year} {2014}{\natexlab{a}})}\BibitemShut
  {NoStop}%
\bibitem [{\citenamefont {Abe}\ \emph {et~al.}(2010)\citenamefont {Abe} \emph
  {et~al.}}]{Abe:2009aa}%
  \BibitemOpen
  \bibfield  {author} {\bibinfo {author} {\bibfnamefont {S.}~\bibnamefont
  {Abe}} \emph {et~al.} (\bibinfo {collaboration} {KamLAND Collaboration}),\
  }\href {\doibase 10.1103/PhysRevC.81.025807} {\bibfield  {journal} {\bibinfo
  {journal} {Phys. Rev.}\ }\textbf {\bibinfo {volume} {C81}},\ \bibinfo {pages}
  {025807} (\bibinfo {year} {2010})},\ \Eprint {http://arxiv.org/abs/0907.0066}
  {arXiv:0907.0066 [hep-ex]} \BibitemShut {NoStop}%
\bibitem [{\citenamefont {Wen}\ \emph {et~al.}(2006)\citenamefont {Wen},
  \citenamefont {Cao}, \citenamefont {Luk}, \citenamefont {Ma}, \citenamefont
  {Wang},\ and\ \citenamefont {Yang}}]{Wen:2006hx}%
  \BibitemOpen
  \bibfield  {author} {\bibinfo {author} {\bibfnamefont {L.-j.}\ \bibnamefont
  {Wen}}, \bibinfo {author} {\bibfnamefont {J.}~\bibnamefont {Cao}}, \bibinfo
  {author} {\bibfnamefont {K.-B.}\ \bibnamefont {Luk}}, \bibinfo {author}
  {\bibfnamefont {Y.-q.}\ \bibnamefont {Ma}}, \bibinfo {author} {\bibfnamefont
  {Y.-f.}\ \bibnamefont {Wang}}, \ and\ \bibinfo {author} {\bibfnamefont
  {C.-g.}\ \bibnamefont {Yang}},\ }\href {\doibase 10.1016/j.nima.2006.04.047}
  {\bibfield  {journal} {\bibinfo  {journal} {Nucl. Instrum. Meth.}\ }\textbf
  {\bibinfo {volume} {A564}},\ \bibinfo {pages} {471} (\bibinfo {year}
  {2006})},\ \Eprint {http://arxiv.org/abs/hep-ex/0604034}
  {arXiv:hep-ex/0604034 [hep-ex]} \BibitemShut {NoStop}%
\bibitem [{\citenamefont {Li}\ and\ \citenamefont {Beacom}(2015)}]{Li:2015lxa}%
  \BibitemOpen
  \bibfield  {author} {\bibinfo {author} {\bibfnamefont {S.~W.}\ \bibnamefont
  {Li}}\ and\ \bibinfo {author} {\bibfnamefont {J.~F.}\ \bibnamefont
  {Beacom}},\ }\href {\doibase 10.1103/PhysRevD.92.105033} {\bibfield
  {journal} {\bibinfo  {journal} {Phys. Rev.}\ }\textbf {\bibinfo {volume}
  {D92}},\ \bibinfo {pages} {105033} (\bibinfo {year} {2015})},\ \Eprint
  {http://arxiv.org/abs/1508.05389} {arXiv:1508.05389 [physics.ins-det]}
  \BibitemShut {NoStop}%
\bibitem [{\citenamefont {Tilley}\ \emph {et~al.}(2004)\citenamefont {Tilley},
  \citenamefont {Kelley}, \citenamefont {Godwin}, \citenamefont {Millener},
  \citenamefont {Purcell}, \citenamefont {Sheu},\ and\ \citenamefont
  {Weller}}]{Tilley:2004zz}%
  \BibitemOpen
  \bibfield  {author} {\bibinfo {author} {\bibfnamefont {D.~R.}\ \bibnamefont
  {Tilley}}, \bibinfo {author} {\bibfnamefont {J.~H.}\ \bibnamefont {Kelley}},
  \bibinfo {author} {\bibfnamefont {J.~L.}\ \bibnamefont {Godwin}}, \bibinfo
  {author} {\bibfnamefont {D.~J.}\ \bibnamefont {Millener}}, \bibinfo {author}
  {\bibfnamefont {J.~E.}\ \bibnamefont {Purcell}}, \bibinfo {author}
  {\bibfnamefont {C.~G.}\ \bibnamefont {Sheu}}, \ and\ \bibinfo {author}
  {\bibfnamefont {H.~R.}\ \bibnamefont {Weller}},\ }\href {\doibase
  10.1016/j.nuclphysa.2004.09.059} {\bibfield  {journal} {\bibinfo  {journal}
  {Nucl. Phys.}\ }\textbf {\bibinfo {volume} {A745}},\ \bibinfo {pages} {155}
  (\bibinfo {year} {2004})}\BibitemShut {NoStop}%
\bibitem [{\citenamefont {Nyman}\ \emph {et~al.}(1990)\citenamefont {Nyman}
  \emph {et~al.}}]{Nyman1990189}%
  \BibitemOpen
  \bibfield  {author} {\bibinfo {author} {\bibfnamefont {G.}~\bibnamefont
  {Nyman}} \emph {et~al.},\ }\href {\doibase 10.1016/0375-9474(90)90236-F}
  {\bibfield  {journal} {\bibinfo  {journal} {Nucl. Phys. A}\ }\textbf
  {\bibinfo {volume} {510}},\ \bibinfo {pages} {189 } (\bibinfo {year}
  {1990})}\BibitemShut {NoStop}%
\bibitem [{\citenamefont {Bjornstad}\ \emph {et~al.}(1981)\citenamefont
  {Bjornstad} \emph {et~al.}}]{Bjornstad:1981ad}%
  \BibitemOpen
  \bibfield  {author} {\bibinfo {author} {\bibfnamefont {T.}~\bibnamefont
  {Bjornstad}} \emph {et~al.} (\bibinfo {collaboration} {ISOLDE
  Collaboration}),\ }\href {\doibase 10.1016/0375-9474(81)90522-4} {\bibfield
  {journal} {\bibinfo  {journal} {Nucl.Phys.}\ }\textbf {\bibinfo {volume}
  {A366}},\ \bibinfo {pages} {461} (\bibinfo {year} {1981})}\BibitemShut
  {NoStop}%
\bibitem [{\citenamefont {Liu}\ \emph {et~al.}(2015)\citenamefont {Liu},
  \citenamefont {Carr}, \citenamefont {Dwyer}, \citenamefont {Gu},
  \citenamefont {Li}, \citenamefont {McKeown}, \citenamefont {Qian},
  \citenamefont {Tsang}, \citenamefont {Wu},\ and\ \citenamefont
  {Zhang}}]{Liu:2015cra}%
  \BibitemOpen
  \bibfield  {author} {\bibinfo {author} {\bibfnamefont {J.}~\bibnamefont
  {Liu}}, \bibinfo {author} {\bibfnamefont {R.}~\bibnamefont {Carr}}, \bibinfo
  {author} {\bibfnamefont {D.~A.}\ \bibnamefont {Dwyer}}, \bibinfo {author}
  {\bibfnamefont {W.~Q.}\ \bibnamefont {Gu}}, \bibinfo {author} {\bibfnamefont
  {G.~S.}\ \bibnamefont {Li}}, \bibinfo {author} {\bibfnamefont {R.~D.}\
  \bibnamefont {McKeown}}, \bibinfo {author} {\bibfnamefont {X.}~\bibnamefont
  {Qian}}, \bibinfo {author} {\bibfnamefont {R.~H.~M.}\ \bibnamefont {Tsang}},
  \bibinfo {author} {\bibfnamefont {F.~F.}\ \bibnamefont {Wu}}, \ and\ \bibinfo
  {author} {\bibfnamefont {C.}~\bibnamefont {Zhang}},\ }\href {\doibase
  10.1016/j.nima.2015.07.003} {\bibfield  {journal} {\bibinfo  {journal} {Nucl.
  Instrum. Meth.}\ }\textbf {\bibinfo {volume} {A797}},\ \bibinfo {pages} {260}
  (\bibinfo {year} {2015})},\ \Eprint {http://arxiv.org/abs/1504.07911}
  {arXiv:1504.07911 [physics.ins-det]} \BibitemShut {NoStop}%
\bibitem [{\citenamefont {An}\ \emph {et~al.}(2013)\citenamefont {An} \emph
  {et~al.}}]{CPC}%
  \BibitemOpen
  \bibfield  {author} {\bibinfo {author} {\bibfnamefont {F.~P.}\ \bibnamefont
  {An}} \emph {et~al.} (\bibinfo {collaboration} {Daya Bay Collaboration}),\
  }\href@noop {} {\bibfield  {journal} {\bibinfo  {journal} {Chin. Phys. C}\
  }\textbf {\bibinfo {volume} {37}},\ \bibinfo {pages} {011001} (\bibinfo
  {year} {2013})}\BibitemShut {NoStop}%
\bibitem [{\citenamefont {Gu}\ \emph {et~al.}(2015)\citenamefont {Gu},
  \citenamefont {Cao}, \citenamefont {Chen}, \citenamefont {Ji}, \citenamefont
  {Li}, \citenamefont {Ling}, \citenamefont {Liu}, \citenamefont {Qian},\ and\
  \citenamefont {Wang}}]{Gu:2015inc}%
  \BibitemOpen
  \bibfield  {author} {\bibinfo {author} {\bibfnamefont {W.~Q.}\ \bibnamefont
  {Gu}}, \bibinfo {author} {\bibfnamefont {G.~F.}\ \bibnamefont {Cao}},
  \bibinfo {author} {\bibfnamefont {X.~H.}\ \bibnamefont {Chen}}, \bibinfo
  {author} {\bibfnamefont {X.~P.}\ \bibnamefont {Ji}}, \bibinfo {author}
  {\bibfnamefont {G.~S.}\ \bibnamefont {Li}}, \bibinfo {author} {\bibfnamefont
  {J.~J.}\ \bibnamefont {Ling}}, \bibinfo {author} {\bibfnamefont
  {J.}~\bibnamefont {Liu}}, \bibinfo {author} {\bibfnamefont {X.}~\bibnamefont
  {Qian}}, \ and\ \bibinfo {author} {\bibfnamefont {W.}~\bibnamefont {Wang}},\
  }\href@noop {} {} (\bibinfo {year} {2015}),\ \Eprint
  {http://arxiv.org/abs/1512.00295} {arXiv:1512.00295 [physics.ins-det]}
  \BibitemShut {NoStop}%
\bibitem [{\citenamefont {{J. Zhao, Z.~Y. Yu, J.~L. Liu, X.~B. Li, F.~H. Zhang
  and D.~M. Xia}}(2014)}]{AlphaN}%
  \BibitemOpen
  \bibfield  {author} {\bibinfo {author} {\bibnamefont {{J. Zhao, Z.~Y. Yu,
  J.~L. Liu, X.~B. Li, F.~H. Zhang and D.~M. Xia}}},\ }\href@noop {} {\bibfield
   {journal} {\bibinfo  {journal} {Chin. Phys. C}\ }\textbf {\bibinfo {volume}
  {38}},\ \bibinfo {pages} {116201} (\bibinfo {year} {2014})}\BibitemShut
  {NoStop}%
\bibitem [{\citenamefont {Wojcik}\ and\ \citenamefont
  {Zuzel}(2013)}]{Wojcik:2013}%
  \BibitemOpen
  \bibfield  {author} {\bibinfo {author} {\bibfnamefont {M.}~\bibnamefont
  {Wojcik}}\ and\ \bibinfo {author} {\bibfnamefont {G.}~\bibnamefont {Zuzel}},\
  }\href {\doibase 10.1007/s10967-012-2180-5} {\bibfield  {journal} {\bibinfo
  {journal} {J.~Radioanal.~Nucl.~Chem.}\ }\textbf {\bibinfo {volume} {296}},\
  \bibinfo {pages} {639} (\bibinfo {year} {2013})}\BibitemShut {NoStop}%
\bibitem [{\citenamefont {Shibata}\ \emph {et~al.}(2011)\citenamefont {Shibata}
  \emph {et~al.}}]{JENDL:2011}%
  \BibitemOpen
  \bibfield  {author} {\bibinfo {author} {\bibfnamefont {K.}~\bibnamefont
  {Shibata}} \emph {et~al.},\ }\href {\doibase 10.1080/18811248.2011.9711675}
  {\bibfield  {journal} {\bibinfo  {journal} {J. of Nucl. Sci. and Tech.}\
  }\textbf {\bibinfo {volume} {48}},\ \bibinfo {pages} {1} (\bibinfo {year}
  {2011})}\BibitemShut {NoStop}%
\bibitem [{\citenamefont {Otuka}\ \emph {et~al.}(2014)\citenamefont {Otuka}
  \emph {et~al.}}]{Otuka2014272}%
  \BibitemOpen
  \bibfield  {author} {\bibinfo {author} {\bibfnamefont {N.}~\bibnamefont
  {Otuka}} \emph {et~al.},\ }\href {\doibase
  http://dx.doi.org/10.1016/j.nds.2014.07.065} {\bibfield  {journal} {\bibinfo
  {journal} {Nuclear Data Sheets}\ }\textbf {\bibinfo {volume} {120}},\
  \bibinfo {pages} {272 } (\bibinfo {year} {2014})}\BibitemShut {NoStop}%
\bibitem [{SRI()}]{SRIM}%
  \BibitemOpen
  \href@noop {} {}\bibinfo {howpublished}
  {\url{http://www.srim.org}}\BibitemShut {NoStop}%
\bibitem [{\citenamefont {An}\ \emph {et~al.}(2016{\natexlab{c}})\citenamefont
  {An} \emph {et~al.}}]{An:2016srz}%
  \BibitemOpen
  \bibfield  {author} {\bibinfo {author} {\bibfnamefont {F.~P.}\ \bibnamefont
  {An}} \emph {et~al.} (\bibinfo {collaboration} {Daya Bay}),\ }\href@noop {}
  {\  (\bibinfo {year} {2016}{\natexlab{c}})},\ \Eprint
  {http://arxiv.org/abs/1607.05378} {arXiv:1607.05378 [hep-ex]} \BibitemShut
  {NoStop}%
\bibitem [{\citenamefont {Djurcic}\ \emph {et~al.}(2009)\citenamefont
  {Djurcic}, \citenamefont {Detwiler}, \citenamefont {Piepke}, \citenamefont
  {Foster}, \citenamefont {Miller} \emph {et~al.}}]{Djurcic:2008ny}%
  \BibitemOpen
  \bibfield  {author} {\bibinfo {author} {\bibfnamefont {Z.}~\bibnamefont
  {Djurcic}}, \bibinfo {author} {\bibfnamefont {J.}~\bibnamefont {Detwiler}},
  \bibinfo {author} {\bibfnamefont {A.}~\bibnamefont {Piepke}}, \bibinfo
  {author} {\bibfnamefont {V.~J.}\ \bibnamefont {Foster}}, \bibinfo {author}
  {\bibfnamefont {L.}~\bibnamefont {Miller}},  \emph {et~al.},\ }\href
  {\doibase 10.1088/0954-3899/36/4/045002} {\bibfield  {journal} {\bibinfo
  {journal} {J.Phys.}\ }\textbf {\bibinfo {volume} {G36}},\ \bibinfo {pages}
  {045002} (\bibinfo {year} {2009})},\ \Eprint {http://arxiv.org/abs/0808.0747}
  {arXiv:0808.0747 [hep-ex]} \BibitemShut {NoStop}%
\bibitem [{\citenamefont {Ma}\ \emph {et~al.}(2013)\citenamefont {Ma} \emph
  {et~al.}}]{bib:fr_ma}%
  \BibitemOpen
  \bibfield  {author} {\bibinfo {author} {\bibfnamefont {X.~B.}\ \bibnamefont
  {Ma}} \emph {et~al.},\ }\href {\doibase 10.1103/PhysRevC.88.014605}
  {\bibfield  {journal} {\bibinfo  {journal} {Phys.Rev.}\ }\textbf {\bibinfo
  {volume} {C88}},\ \bibinfo {pages} {014605} (\bibinfo {year}
  {2013})}\BibitemShut {NoStop}%
\bibitem [{\citenamefont {Huber}(2011)}]{HuberAnomaly}%
  \BibitemOpen
  \bibfield  {author} {\bibinfo {author} {\bibfnamefont {P.}~\bibnamefont
  {Huber}},\ }\href@noop {} {\bibfield  {journal} {\bibinfo  {journal} {Phys.
  Rev. C}\ }\textbf {\bibinfo {volume} {84}},\ \bibinfo {pages} {024617}
  (\bibinfo {year} {2011})}\BibitemShut {NoStop}%
\bibitem [{\citenamefont {Mueller}\ \emph {et~al.}(2011)\citenamefont {Mueller}
  \emph {et~al.}}]{Mueller}%
  \BibitemOpen
  \bibfield  {author} {\bibinfo {author} {\bibfnamefont {T.}~\bibnamefont
  {Mueller}} \emph {et~al.},\ }\href@noop {} {\bibfield  {journal} {\bibinfo
  {journal} {Phys. Rev. C}\ }\textbf {\bibinfo {volume} {83}},\ \bibinfo
  {pages} {054615} (\bibinfo {year} {2011})}\BibitemShut {NoStop}%
\bibitem [{\citenamefont {Cao}(2012)}]{Cao:2011gb}%
  \BibitemOpen
  \bibfield  {author} {\bibinfo {author} {\bibfnamefont {J.}~\bibnamefont
  {Cao}},\ }\href {\doibase 10.1016/j.nuclphysbps.2012.09.033} {\bibfield
  {journal} {\bibinfo  {journal} {Nucl. Phys. B. (Proc. Suppl.)}\ }\textbf
  {\bibinfo {volume} {229-232}},\ \bibinfo {pages} {205} (\bibinfo {year}
  {2012})},\ \Eprint {http://arxiv.org/abs/1101.2266} {arXiv:1101.2266
  [hep-ex]} \BibitemShut {NoStop}%
\bibitem [{\citenamefont {Kopeikin}(2012)}]{Kopeikin:2012zz}%
  \BibitemOpen
  \bibfield  {author} {\bibinfo {author} {\bibfnamefont {V.~I.}\ \bibnamefont
  {Kopeikin}},\ }\href {\doibase 10.1134/S1063778812020123} {\bibfield
  {journal} {\bibinfo  {journal} {Phys. Atom. Nucl.}\ }\textbf {\bibinfo
  {volume} {75}},\ \bibinfo {pages} {143} (\bibinfo {year} {2012})},\ \bibinfo
  {note} {[Yad. Fiz.75N2,165(2012)]}\BibitemShut {NoStop}%
\bibitem [{\citenamefont {An}\ \emph {et~al.}(2016{\natexlab{d}})\citenamefont
  {An} \emph {et~al.}}]{An:2015nua}%
  \BibitemOpen
  \bibfield  {author} {\bibinfo {author} {\bibfnamefont {F.~P.}\ \bibnamefont
  {An}} \emph {et~al.} (\bibinfo {collaboration} {Daya Bay Collaboration}),\
  }\href {\doibase 10.1103/PhysRevLett.116.061801} {\bibfield  {journal}
  {\bibinfo  {journal} {Phys. Rev. Lett.}\ }\textbf {\bibinfo {volume} {116}},\
  \bibinfo {pages} {061801} (\bibinfo {year} {2016}{\natexlab{d}})},\ \Eprint
  {http://arxiv.org/abs/1508.04233} {arXiv:1508.04233 [hep-ex]} \BibitemShut
  {NoStop}%
\bibitem [{\citenamefont {An}\ \emph {et~al.}(2014{\natexlab{b}})\citenamefont
  {An} \emph {et~al.}}]{DBPRL2014}%
  \BibitemOpen
  \bibfield  {author} {\bibinfo {author} {\bibfnamefont {F.~P.}\ \bibnamefont
  {An}} \emph {et~al.} (\bibinfo {collaboration} {Daya Bay Collaboration}),\
  }\href@noop {} {\bibfield  {journal} {\bibinfo  {journal} {Phys. Rev. Lett.}\
  }\textbf {\bibinfo {volume} {112}},\ \bibinfo {pages} {061801} (\bibinfo
  {year} {2014}{\natexlab{b}})}\BibitemShut {NoStop}%
\bibitem [{\citenamefont {Hayes}\ \emph {et~al.}(2014)\citenamefont {Hayes},
  \citenamefont {Friar}, \citenamefont {Garvey}, \citenamefont {Jungman},\ and\
  \citenamefont {Jonkmans}}]{Hayes:2013wra}%
  \BibitemOpen
  \bibfield  {author} {\bibinfo {author} {\bibfnamefont {A.~C.}\ \bibnamefont
  {Hayes}}, \bibinfo {author} {\bibfnamefont {J.~L.}\ \bibnamefont {Friar}},
  \bibinfo {author} {\bibfnamefont {G.~T.}\ \bibnamefont {Garvey}}, \bibinfo
  {author} {\bibfnamefont {G.}~\bibnamefont {Jungman}}, \ and\ \bibinfo
  {author} {\bibfnamefont {G.}~\bibnamefont {Jonkmans}},\ }\href {\doibase
  10.1103/PhysRevLett.112.202501} {\bibfield  {journal} {\bibinfo  {journal}
  {Phys. Rev. Lett.}\ }\textbf {\bibinfo {volume} {112}},\ \bibinfo {pages}
  {202501} (\bibinfo {year} {2014})},\ \Eprint {http://arxiv.org/abs/1309.4146}
  {arXiv:1309.4146 [nucl-th]} \BibitemShut {NoStop}%
\bibitem [{\citenamefont {Hayes}\ \emph {et~al.}(2015)\citenamefont {Hayes},
  \citenamefont {Friar}, \citenamefont {Garvey}, \citenamefont {Ibeling},
  \citenamefont {Jungman}, \citenamefont {Kawano},\ and\ \citenamefont
  {Mills}}]{Hayes:2015yka}%
  \BibitemOpen
  \bibfield  {author} {\bibinfo {author} {\bibfnamefont {A.~C.}\ \bibnamefont
  {Hayes}}, \bibinfo {author} {\bibfnamefont {J.~L.}\ \bibnamefont {Friar}},
  \bibinfo {author} {\bibfnamefont {G.~T.}\ \bibnamefont {Garvey}}, \bibinfo
  {author} {\bibfnamefont {D.}~\bibnamefont {Ibeling}}, \bibinfo {author}
  {\bibfnamefont {G.}~\bibnamefont {Jungman}}, \bibinfo {author} {\bibfnamefont
  {T.}~\bibnamefont {Kawano}}, \ and\ \bibinfo {author} {\bibfnamefont {R.~W.}\
  \bibnamefont {Mills}},\ }\href {\doibase 10.1103/PhysRevD.92.033015}
  {\bibfield  {journal} {\bibinfo  {journal} {Phys. Rev.}\ }\textbf {\bibinfo
  {volume} {D92}},\ \bibinfo {pages} {033015} (\bibinfo {year} {2015})},\
  \Eprint {http://arxiv.org/abs/1506.00583} {arXiv:1506.00583 [nucl-th]}
  \BibitemShut {NoStop}%
\bibitem [{\citenamefont {Nunokawa}\ \emph {et~al.}(2005)\citenamefont
  {Nunokawa}, \citenamefont {Parke},\ and\ \citenamefont
  {Zukanovich~Funchal}}]{Nunokawa:2005nx}%
  \BibitemOpen
  \bibfield  {author} {\bibinfo {author} {\bibfnamefont {H.}~\bibnamefont
  {Nunokawa}}, \bibinfo {author} {\bibfnamefont {S.~J.}\ \bibnamefont {Parke}},
  \ and\ \bibinfo {author} {\bibfnamefont {R.}~\bibnamefont
  {Zukanovich~Funchal}},\ }\href {\doibase 10.1103/PhysRevD.72.013009}
  {\bibfield  {journal} {\bibinfo  {journal} {Phys. Rev.}\ }\textbf {\bibinfo
  {volume} {D72}},\ \bibinfo {pages} {013009} (\bibinfo {year} {2005})},\
  \Eprint {http://arxiv.org/abs/hep-ph/0503283} {arXiv:hep-ph/0503283 [hep-ph]}
  \BibitemShut {NoStop}%
\bibitem [{\citenamefont {Abe}\ \emph {et~al.}(2014)\citenamefont {Abe} \emph
  {et~al.}}]{Abe:2014ugx}%
  \BibitemOpen
  \bibfield  {author} {\bibinfo {author} {\bibfnamefont {K.}~\bibnamefont
  {Abe}} \emph {et~al.} (\bibinfo {collaboration} {T2K Collaboration}),\ }\href
  {\doibase 10.1103/PhysRevLett.112.181801} {\bibfield  {journal} {\bibinfo
  {journal} {Phys. Rev. Lett.}\ }\textbf {\bibinfo {volume} {112}},\ \bibinfo
  {pages} {181801} (\bibinfo {year} {2014})}\BibitemShut {NoStop}%
\bibitem [{\citenamefont {Adamson}\ \emph {et~al.}(2014)\citenamefont {Adamson}
  \emph {et~al.}}]{Adamson:2014vgd}%
  \BibitemOpen
  \bibfield  {author} {\bibinfo {author} {\bibfnamefont {P.}~\bibnamefont
  {Adamson}} \emph {et~al.} (\bibinfo {collaboration} {MINOS Collaboration}),\
  }\href {\doibase 10.1103/PhysRevLett.112.191801} {\bibfield  {journal}
  {\bibinfo  {journal} {Phys. Rev. Lett.}\ }\textbf {\bibinfo {volume} {112}},\
  \bibinfo {pages} {191801} (\bibinfo {year} {2014})},\ \Eprint
  {http://arxiv.org/abs/1403.0867} {arXiv:1403.0867 [hep-ex]} \BibitemShut
  {NoStop}%
\bibitem [{\citenamefont {Adamson}\ \emph
  {et~al.}(2016{\natexlab{b}})\citenamefont {Adamson} \emph
  {et~al.}}]{Adamson:2016xxw}%
  \BibitemOpen
  \bibfield  {author} {\bibinfo {author} {\bibfnamefont {P.}~\bibnamefont
  {Adamson}} \emph {et~al.} (\bibinfo {collaboration} {NOvA Collaboration}),\
  }\href {\doibase 10.1103/PhysRevD.93.051104} {\bibfield  {journal} {\bibinfo
  {journal} {Phys. Rev.}\ }\textbf {\bibinfo {volume} {D93}},\ \bibinfo {pages}
  {051104} (\bibinfo {year} {2016}{\natexlab{b}})},\ \Eprint
  {http://arxiv.org/abs/1601.05037} {arXiv:1601.05037 [hep-ex]} \BibitemShut
  {NoStop}%
\bibitem [{\citenamefont {Wendell}(2014)}]{Wendell:2014dka}%
  \BibitemOpen
  \bibfield  {author} {\bibinfo {author} {\bibfnamefont {R.}~\bibnamefont
  {Wendell}} (\bibinfo {collaboration} {Super-Kamiokande Collaboration}),\
  }\href@noop {} {\bibfield  {journal} {\bibinfo  {journal} {XXVI International
  Conference on Neutrino Physics and Astrophysics}\ } (\bibinfo {year}
  {2014})},\ \Eprint {http://arxiv.org/abs/1412.5234} {arXiv:1412.5234
  [hep-ex]} \BibitemShut {NoStop}%
\bibitem [{\citenamefont {Aartsen}\ \emph {et~al.}(2015)\citenamefont {Aartsen}
  \emph {et~al.}}]{Aartsen:2014yll}%
  \BibitemOpen
  \bibfield  {author} {\bibinfo {author} {\bibfnamefont {M.~G.}\ \bibnamefont
  {Aartsen}} \emph {et~al.} (\bibinfo {collaboration} {IceCube
  Collaboration}),\ }\href {\doibase 10.1103/PhysRevD.91.072004} {\bibfield
  {journal} {\bibinfo  {journal} {Phys. Rev.}\ }\textbf {\bibinfo {volume}
  {D91}},\ \bibinfo {pages} {072004} (\bibinfo {year} {2015})},\ \Eprint
  {http://arxiv.org/abs/1410.7227} {arXiv:1410.7227 [hep-ex]} \BibitemShut
  {NoStop}%
\bibitem [{\citenamefont {Choi}\ \emph {et~al.}(2016)\citenamefont {Choi} \emph
  {et~al.}}]{RENO:2015ksa}%
  \BibitemOpen
  \bibfield  {author} {\bibinfo {author} {\bibfnamefont {J.~H.}\ \bibnamefont
  {Choi}} \emph {et~al.} (\bibinfo {collaboration} {RENO Collaboration}),\
  }\href {\doibase 10.1103/PhysRevLett.116.211801} {\bibfield  {journal}
  {\bibinfo  {journal} {Phys. Rev. Lett.}\ }\textbf {\bibinfo {volume} {116}},\
  \bibinfo {pages} {211801} (\bibinfo {year} {2016})},\ \Eprint
  {http://arxiv.org/abs/1511.05849} {arXiv:1511.05849 [hep-ex]} \BibitemShut
  {NoStop}%
\bibitem [{\citenamefont {Ishitsuka}(2016)}]{Ishitsuka:Moriond2016}%
  \BibitemOpen
  \bibfield  {author} {\bibinfo {author} {\bibfnamefont {M.}~\bibnamefont
  {Ishitsuka}} (\bibinfo {collaboration} {{Double Chooz Collaboration}}),\
  }\href {https://indico.in2p3.fr/event/12279/other-view?view=standard}
  {\bibfield  {journal} {\bibinfo  {journal} {51$^{\mathrm{st}}$ Rencontres de
  Moriond, Electroweak Interactions and Unified Theories}\ } (\bibinfo {year}
  {2016})}\BibitemShut {NoStop}%
\bibitem [{\citenamefont {Iwamoto}(2016)}]{Iwamoto:ICHEP2016}%
  \BibitemOpen
  \bibfield  {author} {\bibinfo {author} {\bibfnamefont {K.}~\bibnamefont
  {Iwamoto}} (\bibinfo {collaboration} {T2K Collaboration}),\ }\href@noop {}
  {\bibfield  {journal} {\bibinfo  {journal} {38th International Conference on
  High Energy Physics}\ } (\bibinfo {year} {2016})}\BibitemShut {NoStop}%
\bibitem [{\citenamefont {Adamson}\ \emph {et~al.}(2013)\citenamefont {Adamson}
  \emph {et~al.}}]{Adamson:2013ue}%
  \BibitemOpen
  \bibfield  {author} {\bibinfo {author} {\bibfnamefont {P.}~\bibnamefont
  {Adamson}} \emph {et~al.} (\bibinfo {collaboration} {MINOS Collaboration}),\
  }\href {\doibase 10.1103/PhysRevLett.110.171801} {\bibfield  {journal}
  {\bibinfo  {journal} {Phys. Rev. Lett.}\ }\textbf {\bibinfo {volume} {110}},\
  \bibinfo {pages} {171801} (\bibinfo {year} {2013})}\BibitemShut {NoStop}%
\bibitem [{\citenamefont {Evans}(2016)}]{Evans:Neutrino2016}%
  \BibitemOpen
  \bibfield  {author} {\bibinfo {author} {\bibfnamefont {J.}~\bibnamefont
  {Evans}} (\bibinfo {collaboration} {MINOS and MINOS+ Collaborations}),\
  }\href@noop {} {\bibfield  {journal} {\bibinfo  {journal} {XXVII
  International Conference on Neutrino Physics and Astrophysics}\ } (\bibinfo
  {year} {2016})}\BibitemShut {NoStop}%
\bibitem [{\citenamefont {Vahle}(2016)}]{Vahle:Neutrino2016}%
  \BibitemOpen
  \bibfield  {author} {\bibinfo {author} {\bibfnamefont {P.}~\bibnamefont
  {Vahle}} (\bibinfo {collaboration} {NO$\nu$A Collaboration}),\ }\href@noop {}
  {\bibfield  {journal} {\bibinfo  {journal} {XXVII International Conference on
  Neutrino Physics and Astrophysics}\ } (\bibinfo {year} {2016})}\BibitemShut
  {NoStop}%
\bibitem [{\citenamefont {Moriyama}(2016)}]{Moriyama:Neutrino2016}%
  \BibitemOpen
  \bibfield  {author} {\bibinfo {author} {\bibfnamefont {S.}~\bibnamefont
  {Moriyama}} (\bibinfo {collaboration} {Super-Kamiokande Collaboration}),\
  }\href@noop {} {\bibfield  {journal} {\bibinfo  {journal} {XXVII
  International Conference on Neutrino Physics and Astrophysics}\ } (\bibinfo
  {year} {2016})}\BibitemShut {NoStop}%
\bibitem [{\citenamefont {Koskinen}(2016)}]{Koskinen:Neutrino2016}%
  \BibitemOpen
  \bibfield  {author} {\bibinfo {author} {\bibfnamefont {J.}~\bibnamefont
  {Koskinen}} (\bibinfo {collaboration} {IceCube Collaboration}),\ }\href@noop
  {} {\bibfield  {journal} {\bibinfo  {journal} {XXVII International Conference
  on Neutrino Physics and Astrophysics}\ } (\bibinfo {year}
  {2016})}\BibitemShut {NoStop}%
\end{thebibliography}%

\end{document}